\documentclass[defaultstyle,11pt]{thesis}

\pdfoutput=1
\usepackage{amssymb,amsmath}		
\usepackage{graphicx,color,graphics}		
\usepackage{rotating}
\usepackage{pmat}
\usepackage[hyphens]{url}
\usepackage{hyperref}

\title{Randomized Benchmarking of Clifford Operators}

\author{A.~M.}{Meier}

\otherdegrees{B.S., Rice University, 2006 \\
	      B.A., Rice University, 2006}

\degree{Doctor of Philosophy}		
	{Ph.D., Physics}		

\dept{Department of}			
	{Physics}		

\advisor{}				
	{Emanuel Knill}			

\reader{Dietrich Leibfried}		

\abstract{  \OnePageChapter	
Randomized benchmarking is an experimental procedure intended to demonstrate control of quantum systems.  The procedure extracts the average error introduced by a set of control operations.  When the target set of operations is intended to be the set of Clifford operators, the randomized benchmarking algorithm is particularly easy to perform and its results have an important interpretation with respect to quantum computation.  The aim of the benchmark is to provide a simple, useful parameter describing the quality of quantum control with an experiment that can be performed in a standard way on any prospective quantum computer.  This parameter can be used to fairly compare different experiments or to mark improvement in a single experiment.

In this thesis I discuss first the original randomized-benchmarking procedure and the importance of the Clifford operators for its implementation.  I develop the statistical analysis of the results and the physical assumptions that are required for the simplest analysis to apply.  The original procedure does not extend in an obvious way to benchmarking of more than one qubit, so I introduce a standardized procedure for randomized benchmarking that applies to any number of qubits.  This new procedure also enables the benchmarking of an individual control operation.  I describe two randomized-benchmarking experiments I helped to design: one involved a single qubit and utilized a variation of the original procedure and the second involved two qubits and demonstrated the new procedure.  I conclude with several potential extensions to the original and new procedures that give them reduced experimental overhead, the ability to describe encoded operations, and fairer comparisons between experiments.
}

\dedication[Dedication]{	
To my parents and my wife, this task would have been much harder without you.
	}

\acknowledgements{	\OnePageChapter	
I want to thank a lot of people.  I certainly would not have made it this far on my own.

My research has been made much more interesting and much more enjoyable by some wonderful people at NIST.  I want to thank my adviser, Manny, for being inspirationally good at math and physics.  I want to thank Bryan for distracting me when I needed distracting and being a patient and insightful collaborator when that was the more appropriate thing.  I want to thank Scott, Mike, and Yanbao for being great editors, sounding boards, and friends.  I also want to thank Dave, Didi, Kenton, David, Jonathan, Andrew, John, and many others in the ion-trapping group for giving me wonderful opportunities to be involved with experiments.  Jim Harrington and Ivan Deutsch have also been very kind to me and very supportive of my research, and I appreciate them very much.

I would certainly not have made it to the University of Colorado if not for some wonderful teachers along the way.  I want to thank Wayne Murrah, Mrs.~Harrison, Kael Martin, Paul Stevenson, Nathan Harshman, and many others who challenged me.

I must also thank my parents for being incredible role models and for providing me with support in so many ways and for so many years.  Good friends, especially Jon Carrasco, Anthony Del Porto, and Charlie Sievers, have kept me going during  difficult times.

Finally, I have no idea of where I would be without my wife, Mariel.  So thanks everyone.

	}

\LoFisShort	

\LoTisShort	

\begin{document}

\newtheorem{theorem}{Theorem}
\newtheorem{lemma}{Lemma}
\newtheorem{corollary}{Corollary}

\newcommand{\diff}[2]{\frac{\partial #1}{\partial #2}}
\newcommand{\diffr}[1]{\diff{#1}{r}}
\newcommand{\diffth}[1]{\diff{#1}{\theta}}
\newcommand{\diffz}[1]{\diff{#1}{z}}

\newcommand{\vth}{V_{\theta}}

\newcommand{\twochoices}[2]{\left\{ \begin{array}{lcc}
        \displaystyle #1 \\ \vspace{-10pt} \\
        \displaystyle #2 \end{array} \right. } 

\newcommand{\threechoices}[3]{\left\{ \begin{array}{lcc}
        #1 \\ #2 \\ #3 \end{array} \right. }    

\newcommand{\fourchoices}[4]{\left\{ \begin{array}{lcc}
        #1 \\ #2 \\ #3 \\ #4 \end{array} \right. }      

\newcommand{\twovec}[2]{\left(\begin{array}{c} #1 \\ #2 \end{array}\right)}
\newcommand{\threevec}[3]{\left(\begin{array}{c} #1 \\ #2 \\ #3 \end{array}\right)}
\newcommand{\twomatrix}[4]{\left(\begin{array}{cc} #1 & #2 \\ #3 & #4 \end{array}\right)}


\newcommand{\I}[0]{\imath}
\newcommand{\ket}[1]{| #1 \rangle}
\newcommand{\bra}[1]{\langle #1 |}
\newcommand{\bracket}[2]{\langle #1 | #2 \rangle}
\newcommand{\tr}[1]{\mbox{tr} \left ( #1 \right )}
\newcommand{\Clif}[1]{\mathcal{C}_{#1}}
\newcommand{\Pauli}[1]{\mathcal{P}_{#1}}
\newcommand{\C}[1]{{{\vphantom{#1}}^{C}\hspace{-.3em}{#1}}}
\newcommand{\CX}{\C{X}}
\newcommand{\CY}{{{\vphantom{Y}}^{C}\hspace{-.1em}{Y}}}
\newcommand{\CZ}{{{\vphantom{Z}}^{C}\hspace{-.2em}{Z}}}
\newcommand{\SWAP}[2]{\mbox{SWAP}_{#1 , #2}}

\chapter{Introduction}
\label{Chap:1}
Quantum computing promises to leverage aspects of quantum mechanics to solve some problems that are not tractable with classical computers.  In particular, simulation of quantum mechanics and factoring of large integers seem to be problems that are not efficiently solvable with a classical computer, but are efficiently solvable with a quantum computer.  There is great scientific interest in quantum simulation because it enables the study of exotic condensed-matter systems and the computation of quantities of interest in field theory.  In contrast, the interest in factoring comes from security concerns, as the presumed inability of classical computers to factor large numbers is the basis of many cryptographic protocols, including secure transactions over the internet. Both of these motivations, combined with intriguing questions in fundamental quantum mechanics and information theory, have spurred rapid growth in the study of how quantum computers might be realized.

Significant debate continues on how to isolate the specific aspects of quantum computing that are responsible for its increased power.  However, one aspect of that power is reflected in the generalization from a classical bit, which can only contain one of the states ``0'' or ``1'', to the quantum bit (qubit), which can contain any normalized linear combination of these states.  This is usually represented by denoting the classical states as $\ket{0}$ and $\ket{1}$, and writing the state of a quantum bit as $\ket{\psi} = \alpha \ket{0} + \beta \ket{1}$, where $\alpha,\beta \in \mathbb{C}$ and $|\alpha|^2 + |\beta|^2 = 1$.  In the same notation, one might think of a classical bit as the system restricted to $\alpha,\beta \in \{0,1\}$.  Essentially, the state space of a qubit is larger and more complicated than that of a bit, and so the information contained in a qubit can be manipulated in many ways that the information in a bit cannot.

In order to build a one-qubit quantum computer, one must first identify a physical system possessing a degree of freedom with two states that can be well isolated from any others.  Take as an example a typical light switch; although a light switch can be found in many different positions, the all-the-way up state and the all-the-way down state are stable, and the mechanism of a light switch tends to force the switch into one of these two states.  Second, one must be able to initialize the system into a particular state.  In the case of a light switch, this corresponds to the ability to flip the switch reliably into the off state.  Third, one must be able to make a measurement that distinguishes (perhaps indirectly) between the two isolated states of the system.  Still considering the light switch, one might place a light detector near the light bulb connected to the switch.  If a significant luminosity is detected, one infers that the switch is on, even though the switch has not been measured directly.  If not, one infers that the switch is off.  There are many assumptions required to make even this simple indirect measurement work.  For example, the light bulb must be correctly connected to the switch for this inference to succeed.

A fourth requirement separates a usable qubit from a light switch, and that is the demand that the physical system be well isolated from all other systems when it is not intentionally being measured or manipulated.  Measurements in quantum mechanics project the state of the system into an eigenstate of the measurement operator.  If a state is to persist in a quantum superposition (linear combination) of these eigenstates, it must not be measured.  This requirement can be difficult to formalize, but it concerns requiring the interactions between other physical systems (often called the environment) and the distinguished qubit to be slow on the computation's time scale, so that these other systems do not perform a measurement of the qubit during the computation.  If the only way of detecting the light switch was to observe the readout of the light detector, it would actually satisfy this requirement as well, since one could turn the light detector off and it would never make a measurement.  Unfortunately, the light from a light bulb interacts strongly with many physical systems around it, and this causes inadvertent measurements of the light switch.

In order to find systems which satisfy this fourth requirement, physicists often turn to conventionally ``quantum-mechanical''-scale systems with binary degrees of freedom: atomic electrons that lie in one of two energy levels, photons with one of two polarizations, or nuclei with one of two spin orientations, for example.  Their hope is that, by choosing smaller systems, they reduce the magnitude of the interactions of those systems with the environment.  This is typically at odds with the desire to measure the state of the systems, however, so there must be some controllable way to strengthen the interaction between a detector and the system when one intends to perform a measurement.  In the light switch example, the light bulb strengthens the interaction by taking information about the light switch and broadcasting it widely.

The final requirement for a one-qubit quantum computer is the ability to manipulate the state of the qubit.  An example of such a manipulation is flipping the light switch from on to off.  However, a qubit might be in a superposition of those two states, and the computer is required to be able to change the state with high precision and accuracy from any initial superposition to any final superposition.  The ability to interact precisely with the system while still not inadvertently measuring it in the process is typically the most difficult aspect of finding a usable quantum bit.  These five requirements are formalized in Ref.~\cite{Divincenzo2000} and are frequently referred to as the DiVincenzo criteria.

The requirements for building a multi-qubit quantum computer are the same as those above, but an added complication comes from the difficulty of manipulating the state space of many qubits.  In particular, interactions that couple two or more of the qubits must be present in order to create most of the states required for computation.  The most common such example is the Bell state of two qubits, $\ket{B} = \frac{1}{\sqrt{2}} \left ( \ket{0}_1\ket{0}_2 + \ket{1}_1\ket{1}_2 \right )$, which cannot be created from the initial state $\ket{0}_1\ket{0}_2$ without an interaction or measurement that couples qubits.  In order to create superpositions of many more qubits, one can either introduce interactions that couple them all directly or build up the interaction out of one-qubit and pairwise terms.  This second approach of creating complicated Hamiltonians from simpler, few-body interactions is the gate model of quantum computation, and the component interactions are referred to informally as gates.

As an example, a frequently used qubit in experiments is based on the valence electron of an ion (see Ref.~\cite{Wineland1998} for a comprehensive review).  Two stable electronic energy levels are chosen as the $\ket{0}$ and $\ket{1}$ states.  Preparation of the $\ket{0}$ state can be performed by illuminating the ion with a laser that excites the $\ket{1}$ state to some short-lived state that preferentially decays to the $\ket{0}$ state.  After some time, the ion will be found to have decayed into the $\ket{0}$ state with high probability.  Measurement can be performed in a similar way by illuminating the ion with a laser that excites the $\ket{0}$ state to a short-lived state that decays preferentially to the $\ket{0}$ state.  Such a decay emits a photon that can be detected by a photodetector, indirectly measuring the electron's state; exposure of the ion to the laser for longer periods of time results in the detection of many photons if the ion was in the $\ket{0}$ state and few photons if the ion was in the $\ket{1}$ state.  

Because the dependence of the dipole moment of the ion on the state of the valence-electron is small, the approximation that the environment does not affect the qubit state of the ion is good.  An exception is seen if the ion is illuminated by a laser resonant with a transition for only one of the two states, as in the measurement operation.  To satisfy the final requirement, the state of a single qubit can be manipulated by pulsing the ion with a laser resonant with the $\ket{0}\leftrightarrow \ket{1}$ transition.  By controlling the pulse strength, duration, and phase, arbitrary superpositions of the qubit states can be created.  Two-qubit interactions of ions are considerably more complicated and typically involve using a laser to drive a motional mode of the two ions conditional on the electronic state of one of the ions.  Precise control of the lasers, stray fields, and trapping fields for the ion are some of the main experimental factors affecting the quality of the ion as a computational qubit.

Although resource-requirement estimations for quantum computers are still very rough, the consensus seems to be that to solve an interesting instance of the factoring problem, a quantum computer will need to have thousands of perfectly controllable qubits or, realistically, millions of imperfectly controllable qubits (see Ref.~\cite{Jones2012}, for example).  The number of gates, or fundamental control operations, is expected to be several factors of ten larger than the number of qubits.  In order to build and test such a large system, a modular design that allows engineers to break the computer up into understandable and testable sub-parts is essential.  For this reason, researchers commonly refer to the idea of a scalable quantum computer: this concept denotes a strategy of designing systems and gates in such a way that the method of combining smaller modular sets of systems (or ``registers'') into a large register is straightforward and efficient.  In particular, an important feature of a scalable quantum computer is the ability to create a Bell state of two qubits even when they are contained in different registers at the smallest register size.  Because it is known (see Ref.~\cite{Nielsen2004}) that one and two-qubit gates are sufficient to accomplish any quantum computation, a common program for building up to large-scale quantum computation is to first satisfy the five DiVincenzo criteria for one-qubit computation, then focus on a single interaction between two qubits that can be controlled well.  Finally, an architecture and networking scheme must be designed that allows for pairwise interactions to connect distant systems.

Regardless of any particular computing scheme, improvement of the quality of the gates can always be leveraged to reduce other resource requirements.  The strategies used to protect against and correct for errors in computational gates fall under the broad title of fault-tolerance.  One of the most important results of the study of fault-tolerance is that there is a threshold for gate quality above which extra qubits and extra gates can be used to construct quantum codes that allow for computational gates of arbitrarily high quality.  In this paradigm, by reducing the probability that a gate acts erroneously with no coding, one can reduce the extra coding overheads required.  The quality of gates then becomes of the utmost importance to designing a functioning quantum computer of manageable size.  Likewise, the ability to quantify the quality of gates becomes a crucial tool both for informing the design of quantum-computing architectures and marking progress toward the goal of a functioning quantum computer.

\section{Gate Fidelity \label{Sec:1GateFidelity}}

The most commonly used expression of the quality of a gate is gate (also: process or operator) fidelity, which describes how close the result of the operation implementing a gate is to the idealized result, averaged over all initial states of the qubit it might be applied to.  Before introducing the formal definition of gate fidelity, it is important to understand what gates are and how they might have errors.

The states of a qubit introduced in the previous section are pure states.  For a system consisting of $n$ qubits, there are $2^n$ pure quantum states that correspond to classical states of $n$ bits ( $\ket{0}$ and $\ket{1}$ for a single qubit and $\{\ket{00},\ket{01},\ket{10},\ket{11}\}$ for two qubits, for example).  These states are orthogonal and additively generate all pure quantum states of $n$ qubits.  The space of all pure states on $n$ qubits can then be considered a vector space over the complex numbers of dimension $2^n$ representing the coefficients of these basis states with an additional normalization condition.  For a good, basic resource for states, gates, and other fundamentals of quantum computing, see Ref.~\cite{Nielsen2004}.

Pure states are distinguished from more-general mixed states, which can be written as probabilistic sums (or mixtures) over pure states.  The concept of mixed states is useful when describing systems about which one does not have maximal knowledge; particularly, this arises when there is a possibility that an unwanted interaction has occurred.  Any mixed state of a finite number of qubits can be represented by a density matrix $\rho,$ where $\rho = \sum_{i,j}\rho_{i,j}\ket{i}\bra{j}$ in the typical outer product notation and $\{\ket{i}\}_i$ is any basis for the vector space of pure states, e.g., the basis of classical states described above.  Using this basis of states for the sums in the definition of a density matrix reveals that mixed states can be represented as $2^n \times 2^n$ matrices over complex numbers.  The normalization condition for pure states translates to the requirement that the trace of the density matrix $\tr{\rho}$ is equal to one for these states, and the fact that mixed states are probabilistic sums of pure states preserves this property for all mixed states as well.  Other constraints in this definition force density matrices to be Hermitian and positive semi-definite.  It will be useful later to note that, if a density matrix $\rho$ describes a system in a pure state, then $\rho^2 = \rho$. 

An interaction that takes all pure states to pure states is constrained to have a unitary (or anti-unitary) matrix representation $U$ acting on the vector space of pure states.  Such idealized interactions are called unitary (or anti-unitary) operators.  The result of the same interaction acting on a system represented by a density matrix $\rho$ is given by $U \rho U^{\dag}$.  If a process takes some pure states to mixed states, then it cannot be represented in this way as a unitary matrix.  These more-general transformations are typically called operations, superoperators, or quantum channels, and, because they do not preserve the space of pure states, they can only be represented conveniently using the density-matrix formalism.

For most (but not all) strategies for quantum computing, the desired gates are unitary.  In practice, implementation of these gates will always cause some unwanted interaction of the computational system with the environment (decoherence), so the implemented gates will not actually be unitary.  I use the notation $U$ to denote a desired unitary gate and the notation $\hat{\Lambda}_U$ to denote the experimental implementation of that gate.  Generally, the ``hat'' notation designates a quantum operation or superoperator, which is a linear function on density operators.  The quality of $\hat{\Lambda}_U$ is the degree to which the density operator $\hat{\Lambda}_U(\rho)$ is close to $U\rho U^{\dag}$.  There are various ways to quantify this quality that describe average behavior and worst-case behavior, among others.  For example, worst-case behavior is frequently the quantity of interest for cryptographic protocols.  For a description of a quantum computer that is not under cryptographic attack, however, average-case behavior is usually appropriate.

In order to describe the discrepancy between a density matrix $\sigma$ and a desired density matrix $\rho$, the state fidelity $F(\rho,\sigma) = \tr{\sqrt{\sqrt{\rho}\sigma\sqrt{\rho}}}^2$ is frequently used.  This quantity is sometimes defined without the square, including in Ref.~\cite{Nielsen2004}, but I prefer the definition above and use it consistently in this thesis.  If $\rho$ represents a pure state $\rho=\ket{\psi}\bra{\psi}$, then $\sqrt{\rho}=\rho$ and so $F(\ket{\psi}\bra{\psi},\sigma) = \tr{\sqrt{\ket{\psi}(\bra{\psi}\sigma\ket{\psi})\bra{\psi}}}^2 = \bra{\psi}\sigma\ket{\psi},$ where the last equality uses the facts that the quantity in the inner parentheses is a scalar and so can be removed from the trace and that the trace of a pure state is $1$.  If $\sigma = \ket{\phi}\bra{\phi}$ is also a pure state, then the definition further simplifies to $F(\ket{\psi}\bra{\psi},\ket{\phi}\bra{\phi}) = |\bracket{\psi}{\phi}|^2$, which is familiar as the square of the overlap between the two pure states.  This state fidelity has the following useful operational definition: if one attempts to measure the system described by $\sigma$ in an orthogonal basis containing $\ket{\psi}$, then the result indicating projection onto $\ket{\psi}$ occurs with probability $F(\ket{\psi}\bra{\psi},\sigma)$.  Loosely speaking, state fidelity describes the probability that a measurement of a system reveals that it is in the desired state.

The definition of state fidelity can be extended to a definition of gate fidelity 
\begin{equation} \label{Eq:1GateFidelInt}
F(U,\hat{\Lambda}_U) = \int d\rho \; F(U\rho U^\dag, \hat{\Lambda}_U(\rho))
\end{equation}
that describes the fidelity between the result of the ideal operation and the result of the actual operation averaged with Haar measure $d\rho$ over all pure-state density matrices for the correct number of systems.  The Haar measure (with respect to the unitary group) is a topological measure that formalizes the concept of a uniform distribution over these matrices.  Operationally, this fidelity describes the average probability that a measurement of a system initialized in a random (but known) initial state reveals it to be in the correct final state after the operation.  Similarly, a worst-case gate fidelity can be defined as the minimum state fidelity over all input states.  For a pure state $\rho=\ket{\psi}\bra{\psi}$, the integrand in the gate fidelity can be simplified to 
\begin{equation} \label{Eq:1GateFidelity}
F(U\rho U^\dag, \hat{\Lambda}_U(\rho)) = \bra{\psi}U^{\dag}\hat{\Lambda}_U \left (\ket{\psi}\bra{\psi} \right )U\ket{\psi}.
\end{equation}
If $\hat{\Lambda}_U(\rho) = U\rho U^\dag$, i.e. $\hat{\Lambda}_U$ exactly implements the unitary gate $U$, then this equation reduces to $1$ .

The form of Eq.~\ref{Eq:1GateFidelity} suggests a hypothetical experiment in which one first prepares a system in state $\ket{\psi}$, then performs the operation $\hat{\Lambda}_U$ on it, then performs an ideal inverse gate $U^\dag$, and finally measures in an orthogonal basis containing $\ket{\psi}$.  The probability of the measurement result indicating projection onto $\ket{\psi}$ averaged over all pure states $\ket{\psi}$ would reveal the gate fidelity $F(U,\hat{\Lambda}_U))$.  There are two principle difficulties in performing such an experiment: first, averaging over all pure states is not experimentally possible, and, second, no ideal inverse gate exists.  There are (at least) two potential solutions to the first problem.  If the integral over all states in the gate fidelity definition is formally equal to a sum over a finite set of states, then sufficient resources might allow the hypothetical experiment above to be implemented for each of the states in that set.  This is the approach of quantum process tomography, which is detailed in the next section.  Another solution is to perform the hypothetical experiment for many randomly chosen initial states and hope that this Monte-Carlo-integral approach converges quickly to the actual integral.  This approach is employed in the bulk of this thesis.

The second problem with the hypothetical experiment described in the last paragraph is its reliance on an ideal gate to invert the faulty desired operation.  If the inverting gate $U^{\dag}$ is actually a faulty operation $\hat{\Lambda}_{U^{\dag}}$ as well, as it must be in a real experiment, then one must attempt to distinguish between the errors of the operations $\hat{\Lambda}_{U^{\dag}}$ and $\hat{\Lambda}_{U}$ in the data analysis.  This is not a particularly difficult problem when the former operation has much higher fidelity than the latter, but this is frequently not the case.  The task can also be made much simpler if $\hat{\Lambda}_{U^{\dag}}$ can be well characterized; however, in the next section I demonstrate that the typical protocol for characterizing gates, quantum process tomography, often suffers from the same need to assume perfect inverting gates.

\section{Quantum Process Tomography} \label{Sec:1QPT}

The now-standard experiment for revealing a complete description of a gate is quantum process tomography (QPT).  The goal of process tomography is slightly different than the experiment imagined in the previous section in that tomography seeks a complete description of a process instead of an overall fidelity.  This is most obviously useful when the goal of the experiment is to discover the nature of some unknown process.  For example, quantum process tomography is used in Ref.~\cite{Chow2009} as a diagnostic tool to describe the identity operation (and several non-identity operations); by investigating the transformation of quantum states when no qubit dynamics are intended, systematic errors such as stray fields and unwanted couplings can be identified and corrected.  QPT is also frequently used, in Refs.~\cite{Home2009} and~\cite{Riebe2006} for example, to demonstrate that a new, interesting gate is performing as it should and to provide information on the failure mechanisms if it is not.

A description of the basic QPT procedure is given in Ref.~\cite{Mohseni2008} and is summarized here for faster reference and uniform notation.  If the qubits are initially decoupled from their environment, any quantum operation  $\hat{\varepsilon}(\rho)$ can be written as $\hat{\varepsilon}(\rho) = \sum_i A_i \rho A_i^{\dag}$ for a (not-uniquely defined) set of $2^n \times 2^n$ matrices $A_i$.  In turn, the set of Pauli matrices $\{P_j\}_j$ on $n$ qubits form an additive basis (with complex coefficients) for any $2^n \times 2^n$ matrix, so each $A_i$ can be decomposed into $A_i = \sum_j \beta_{i,j} P_j$.  For a more complete description of the Pauli matrices, see Secs.~\ref{Sec:2CliffordFacts} and \ref{Sec:APauliOps}.  Together, these decompositions imply that
\begin{equation} \label{Eq:1QPT}
\hat{\varepsilon}(\rho) = \sum_{j,k} \chi_{j,k} P_j \rho P_k^{\dag},
\end{equation}
where $\chi$, the process matrix, is uniquely defined given the standard set of Pauli matrices.  The process matrix completely describes any quantum process on a fixed number of qubits; for $n$ qubits, there are $4^n$ Pauli matrices of this form, so the process matrix has dimensions $4^n \times 4^n$.

The process matrix contains, among other things, information about gate fidelity.  If the ideal gate is the identity, then $\hat{\varepsilon}(\rho)$ ideally takes $\rho \rightarrow I \rho I$, which is commonly identified with the $\chi_{0,0}$ term in Eq.~\ref{Eq:1QPT}.  Ref.~\cite{Nielsen2002} provides a simple expression for the gate fidelity in terms of the matrices $A_i$ above:
\begin{equation} \label{Eq:1NielsenForm}
F(I,\hat{\varepsilon}) = \frac{2^n + \sum_i |\tr{A_i}|^2}{2^n(2^n+1)}
\end{equation}
Noting that $\tr{A_i} = 2^n\beta_{i,0}$ and that $\chi_{0,0} = \sum_i |\beta_{i,0}|^2$, this equation can be simply rewritten as
\begin{equation} \label{Eq:1Chi00}
F(I,\hat{\varepsilon}) = \frac{1 + 2^n \chi_{0,0}}{(1+ 2^n)},
\end{equation}
which provides a simple relationship between the process matrix and the gate fidelity of the process with the identity.

If the gate fidelity with some other unitary $U$ is desired, substitution reveals that $
F(U \rho U, \hat{\varepsilon}(\rho)) = F(I \rho I, U^{\dag} \hat{\varepsilon}(\rho) U)$.  If $U = \sum_a \beta_a P_a$, then
\begin{equation} \label{Eq:1QPT2}
U^{\dag} \hat{\varepsilon}(\rho) U = \sum_{j,k} \chi_{j,k} (\sum_a \beta_a^{*} P_a^{\dag}) P_j \rho P_k^{\dag}  (\sum_b \beta_b P_b) = \sum_{j,k} \chi_{j,k}' P_j \rho P_k^{\dag},
\end{equation}
where $\chi_{0,0}' = \sum_{a,b} \chi_{a,b} \beta_a^{*} \beta_b$ can be used in Eq.~\ref{Eq:1Chi00} to calculate the gate fidelity of $U^{\dag}\hat{\varepsilon}(\rho)U$ with $I$ or, equivalently, the process fidelity of $\hat{\varepsilon}(\rho)$ with $U$.  In this way, one can calculate the fidelity of a process with any ideal gate once the process matrix has been constructed.

In order to extract the elements $\chi_{j,k}$ of the process matrix, the following experiment suffices: First, establish a linearly independent basis for the space of density matrices considered as a $2^{2n}$ dimensional vector space.  The set $\{\ket{l}\bra{m}\}_m$ is such a basis (the $n$-qubit Pauli matrices are another).  Second, find a way to rewrite each $\ket{l}\bra{m}$ as a linear combination of pure states: $\ket{l}\bra{m} = \sum_r \alpha_{r,l,m} \ket{\psi_{r,l,m}}\bra{\psi_{r,l,m}}$.  Now experimentally prepare each $\ket{\psi_{r,l,m}}$ as well as possible and perform the operation $\hat{\varepsilon}$ on the state to obtain the output state $\rho_{r,l,m}' = \hat{\varepsilon}(\ket{\psi_{r,l,m}}\bra{\psi_{r,l,m}})$.  Finally, perform quantum state tomography (see \cite{Nielsen2004}, for example) on each $\rho_{r,l,m}'$ in order to provide a complete description of these output states.  Each output is also considered as an element of a $2^{2n}$-dimensional vector space, and so its tomography requires a commensurate number of different measurements. 

To predict $\hat{\varepsilon}(\rho)$ for an arbitrary state $\rho$, one can now simply decompose $\rho$ into the basis $\{\ket{l}\bra{m}\}_m$ and then use linearity to recompose the various $\rho_{r,l,m}'$ with appropriate weights.  In order to calculate the elements of $\chi$, one must invert the results of the state tomography for the output states.  This inversion process can itself be a difficult computational task (see supplementary information of Ref.~\cite{Mallet2011} or Ref.~\cite{Merkel2012}, for example).

Because it requires determining the elements of a complex $4^n \times 4^n$ matrix that are experimentally independent, quantum process tomography fundamentally requires $O(2^{4n})$ applications of the operation of interest.  The order notation here hides the fact that many applications might be required to determine each matrix element with sufficiently small noise.  Fundamentally, this scaling implies that QPT is not viable for investigating processes on more than ten or so qubits.  Even when the experiments are feasible, the computational inversion step can be extremely taxing and much of the data might not lend itself to useful interpretation in any case.  For these reasons, much theoretical attention has been directed toward the task of finding less-intensive experimental methods for directly extracting certain information contained in the process matrix  (e.g., gate fidelity with an ideal gate or set of gates) without its full determination.

A second significant challenge of QPT lies in the preparation and measurement steps of its experimental implementation.  At the present, most QPT experiments seek to describe the same set of gates that are used to manipulate quantum information in the experiment.  That is, the set of gates used to prepare and measure the quantum states as required by QPT is often not well characterized and sometimes even includes the gate(s) under investigation.  The authors of Ref.~\cite{Merkel2012} propose a strategy based on maximum likelihood that allows QPT to be performed with only weak assumptions on the gates involved; however, they note that the time required for the procedure is ``ridiculous'' for investigations of more than two qubits.

\section{Randomized Benchmarking} \label{Sec:1RB}

An alternative approach to fidelity calculation is to use Monte-Carlo integration over the set of quantum states \cite{Emerson2007, Emerson2005,Levi2007,Lopez2010}.  In this approach, instead of preparing a special set of initial states and measuring in a special set of bases, random initialization and random measurements are used.  The advantages of randomized approaches to integration frequently lie in their faster convergence rates, simplicity, and immunity to certain adversarial noise models.  The disadvantages are directly related to the non-structured nature of the experiment: the data collected from a randomized fidelity-estimation experiment is not obviously useful for any other task.  For example, one cannot simply manipulate the information to reveal the fidelity of the process with some other ideal process, as is possible with QPT.

Other than poor scaling, the major problem with QPT mentioned in the previous section was its reliance on imperfect and ill-characterized gates to perform the preparation and measurement of the states.  This same problem manifests itself in randomized fidelity estimation if it is approached in the manner suggested in Sec.~\ref{Sec:1GateFidelity}.  That thought experiment relied on a perfect implementation of the inverse of the ideal gate, which is never available in an experiment except in an approximate sense.  When the experiment can be constructed so that the inverting gate has very high fidelity relative to the gate of interest, this approximation can be acceptable.  Another approach to this difficulty might be to try to ensure that the preparation and measurement steps are consistently faulty in a simple way.  Then the faults might be estimated and eliminated in the analysis.  This is the solution employed by randomized benchmarking.

As originally formulated in Ref.~\cite{Knill2008}, randomized benchmarking is a procedure that extracts the average gate fidelity over a set of gates by performing fidelity estimations on randomized sequences of gates from that set.  The elegance of this approach comes from its implementation of either unitary or Clifford twirling (discussed in Secs.~\ref{Sec:2Twirling} and \ref{Sec:2Clifford}).  The twirling randomizes correlations between errors, effectively decoupling errors from each other, and averages the errors themselves.  For example, the preparation and measurement aspects of the experiment are largely decoupled from the process in question.  Twirling ensures that these average processes have a mathematically tractable form as well, so the analysis of randomized benchmarking is simpler in comparison with QPT.

One peculiarity of the standard randomized benchmarking procedure, however, is that it produces an average gate fidelity over a set of gates (which has size necessarily greater than 1) instead of a fidelity for a single gate.  While this number is worthwhile, it has a slightly different purpose than a single gate fidelity: the result of a randomized-benchmarking experiment most effectively describes the aggregate quality of the quantum control, instead of the quality of any particular gate.  This makes standard randomized benchmarking, a poor tool for diagnosing specific errors but an excellent tool for describing overall performance.  In Chap.~\ref{Chap:3} I describe an extended randomized benchmarking protocol that overcomes this limitation.

I specifically discuss Clifford randomized benchmarking, where the set of gates of interest is the set of Clifford operators, which is a finite-size subset of the unitary operators.  Use of the Clifford operators instead of the full set of unitaries further simplifies the design and analysis of randomized-benchmarking experiments.  While the Clifford operators are a small subset of unitary quantum operators, the average fidelity over Clifford operators estimated by the benchmarking analysis is a worthwhile figure of merit in the study of fault-tolerant quantum computing, which deals almost exclusively with Clifford operators.

In the chapters that follow, I describe my contributions to expanding the utility and soundness of Clifford randomized benchmarking as a fidelity-estimation tool for quantum computing.  
\begin{itemize}
\item
In Chapter \ref{Chap:2}, I describe the original randomized-benchmarking procedure in detail, noting both its theoretical and experimental aspects.  I also present an experiment I helped to design that applied this procedure in a fairly straightforward way, but with improvements to the statistical analysis.  

\item
In Chapter \ref{Chap:3}, I discuss the problems with performing randomized benchmarking for experiments with more qubits and with different natural sets of gates.  I describe a new procedure for randomized benchmarking that solves these problems and the new difficulties and opportunities it introduces.  Finally, I present a second experiment I helped to design that implemented the new procedure.

\item
In Chapter \ref{Chap:4}, I introduce several as-yet-unrealized extensions to the randomized-benchmarking procedure.  These include proposals for how to perform randomized benchmarking with limited sets of gates and on information that has already been encoded in a quantum code.  I also attempt to improve upon the analysis of the original procedure.

\item
In Chapter \ref{Chap:5}, I offer some conclusions and perspective on the uses of randomized benchmarking in the overall effort to build a large-scale quantum computer.

\item
Appendix \ref{App:A} contains details about the Clifford and Pauli operators and is intended to be used as a reference for these topics throughout the first five chapters.  It also contains a description of a computer program called a Clifford simulator that is the main tool I used to enable my design and investigations of randomized-benchmarking experiments.
 
\item
Appendix \ref{App:B} is a table that collects information about all of the randomized-benchmarking experiments that have been performed so far. 

\item
Appendix \ref{App:C} is contains notational conventions for the Clifford operators, including a table cataloging and defining those that are used frequently in this thesis.
\end{itemize}

\chapter{Standard Randomized Benchmarking}
\label{Chap:2}

Randomized benchmarking (RB) is an experimental procedure that extracts the average operator fidelity for a set of operators.  This average fidelity is intended to describe how well a quantum-computing architecture will perform when it is asked to compute random-looking circuits. Particularly, by multiplying one minus the average fidelity (called the error per step) by the number of gates in such a circuit, one obtains a first-order estimate of the probability that the measurements at the end of the circuit will return the incorrect results.  The error per step is also useful in relation to fault-tolerance thresholds, which are a common goal of quantum control and are discussed in more depth in Sec.~\ref{Sec:2CliffRand}. 

Randomized benchmarking involves long sequences of gates, so the error per step inevitably reflects the consistency of the gates, that is, how well they perform in a scenario where they might be affected by time dependence of the apparatus and cannot be constantly recalibrated for optimal performance.  This emphasis on consistency enhances the benchmark's appeal as a prediction of the performance of actual, long quantum computations.  In contrast, quantum process tomography is designed to provide information only about an isolated gate.  Alternatively, calibration experiments, e.g. a Rabi-oscillation experiment, provide information about consistency of gates, but only in a very particular, non-random way, and so they are blind to certain types of gate interactions and error types that are likely to occur in a more realistic computation.  Another advantage to benchmark-type experiments that use long sequences of gates is that the repetition amplifies the errors in gates, making them easier to detect when the gates have high fidelity.

The theoretical underpinnings of randomized benchmarking (RB) were developed originally by Refs.~\cite{Emerson2005,Dankert2009}.  The first RB experiment was performed by the authors of Ref.~\cite{Knill2008} in an ion-trap architecture.  The procedure has grown quickly in popularity since then, in large part because it provides a simple and useful way to quantify the quality of quantum control experiments.  Because many gates are used in each instance of an RB experiment, the errors of high-fidelity gates are amplified by repetition and made easier to distinguish from noise.  Most RB experiments have directly copied the experimental procedure of Ref.~\cite{Knill2008}, although recent advances discussed in Chap.~\ref{Chap:3} have led to new procedures.  A complete table of published randomized-benchmarking results is given in App.~\ref{App:B}; the two experiments to which I made contributions are discussed at length in Secs.~\ref{Sec:2Exp1} and \ref{Sec:3Exp2}.  In this chapter I lay out the mathematical and physical foundations for randomized benchmarking, discuss the original experimental procedure and how it is adapted for different experiments, present my approach to the statistical analysis of the experiments, and detail the ion-trap experiment published as Ref.~\cite{Brown2011}, which utilized these experimental and analytical procedures.

\section{Theory of Randomized Benchmarking} \label{Sec:2Theory} 

The fundamental theoretical tool that underlies randomized benchmarking is whimsically called twirling; this process transforms a large class of quantum  superoperators into depolarizing channels, which are superoperators with a particular, simple form.  In this section I introduce twirling and depolarizing channels before deriving the equation that describes the results of an ideal randomized-benchmarking experiment.  These derivations are condensed versions of those contained in Refs.~\cite{Emerson2005,Dankert2009,Nielsen2002}.

\subsection{Twirling} \label{Sec:2Twirling}

Consider any memory-less quantum operation (superoperator) $\hat{\Lambda}$ on $n$ qubits.  $\hat{\Lambda}$ might be a perfect unitary gate, a process of decoherence, or an experimental, imperfect attempt at a unitary gate.  The only requirement is that the output $\hat{\Lambda}(\rho)$ depends only on the input state $\rho$.  As discussed in Sec.~\ref{Sec:1QPT}, the action of any such operation (or superoperator) on an operator $\rho$ can be written (not uniquely) as a sum $\hat{\Lambda}(\rho) = \sum_i A_i \rho A_i^{\dag}$, called a Kraus decomposition, where each $A_i$ is a $2^n \times 2^n$ matrix and $\sum_i A_i A_i^{\dag} \leq I$.  It is sometimes convenient to refer to a superoperator using these Kraus operators by writing $\hat{\Lambda} = \sum_{i} A_i \otimes A_i^{\dag}$, where the action of a tensor product term in the sum on an operator $X$ is $(A_i \otimes A_i^{\dag})(X) = A_i X A_i^{\dag}$. In Eq.~\ref{Eq:1QPT} I showed that, if an additive, orthogonal basis of $2^n \times 2^n$ matrices is fixed, then $\hat{\Lambda}$ can be fully, uniquely specified by a corresponding $4^n \times 4^n$ matrix of complex numbers called a process matrix.

One of the most difficult problems of the analysis of quantum control experiments is that the process matrix is very large and difficult to interpret meaningfully for $n>2$.  Chap.~\ref{Chap:1} argued that one of the most useful single-number descriptions of an operation is its gate (or operator) fidelity with a desired unitary.  The goal of twirling is to manipulate the quantum operation so that a simple experiment can be used to extract this gate fidelity in a more efficient way than through QPT.

The general notion of twirling is to transform the superoperator $\hat{\Lambda}$ into the superoperator $\int d\mu(U) \hat{U} \circ \hat{\Lambda} \circ \hat{U^{\dag}}$ where $d\mu(U)$ is some probability distribution over unitaries, i.e.~$\int d\mu(U) = 1$.  The "$\circ$" notation indicates composition of the operations, or the first operation applied to the result of the second.  The notation $\hat{U}$ is meant to indicate the superoperator that acts as $\hat{U}(\rho) = U \rho U$.  In many cases of interest, the distribution $d\mu(U)$ has support only on a finite set of unitaries, so the integral can be written as a sum $\sum_i p(U_i) \hat{U_i} \circ \hat{\Lambda} \circ \hat{U_i^{\dag}}$.  A specific instance of twirling can be referred to by the set of unitary operators with nonzero support (e.g. a Clifford operator twirl) when the distribution over the operators is obvious in context.  Because twirling transforms a superoperator into a new superoperator, twirling is an example of a super-superoperator, or linear mapping in superoperator space.  

The first twirl of interest is the exact Haar twirl, where the integral is over all unitaries (on a specified system) and $\mu$ is the Haar measure on unitaries.  In order to understand the action of the exact Haar twirl on a superoperator, it is convenient to first introduce some basic results commonly derived in the representation theory of groups.  The general lemma that follows can be used to understand this twirl by considering the specific case of the unitary group of operators on $n$ qubits applied to the vector space of linear superoperators on $n$ qubits.

\begin{lemma} \label{Lem:2Proj}
Suppose $\phi: G \rightarrow GL(V)$ is a homomorphism of a compact group $G$ into the group $GL(V)$ of linear operators on a vector space $V$.  Suppose further that $W$ is the maximal subspace of $V$ such that $\phi(g)(w) = w$ for all $g\in G, w \in W$, i.e., $W$ is the maximal subspace of vectors in $V$ stabilized by the action of $\phi(G)$.  Finally, let $\mu$ be the left-invariant Haar measure on $G$.  Then, first, $\int_{g \in G} d\mu(g) \phi(g)(v) \in W$ for any $v \in V$.  Second, $\int_{g \in G} d\mu(g) \phi(g)(w) = w$ for any $w \in W$.

\paragraph{Proof:} The crucial property of the measure $\mu$ is that $d\mu(g) = d\mu(hg)$ for all $g,h \in G$.  For any vector $v \in V$ consider the vector $\int_{g \in G} d\mu(g) \phi(g)(v)$.  To see that this vector is contained in $W$, observe that for any $h \in G$
\begin{eqnarray}
\phi(h) \left (\int_{g \in G} d\mu(g) \phi(g)(v) \right ) &=& \int_{g \in G} d\mu(g) \phi(h)\phi(g)(v) \nonumber \\
&=&  \int_{g \in G} d\mu(g) \phi(hg)(v) \nonumber \\
&=&  \int_{g \in G} d\mu(hg) \phi(hg)(v) \nonumber \\,
\end{eqnarray}
where the second equality relies on the definition of a group homomorphism.  At this point, because the left multiplication action of $h$ maps $G$ onto itself (that is hG = G), the integral over all $g \in G$ of $d\mu(hg) \phi(hg)(v)$ can be seen to be the same as the integral over all $g \in G$ of $d\mu(g) \phi(g)(v)$, so
\begin{equation} \label{Eq:2LemmaResult}
\phi(h) \left (\int_{g \in G} d\mu(g) \phi(g)(v) \right ) = \int_{g \in G} d\mu(g) \phi(g)(v).
\end{equation}
Because $\int_{g \in G} d\mu(g) \phi(g)(v)$ is stabilized by any element of $G$, it must be contained in $W$.

The second result holds simply because $\phi(g)(w) = w$ for any $w \in W$, so the integrand does not depend on $g$.  Because the Haar measure of the entire group is always $1$, $\int_{g \in G} d\mu(g) \phi(g)(w) = \left (\int_{g \in G} d\mu(g) \right ) w = w$. $\square$
\end{lemma}

The integral in this lemma acts as a projector onto the subspace that is stabilized by the group action.  
For the case needed here based on the group of unitary operators and the space of linear superoperators, the homomorphism of interest takes a unitary $U$ to a super-superoperator $\phi(U) = \hat{U} \otimes \hat{U^{\dag}}$, which acts on a superoperator $\hat{\Lambda}$ by $\phi(U)(\hat{\Lambda}) = \hat{U} \circ \hat{\Lambda} \circ \hat{U^{\dag}}$.  This looks like the integrand of the Haar twirl.  Finally, the resulting superoperator acts on an operator as $\phi(U)(\hat{\Lambda})(\rho) = U\hat{\Lambda}(U^{\dag}\rho U)U^{\dag}$, where the $U$ are now once again thought of as simply operators (or matrices) on the right hand side of the equation.

The next task is to understand the stabilized subspace of superoperators under this group action.  
Ref.~\cite{Emerson2005} implies that there is a two-dimensional space of superoperators such that $\phi(U)(\hat{\Lambda}) = \hat{\Lambda}$ for all unitaries $U$, and it is spanned by the superoperators 
\begin{eqnarray}
\hat{\Lambda}_1 = \sum_{i,j} \ket{i}\bra{i} \otimes \ket{j}\bra{j} = I \otimes I, \\
\hat{\Lambda}_2 = \sum_{i,j} \ket{i}\bra{j} \otimes \ket{j}\bra{i} = \frac{1}{D} \sum_{P_a \in \mathcal{P}_n} P_a \otimes P_a, 
\end{eqnarray}
where $I$ is the identity operator and I have reverted to the bra-ket notation for matrices the middle expressions.  The set $\mathcal{P}_n$ of Pauli operators is described in depth in Sec.~\ref{Sec:APauliOps}.  Lemma \ref{Lem:2Proj} implies that any superoperator in the stabilized subspace of the action $\phi(U)$ for all $U$ can be written as $b\hat{\Lambda}_1 + c\hat{\Lambda}_2$.   

Acting with a superoperator of the form $\hat{\Psi} = p_1\hat{\Lambda}_1 + \frac{p_2}{D}\hat{\Lambda}_2$ on an operator $X$ gives
\begin{equation} \label{Eq:2DepXold}
\hat{\Psi}(X) = p_1 X + \frac{p_2}{D}\tr{X}I.
\end{equation}
If such a superoperator is known also to preserve (trace-one) $D \times D$ density matrices, e.g., if it is a quantum operation on a $D$-level system ($D=2^n$ for $n$ qubits), then checking the trace of an arbitrary density matrix before and after this transformation reveals that $p_1 = 1-p_2$.

Consider a basis for operator space $\{E_k\}_k$ that is orthonormal with respect to the trace norm (i.e. $\tr{E_iE_j} =\delta_{i,j}$).  Ref.~\cite{Emerson2005} defines a matrix $M(\hat{\Lambda})$ with $(l,m)$ entry $[M(\hat{\Lambda})]_{l,m} \equiv \tr{E_l^{\dag} \hat{\Lambda}(E_m)}$.  Then the authors define a superoperator trace $\tr{\hat{\Lambda}} = \tr{M(\hat{\Lambda})}$, so that $\tr{\hat{\Lambda}_1} = D^2$ and $\tr{\hat{\Lambda}_2} = D$.  For a general $\hat{\Lambda} = \sum_a A_a \otimes A_a^{\dag}$; this trace can also be written as $\tr{\hat{\Lambda}} = \sum_{a,k} \tr{E_k^{\dag} A_a E_k A_a^{\dag})} = \sum_a |\tr{A_a}|^2$.  For the rest of this section, I consider the basis $E_{i,j} = \ket{i}\bra{j}$.  Note that $M(\hat{\Lambda})$ and the process matrix for $\hat{\Lambda}$ have the same size but different meanings.   Finally, for
\begin{equation} \label{Eq:2DepX} 
\hat{\Psi}(X) = (1-p_d) \hat{\Lambda}_1(X) + \frac{p_d}{D} \hat{\Lambda}_2(X) = (1-p_d) X + \frac{p_d}{D}\tr{X}I,
\end{equation}
once can express $p_d$ as a function of $\tr{\hat{\Psi}}$:
\begin{equation} \label{Eq:2DepStr}
p_d = \frac{D^2 - \tr{\hat{\Psi}}}{D^2-1}.
\end{equation}

Lemma~\ref{Lem:2Proj} indicates that $\hat{\Lambda}_{t} = \int_U dU \hat{U} \circ \hat{\Lambda} \circ {U^{\dag}}$ is in the stabilized subspace described in the preceding paragraphs for any $\hat{\Lambda}$.  I assume for now that $\hat{\Lambda}$ is trace preserving, so that $\tr{\hat{\Lambda}(\rho)} = \tr{\rho},$ and discuss violations of this assumption in Sec.~\ref{Sec:2Memory}.  Eq.~\ref{Eq:2DepStr} shows that $\hat{\Lambda}_{t}$ can be parametrized by the single parameter $p_d$ that is a function of $\tr{\hat{\Lambda}_t}$.  This parameter can also be seen as a function of $\tr{\hat{\Lambda}}$ because
\begin{equation} \label{Eq:2TraceEquiv}
\tr{\hat{\Lambda}_t} = \tr{\int_U dU \hat{U} \circ \hat{\Lambda} \circ \hat{U^{\dag}}} = \int dU \tr{\hat{U} \circ \hat{\Lambda} \circ \hat{U^{\dag}}} = \tr{\hat{\Lambda}}.
\end{equation}
Putting these expressions together gives 
\begin{equation} \label{Eq:2DepFormGen}
\hat{\Lambda}_t = (1-p_d) \hat{\Lambda}_1 + \frac{p_d}{D} \hat{\Lambda}_2,
\end{equation}
where $p_d = \frac{D^2-\tr{\hat{\Lambda}}}{D^2 - 1}$.  The last equality of Eq.~\ref{Eq:2TraceEquiv} relies on the fact that if $\hat{\Lambda} = \sum_k A_k \otimes A_k^{\dag}$, then $\hat{U} \circ \hat{\Lambda} \circ \hat{U^{\dag}} = \sum_k UA_kU^{\dag} \otimes UA_k^{\dag}U^{\dag}$, so $\tr{\hat{U} \circ \hat{\Lambda} \circ \hat{U^{\dag}}} = \sum_k |\tr{U A_k U^{\dag}}|^2 = \sum_k |\tr{A_k}|^2 = \tr{\hat{\Lambda}}$.  

To see how this twirl might be useful for determining gate fidelities, consider again Eq.~\ref{Eq:1GateFidelInt} and \ref{Eq:1GateFidelity}.  Suppose $\hat{\Lambda}_U(X) = \sum_{i} A_iXA_i^{\dag}$, and rewrite this as $\hat{\Lambda}_U(X) = \sum_{i} (A_iU^{\dag})UXU^{\dag}(UA_i^{\dag})$ by inserting the identity twice.  Now define a new operator
\begin{equation} \label{Eq:2ErrorOp}
\hat{\Lambda}_{U,e} = \sum_{i} A_iU^{\dag} \otimes UA_i^{\dag} = \sum_{i} A_iU^{\dag} \otimes (A_iU^{\dag})^{\dag},
\end{equation}
so that $\hat{\Lambda}_U(X) = \hat{\Lambda}_{U,e}(UXU^{\dag})$.  Essentially, this operator can be used in the context of gate fidelity to encapsulate the error that differentiates a superoperator $\hat{\Lambda}_U$ from the intended (unitary) superoperator $\hat{U}$.  By use of this identity and the introduction of a trace, Eq.~\ref{Eq:1GateFidelity}, can be rewritten as
\begin{equation} \label{Eq:2GateFidelity}
F \left (U\ket{\psi}\bra{\psi} U^{\dag}, \hat{\Lambda}_U(\ket{\psi}\bra{\psi}) \right ) = F \left ( (U\ket{\psi})(\bra{\psi} U^{\dag}), \hat{\Lambda}_{U,e} \left ((U\ket{\psi})(\bra{\psi}U^{\dag}) \right ) \right ),
\end{equation}
which is the fidelity of the state obtained by applying the identity operator  to the pure state $U\ket{\psi}$ with the state obtained by applying the (error) superoperator $\hat{\Lambda}_{U,e}$ to the pure state $U\ket{\psi}$.  

By inserting this re-definition into the gate-fidelity integral and using the invariance of the Haar measure to rename the variable of integration from $U\ket{\psi}$ to $\ket{\phi}$,
one finds that
\begin{equation} \label{Eq:2FidelWithIden}
F(\hat{U},\hat{\Lambda}_U) = F(\hat{I},\hat{\Lambda}_{U,e}).
\end{equation}
In other words, the fidelity of $\hat{\Lambda}_U$ with the ideal superoperator $\hat{U}$ is equal to the fidelity of $\hat{\Lambda}_{U,e}$ with the identity superoperator.

Next, Ref.~\cite{Nielsen2002} provides a simple proof that the fidelity of an operator $\hat{\Lambda}$ with the identity is the same as the fidelity of its exact Haar twirl $\int_U dU U^{\dag}\hat{\Lambda} U$ with the identity.  In short,
\begin{eqnarray} \label{Eq:2FidEquivalence}
F \left (\int_U dU \hat{U^{\dag}} \circ \hat{\Lambda} \circ \hat{U},\hat{I} \right ) &=&  \int_{\psi} \tr{\bra{\psi} (\int_U d\mu(U) U^{\dag} \hat{\Lambda} (U(\ket{\psi}\bra{\psi})U^{\dag})U \ket{\psi}} \nonumber \\
 &=& \int_U d\mu(U) \int_{\psi} d\psi \tr{(\bra{\psi}U^{\dag})\hat{\Lambda} ((U\ket{\psi})(\bra{\psi})U^{\dag}))(U \ket{\psi})} \nonumber \\
 &=& \int_{\psi} \tr{\bra{\psi}\hat{\Lambda} \left (\ket{\psi}\bra{\psi} \right )\ket{\psi}} \nonumber \\
 &=& F(\hat{\Lambda},I),
\end{eqnarray}
using again that the measure $d(U\ket{\psi}) = d(\ket{\psi})$ for each unitary $U$.

Finally, consider the average over all $U$ of the gate fidelity of $\hat{\Lambda}_U$ with $\hat{U}$:
\begin{eqnarray}
\int_U d\mu(U) F \left (\hat{\Lambda}_U,U \right ) &=& \int_U d\mu(U) F \left (\hat{\Lambda}_{U,e},I \right ) \nonumber \\
&=& \int_U d\mu(U) F \left (\int_{U'} dU' \hat{U'}^{\dag} \circ \hat{\Lambda}_{U,e} \circ \hat{U'},I \right ) \label{Eq:2TwirlStep} \\
&=& \int_U d\mu(U) F \left ((1-p_{d,U})\hat{\Lambda}_1 + \frac{p_{d,U}}{D} \hat{\Lambda}_2,I \right ) \nonumber \\
&=& \int_U d\mu(U) \int_{\psi} \tr{\bra{\psi} \left ( (1-p_{d,U}) \ket{\psi}\bra{\psi} + \frac{p_{d,U}}{D} I \right ) \ket{\psi}} \nonumber \\
&=& \int_U d\mu(U) \left ( (1-p_{d,U}) + \frac{p_{d,U}}{D} \right ) \nonumber \\
&=& 1- \frac{D-1}{D}\overline{p_{d}} \label{Eq:2TwirlIntroInfid}\\
&=& \left (1-\frac{D-1}{D} \right ) + \frac{D-1}{D}(1-\overline{p_{d}}) \label{Eq:2TwirlIntro},
\end{eqnarray}
where $\overline{p_{d}} = \int_U dU p_{d,U}$ is the average strength of the twirled errors on the set of unitaries $U$.  Because it shows up frequently, I define $\alpha_n \equiv \frac{2^n-1}{2^n} = \frac{D-1}{D}$.  The quantity $\alpha_n\overline{p_d}$ is equal to the average gate infidelity (one minus the average gate fidelity).  The strategy of randomized benchmarking is then to intentionally twirl all errors in order to simplify the experiment, calculate the average strength of these twirled errors (Eq.~\ref{Eq:2TwirlStep} through Eq.~\ref{Eq:2TwirlIntro}), and then use this parameter to solve for the average gate fidelity of the original gates.

\subsection{Depolarizing Channels} \label{Sec:2Dep}

Superoperators of the form $(1-p)X + \frac{p\tr{X}}{D}I$ are called depolarizing channels.  They are parametrized by a single real number $p$, which I refer to as the strength of the depolarizing channel.  A depolarizing channel of strength $0$ is the identity superoperator and does not change the operator to which it is applied; a depolarizing channel of strength $1$ replaces the operator with the identity operator (properly normalized).  Because depolarizing channels are the basis for most of the theory of randomized benchmarking it is worth touching upon some of their features.
\begin{itemize}
\item 
The composition (or product) of two depolarizing channels is a depolarizing channel.  If $\hat{\Lambda}_{a}$ has strength $p_a$ and $\hat{\Lambda}_{b}$ has strength $p_b$, then
\begin{eqnarray}
\left (\hat{\Lambda}_{a} \circ \hat{\Lambda}_{b} \right )(X) &=& \hat{\Lambda}_{a}\left (\hat{\Lambda}_{b} (X) \right ) \nonumber \\
&=& (1-p_a) \left( (1-p_b) X + \frac{p_b\tr{X}}{D} I \right ) + \frac{p_a\tr{(1-p_b)X + \frac{p_b\tr{X}}{D}I}}{D} I \nonumber \\
&=& (1-p_a)(1-p_b) X + \left(1- (1-p_a)(1-p_b) \right)\frac{\tr{X}}{D}I. \label{Eq:2DepProduct}
\end{eqnarray}
This resulting superoperator is a depolarizing channel with strength $\left(1- (1-p_a)(1-p_b) \right)$.  Conceptually, the quasi-probability of not depolarizing (one minus the depolarizing strength) is the product of the quasi-probabilities of not depolarizing for the two individual channels.  Note that, if $p_a$ and $p_b$ are both less than one, then so is $\left(1- (1-p_a)(1-p_b) \right)$, the strength of the composition of the channels.
\item
Given the two depolarizing channels $\hat{\Lambda}_a, \hat{\Lambda}_b$ defined in the previous paragraph, the probabilistic mixture $\hat{\Lambda}_{a+b} = (1-q)\hat{\Lambda}_{a} + q\hat{\Lambda}_{b}$ is a depolarizing channel with strength $(1-q)p_a + q p_b$.  Consequently, any probabilistically weighted sum or integral over depolarizing channels is a depolarizing channel with strength equal to the weighted sum or integral of the strengths of the individual terms.
\item
Because the totally mixed state can be written as $\rho_{\mbox{mix}} = \frac{1}{D} I$ for any basis of qubit states, it is convenient and accurate to think of the action of a depolarizing channel as mapping a state to a sum of the state and the totally mixed state.  If the depolarizing strength is one, then the channel takes any state to the totally mixed (information-less) state.  It is actually possible for the depolarizing strength to be greater than one.  For example, the depolarization (defined below) of any non-identity Pauli superoperator has depolarizing strength $\frac{D^2}{D^2-1} > 1$.  For randomized benchmarking one is typically concerned with error superoperators that are near the identity (and therefore have strength $<1$); the superoperator composition of the depolarizations of these errors will never have strength more than one, so the depolarizing strength can be loosely thought of as if it was a probability.
\item
For any trace-preserving superoperator $\hat{\Lambda}$, the superoperator $\int_U dU \hat{U} \circ \hat{\Lambda} \circ{U^{\dag}}$ is a depolarizing channel referred to as the depolarization of $\hat{\Lambda}$, and its strength is $p_{d,U} = \frac{D^2 - \tr{\hat{\Lambda}}}{D^2-1}$.  Further, the fidelity of $\hat{\Lambda}$ with the identity operator $I$ is $F(\hat{\Lambda},I) = 1 - \alpha_n p_{d,U}$, where $\alpha_n = \frac{D-1}{D}$, as before.  Note that this fidelity is strictly bounded from below by $\frac{1}{D+1}$, but for any composition of depolarizations of near-identity superoperators, it is bounded from below by $\frac{1}{D}$.
\end{itemize}

\subsection{Gate Error Model} \label{Sec:2GateError}

If $\hat{\Lambda}_U$ is a general, trace-preserving superoperator, then the discussion surrounding Eq.~\ref{Eq:2ErrorOp} demonstrated a decomposition $\hat{\Lambda}_U = \hat{\Lambda}_{U,e} \circ \hat{U}  $ that splits $\hat{\Lambda}_U$ into a perfect part implementing the desired unitary and an error superoperator.  In this case, the error superoperator is seen to act after the ideal operator, although an error operator that acts before the ideal operator is just as easy to define.  In the case of randomized benchmarking, the main quantity of interest is the gate fidelity of $\hat{\Lambda}_U$ with $\hat{U}$, averaged over all $U$.  In order for the derivation in the previous section to apply, I must make several assumptions about the gate error model that I collect here for reference.  In particular, I describe problems with the gate error model that can cause the derivation to fail.

An experimental process $\hat{\Lambda}_U$ might change in time as (for example) stray fields in the lab shift qubit resonances.  The process might also be a function of the number and type of gates that have been performed since the beginning of a particular experiment, especially if the control apparatus or the qubit have heating processes that are reset before each experiment.  Such variations are discussed in some more depth in Secs.~\ref{Sec:2TimeDep},~\ref{Sec:2Systematics}, and~\ref{Sec:3ExpResults}.  Time-dependent errors of these kinds are inevitable in experiment, but they must be small if any quantum computation is to succeed.  The derivation of the behavior of the randomized-benchmarking experiment that follows relies on the assumption that $\hat{\Lambda}_U$ is a function only of the gate $U$ and not time.

A separate issue is whether $\hat{\Lambda}_{U,e}$ varies for different unitaries $U$.  For the derivation that follows, I assume that $\hat{\Lambda}_{U,e} = \hat{\Lambda}_e$ is independent of the intended unitary $U$.  This is almost always an unreasonable assumption, and Refs.~\cite{Magesan2011,Magesan2011B,Magesan2012B} have dealt with relaxations of this assumption in depth.  In particular, they show that the average depolarizing strength of the errors is obtained correctly from the benchmark experiment as long as the error superoperators differ from their weighted average by only a small distance according to a particular norm.  Sec.~\ref{Sec:2GateDep} describes two special error models that do not have this property for which the same simple results can be proven.  

If there are states of the system other than the two distinguished qubit states, then superoperators on the system need not preserve trace when considered to act only on the qubit states.  This is easily seen for a three-level system: A unitary operator may swap the $\ket{0}$ and $\ket{2}$ states.  Looked at as an operator on the $\ket{0}$ and $\ket{1}$ states only, such an operator does not conserve the sum of squared state amplitudes.  Operations of this sort are also common in experiments, e.g. when the electronic state of an ion is accidentally excited from the ground state to an unwanted state.  In photonic experiments, photon loss events are also modeled by trace-non-preserving operations and are a dominant error source.  Again, however, I must assume that all error operations are trace-preserving.  The effects of violation of this assumption are briefly considered in Sec.~\ref{Sec:2Memory}.  

\subsection{Random Sequences} \label{Sec:2Sequences}

Consider a long sequence of ideal unitaries $U_{tot}^{\dag}, U_l, \cdots U_1$, where $U_{tot}= U_l \cdots U_1$.  The product of all of these matrices (taken in the order given in the sequence) is the identity.  Fix $m \in \{1,l\}$.  Given a large set of such sequences, consider only the subset, indexed by $k \in [1,N]$, such that the $m^{th}$ unitary in the sequence, $U_{m,k}$ is the same distinguished unitary $U$.  A probabilistic mixture of the superoperators implied by all such sequences is $\frac{1}{N}\sum_{k} \hat{U}_{tot,k}^{\dag} \circ \hat{U}_{l,k} \circ \cdots \hat{U} \circ \cdots \hat{U}_{1,k}$.  Now consider experimental implementations of these ideal sequences where all superoperators are ideal except for $\hat{\Lambda}_U = \hat{\Lambda}_{U,e} U$.  Then the superoperator mixture above becomes
\begin{equation} \label{Eq:2SeqSum1}
\frac{1}{N}\sum_{k} \hat{U}_{tot,k}^{\dag} \circ \hat{U}_{l,k} \circ \cdots \hat{\Lambda}_{U,e} \hat{U} \circ \cdots \hat{U}_{1,k}
\end{equation}

Denote the product of the $b^{th}$ - $a^{th}$ consecutive ideal unitaries in sequence $k$ by $U_{[b,a],k}$.  For these purposes, I also denote $U_{l+1} \equiv U_{tot}$.  Because the product of all of the unitaries in each sequence is the identity, $U_{[l+1,b],k} = U_{[b,1],k}^{\dag}$ for any $b$. Then the expression of Eq.~\ref{Eq:2SeqSum1} is equal to
\begin{equation} \label{Eq:2SecSum2}
\frac{1}{N}\sum_{k} U_{[m-1,1],k}^{\dag}\hat{\Lambda}_{U,e}U_{[m-1,1],k}
\end{equation}

Now consider instead a distribution over all length-$l$ sequences of unitaries according to the Haar measure of each unitary in the sequence, fixing the $m^{th}$ superoperator to be $\Lambda_U$ as before.  Then consider the $(l-1)$-fold integral over this distribution where each integral varies a different $U_i$ with Haar measure.  The left-invariance of the Haar measure in each integral guarantees that this is equivalent to the integral 
\begin{equation} \label{Eq:2SecInt}
\overline{\hat{\Lambda}}_{m,U} \equiv \int d\mu(U)' U'^{\dag} \hat{\Lambda}_{U,e} U'.
\end{equation}
This superoperator has the form of a depolarizing channel with strength $p_{U,e} = \frac{D^2 - \tr{\hat{\Lambda}_{U,e}}}{D^2-1}$.  

Allowing only the $m^{th}$ unitary to have errors, but integrating now over all possible ideal unitaries in the $m$ location, gives $\overline{\hat{\Lambda}}_m \equiv \int_U d\mu(U) \overline{\hat{\Lambda}}_{m,U}$.  This superoperator is also a depolarizing channel, and its strength is $\overline{p}_e \equiv \int_U d\mu(U) p_{U,e} = \frac{D^2 - \int_U d\mu(U) \tr{\hat{\Lambda}_{U,e}}}{D^2-1}$.  The important insight to take is that the error superoperator associated with the implementation of any single unitary, when integrated over all possible surrounding unitary sequences, becomes a depolarizing channel.

At this point, I leap to a conclusion that is proven in detail in Ref.~\cite{Magesan2012B}  and elsewhere.  Let the implementations of all $U$ be error-prone, so that $\hat{\Lambda}_U = \hat{\Lambda}_{U,e} \circ  \overline{U}$ for all $U$.  Further, assume that $\hat{\Lambda}_{U,e} = \hat{\Lambda}_e$ does not depend on the unitary $U$.  Then the integral over all self-inverting unitary sequences is
\begin{equation} \label{Eq:2SeqIntFin}
\int_{U_1, \cdots U_l} dU_1 \cdots dU_l \hat{\Lambda}_{U_{tot}^{\dag}} \circ \hat{\Lambda}_{U_l} \cdots \hat{\Lambda}_{U_1} = \hat{\Lambda}_e \circ (\hat{\Lambda}_d)^l ,
\end{equation}
where
\begin{equation} \label{Eq:2DepOp}
\hat{\Lambda}_d = \int_U dU \hat{U^{\dag}} \circ \hat{\Lambda}_e \circ \hat{U} = (1-p_e)\hat{\Lambda}_1 + p_e\hat{\Lambda}_2
\end{equation}
is the depolarization of $\hat{\Lambda}_e$.  Intuitively, each ``step'' of the sequences, with the exception of the inverting step, becomes depolarized by the integral of unitaries around it as in the single-error example above.

There is an interpretation for the integral over all unitary sequences that connects more directly with the experimental implementation of randomized benchmarking.  For each index in the overall sequence integral one can think of applying the operator corresponding to $U_i$ with (infinitesimal, Haar) probability $d\mu(U_i)$.  Then the superoperator represented by the integral over all sequences can be thought of as the composition of $l$ superoperators each chosen randomly with Haar distribution and a final superoperator that is chosen deterministically to invert the others. 

The goal of a randomized-benchmarking experiment is to Monte-Carlo approximate the integral 
\begin{eqnarray} 
\int_{\{U_1 \cdots U_l\}}  &dU_1& \cdots dU_l \tr{\ket{\overline{0}}\bra{\overline{0}} \hat{\Lambda}_{U_{tot}} \circ \hat{\Lambda}_{U_l} \circ \cdots \hat{\Lambda}_{U_1}  (\ket{\overline{0}}\bra{\overline{0}})} \nonumber \\ 
&=& \tr{\ket{\overline{0}}\bra{\overline{0}} \int_{\{U_1 \cdots U_l\}} dU_1 \cdots dU_l \hat{\Lambda}_{U_{tot}} \circ \hat{\Lambda}_{U_l} \circ \cdots \hat{\Lambda}_{U_1} (\ket{\overline{0}}\bra{\overline{0}})} \nonumber \\
&=& \tr{\ket{\overline{0}}\bra{\overline{0}} \hat{\Lambda}_{e} \circ \hat{\Lambda}_d^l \left (\ket{\overline{0}}\bra{\overline{0}} \right )} \nonumber \\
&=& \tr{\ket{\overline{0}}\bra{\overline{0}} \hat{\Lambda}_{e} \left ((1-p_e)^l \ket{\overline{0}}\bra{\overline{0}} + \frac{1-(1-p_e)^l}{D} I \right )} \label{Eq:2RBIntegral}
\end{eqnarray}
by sampling at random from the possible unitary sequences and averaging the resulting fidelities of their implementations (called sequence fidelities).  For the $i^{th}$ chosen sequence, this experiment involves preparing the $\ket{\overline{0}}$ state (where each qubit is in the $\ket{0}$ state), applying a sequence of implementations of unitaries (including the inverting unitary intended to return the system to the $\ket{\overline{0}}$ state), and then measuring the probability that the result lies in the $\ket{\overline{0}}$ state.  From the results of these experiments, one hopes to be able to extract the average strength of the depolarizations of the error operators.
 
\subsection{Measurement Step} \label{Sec:2Measurement}

The main quantity of interest in a randomized-benchmarking experiment is the depolarizing strength of $\hat{\Lambda}_d$, as Eq.~\ref{Eq:2TwirlIntroInfid} showed it to be the mean infidelity of the operations in the sequence.  If not for the remaining $\hat{\Lambda}_e$ term in Eq.~\ref{Eq:2RBIntegral}, solving for this quantity would seem straightforward:  As established in Sec.~\ref{Sec:2Dep}, $\hat{\Lambda}_d^l(X) = (1-p_d)^l X + \frac{(1-(1-p_d)^l)\tr{X}} {D}I$, so 
\begin{equation} \label{Eq:2EasyExtraction}
\tr{\ket{\overline{0}}\bra{\overline{0}} \hat{\Lambda}_d^l \ket{\overline{0}}\bra{\overline{0}}} = \tr{\ket{\overline{0}}\bra{\overline{0}} \left ((1-p_d)^l \ket{\overline{0}}\bra{\overline{0}} + \frac{1-(1-p_d)^l}{D}I \right )} = (1-p_d)^l + \frac{1-(1-p_d)^l}{D}.
\end{equation}
As the number of sequences considered increases, improving the Monte-Carlo approximation to the integral in Eq.~\ref{Eq:2RBIntegral}, the average of the sequence fidelities  approaches $(1-\alpha_n) + \alpha_n (1-p_d)^l $.

The problem is that the inverting step has errors that may not behave as depolarizing channels under this averaging.  This problem is related to the problem of imperfect inversion operators introduced in Sec.~\ref{Sec:1GateFidelity}.  In fact, there can also be errors in the measurement of the $\ket{\overline{0}}$ state, which might be modeled as $\hat{\Lambda}_{m}$ before a perfect measurement.  Because these two errors occur at the same place in the operator sequence, their composition is the only relevant operator, and I refer only to $\hat{\Lambda}_{m}$.  Errors are also likely in the preparation of the $\ket{\overline{0}}$ state, and I model these by $\hat{\Lambda}_{p}$ acting after a perfect preparation.  Assume that however these errors behave, they are time- and gate-independent and trace-preserving.  

Following the careful derivation of Refs.~\cite{Emerson2005} and \cite{Magesan2012B}, these errors together contribute a constant error probability $p_m$.  Errors of this sort are called state preparation and measurement (SPAM) errors, so $p_m$ is the SPAM probability of error.  SPAM errors can also contribute an additive (as opposed to multiplicative) offset to the sequence fidelities, although the measurement technique described in the next section is designed to avoid this contribution.  As long as the SPAM errors are independent of the gate sequence, the error for each sequence experiment can be interpreted as the product of the errors for the sequence of gates and the errors for the SPAM processes.  Then, even if the SPAM errors vary, the average error over all experiments is a product of the average error for the sequences of gates and the average SPAM error.

Such a constant, multiplicative error contribution has the same effect in the description of the randomized-benchmarking experiment as a depolarizing channel of strength $p_m$, so the same notation is used even though there is no reason to believe that these errors are actually depolarizing.  This is possible because the input state and the measurement basis are the same for every experiment, and complicated errors have simplified behavior in this case.  In the next section, I discuss one common experimental scenario in which the measurement error appears to depend on the sequence and two remedies for the problem.   It is worth reexamining the validity of assumptions made for the SPAM errors for each experiment.

\subsection{Measurement Randomization} \label{Sec:2MeasRand}
A likely failure of the assumptions of the previous section is that the probability of error in the measurement will depend on the state before the measurement.  This occurs particularly if the probability $p_{01}$ of mistakenly measuring $\ket{0}$ as $\ket{1}$ is not equal to the probability $p_{10}$ of mistakenly measuring $\ket{1}$ as $\ket{0}$.  A careful experiment might try to balance these two errors, but they are never exactly equal.  Consider the case where $p_{01} = 0$ and $p_{10} > 0$ for a single qubit.  Let the probability of measuring the final state of some sequence to be in the state $\ket{0}$ with a perfect measurement be $p_0$ and the corresponding probability for the $\ket{1}$ state be $p_1$. Then the probability of measuring the system in the state $\ket{0}$ using this imperfect measurement is $p_0 +p_{10}p_1 = (1-p_{10})p_0 + p_{10}$, and the probability of measuring it in the state $\ket{1}$ is $(1-p_{10})p_1 = (1-p_{10})(1-p_0)$.  

If the system is in the pure $\ket{0}$ state at the end of the sequence, the measurement operator acts like a perfect measurement.  However, if the system is in the totally mixed state, as it will be after very long sequences, then the measurement operator acts imperfectly, measuring the state $\ket{1}$ with probability $\frac{1}{2}(1+p_{10}) \neq \frac{1}{2}$.  If the measurement is to look like a depolarizing error, it must act trivially on the totally mixed state, and this error does not.

It is possible to detect and model such an issue if very long sequences are used in the experiment because the experimental probability of measuring the correct result will not decay to the expected value $\frac{1}{2}$ (or  $(1-\alpha_n)$ for $n$ qubits).  This is the approach implied by the authors of Ref.~\cite{Magesan2012B}.  However, it is possible to avoid this particular problem entirely if the state of each qubit is flipped with a probability $\frac{1}{2}$ immediately before the measurement.  Ignoring for now how this might be implemented, consider its effect on this unbalanced measurement.  For each qubit, the measurement measures the un-flipped state as the correct $\ket{0}$ state with probability $p_0 + p_{10}p_1 = p_0 + p_{10}(1-p_0)$ as before.  It measures the flipped state as the correct $\ket{1}$ state with probability $(1-p_{10})p_0$.  Then the average probability of measuring the correct result is $p_0 + \frac{1}{2}(p_1-p_0)p_{10} = (1-p_{10})p_0 + \frac{1}{2}p_{10}$.  Such an error does act like a depolarizing error; this can be seen particularly in the probability $\frac{1}{2}$ of detecting the correct result if the system ended in the totally mixed state.

In order to implement this random bit flip, one might add an additional gate (a $\pi$ rotation about the $X$ axis would suffice).  However, this additional gate might introduce additional error, changing the effective error of the measurement step conditional on whether the flip was applied or not.  Instead, Ref.~\cite{Knill2008} introduced a convention whereby the last unitary operator in the sequence is chosen such that it acts as the product of the unitary needed to invert the aggregate unitary of the sequence times a unitary chosen to act randomly and independently on each qubit as the $X,Y,Z,$ or $I$ Pauli operators.  Although this operator is computed as a product, it is implemented as any other superoperator would be, in a single step.  The unitary chosen is marginally Haar distributed, and its superoperator has the same average error as the superoperator for the exact inverse unitary would have had, but it implements the random flip described in the previous paragraph.

The analysis of an experiment with this measurement randomization includes one additional complication.  When determining the fidelity of an implementation of a sequence, each measurement result of the experiment is compared to the ideal result of the sequence (assuming no errors occurred), which may correspond to either $\ket{0}$ or $\ket{1}$, depending on the random flip.  In contrast, the fidelity of an implementation of the non-randomized sequences is always the probability that all qubits are detected in the $\ket{0}$ state.

Whether it is better to model such an unbalanced measurement error or to remove it from the analysis  by randomizing is an experimental choice.  However, the same randomization that allows the measurement imbalance to be ignored also makes the procedure immune to an adversarial error model discussed in Ref.~\cite{Magesan2012B} that otherwise fools RB experiments.  This is an error model where the error on each superoperator is the exact inverse of the ideal unitary being implemented, so that each experimental superoperator acts exactly as the identity.  In the formulation without randomization before measurement, the state is always in the $\ket{0}$ state and the sequences are always found to have fidelity $1$, even though the errors are quite large.  If the measurement randomization is added, then the measurement is found to be incorrect exactly $\frac{1}{2}$ of the time, correctly identifying the failure of the error assumptions and the high strength of the errors.

\subsection{Standard Equation} \label{Sec:2MainEquation}

If the many assumptions in the previous sections are valid, then the average length-$l$ sequence is described by a product of $l$ depolarizing channels of the same strength ($p_s$) and one additional depolarizing channel of different strength ($p_m$) corresponding to the SPAM errors.  Such a superoperator is a depolarizing channel of strength $1-(1-p_m)(1-p_s)^l$.  The average sequence fidelity of all length $l$ sequences is then
\begin{equation} \label{Eq:2Main}
F_l = (1-\alpha_n) + \alpha_n \left (1-\frac{\epsilon_m}{\alpha_n} \right ) \left (1-\frac{\epsilon_s}{\alpha_n} \right )^l,
\end{equation}
where $\alpha_n = \frac{2^n - 1}{2^n}$ for $n$ qubits, and $\frac{\epsilon_{s/m}}{\alpha_n}$ may be identified with the depolarizing strength of the corresponding superoperator.  This scalar multiple of the depolarizing strength is identified, for convenience, because $\epsilon_s$ is the probability that the step introduces an error that causes a measurement of all $n$ qubits to reveal at least one error.  As the number of sequences included in the experiment increases, the result of Monte-Carlo approximation using the finite number of sequences approaches Eq.~\ref{Eq:2Main}.

In many ways, this model is more universal than the above analysis indicates.  As long as the gate error assumptions introduced in Sec.~\ref{Sec:2GateError} are not violated strongly or in an adversarial way, a RB experiment is expected to behave in this way.  For this reason, if experimental results are well described by Eq.~\ref{Eq:2Main}, as quantified by goodness-of-fit tests, for example, then the error per step is a useful experimental parameter even if the degree to which the assumptions are satisfied individually is unclear.  In taking this approach, I differentiate the use of RB as a benchmark from the use of RB as a tool to uncover a particular parameter of the physical control errors.

\subsection{Many-Parameter Models} \label{Sec:2EqExten}
In some cases it might be that detectors are not be sufficiently calibrated to allow for convincing discrimination of the $\ket{0}$ and $\ket{1}$ states.  This happens especially in ensemble experiments or experiment using photon counters for which the number of counts expected from the ``bright'' and ``dark'' states cannot be calibrated because these states cannot be prepared with high fidelity.  In these cases, if no measurement randomization is used, Eq.~\ref{Eq:2Main} can be modified to
\begin{equation} \label{Eq:2RB3}
N_l =C(1-\alpha_n) + C\alpha_n \left (1-\frac{\epsilon_m}{\alpha_n} \right ) \left (1-\frac{\epsilon_s}{\alpha_n} \right )^l,
\end{equation}
where the parameter $C$ (with units of detector clicks per experiment), which is unity in the standard equation, is allowed to vary and $N_l$ is the mean number of clicks at the detectors for sequences of length $l$.  As long as the depolarization is effective, each qubit will eventually decay to the totally mixed state, and this state will be observed in the detector with a number of clicks which is the mean of the number of clicks for the bright $\ket{1}$ and dark $\ket{0}$ states.  In this way, the long-sequence data effectively calibrate the detector counts during the fitting procedure.  For this procedure to be effective, the detector efficiency and dark count rate must not drift significantly during the randomized benchmark experiment.  An equation of this same form can be used (with different parameter interpretations) to model a RB experiment with unbalanced measurement errors and no measurement randomization.

The authors of Ref.~\cite{Magesan2011,Magesan2011B,Magesan2012B} have developed a model that accounts for gate- and time-dependent errors as long as their variation from a mean error is small.  They use a perturbative approach to derive a first order correction to the model and bound the magnitude of higher order corrections.  A somewhat restricted form of their result is
\begin{equation} \label{Eq:2Magesan}
F_l = a + \alpha_n (1- \frac{\epsilon_m}{\alpha_n}) (1-\frac{\epsilon_s}{\alpha_n})^l + b(l-1) (1-\frac{\epsilon_s}{\alpha_n})^{l-2}.
\end{equation}

When using equations with more parameters to fit to the experimental data, one must take care to avoid over-fitting.  In particular, I demonstrate in Sec.~\ref{Sec:2ExpResults1} that if data points for sequences with a high probability of error are not available, then Eq.~\ref{Eq:2RB3} uses three parameters to fit data that is well described by a line, leading to some degeneracy of the fitting parameters.  Additionally, in Sec.~\ref{Sec:3Exp2} I show that using fits of more than 2 parameters on 6 noisy data points causes a similar problem.  Only experiments with larger range of $l$ and smaller variance on the individual data points are amenable to fitting to such models.

\section{Use of Clifford Operators} \label{Sec:2Clifford} 

I show in this section that the exact Haar twirl is not necessary to achieve the depolarization of errors that is the basis of the randomized-benchmarking procedure.  Among others, the twirl over Clifford operators, which form a finite-size subgroup of unitary operators, is equal to the exact Haar twirl as a super-superoperator.  App.~\ref{App:A} contains a more-detailed overview of Clifford operators and should be used as a reference for this section.  There are several appealing reasons to try to implement a randomized benchmark using Clifford operators instead of arbitrary unitaries:
\begin{itemize}
\item
Because the Clifford operators are finite in number, it is possible to calibrate them all for small qubit number $n$.  Failing that, each Clifford operator can be exactly constructed as the finite product of a small number of generating gates, each of which can be calibrated.  This compares favorably to the full set of unitary operators, which do not have a generating set of this kind and so cannot be constructed exactly from a finite set of calibrated gates.  This issue is particularly relevant for the encoded benchmarks discussed in Sec.~\ref{Sec:4EncodedBenchmark}, as generating arbitrary unitaries in fault-tolerant architectures is quite complicated.
\item
As described in more depth in Sec.~\ref{Sec:2CliffRand}, the Clifford operators are the basis of almost all practical quantum error-correction protocols.  Quantifying their gate fidelity is an essential step in deciding whether error correction is possible for a given architecture.
\item
In order to perform the inverting operator at the end of a sequence of random unitaries, it is necessary to calculate the inverse of the product of the preceding unitaries.  This (classical) computation can take a very long time, making the design of RB experiments prohibitively difficult when $n$ is large.  Because general unitaries are represented as $2^n \times 2^n$ size matrices, inversion takes time of order $O(2^{3n})$ using typical inversion algorithms.  Even the best inversion algorithms take time that scales exponentially in $n$.  For $n$ of a modest size (certainly $n \gtrsim 100$), this calculation is impossible using current computing technology, and so randomized-benchmarking cannot be performed as described so far.  In contrast, the inversion algorithms for Clifford operators described in Sec.~\ref{Sec:AGeneration} require time of order $O(n^3)$ and are therefore tractable even for very large $n$.
\item
In contrast to the infinite number of possible unitary sequences of fixed length, the number of possible Clifford sequences of fixed length is finite.  All integrals over unitaries in the analysis are replaced by sums.  It is easier to analyze the degree to which the sum of a small number of sequence fidelities approximates the sum over all sequence fidelities (in comparison to analyzing the convergence of a finite sum of sequences to the integral over all unitary sequences).  As addressed in Sec.~\ref{Sec:4StepDef}, if some operators are never performed (this is always the case when randomizing over an infinite group), then the analysis error in reporting the average over all unitaries using a finite number of sequences is poorly constrained unless some assumptions regarding the physical gate error models are employed.
\end{itemize}
\subsection{Clifford Operator Group} \label{Sec:2CliffordFacts}
The Clifford operators form a finite subgroup of the unitary operators that has played a prominent part in the theory of quantum error correction.  Many aspects of the Clifford (and Pauli) operators and related definitions are treated in App.~\ref{App:A}.  In order to show that the Clifford twirl depolarizes superoperators, I use the following facts.
\begin{itemize}
\item
The Pauli matrices with initial phase $\imath^{\zeta} = 1$ on $n$ qubits (as defined in Eq.~\ref{Eq:APauliDef}) form an orthogonal additive basis for $2^n \times 2^n$ matrices, where the inner product is $\tr{P_iP_j} = \delta_{i,j}D$.  Therefore, for any matrix $T$, $T = \sum_i \beta_i P_i$ for $\beta_i = \frac{\tr{P_iT}}{D}$.  If $T$ is a density matrix and $P_0$ is identified with the identity operator, then $1=\tr{T} = D\beta_0$.  Because the overall phase of a quantum operator does not have a physical meaning, let the quotient of the group of Pauli matrices (under multiplication) with the subgroup $\{\imath^{\zeta} I\}$ for $\zeta \in \{0,1,2,3\}$ define the group of Pauli operators, which I denote with $\mathcal{P}_n$.  As a representative matrix for each Pauli operator, I choose the coset member with $\zeta = 0$ in Eq.~\ref{Eq:APauliDef}.  This definition of the Pauli operators as distinct from the Pauli matrices is not standard, but enables one to avoid discussing the initial phase of the operators, which are frequently not important.
\item
The Clifford unitaries on $n$ qubits are the matrices $C$ such that $\hat{C}_i(P_j) \equiv CPC^{\dag} = \pm P'$, for all $P \in \mathcal{P}_n$ and some (not fixed) $P' \in \mathcal{P}_n$.  As with the Pauli operators, I define the group of Clifford operators to be the quotient of the group of Clifford matrices with the subgroup $\{e^{\imath \phi} I\}$ for $\phi \in \mathbb{R}$.  The Pauli and Clifford operator groups defined in this way are finite.
\item
A Clifford operator can be uniquely specified by one of its unitary matrix representations or by its action on the Pauli matrices (see Lem.~\ref{Lem:AClifAction}).  I write this action as $\hat{C}(P) \equiv CPC^{\dag}$ and refer to the action on each $P$ as the transformation rule for $C$ and $P$.

\item
The Pauli operators are a normal subgroup of the Clifford operators.  For a fixed Pauli operator $P$ considered as a Clifford operator, the Clifford action on any Pauli matrix $P'$ gives $P(P') = \pm P'$ (depending on whether $P$ commutes or anti-commutes with $P'$).  

Consider the quotient $\overline{\mathcal{C}}_n = \mathcal{C}_n/\mathcal{P}_n$ of the group of Clifford operators with the subgroup of Pauli operators.  Its elements are equivalence classes (or cosets) of Clifford operators, all of which have the same action on each given Pauli matrix up to sign of the result.  It is possible to pick a representative of each coset that, for all qubit indices $i$, maps the Pauli matrices $X_i$ and $Z_i$ (with $\zeta = 0$) to Pauli matrices with $\zeta= 0$.  Each Clifford operator can be decomposed into the product of this representative of its coset and a Pauli operator that changes the signs in the Clifford action.  I denote the coset of $C$ with $[C]$ and the distinguished representative of this coset with $\tilde{C}$.  Notably, the Clifford operator $I$ is the distinguished representative $\tilde{I}$ for its coset.

\end{itemize}

\subsection{Pauli Twirl} \label{Sec:2PauliTwirl}
Let $A = \sum_{i,j} \beta_{i,j} P_i \otimes P_j$ be a general superoperator on $n$ qubits, where $P_i,P_j \in \mathcal{P}_n$.  Consider the super-superoperator which takes $A \rightarrow \frac{1}{|\mathcal{P}_n|}\sum_{P_k \in \mathcal{P}_n} \hat{P}_k(A) = \frac{1}{|\mathcal{P}_n|} \sum_{P_k \in \mathcal{P}_n} \sum_{i,j} \beta_{i,j} P_k P_i P_k \otimes P_k P_j P_k$, which is called the Pauli twirl.  As seen in Lemma~\ref{Lem:2Proj}, because the Pauli operators are a group, the Pauli twirl projects $A$ onto the space of superoperators $B$ that are stabilized by the action $P_k(B)$ for each $P_k$. 


Let $B = \sum_{i,j} \gamma_{i,j} P_i \otimes P_j$ be such that $P_k(B) = B$.  Then 
\begin{eqnarray} \label{Eq:2PauliTwirl}
\sum_{i,j} \gamma_{i,j} P_i \otimes P_j & = & \sum_{i,j} \gamma_{i,j} P_k P_i P_k \otimes P_k P_j P_k \\
& = & \sum_{i,j} \gamma_{i,j} (-1)^{f(i,k)} P_i \otimes (-1)^{f(j,k)} P_j,
\end{eqnarray}
where
\begin{equation} \label{Eq:2ComFunction}
 f(a,b) = 
\begin{cases}
1  \mbox{ if } P_a,P_b \mbox{ commute}\\
-1  \mbox{ if } P_a,P_b \mbox{ anti-commute}
\end{cases}.
\end{equation}
Matching up tensor terms in Eq.~\ref{Eq:2PauliTwirl}, shows that either $f(i,k) = f(j,k)$ or $\gamma_{i,j} = 0$ must hold for each $(i,j)$ in order for the equality to hold.  If this equality is to hold for all Pauli elements $P_k$, then it can only hold on tensor term $(i,j)$ if either $f(i,k) = f(j,k)$ for all $k$ or $\gamma_{i,j}=0$.  Inspection of the Pauli group reveals that two Pauli elements only share the same commutation relations with all other elements if they are the same.  Therefore, $\gamma_{i,j}$ must equal $0$ for all $i\neq j$.  I conclude that the Pauli twirl projects the superoperator $A$ onto the space of superoperators of the form $\sum_i \gamma_i P_i \otimes P_i$; this set of superoperators are called (stochastic) Pauli channels, and are discussed further in Secs.~\ref{Sec:AError} and~\ref{Sec:4AppDep}  

\subsection{Clifford Twirl} \label{Sec:2ClifTwirl}
Consider again a general superoperator $A = \sum_{i,j} \beta_{i,j} P_i \otimes P_j$ and an arbitrary Clifford operator $C$.  Because $\hat{C}(A) \equiv \sum_{i,j} \beta_{i,j} CP_iC^{\dag} \otimes CP_jC^{\dag}$, a Clifford twirl super-superoperator can be defined that takes $A \rightarrow \frac{1}{|\mathcal{C}_n|}\sum_{C \in \mathcal{C}_n} \sum_{i,j} \beta_{i,j} CP_iC^{\dag} \otimes CP_jC^{\dag}$.  As before, this super-superoperator projects onto the space of superoperators that are stabilized by each $\hat{C}$.  The structure of the Clifford group described in Sec.~\ref{Sec:2CliffordFacts} is such that for each Clifford $C$ the operators $CP_l$ are distinct for distinct Pauli operators $P_l$.  These operators form a coset $[C]$ of $C$ in the Clifford quotient $\overline{\mathcal{C}}_n$.  I identify the representative $\tilde{C}$ of each of these equivalence classes, so that 
\begin{equation} \label{Eq:2TwirlDecomp}
\frac{1}{|\mathcal{C}_n|}\sum_{C \in \mathcal{C}_n} \hat{C}(A) = \frac{1}{|\mathcal{C}_n|}\sum_{[C] \in \overline{\mathcal{C}}_n} \hat{\tilde{C}} \left ( \sum_{P_l \in \mathcal{P}_n} \hat{P}_l(A) \right ).
\end{equation}

Eq.~\ref{Eq:2TwirlDecomp} decomposes the Clifford twirl into the composition of a twirl over representatives of each coset in $\overline{\mathcal{C}_n}$ and a Pauli twirl.  The Pauli twirl was already shown to project onto operators of the form $B = \sum_i \gamma_i P_i \otimes P_i$, so superoperators stabilized by the full Clifford twirl must have this form as well, as can be seen by considering those stabilized by all the Cliffords in $[I]$.  For such a superoperator $B$, consider the action $\hat{C}(B) = \hat{\tilde{C}}(\hat{P}(B)) = \hat{\tilde{C}}(B)$.  Define $P_{i,\tilde{C}} \equiv \tilde{C} P_i \tilde{C}^{\dag}$.  If $\hat{C}(B) = B$, then
\begin{eqnarray} \label{Eq:2ClifTwirlWork}
\sum_i \gamma_i P_i \otimes P_i & = & \sum_i \gamma_i \tilde{C} P_i {\tilde{C}}^{\dag} \otimes \tilde{C} P_i {\tilde{C}}^{\dag} \\
& = & \sum_i \gamma_i P_{i,\tilde{C}} \otimes P_{i,\tilde{C}}.
\end{eqnarray}
Because conjugation by a particular Clifford is an isomorphism on Pauli operators, the terms in the last sum must all be distinct.  For the equation to hold, the coefficients of like tensored Pauli terms must agree.  Calling $\gamma_i'$ the coefficient for the $P_{i,\tilde{C}} \otimes P_{i,\tilde{C}}$ term in the first expression, the requirement is $\gamma_i = \gamma_i'$.  It is possible to find a Clifford matrix $C_{i,i'}$ for which $C_{i,i'} P_i C_{i,i'}^{\dag} = P_i'$ for any $P_i \neq P_0, P_{i'} \neq P_0$, where $P_0 = I$ (see Lem.~\ref{Lem:ACliffordSym}).  Therefore, if $B$ is to be stabilized by the action of each Clifford element, it must satisfy $\gamma_i = \gamma$ for all $i \neq 0$.

The Clifford twirl then maps a superoperator onto a superoperator of the form $\hat{\Lambda}(\gamma_0,\gamma) = \gamma_0 P_0 \otimes P_0 + \sum_{i\neq 0} \gamma P_i \otimes P_i$.  Recall that for a general operator (matrix) $X = \sum_j \chi_j P_j$ one finds $\tr{X} = \chi_0 D$.  The techniques used in the derivation of the Pauli twirl can be used again to show that $\frac{1}{|\mathcal{P}_n|}\sum_i P_i X P_i = \chi_0 I = \frac{\tr{X}}{D} I $.  Then
\begin{equation} \label{Eq:2ClifDep}
\hat{\Lambda}(\gamma_0,\gamma)(X) = \left ( \gamma_0 - \gamma \right ) X + \frac{\gamma \tr{X}}{D} I.
\end{equation}
Assuming $\hat{\Lambda}(\gamma_0,\gamma)$ is trace-preserving 
forces $\gamma_0 = 1$, so, for any trace-preserving superoperator $\hat{\Lambda}$,
\begin{equation} \label{Eq:2ClifDepFin}
\frac{1}{|\mathcal{C}_n|} \sum_{C_i \in \mathcal{C}_n} \hat{C}_n \circ \hat{\Lambda} \circ \hat{C}_n^{\dag}(X) = (1-\gamma) X + \gamma \frac{\tr{X}}{D}I.
\end{equation}
This is of the same form as Eq.~\ref{Eq:2DepX}, so the result of applying the Clifford twirl to a superoperator is a depolarizing channel.  Repeating the arguments of that section shows that the strength of the depolarizing channel is $\gamma = \frac{D^2-\tr{\hat{\Lambda}}}{D^2-1}$.

Because the Clifford twirl acts the same as the exact Haar twirl, all of the results that relied only on the depolarizing nature of integrals of sequences of unitary operators hold as well for mixtures of sequences of uniformly random Clifford operators.  The same assumptions about gate errors must be made in this case as well.  With these assumptions, the mean sequence fidelity over all Clifford sequences of length $l$ behaves the same as the mean sequence fidelity over all unitary sequences described in Eq.~\ref{Eq:2SeqIntFin}.  The superoperator representing this mean is 
\begin{equation}
\label{Eq:2SeqIntClif}
\frac{1}{|\mathcal{C}_n|^l}\sum_{C_1 \cdots C_l}\hat{\Lambda}(C_{[l,1]}^{\dag}) \circ \hat{\Lambda}(C_l) \cdots \hat{\Lambda}(C_1) = \hat{\Lambda}_e (\hat{\Lambda}_d)^l,
\end{equation}
where $\hat{\Lambda}_e$ is again the common error operator for the gates and $\hat{\Lambda}_d$ is its depolarization.

For this reason, all of the results of Secs.~\ref{Sec:2Measurement} and \ref{Sec:2MainEquation} apply as well, and a Clifford randomized-benchmarking experiment is described by the same equations as a unitary randomized-benchmarking experiment.  One key difference lies in the meaning of the benchmark.  Instead of being a description of the error superoperator associated with all unitary operators, $\epsilon_s$ now describes the error operator associated with all Clifford operators.  Although the constructions are similar, the practical meaning of these two benchmarks can be quite different.

\subsection{Case for Unitary Randomization} \label{Sec:2UnitaryRand}

In the rest of this thesis, I focus on Clifford randomized benchmarking, but there is also a case to be made for unitary randomized benchmarking.  The advantages noted for Clifford benchmarking earlier in this section largely concern the classical computation and approximation difficulties with designing unitary benchmarking experiments.  The biggest computational advantage of Clifford benchmarking is the ability (discussed at length in Sec.~\ref{Sec:AGeneration}) to calculate efficiently the inverse of the product of the first $l$ operators.  This issue could be avoided by exactly reversing and inverting the first $l$ operators, as long as the first $l$ operators were composed of easily invertible gates; however, the analysis of this variation of benchmarking is quite different.  All advantages of Clifford RB based on the efficient computation of Clifford computations are only crucial when the number of qubits $n$ is large.

In contrast, there are two related advantages to unitary benchmarking.  The first is that the benchmark can be used to estimate the experimental fidelity of any unitary gate.  For quantum-computing strategies that do not particularly rely on stabilizer coding, there is no reason to think that Clifford fidelity is a particularly useful figure of merit; the average unitary fidelity might be a more appropriate starting point for benchmarking in this case.  The second  advantage comes from a technique described in Sec.~\ref{Sec:3GateBenchmark}.  In that section I describe how to alter the randomized-benchmarking experiment to learn about the depolarizing strength of the error on a particular gate.  In order for some of the advantages of Clifford randomized benchmarking to be used, this technique must be limited to investigating Clifford gates.  However, if unitary benchmarking is used, the depolarizing strength of any gate of interest might be investigated.  In any quantum-computing strategy, some non-Clifford gates are necessary; unitary benchmarking could be a useful tool for benchmarking these gates.

\subsection{Fault-Tolerance and the Case for Clifford Randomization} \label{Sec:2CliffRand}

The reason that Clifford fidelity is a useful benchmark is tied to the importance of stabilizer codes for quantum error correction.  In order to perform an interesting instance of Shor's factoring algorithm for a plausible architecture, Ref.~\cite{Jones2012} estimates that $>10^8$ gates of very high quality are needed.  To a crude approximation, each of these gates must have a probability of introducing an error that is smaller than $10^{-8}$.  Except perhaps in topological quantum computers of the sort described in Ref.~\cite{Nayak2008}, this seems to be an unachievable goal, and algorithms requiring more qubits would require even better gates.  The detour around this roadblock is provided by quantum coding, which allows quantum information to be encoded in many qubits in such a way that implementation errors in gates can be corrected at the cost of a large multiplicative overhead in the number of (unencoded) physical qubits required.

Partially for historical reasons and partially because of their relative simplicity, so-called stabilizer quantum codes have become the preeminent strategy for quantum error correction.  These codes encode some number $k$ of qubits into a subspace of the Hilbert space of $n$ qubits specified by $2^{n-k}$ $n$-qubit Pauli operators of which the subspace is an eigenspace.  Because a  quantum state can lie simultaneously in the eigenspace of $2^n$ distinct, commuting Pauli operators, this leaves a subspace or ``code space'' of dimension $2^k$ of states which satisfy the conditions implied by the stabilizing operators.

This code space is decomposed in such a way that $k$ ``logical'' or encoded qubits are distinguished.  The logical qubits are typically chosen in such a way that the most common experimental errors do not change the logical part of the state.  These errors can then be measured (without disturbing the encoded information), detected, and sometimes corrected, depending on the structure of the code and the goal of the coding.

For more information about stabilizer operators, see Sec.~\ref{Sec:AStabilizer}, and, for more information about stabilizer codes, see \cite{Nielsen2004} and \cite{Gottesman1997}.  For current purposes, there are two important facts about stabilizer codes.  First, their encoding, decoding, and error-detection algorithms can all be performed using only Clifford operations, i.e.~preparation and measurement of states in the $\ket{0}/\ket{1}$ basis and Clifford unitary operators.  Second, there exist threshold error probabilities for Clifford gates such that stabilizer error correction is useful.  In other words, if physical implementations of Clifford gates on bare (unencoded) states have errors less than some thresholds, then coding can be used to create encoded gates that have arbitrarily low error probabilities (at the cost of an increasingly large overhead in number of qubits).  If the physical gates have error probabilities above these thresholds, then quantum coding will not improve the situation.  Sets of error-type assumptions and error-magnitude bounds of this kind for each Clifford gate used are called fault-tolerance thresholds and have been the subject of a great deal of research (see Refs.~\cite{Knill2005},~\cite{Raussendorf2007}, or~\cite{Steane2003} for examples).

Fault-tolerance thresholds are given frequently with the assumption that Clifford gates have depolarizing errors. In order to further simplify the statement of the thresholds, one gate (usually the $\CX$ gate) is assumed to have greater error probability than all other implemented gates, so that the error threshold for the $\CX$ gate is used as an upper bound for the allowable error rates on all gates.  Depending on the error and architecture assumptions made, fault-tolerance thresholds vary from allowing probabilities of $\CX$ errors of order approximately $10^{-2}$ \cite{Knill2005,Raussendorf2007} in the best case to $10^{-4}$ in a more typical case \cite{Steane2003}.  The overhead in gates and qubits for fault-tolerant quantum computing near these thresholds is unmanageable; however, this overhead tends to decay very rapidly as the error probability decreases below the threshold.  Achieving consistent 1- and 2-qubit gates with quality surpassing the fault-tolerance threshold for a practical architecture is a major goal of every scalable quantum-computing initiative.  

Clifford randomized benchmarking may then be used to describe the performance of stabilizer-quantum-code-based quantum computations in two ways.  First, the average fidelity over all Clifford operators on $n$ qubits provides an estimate of how well any stabilizer coding task on $n$ qubits might be expected to perform on average.  Second, by using techniques described in Sec.~\ref{Sec:3GateBenchmark}, it is possible to isolate the depolarizing strengths of particular gates.  This information can be compared directly to the error bounds in fault-tolerance thresholds.  In addition, if the error probabilities are below thresholds, studies such as Ref.~\cite{Steane2003} can be used to estimate how much coding overhead will be required to accomplish various quantum algorithms.

\section{Approximate Twirling} \label{Sec:2AppTwirl}

In the first randomized-benchmarking experiment (\cite{Knill2008}),  Knill, et al., did not actually compose their sequences from uniformly random Clifford operators.  To generalize the idea of a single random operator occurring at each index of a sequence, I use the terminology ``step'' and refer to the composition of the (potentially many) gates in each step as the aggregate operator of the step.  Each step in this experiment consisted of the composition of two gates chosen with probabilistic weights from subsets of the Clifford operators.  The aggregate operators that could occur in a step with non-zero probability did not include all Clifford operators on one qubit, but they did include generators for the Clifford group on one qubit.  In principle, any particular step in the middle of such a sequence is surrounded by sufficiently many of these randomly chosen generating operators that the composition of these operators appears to be chosen uniformly at random from the Clifford group.  This argument is made more carefully below.  Under this assumption, the error of a step is Clifford twirled by the aggregate Clifford operators before and after it, so it should appear as a depolarizing error in the analysis.  This strategy for randomized benchmarking has significant advantages and disadvantages, especially with respect to the generation of the sequences and the analysis of the results.  I introduce these issues in this section and attempt to characterize better and mitigate some of the disadvantages in Sec.~\ref{Sec:4AppTwirl}.

Let $q_i(C)$ be the probability that the composition of $i$ sequential Clifford steps is $C$ when the aggregate operator in each step is chosen independently from the distribution $q_1(C)=q(C)$.  Then the aggregate of the first $i$ operators in a sequence can be thought of as a random variable $C_{(i)}$, and $C_{(1)}, C_{(2)}, \cdots$ is a Markov chain.  As long as the operators with $q_1(C) \neq 0$ generate the Clifford group, $q_m(C) > 0$ for all  $C \in \mathcal{C}_n$ for some $m$ sufficiently large, unless there is some periodicity to the Markov chain.  In fact, because $q_{\infty}(C)=\frac{1}{|\mathcal{C}_n|}$ is a stationary distribution for this process, basic results of the theory of Markov chains dictate that the sequence of distributions $q_1, q_2, \cdots$ is either periodic or converges to $q_{\infty}$.  Therefore, for randomized-benchmarking experiments with sufficiently long (aperiodic) sequences, one can treat steps with distant indices as if they are separated by a uniformly random Clifford operator.

The first crucial distinction between RB with exact and approximate twirls is the difference in meaning of the resulting benchmark.    The exact Clifford twirl corresponds to the case where $q(C) = \frac{1}{|\mathcal{C}_n|}$ for all $C \in \mathcal{C}_n$, i.e.~$q(C)$ is the uniform distribution.  If all errors $\hat{\Lambda}_g$ on operators in a RB experiment are depolarizing channels, then Eq.~\ref{Eq:2Main} models the mean sequence fidelities correctly, and the resulting benchmark parameter $\frac{\epsilon_s}{\alpha_n}$ describes the weighted average of the depolarizing strengths of the errors, where the weighting is also $q(C)$.  As a consequence, the results of randomized-benchmarking experiments based on different distributions $q(C)$ are not obviously comparable.

The discrepancy between benchmarks provided by experiments based on different $q(C)$ is particularly important when the assumption that all gate errors are identical is relaxed to allow gate errors that are depolarizing channels of different strengths.  I treat this scenario explicitly in Sec.~\ref{Sec:4StepDef}.  There are two notable, common scenarios for which this relaxation is a better approximation: First, if one of the gates implements the identity by doing ``nothing'', then this gate would typically have a much lower error than all of the others.  Second, if one gate is a two-qubit interaction and the others are all one-qubit interactions, the two-qubit gate would typically have a significantly higher error rate.  The error per step describing such experiments depends strongly on the probability with which each step contains the higher-error gate or gates.  In order to use the benchmark for one step distribution to rigorously (upper) bound the benchmark result of a different step distribution on the same apparatus, one must make some adversarial and penalizing assumptions about the worst case errors for gates which are represented with different weights.

A second, more complicated, distinction relates to the imperfect twirl that occurs between errors in an approximately twirled sequence.  Unitary errors can add coherently in the following sense: If an implementation of an operator has a small unitary error $U$, call the strength of the depolarization of this error $p_U$.  This strength is proportional to the infidelity of the implementation.  If two such errors act consecutively, then the aggregate error is $U^2$, and the strength of the depolarization of $U^2$ is approximately $4 p_U$.  This result is derived for a specific physical scenario in Sec.~\ref{Sec:2Systematics}. If, instead, two consecutive unitary errors are rotations of the same angle about opposite axes, then the depolarization of their product will have strength 0.  In contrast, one expects from an ideal RB experiment that all errors are depolarized.  The composition of two depolarizing channels with small strengths is always the sum of their strengths.  One might say that these errors add incoherently.  When approximate twirls are used, unitary errors do not get depolarized exactly, so the composition of errors can behave partly coherently and partly incoherently, and Eq.~\ref{Eq:2Main} is no longer necessarily a good model.  However, to the degree that two errors are separated by enough steps for the aggregate Clifford between them to be uniformly distributed, the errors will add incoherently and the depolarized error model will hold.

In Sec.~\ref{Sec:4AppDep}, I describe some attempts to quantify the difference between the reported error per step for an approximately twirled benchmark and the true error per step (based on an exact, ideal benchmark).  A large difference, especially for short, high-fidelity sequences, occurs when Pauli errors are not transformed into depolarizing errors correctly before they are measured.  This effect is simpler to quantify than the general failure of depolarization described in the previous paragraph, but it is expected to show some of the same characteristics.  Until that section, I use the same simplifying approximation made in experiments so far: for sufficiently long sequences, nearly all of the errors are separated from each other and from the measurement by operators that look sufficiently close to a uniformly random Clifford for the independent depolarizing channel behavior to apply approximately.  However, one must then be careful in these experiments with short sequences and with the errors near the end of each sequence.

\section{Assumptions and Failure Signatures} \label{Sec:2Assumptions} 

In this section, I collect some of the assumptions required for idealized randomized-benchmarking that are most likely to be violated in experiments.  I discuss briefly the signatures of failure that might be observed for each assumption and how to structure the experiment to correct for their failure.

\subsection{Approximate Twirling} \label{Sec:2AppTwirlFail}

As discussed in Secs.~\ref{Sec:2AppTwirl} and~\ref{Sec:4AppTwirl}, for experiments that rely on approximate twirling, the depolarization of each error applies only approximately even in the limit of very many different random sequences.  Two issues with approximate twirling are discussed there, but the first, regarding the meaning of error per step, is an interpretational issue and does not cause failure of the randomized-benchmarking experiment.  The second issue, which explicitly deals with the imperfect depolarization of errors, causes errors in each step to not be independent with respect to each other except when the steps are very far removed. Such a failure mode has not been rigorously investigated to date.  

In long sequences, where almost all steps are sufficiently far from each other for the depolarized model to hold, one might expect approximate twirling to behave much like exact twirling.  Therefore, failure of the behavior of Eq.~\ref{Eq:2Main} due to approximate twirling would be most likely when considering  short sequences or using significantly non-uniform gate distributions.  For this reason, it is desirable to perform sequences that are as long as possible and with gate distributions as uniform as possible.  If many sequence lengths are available, it is worthwhile to investigate how the benchmark result changes if short sequence lengths are not included in the analysis.  However, one must be careful to avoid biasing the results by discarding data selectively.

\subsection{Time Dependence} \label{Sec:2TimeDep}

Two somewhat-unrelated issues with time dependence of errors are likely to occur in any experimental apparatus.  First, gate quality is likely to vary over many different time scales as stray fields, apparatus heating, alignment drift, etc. occur.  If these drifts have short time scales (compared to the time scale of a step), then they are not of any great concern, as they will be roughly independent from step to step.  If these drifts have characteristic time scales longer than an average sequence length and are not strongly periodic, then the main concern is that the experiment is recalibrated in between experiments to correct these drifts to an acceptable level.  In addition, the order of sequences can and should be randomized so that the short sequences do no occur under a consistent and different environment than the longer sequences.  Time variation on approximately the same time scale as a sequence is the most difficult and can be considered together with the issue discussed next.

\begin{figure}[tbhp] \label{Fig:2TimeDep}
    \begin{center}
	\includegraphics[width=140mm]{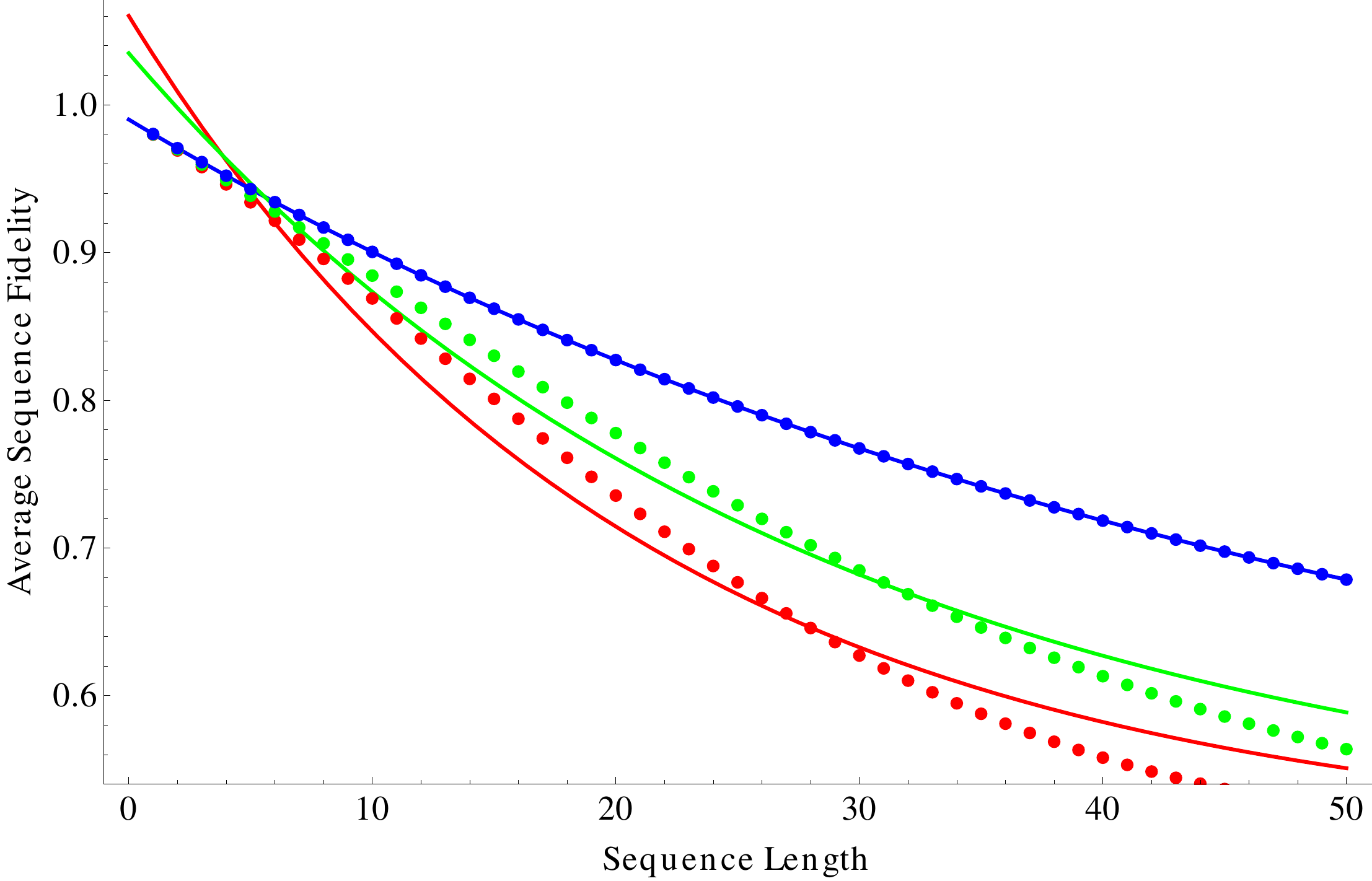}
    \end{center}
    \caption[Model of Time Dependence in Benchmark Results]{
Depicts artificial data for which the gate infidelity increases linearly as a function of step number as described by Eq.~\ref{Eq:2LinVary}.  The three data sets correspond to $\gamma = 0$ (top, blue), $\gamma = .02$ (middle, green), and $\gamma = .04$ (bottom, red).  Also shown are least-squares exponential decay fits to each data set.  The exponential models are increasingly poor descriptions of the data as $\gamma$ increases.
	}
\end{figure}

A more pernicious effect is likely to occur whenever the experiment is allowed to cool or recalibrate in between sequences.  I stress that recalibrating and cooling frequently during the course of a RB experiment is generally a good idea, as it negates long-term drifts and improves the reported benchmark.  However, errors will likely increase in magnitude as a function of the number of steps since the last recalibration or cooling process, creating a relationship between the step index in a sequence and the observed error.  Exactly such a scenario can be modeled by modifying Eq.~\ref{Eq:2Main} to
\begin{equation}\label{Eq:2TimeVary}
F_l = \alpha(n) \left (1-\frac{\epsilon_m}{\alpha(n)} \right ) \prod_{t=1}^l\left (1-2\frac{\epsilon_s(t)}{\alpha(n)} \right ) + (1-\alpha(n)).
\end{equation}
If the dominant error source has depolarizing strength that increases linearly from some initial strength, then the experiment behaves according to 
\begin{equation}\label{Eq:2LinVary}
F_l = \alpha(n) \left (1-\frac{\epsilon_m}{\alpha(n)} \right ) \prod_{t=1}^l\left (1-2\frac{(1+\gamma t)\epsilon_s}{\alpha(n)} \right ) + (1-\alpha(n)).
\end{equation}
This equation is plotted for various $\gamma$ in Fig.~\ref{Fig:2TimeDep}.  Notably, the behavior is poorly approximated by exponential decay if $\frac{1}{\gamma}$ is large compared to the longest sequence lengths, so time dependence can be detected by testing the goodness of fit of an exponential, as detailed in Sec.~\ref{Sec:2Goodness}.

In more realistic error models, time dependence of the error per step is likely to plateau as the apparatus reaches a thermal equilibrium.  If this is the case, then performing sequences of sufficient length and disregarding short sequences in the analysis recovers a consistent, valid error per step, as the short term behavior is incorporated into the $\epsilon_m$ parameter.  As in the previous section, the procedure of discarding sequence lengths during an experiment is fraught with the potential for experimenter bias and should be justified explicitly, before the final data is taken if possible.

\subsection{Gate Dependence} \label{Sec:2GateDep}

In order for the idealized derivation of the behavior of randomized sequences to proceed in Sec.~\ref{Sec:2Sequences}, I assume that the implementation $\hat{\Lambda}_U$ of every unitary $U$ can be decomposed as $\hat{\Lambda}_U = \hat{\Lambda}_e \circ \hat{U}$, where $\hat{\Lambda}_e$ is an error operator that does not depend on $U$.  This assumption is almost surely violated in practice, e.g. because a two-qubit gate does not have the same type of error as a one-qubit gate.  Refs.~\cite{Magesan2011,Magesan2011B,Magesan2012B} investigate this problem in detail, using a perturbative approach that predicts a next-order correction term that can be added to Eq.~\ref{Eq:2Main}.  A reduced version of their equation is given by Eq.~\ref{Eq:2Magesan}.  The perturbation in this approach is quantified by an operator measure of the distance between the mean error and the error of each gate; this distance is not actually likely to be small in experiments.

There are several cases for which Eq.~\ref{Eq:2Main} can be shown to hold exactly even when gate errors are non-identical.  The first and clearest case is when $\hat{\Lambda}_{U,e}$ are all depolarizing errors of different strengths, so twirling is not actually necessary for depolarization.  In this case, the average of the depolarizing errors is a depolarizing error with average strength, and the analysis proceeds in a straightforward fashion.    A second case where Eq.~\ref{Eq:2Main} holds is when the errors are all stochastic Pauli channels, as described in Sec.~\ref{Sec:AError}.  
In this case the twirl over $\overline{\mathcal{C}_n}$ is still effective even with Pauli errors in its implementation, and it correctly depolarizes the errors.  It seems reasonable to predict that gate-dependent errors are more likely to cause deviations to the expected RB behavior when they are strongly coherent, i.e.~systematic unitary errors.  This behavior is demonstrated in numerical simulations in Ref.~\cite{Laforest2008}.

Gate error dependence has not been seen clearly so far in experiments, but it is unknown whether this is because the RB procedure is more robust to this dependence than proofs so far have indicated or whether noisy experiments and coincidence have prevented it from appearing.  This problem is in need of more research, both through mathematical attempts to clarify when Eq.~\ref{Eq:2Main} holds exactly and through simulations that model cases where it might fail.  In any case, the improved RB procedure described in Chap.~\ref{Chap:3} should be more robust to violations of this assumption for reasons that are described in Sec.~\ref{Sec:3ExactSimple}.  

\subsection{Environmental Memory and Trace Non-preserving Errors} \label{Sec:2Memory}

The depolarization of errors by twirling relies on the assumption that the errors are trace-preserving quantum processes on $n$ qubits.  While any quantum process is expected to be trace preserving on a full Hilbert space, there are several ways this assumption might be violated on the $n$-qubit subspace.  In ion traps, for example, a state can be driven off-resonance from the energy level represented by the $\ket{0}$ state to an excited state that is not the $\ket{1}$ state.  On a larger Hilbert space, this process is trace preserving, but the twirling operator is only designed to depolarize errors on the $n$ qubit subspace, so no guarantees can be made about the depolarization of such a process.

A related assumption requires that errors be memoryless.  Errors that temporarily remove the state from the qubit subspace might appear to have a memory: a gate applied to a system not in the qubit subspace will appear to perform incorrectly, so the interpretation of the gate implementation is conditioned on previous errors.  If the extra-qubit states have a finite lifetime, for example if they decay back into the qubit states with high probability after some characteristic time, then this lifetime is roughly the same as the perceived memory lifetime.  If the memory lifetime is less than the time required to perform a step of the randomized-benchmarking experiment, then the errors are well described by decoherence processes and do not cause a problem for the analysis.  At the time of this thesis, I am aware of no research into the effects of general trace non-preserving errors on randomized-benchmarking experiments.

The randomized inversion step originally used in Ref.~\cite{Knill2008} and described in Sec.~\ref{Sec:2MeasRand} has a convenient property for detecting extra-qubit states.  Because a Pauli randomization is performed shortly before the measurement, the ideal sequence is equally likely to return the qubit to the $\ket{0}$ or $\ket{1}$ state, independent of the sequence of Clifford gates in the randomized steps.  If the measurement is calibrated so that it has an equal probability of correctly identifying the $\ket{0}$ state as $\ket{0}$ and $\ket{1}$ as $\ket{1}$, then the results should show that equal length sequences have the same average fidelity regardless of whether they end in the $\ket{0}$ or $\ket{1}$ state.  However, an extra-qubit state is preferentially detected as one of the two states in many measurement schemes, and this provides a signature for the error.  

For example, in the ion-trapping experiment described in Sec.~\ref{Sec:2ExpBasics}, the laser used for measurement does not excite a transition for an electron in any relevant extra-qubit state.  Because the absence of scattered photons from this beam is taken as an indication of the $\ket{1}$ state, extra-qubit states are measured as if they were in the $\ket{1}$ state.  In this case, if the measurement error is found to be much higher for sequences that end with the qubit ideally in the $\ket{0}$ state compared to those that end with it in the $\ket{1}$ state, this might be an indication that the system is being excited to an extra-qubit state.

\section{Experiment Design} \label{Sec:2ExpDesign} 

In this section, I describe some of the experimental design choices involved in performing a randomized benchmark following the prototype of Knill, et al.~\cite{Knill2008}.  The experiment consists of performing many individual instances (or shots) of many randomly chosen Clifford sequences and analyzing the fidelities of the sequences as a function of their length.  The primary concerns of the experiment design discussed here are, first, to decide on a probability distribution to describe which gates to perform in a step and, second, to distribute the available experimental time such that parameters extracted from the fit of Eq.~\ref{Eq:2Main} to the experimental data have good precision and accuracy.

\subsection{Gate Choices} \label{Sec:2GateChoice}

The first consideration in deciding which Clifford gates to implement in a randomized-benchmarking experiment is to catalog the interactions that are experimentally controllable.  From these interactions, it is necessary to choose at least a generating set of Clifford gates (I relax this requirement in Sec.~\ref{Sec:4SmallDesigns}).  Many sets of generating gates are known, but the most canonical is the set that consists of one-qubit $H$ (Hadamard) and $S$ ($Z \left (\frac{\pi}{2} \right )$ rotation) gates on each qubit and $\CX$ gates between all pairs of qubits.  See App.~\ref{App:C} for definitions of these, and other, Clifford gates.  The one-qubit gates in this set generate the group of all tensor products of one-qubit Clifford operators on all qubits.  This same condition can be satisfied by replacing $H$ with the $X \left (\frac{\pi}{2} \right )$ gate.  Given that the one-qubit gates generate the tensor products of one-qubit Clifford gates, the $\CX$ gates between pairs of qubits are sufficient to complete the generating set.
The $\CX$ gates in this part of the construction may be replaced by $\CZ$ gates or M{\o}lmer-S{\o}rensen gates, among others; the construction is flexible.  If it is unclear whether a gate set generates the Clifford gates, one can use the ideas presented in Sec.~\ref{Sec:ADecomp} to check whether each canonical generating gate can be decomposed into gates from the generating set in question.

Once a set of gates has been identified that generates the Clifford group, the next task is to decide the probabilities with which each will be implemented in a given step.  For reference, the standard one-qubit experiment of Ref.~\cite{Knill2008} defines each step to be constructed from a Pauli gate ($\pm X,\pm Y, \pm Z,$ or $ \pm I$) chosen uniformly at random followed by a ``computational'' gate $X \left (\pm \frac{\pi}{2} \right ),Y \left (\pm \frac{\pi}{2} \right )$ chosen uniformly at random.  This pairing of Pauli and Clifford gates is intended to separate the Pauli twirl aspect of the randomization from the Clifford twirl; it is not necessary, but the separation allows the two twirls to be modified separately if desired and allows one to pinpoint where the Pauli twirl occurs.  As indicated in Sec.~\ref{Sec:2GateDep}, there are two concerns to address when optimizing the probability distribution of gates defining a step: the desired interpretation of the resulting benchmark and the degree to which the approximate twirl approximates the exact Clifford twirl.  

The step distributions in Ref.~\cite{Knill2008} and subsequent papers (e.g. Refs.~\cite{Biercuk2009,Olmschenk2010,Brown2011}) were partially defined for interpretation reasons.  Because $Z$ rotations are not implemented in the same fundamental way as $X$ and $Y$ rotations in those experiments, the gate $Z(\frac{\pi}{2})$ was not allowed in the computational part of the step.  This gate would have been relatively easy to perform, but the authors wished to be able to interpret the error per computational step in a uniform way, i.e.~without worrying about it being an average over fundamentally different gates.  For experiments that do not share this $Z$ rotation implementation, this gate distribution is somewhat unnatural.

Define $p_j(g)$ to be the probability that the composition of the gates in $j$ consecutive steps is the operator $g$, so that $p_1(g) = p(g)$.  In order to have the chosen distribution over gates depolarize errors as effectively as possible, it is desirable for $p_j(g) - \frac{1}{|\mathcal{C}|}$ to be as small as possible for each $g \in \mathcal{C}$.  If the generating gate set is small, then the sum of these differences is large for small values of $j$.  To the extent that the difference is ``small'', then errors that are separated by $j$ steps are decoupled from each other so that they may be approximated as independent depolarizing channels.  Therefore, it is desirable to attempt to minimize $p_j(g) - \frac{1}{|\mathcal{C}|}$ for each $j$ and $g$.  In experiments this optimization has so far not been investigated.  As a simple tool, consider the total variation distance $v_j = \frac{1}{2}\sum_{g \in \mathcal{C}} |p_j(g) - \frac{1}{|\mathcal{C}|}|$.  The theory of Markov chains indicates that $v_j \rightarrow 0$ exponentially fast in the asymptotic limit as long as the sequence $p_j(g)$ is aperiodic and the gate set generates the Clifford group.  A coarse goal of such an optimization might be to maximize the exponential decay rate of $v_j$.

\begin{figure} [tbh]
    \begin{center}
	\includegraphics[width=140mm]{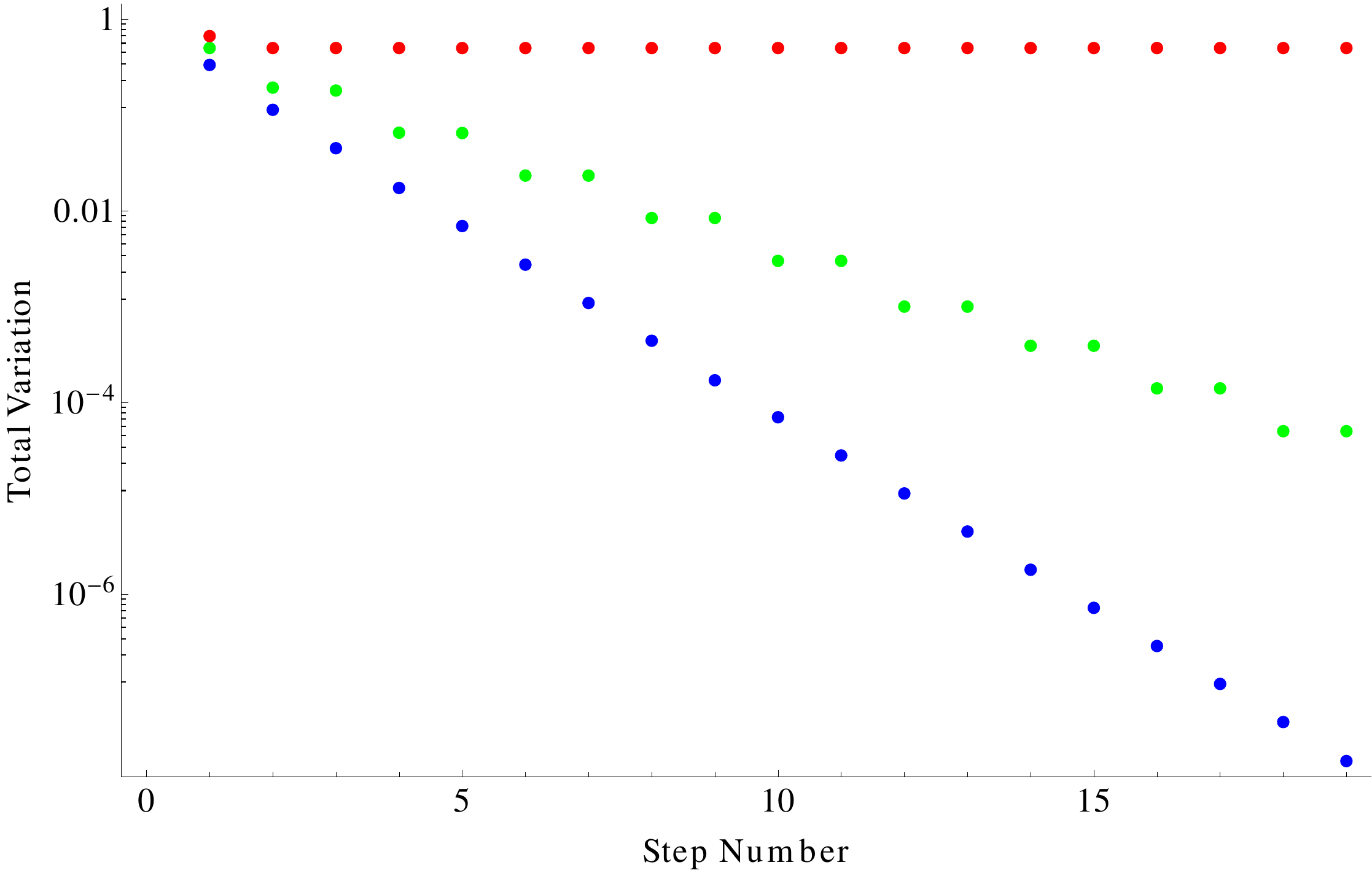}
    \end{center}
    \caption[Total Variation Decay]{
Log plot of the total variation decay of the composition of steps as a function of number of steps for several gate probability distributions for one qubit.  The top data sequence is for the distribution of gates used by Ref.~\cite{Knill2008}.  This distribution does not decay toward $0$ for reasons noted in the text.  The middle data sequence is for the distribution used for the one-qubit benchmarks in Sec.~\ref{Sec:3ExpDesign}.  The lowest sequence is for a computational-gate distribution in which $\frac{\pi}{2}$ rotations about the $X,Y,$ and $Z$ axes are performed with probability $0.2$ and the identity gate is performed with probability $0.4$.      \label{Fig:2TotalVariation} }
\end{figure}

For example, consider the distribution used to define a step in Ref.~\cite{Knill2008}.  The total variation as a function of $j$ for this distribution is plotted in Fig.~\ref{Fig:2TotalVariation}.  It is clear that total variation does not decay to zero for this gate distribution.  In fact, the Markov chain associated with this distribution is periodic.  This can be seen by considering the computational operators available after step numbers of even and odd parities: for example $p_{\mbox{odd}}(I) = 0$, while $p_{\mbox{even}}(I) \neq 0$. While this periodicity has the potential to cause problems for the randomized benchmark, it only does so for particular error models, which are somewhat pathological.  In any case, Fig.~\ref{Fig:2TotalVariation} also depicts the total variation decay for two other gate distributions.  These distributions do not suffer from the periodicity concern and demonstrate the sort of differences in total variation decay rate that can be expected.

\subsection{Number and Lengths of Sequences} \label{Sec:2Lengths}

Consider an experiment that has a total allowable time interval of $T_t$ and for which one average step requires time $t_s$.  Suppose also that the gates in the experiment are expected to require recalibration after time $T_c$ to maintain acceptable time-independence.  Finally, assume that preliminary experiments have roughly estimated that the average step has an error with depolarizing strength $p_{d,\mbox{est}}$.  Given these estimates, one must decide the following parameters: the lengths of the sequences, the number of random sequences of each length, and the number of shots, or individual experiments, that will be performed for each sequence.

Most of the sequence lengths should be long enough that errors are observable with high probability.  This means that most of the sequences should have lengths of order approaching, but not significantly exceeding, $l_{\mbox{max}} \equiv \frac{1}{p_{d,\mbox{est}}}$.  In order to observe short-sequence effects, it is usually desirable to perform some sequences of very short length, even if the data for these sequences is not used in the actual analysis of the experiment (see Sec.~\ref{Sec:2GateDep}).  Therefore, sequences of lengths from $1-l_{\mbox{max}}$ should be performed.  

No experiment yet has attempted to fully optimize the analysis by picking sequences lengths from that interval to minimize the parameter uncertainties from the curve fitting given the time constraint. Typical sequence length choices have either been linearly distributed as $l_i = ki$, for simplicity, or Fibonacci distributed as $l_i = l_{i-2} + l_{i-1}$.  This second choice (which has roughly exponential spacing) emphasizes short sequences, which might display  effects from imperfect depolarization, and de-emphasizes long sequences, which take much longer and might have signals overwhelmed by noise.  The choice of how many different lengths to use has also not been optimized.  A loose criterion indicates that many more data points are needed than there are parameters in the fit equation, but there is no obvious bound to the maximum number of different lengths that should be used.

After the sequence lengths have been chosen, the next decision to be made is how many different sequences of each length should be generated.  This is related to the question of how many shots of each experiment should be performed, so I treat them at the same time.  Generally, it is desirable to perform as many different sequences of each length as possible, as this reduces the probability that atypical sequences will be over- or under-represented in the mean. The issue of analyzing how many different sequences are required for the mean sequence fidelity to converge reliably to its asymptotic value is difficult and is treated in depth in Refs.~\cite{Magesan2012B} and \cite{Emerson2005B}.  This is another area in which the analysis of RB for experiments seems to be better behaved than expected.

In practice, there are two reasons experimenters have decided to restrict the number of sequences.  First, in many experiments an appreciable amount of time must be allocated to programming the apparatus to perform a new sequence.  At the point where more time is spent re-programming than is spent on actual performance of experiments, the number of different sequences should be reduced and the number of shots per sequence increased.  Second, it has often been interesting to look at the distribution of fidelities for different sequences (see \cite{Knill2008,Gaebler2012}, for example).  In order to learn anything about an individual sequence, it is necessary to perform many shots of the experiment to reduce the standard error of the estimate of its fidelity.  Most experiments so far have used an equal number of sequences $n_l$ at each length; this can simplify the analysis, but no other justification for the choice is evident.  In a similar way, experiments have used the same number of shots $n_e$ for each sequence of each length, and this choice has also not been well justified.  A reasonable choice might be to let $n_e \approx T_c/(l*t_s)$, so that calibration need only be performed after all of the shots of each sequence.

It is desirable to perform long sequences so that, first, any initial time-dependent ramp-up effects in error behavior have minimal influence and, second, most of the steps are sufficiently well-separated for the assumption of independence to be approximately valid.  In contrast, longer sequences clearly take more time, and they also might not contribute strongly to the statistical analysis.  If the expected standard deviation of the mean of the average fidelity $F_{l}$ for sequences of length $l$ is $s_{l,n_e,n_l}$ after $n_e$ shots each of $n_l$ different sequences, then a rough bound on useful sequence lengths is established by 
\begin{equation} \label{Eq:2SeqLength}
s_{l,n_e,n_l} < F_{l} - (1-\alpha_n) \approx \alpha_n(1-\frac{\epsilon_{est}}{\alpha_n})^l,
\end{equation}
where $n$ is the number of qubits, $\alpha_n$ equals $\frac{2^n-1}{2^n}$ as in Sec.~\ref{Sec:2MainEquation}, and SPAM errors are disregarded for simplicity.  

The noise $s_{l,n_e,n_l}$ can come from statistical noise due to a finite number of sequences and shots.  Essentially, this equation describes the point where the noise overwhelms the small signal above the $1-\alpha_n$ background.  This cutoff does not need to be estimated very precisely; there is some value in intentionally trying sequences whose length exceeds this bound so that the asymptotic fidelity limit of $1-\alpha_n$ is clearly observed.  However, solving the equation for $l$ gives a better idea of how long the longest sequences should be.  If the expected sequence fidelity for a short length $l$ is expected to be within the noise bound $s_{l,n_e,n_l}$ of $1$, then a similar argument can be made that this length is not useful; this criterion is less common in practice because the SPAM error usually separates the fidelities sufficiently from $1$.  These bounds can be used to further refine the choices of sequence lengths.

If $n_e$ shots are to be performed for each of $n_l$ sequences for each of the lengths $l_i$, then the total time required by the experiment, neglecting the programming, recalibration, measurement, and preparation times, is $T_t' = n_e n_l \sum_i l_i$.  Given a fixed constraint $T_t$ on the available time, it is desirable to saturate this time by maximizing $n_l$.  The parameter choices for two actual experiments are shown in Tables~\ref{Tab:2Exp1} and~\ref{Tab:3Exp2}.

In principle, a formal optimization could be performed by minimizing the Fisher information (expected variance of the estimate) of the model parameters with respect to some estimates of the physical parameters of the system.  Adding experimental constraints into this procedure would make it complicated, but such an analysis might point toward better general strategies or provide an idea of the how much there is to gain by further study into this issue.

\subsection{Measurement Step} \label{Sec:2MeasChoice}

As discussed in Secs.~\ref{Sec:2Measurement} and~\ref{Sec:AGeneration}, several different inversion techniques have been devised to return the state at the end of a sequence to a state with a conveniently measurable Pauli operator.  While the most straightforward approach is to find the Clifford operator that exactly inverts all of the previous steps of the sequence, this approach is undesirable for sequences utilizing the approximate twirl.  If the gate distribution is limited so that, for example, some Clifford operators are only constructible after $k$ steps, then, for sequences of length less than $k$, the measurement operators are not randomly distributed and differ on average from the average measurement operator for larger sequences.  If the possible inversion operators vary between sequence lengths, then $\epsilon_m$ is not likely to be a constant as desired.  

A common situation of concern is as follows: Suppose there is only one entangling gate $G$ in the generating gate set and $G$ has significantly worse fidelity than any other gate in the generating set.  A short sequence containing a small number $m < n^2$ of $G$ gates is likely to require $m$ $G$ gates to invert because no other gate can undo the entanglement.  Thus a short sequence with $m$ random $G$ gates before the inversion step is likely to display a factor of four higher error per step compared with a sequence of the same length but with $\frac{m}{2}$ $G$ gates instead of the expected factor of two.  This can be thought of as a violation of the assumption that the SPAM error is constant, and it can cause an overestimate of the infidelity by a factor of two.

One solution to the problem described in the previous paragraph might be to disregard sequences with length shorter than some small multiple of $k$.  However, because sequences of short length are often interesting for analysis, another technique was introduced in Ref.~\cite{Knill2008} and has been used with modifications in many papers since.  Instead of exactly inverting the Clifford operator or inverting it modulo a Pauli operator, the authors only return the system to a state which has (at least) one stabilizing operator made up of a tensor product of Pauli $Z$ and $I$ gates on each qubit.  In that experiment, only one qubit was involved, so the process looked trivial, but it can be extended directly to more qubits.  

The following describes an algorithm that finds an inversion operator that is composed of a tensor product of one-qubit operators (extending the one-qubit procedure in the previous paragraph): Using the stabilizer matrix formalism introduced in Sec.~\ref{Sec:AStabMat}, choose a random stabilizer of the state prior to the inversion step.  For each qubit index $i$ for which the tensored Pauli part of this stabilizer is not the identity, perform experimentally a one-qubit Clifford gate that transforms the Pauli operator on that qubit to $Z$.  At this point, the state is stabilized by a Pauli operator with only tensor parts of $Z$ and $I$.  Then measure the $Z_i$ operator on each qubit whose tensor part in the stabilizer is $Z$ to find measurement result $m_i$.  Although the individual binary measurement results $m_i$ are not necessarily deterministic, the sum $(\sum_i m_i \mod{2})$ is deterministically equal to the measurement result that would be obtained by measuring this stabilizer directly.  Because the measurements are potentially destructive (each $Z_i$ is not individually a stabilizing operator), this is usually the only bit of information that measurement reveals.

The incomplete inversion of the state described in the previous paragraph represents a single binary measurement of the system, instead of the $n$ deterministic measurements possible after exact inversion.  If all errors are assumed to be depolarizing channels, then this single binary measurement of the state detects the ideal result with probability 
\begin{equation}\label{Eq:2MainApp}
F_l = \frac{1}{2} + \frac{1}{2} \left (1-\frac{1}{\alpha_n}\epsilon_m \right ) \left  (1-\frac{1}{\alpha_n}\epsilon_s \right )^l,
\end{equation}
using the same notation as Eq.~\ref{Eq:2Main}.  The two equations differ when $n\neq 1$.
  
Refs.~\cite{Knill2008,Gaebler2012L} describe several related options for inexact inversion steps that might be chosen.  The goal of these alternative techniques is to decouple to some degree the choice of inverting unitary from the random gates that precede it. To the degree that the inversion step is independent of the sequence, the SPAM error can appear independent of the errors of the random steps of the sequence, avoiding concerns about certain adversarial error models.  On the other hand, it is important to note that when the operation in the inversion step has a different marginal distribution than the operations in other steps of the sequence less can be predicted about the relationship between $\epsilon_s$ and $\epsilon_m$.

\section{Analysis} \label{Sec:2Analysis} 

The goal of the statistical analysis of randomized-benchmarking experiments is to extract the benchmark parameters $\epsilon_m$ and $\epsilon_s$ from the raw data representing measurement results for each experimental shot of each sequence.  There are several layers to this analysis: First, I discuss the expected distribution of the shots for each sequence.  Second, I discuss the expected distribution of sequences for a given length.  Third, I discuss how to fit Eq.~\ref{Eq:2Main} or Eq.~\ref{Eq:2MainApp} to the data.  Fourth, I discuss the use of resampling techniques to estimate confidence regions for the inferred parameters.  Finally, I discuss the use of goodness-of-fit tests to express confidence that the benchmark parameters are valid. 

\subsection{Sequence-Fidelity Statistics} \label{Sec:2FidelStats}

Assume that $\rho_{i,l}$ describes the mixed state present immediately before the measurement at the end of sequence $i$ of length $l$.  Assume that any error in the measurement has already been incorporated into the information about the state.  I treat explicitly the case where only a single binary measurement is made that reveals the state to be either in the ``1''($-1$)-eigenvalue eigenspace of the measurement operator or in its orthogonal space, corresponding to the ``0''($+1$) eigenvalue.  Let $\Pi_1$ be the projector onto the first subspace and $\Pi_0$ be the projector onto the second.  The probability of measuring the state with the ``1'' result is then $p_{1,i,l} =\tr{\Pi_1 \rho_{i,l}}$ and the probability of measuring the ``0'' result is $p_{0,i,l}=\tr{\Pi_0 \rho_{i,l}}$.  The sum of these projectors is the projector onto the whole space (since the measurement error is assumed to preserve trace), and $\tr{\rho_{i,l}} = 1$ is assumed, so $p_{0,i,l} + p_{1,i,l} = 1$.  Therefore, the random variable describing the result of a single experiment with correct result ``0'' follows a Bernoulli distribution with a probability $p_{0,i,l}$ of success.  

The sum of the binary results of $n_e$ such experiments then follows a binomial distribution with success probability $p_{0,i,l}$.  The mean value of this distribution is $n_e p_{0,i,l}$, so dividing the number of successful measurements of the sequence $n_{i,l}$ by $n_e$ gives an estimator $\hat{p}_{0,i,l}$ for $p_{0,i,l}$.  The variance of such a distribution is $n_e p_{0,i,l}(1-p_{0,i,l})$, so an unbiased estimate of the variance of $\hat{p}_{0,i,l}$ is $\frac{\hat{p}_{0,i,l}(1-\hat{p}_{0,i,l})} {n_e - 1}$.  If $n_e$ is large and the mean is far from the boundaries at $1$ and $0$, then this distribution approximates a normal distribution and Gaussian statistics can be applied; however, these conditions might not be satisfied, typically because $p_{0,i,l}$ is too close to the upper bound of $1$.

In contrast, little can be said with certainty about the true distribution of the $p_{0,i,l}$ over different sequences $i$ unless the gate-independent-error assumption holds exactly.  In this case, $p_{0,i,l}$ is independent of the sequence $i$.  However, if this assumption does not hold and, for example, all error is concentrated in one gate $G$, then the distribution of the $p_{0,i,l}$ is closely related to the distribution in number of $G$ gates in the different sequences.  Another example of use concerns the case for which all gates have equal, depolarizing error but the number of gates per step varies.  Then the distribution of $p_{0,i,l}$ is directly related to the distribution of total gate number per step.  Other error models lead to even more complicated distributions.  

Due to the central limit theorem, the mean $F_l \equiv \frac{1}{n_l}\sum_i p_{0,i,l}$ of the success probabilities of the sequences performed converges to the true mean over all sequences as $n_l$ grows large.  Similarly, the sample variance $\sigma_{l,s}^2$ of these probabilities converges to the true variance $\sigma_{l}^2$, and $\frac{\sigma_{l,s}^2}{n_l - 1}$ becomes a good estimate of the variance of the mean.  Making $n_l$ as large as experimental constraints allow improves this approximation.  Sec.~\ref{Sec:2Resampling} describes a useful approach to the analysis in the case where $n_l$ is not large.  Figs.~\ref{Fig:2SeqDistKen} and \ref{Fig:3SeqDistGaeb} depict the actual sequence fidelity distributions for two experiments.

\subsection{Fitting the Benchmark Parameters} \label{Sec:2Fitting}

Given the sample mean $F_l$ and sample variance $\sigma_l^2$ for the sequence fidelities at each length $l$, the next goal is to find parameters $\epsilon_m$ and $\epsilon_s$ such that Eq.~\ref{Eq:2Main} best fits the $F_l$.  Because the data points should not have equal weight (especially in cases where $n_l$ or $n_e$ is allowed to vary), I choose to use a weighted, non-linear least-squares fit to the data (for example the \textbf{leasqr()} function provided by OCTAVE \cite{leasqr}).    The weight input into the function for each length $l$ is the inverse square root of the sample variance $\sigma_l^2$.  It is possible for constraints to be placed on the parameters $\epsilon_m$ and $\epsilon_s$, to reflect the maximum and minimum possible depolarizing strengths, for example.  However, I find that unexpected parameter values, especially $\epsilon_m \geq 1$ and $\epsilon_s \leq 0$, occur when the gate error model violates some of the assumptions of Sec.~\ref{Sec:2GateError} and can be useful diagnostics. In addition, constrained fitting procedures can be significantly biased.  For these reasons, I do not utilize the constraint options.

Such a fitting routine can also be used to test some of the predicted behaviors given  in Eq.~\ref{Eq:2Magesan}, Ref.~\cite{Magesan2011B} and Eq.~\ref{Eq:2RB3}.  Potential degeneracies between fitting parameters are noticeable in the parameter covariance matrix provided by \textbf{leasqr()}, where the covariances can be quite large for parameters allowing degenerate fits.  I describe $\chi^2$-goodness-of-fit tests that can used to more-rigorously justify the choice of models in Sec.~\ref{Sec:2Goodness}.

\subsection{Semi-Parametric Bootstrap Resampling} \label{Sec:2Resampling}

Although the covariance matrix returned by the \textbf{leasqr()} procedure provides an estimate of the errors on the estimates of $\epsilon_m$ and $\epsilon_s$, these estimates are partly based on an assumption of normality of the distribution of sequence fidelities at each length, which is not expected to be generally valid for experiments with small $n_l$. In order to estimate the variances of the parameter statistics, establish confidence intervals, and investigate statistical bias in the analysis procedure, I use the technique of bootstrap resampling.  For statistical justifications of bootstrapping and resampling and a thorough analysis of the procedures, see Ref.~\cite{Efron1994}.  In this section I summarize a particular variation of the technique that I call semi-parametric bootstrap resampling and emphasize some of the features it can reveal.

The basic insight of non-parametric bootstrapping is that, in the absence of a priori knowledge about the true distribution $p(X)$ of some random variable $X$ (in this case the sequence fidelities), the best available estimate is to assume that $p(X)$ is exactly the same as the distribution observed in the experiment.  That is, if $p_{exp}(X=x)$ is the frequency with which $x$ was observed in a limited experiment, one estimates that $p(X=x) = p_{exp}(X=x)$.  Given this assumption, one can estimate the true value of any statistic of the random variable by applying it to the distribution $p_{exp}(X)$.  For example, in the absence of other knowledge, the true mean of $X$ is estimated by the sample mean, and the true variance of $X$ is estimated by the sample variance.

Parametric bootstrapping uses the same idea but assumes some prior knowledge of the distribution of $X$.  For example, if $X$ is assumed to be a binomial random variable and experimental results find it to have a sample mean of $\mu$, then parametric bootstrapping would assume that $X$ is distributed according to a binomial distribution with mean parameter $\mu$.  Note that one could alternatively estimate the true value of $\mu$ by using the sample variance of the experiment and the relationship between the variance and mean for a binomial distribution, and the results would likely differ.

In order to use all of the information available in a randomized-benchmarking experiment, I first assume that the true distribution of sequence fidelities is the same as the observed distribution.  Since each sequence fidelity is calculated through $n_e$ trials of a binary process, I assume that the variable representing the number of correct measurement results observed follows a binomial distribution with mean $n_e p_{0,i,l}$, where $n_e p_{0,i,l}$ is the number of correct measurements observed over $n_e$ shots of sequence $i$ of length $l$.  Because this technique combines parametric and non-parametric bootstraps, I call it semi-parametric bootstrapping.

Given this assumed form of the true distribution of sequence results, the technique of resampling can be employed.  In resampling, one creates data sets that mimic the observed experimental data by drawing random variables according to an assumed distribution.  By creating many such resampled data sets and performing the same analysis used on the actual data set, one can predict the statistics that might be observed if the actual experiment were repeated many more times.  In the case of randomized benchmarking, I create and analyze $N_{\mbox{rs}}$ resampled data sets (indexed by $j$) as follows: 
\begin{itemize}
\item
First, choose $n_l$ sequences $\{S_{i,l,j}\}_j$ (with replacement) from the $n_l$ actual sequences of length $l$.  This means that a sequence may be repeated multiple times, but the expected frequency with which any one sequence shows up is the same as in the actual experiment.  This resampling uses the non-parametric bootstrap. 
\item
Now, for each sequence $S_{i,l,j}$, (parametrically) sample the number $M_{i,l,j}$ of correct measurement results according to the binomial distribution with mean $n_e p_{0,i,l}$.  This creates a data set of the same size and type as the actual data set.  
\item
Perform the fitting procedure on the resampled data $M_{i,l,j}$ to extract $\epsilon_{m,j}$ and $\epsilon_{s,j}$ (or the parameters in any other model for the data).

\end{itemize}

At this point, the available experimental results have been used to simulate what might happen if the experiment was repeated $N_{\mbox{rs}}$ times.  This information can be used to estimate the variances of the estimates of the benchmark parameters by observing the variances of the resampled parameters.  One can also estimate confidence regions for the benchmark parameters by finding regions of parameter space that contain the appropriate fraction of the resampled results.  Such a confidence region estimate is depicted later in Fig.~\ref{Fig:2Resamples}.

Plots like Fig.~\ref{Fig:2Resamples} also provide a clear picture of the relationship between the fitting parameters.  In this case, it is clear from the tilt of the Gaussian spread of the fit parameters that there is a non-zero covariance between $\epsilon_m$ and $\epsilon_s$.  In more extreme examples, such as when the fitting parameters are under-constrained, such a plot can look quite non-Gaussian, encouraging the use of confidence intervals instead of standard errors to qualify the reported parameters.  In less extreme cases, it is still desirable to note large covariances between parameters, as typical error reporting carries the implicit assumption of independence.

Finally, an estimate of the bias of the fitting procedure can be obtained. Suppose that an experiment satisfies all of the assumptions necessary for Eq.~\ref{Eq:2Main} to hold exactly with parameters $\epsilon_m'$ and $\epsilon_s'$.  After performing the experiment with a limited number of sequences and using the fitting routine, the experimental estimates $\epsilon_m$ and $\epsilon_s$ are extracted.  When the resampling procedure above is performed, suppose that the mean of the resampled parameters are $\overline{\epsilon_{m,j}}$ and $\overline{\epsilon_{s,j}}$.  If $b_m = \overline{\epsilon_{m,j}} - \epsilon_m$ or $b_m = \overline{\epsilon_{s,j}} - \epsilon_s$ is non-zero, then the fitting procedure applied to a simulated experiment with simulated true parameters $\epsilon_m,\epsilon_s$ is finding biased estimates of those parameters with biases $b_m, b_s$, respectively.  The experimenter is forced to assume that the same bias has been applied in the estimation of $\epsilon_s'$ by $\epsilon_s$, and so the bias-corrected estimate for $\epsilon_s'$ is $\epsilon_s-b_s$ and likewise for $\epsilon_m'$.  In a Gaussian setting, the bias is only considered significant to the extent that it is large compared to the standard error $e_s$ of the estimate of the parameter (Ref.~\cite{Efron1994} suggests the significance criterion $b > .25 e_s$).  If bias correction is necessary, it should be explicitly stated and explained.

\subsection{Goodness of Fits} \label{Sec:2Goodness}

Along with the presentation of the fit parameters, it is important to report the degree to which the data agrees with the chosen model.  I choose to use $\chi^2$-goodness-of-fit-tests to evaluate the chosen models.  The $\chi^2$ test essentially looks at the squares of each residual of the fitting (i.e.~the difference $r_l = F_{l} - F_{l,\mbox{mod}}$ between the observed values $F_{l}$ and the values $F_{l,\mbox{mod}}$ of the model given the fit parameters) and compares them to the variance expected from the model. 
\begin{equation} \label{Eq:chisq}
X_{\chi} = \sum_l \frac{(r_l)^2}{\sigma_l^2}
\end{equation}
is the statistic of interest.  If a parameter-less model describes the data perfectly, then each term in the sum has expected value $1$, and the expected value of the statistic is equal to the number of different sequence lengths $l$ used in the experiment.  For each parameter of a model that is fit using the data, one term of the sum can be reduced exactly to zero (assuming the model is good), so the expected value of the sum for a perfect model is the number of data points minus the number of parameters in the fit; this difference is often called the number of degrees of freedom.  As stated in previous sections, the sample variance $s_l^2$ for the fidelities at length $l$ is used as a estimator for the true variance $\sigma_l^2$, since nothing is known in general about the distribution of sequence fidelities.  Notably, the applicability of the $\chi^2$ distribution relies on the assumption that the residuals should be normally distributed, which is not always valid in benchmarking analysis when small numbers of sequences are used; it might be worthwhile to apply a more sophisticated statistic that does not require this assumption.

The least-squares fitting procedure is intended to minimize the $X_{\chi}$ statistic given above, so the parameters found in the fitting procedure specify the model giving the best value of $X_{\chi}$.  A useful interpretation of this value is given by quoting the probability that $X_{\chi}$ would be greater than or equal to the observed value if the model were exactly correct.  Instead of reporting this probability directly, typical practice involves reporting whether it is above or  below a significance threshold; I use $.95$.  In this case, being below the threshold is seen as an indication that one need not reject the null hypothesis that the model is valid.  If the value is above the threshold, then it is significant in that it indicates that the model is a worse fit to the data than expected.  Tables of threshold values for a $\chi^2$ statistic as a function of the degrees of freedom are widely available.

There is much statistical debate over the proper use of goodness-of-fit tests.  
One common concern comes from experimenter bias of the sort in which the experimenter decides which tests and models to use after the data has been observed.  This is a very difficult bias to avoid in practice; a good approach is to explicitly decide on a set of models, goodness-of-fit-tests, and significance thresholds to test in advance.  In particular, this problem arises in RB in two ways: First, there are many corrections to Eq.~\ref{Eq:2Main} that might be used (see Eq.~\ref{Eq:2RB3} or \ref{Eq:2Magesan} for example).  If one searches after the experiment for the model that reveals the best fit, one is in danger of reporting results with incorrect significance.  As a toy example, imagine testing many different models at the $.95$ significance level; it is not improbable that one of the models appears valid by a statistical coincidence even if they are all actually insignificant at that level.  In order to avoid this danger, I attempt to isolate a small number of models before the analysis that are expected to be appropriate.  If many models are used, an explicit note should be made about the interpretation of their significance.

A second, related problem arises with the choice of data points to use in analysis.  Experimental practice sometimes includes discarding outliers or repeating experiments if unexpected results are observed.  Both of these practices have the potential to introduce significant bias.  However, I show in Secs.~\ref{Sec:2Assumptions} and \ref{Sec:4AppDep} that sequence fidelities for short sequences are not necessarily well described by the benchmarking model, so discarding these points in analysis can be justified.  The most satisfactory resolution to this conflict is again to establish before the experiment the range of sequence lengths whose fidelities will be included in the fitting procedure.  In the other dangerous scenarios, a stringent threshold for how far out a data point or experiment must lie before it is considered an outlier should ideally also be established in advance of the experiment.  Finally, it is always desirable to repeat the experiment after the final analysis has already been applied once to see if results are consistent.

\section{One-Qubit Experiment Example} \label{Sec:2Exp1} 

In this section I summarize an experiment published as Ref.~\cite{Brown2011} with whose planning and analysis I assisted.  This experiment demonstrates many of the experiment-design and analysis choices described in Sec.~\ref{Sec:2ExpDesign} and Sec.~\ref{Sec:2Analysis}, respectively.  The goal of the experiment was to demonstrate quantum control of a single qubit with an error per gate better than the commonly cited fault-tolerance threshold of $10^{-4}$.  The results achieved this goal, with an estimated error per step of $\epsilon_s = 2.0 * 10^{-5}$, where each step consisted of two gates.

\subsection{Experiment Basics} \label{Sec:2ExpBasics}

The physical system constituting the qubit in our experiment consists of a trapped and laser-cooled ${}^9\mbox{Be}^{+}$ ion whose single valence electron lies in either the $\ket{F=2,m_f=0} \equiv \ket{0}$ or $\ket{F=1,m_f=0} \equiv \ket{1}$ hyperfine states of the $2s {}^2S_{1/2}$ energy level.  The ion is initialized into the $\ket{0}$ state by first pumping it with a laser into the $\ket{F=2,m_f=0}$ state and then using microwave pulses to transfer it into the $\ket{F=1,m_f=-1}$ and then $\ket{0}$ states. This transfer and the qubit transitions are driven by a microwave line integrated into the ion trap at around $1.25$ GHz.  An $X(\pm \frac{\pi}{2})$ pulse, which takes the qubit from the $\ket{0}$ state to the $\frac{1}{\sqrt{2}}\left ( \ket{0} \pm \ket{1} \right )$ state, requires $21 \; \mu$s.  A $Y(\pm \frac{\pi}{2})$ gate can be implemented in the same way by adjusting the phase of the pulse.  $X(\pm \pi)$ and $Y(\pm \pi)$ gates can be implemented by performing two $\frac{\pi}{2}$ pulses of the appropriate type in a row.  The microwave control replaces two lasers in a Raman configuration, which are used in typical ion-trap experiments with this kind of transition.  If well controlled, the microwave pulses have smaller intensity fluctuations and less probability of exciting extra-qubit states than lasers, potentially allowing for higher fidelity gates over longer time intervals.  The trap itself is a surface-electrode, linear Paul trap sitting at $4.2$ K in a cryostat.  Many aspects of the trap and transitions used are described in greater depth in Ref.~\cite{Brown2011B}.

Cooling, state preparation, and measurement of the ion are performed using lasers.  A cycling transition from the $\ket{{}^2S_{1/2},F=2,m_f=-2}$ level to the $\ket{{}^2P_{3/2},F=3,m_f=-3}$ level at $313$ nm is the essential tool for both cooling and measurement; laser intensity fluctuations in these operations are not very detrimental.  At the end of the sequence, two microwave pulses are used to transfer the qubit from the $\ket{0}$ state to the $\ket{{}^2S_{1/2},F=2,m_f=-2}$ state and another pulse transfers the $\ket{1}$ state to the $\ket{{}^2S_{1/2},F=1,m_f=1}$ state.  Then the cycling transition is driven for $400 \; \mu$s; if the qubit had been in the $\ket{0}$ state, a detector registers an approximately Poissonian distribution of scattered photons with mean $\sim 13$.  If the qubit had been in the $\ket{1}$ state, then the laser is far off resonance from any relevant transition and a mean of $\sim .14$ scattered photons are detected, mostly due to stray laser light scattered off of the trap.  Because these two scattered photon distributions are well separated, it is straightforward to introduce a photon number cutoff of $1.5$ to convert the detected photon number into a binary measurement of the $\ket{0}/\ket{1}$ states.  The probability of recording the incorrect state in a measurement was estimated loosely at $\sim .01$ and was thought to be dominated by imperfect laser polarization..

Although the $X(\pm \frac{\pi}{2})$ and $Y(\pm \frac{\pi}{2})$ gates are enough to generate the one-qubit Clifford group, two other gate types, the identity $I$ and the $Z$-axis rotations $Z(\pm \pi)$ are used to more closely replicate the RB step definition given by Ref.~\cite{Knill2008}.  That the $Z$ and $I$ gates are of a fundamentally different type than the $X$ and $Y$ gates in the experiment informs the experiment design.  The identity gate is implemented without a pulse by simply doing nothing to the qubit for $42 \; \mu$s, the same duration as two $\frac{\pi}{2}$ pulses.  

The $Z(\pm \pi)$ gates are also implemented by waiting for this interval.  However, after a $Z(\pm \pi)$ gate, the phase of every subsequent pulse is adjusted by $\pm \pi$:  Consider the unitary for the $X(\frac{\pi}{2})$ gate, $e^{ -i \frac{\pi}{4} X} = \frac{1}{\sqrt{2}} (I + iX)$.  If this gate is preceded by a $Z(\pi) = Z$ gate, then $Z(\pi)X(\frac{\pi}{2}) = Z + iZX = Z - iXZ = X(-\frac{\pi}{2})Z(\pi)$, so the $Z$ gate may simply be propagated forward through the sequence, with the phase of all other Clifford pulses updated along the way.  The same technique applies to propagation past $Y(\frac{\pi}{2})$ gates.  An extra $Z$ gate immediately preceding a measurement in the $\ket{0}/\ket{1}$ basis does not change the measurement outcome, so it may be ignored when it appears in this location.  

A valid objection to this strategy for the $Z$ gates is raised in Ref.~\cite{Ryan2009}.  The authors note that this $Z$ gate might not commute with an error, so the propagation through the sequence does not work exactly as described.  They remark that this leads to imperfect depolarization of some errors and could lead to deviations from the model of Eq.~\ref{Eq:2Main}.  To avoid this potential problem and simplify the procedure, it might be preferable in future experiments to implement $Z$ gates as physical gates (for example as a composition of $X$ and $Y$ gates). However, the current protocol better describes how an actual computation would be performed using this hardware.

\subsection{Experiment Design} \label{Sec:2ExpDesign1}

At the time of this experiment, all previous randomized-benchmarking experiments with the exception of that of Ref.~\cite{Ryan2009} had been performed using a procedure almost identical to that in Ref.~\cite{Knill2008}.  This one-qubit procedure calls for sequences for which each step is composed of a Pauli gate $\pm X,\pm Y,\pm Z,$ or $\pm I$, each chosen with probability $\frac{1}{8}$, followed by a Clifford $X(\pm \frac{\pi}{2})$ or $Y(\pm \frac{\pi}{2})$ gate, each with probability $\frac{1}{4}$.  Measurements used in different experiments have varied slightly; we chose not to exactly invert the sequence, instead only performing a randomized Pauli operator and another $X(\pm \frac{\pi}{2})$ or $Y(\pm \frac{\pi}{2})$ gate, if necessary to return the state to the measurement basis.  As described in the previous section, $X(\pi)$ and $Y(\pi)$ rotations were implemented by two successive $\frac{\pi}{2}$ rotations, while the $Z$ and $I$ rotations required no pulses, so each step contained $2$ $\frac{\pi}{2}$ pulses on average.
 
Because the target gate fidelity was $10^{-4}$, it was expected that approximately $10^4$ gates might be required before an error could be observed.  Therefore sequence lengths were upper bounded by approximately $10^4$, and it would have been desirable to investigate sequences with lengths approaching this value.  However, the experiment was constrained by the control software and hardware that program the pulse sequences to be performed in each shot of the experiment.  These limitations put an upper bound of $\sim 1.3*10^3$ on the number of steps per sequence, and pulse-sequence loading times contributed to the time constraints of the experiment, as discussed below.  

Each step of the experiment requires three $\frac{\pi}{2}$ gates (or equivalent wait times), so a characteristic step time is $\sim 100 \; \mu$s.  Given a time frame of hours for the experiment, of order $10^9$ steps could be performed.  For sequences of length $1000$, this allowed $\sim 10^6$ individual sequence experiments to be performed; however, measurement and re-preparation times contributed significantly to sequence time, proportionally much more for short sequences.  A final consideration was the sequence load time, which we approximated to be $360 \; \mu$s, roughly independent of the sequence length.  For consistency with previous experiments, we decided to perform each sequence $n_e = 100$ times after it had been loaded. Due to the extra time required for measurements, cooling, sequence loading, and recalibration, we decided to use $n_l = 100$ sequences for each length.  Finally, in following with the sequence lengths in \cite{Olmschenk2010}, we decided to choose Fibonacci number sequence lengths $\{1, 3, 8, 21, 55, 144, 233, 377, 610, 987\}$, judging that they were unlikely to hide any plausible error models, that might, for example, show different effects for even vs.~odd sequence lengths.  These parameter choices are summarized in Tab.~\ref{Tab:2Exp1}.

\begin{table} \label{Tab:2Exp1}
\begin{center}
\begin{tabular}{|c|c|c|c|c|}
\hline
$T_t$ & $t_s$ & $n_e$ & $n_l$ & $\{l_i\}$ \\
\hline
1 hr & $20 \; \mu$s & $100$ & $100$ & $\{1, 3, 8, 21, 55, 144, 233, 377, 610, 987\}$ \\
\hline
\end{tabular}
\end{center}
\caption[Experiment-Design Parameters for the One-Qubit Benchmark]{Presented are the experimental design parameters chosen for the one-qubit benchmark of Ref.~\cite{Brown2011}.  Notation for the parameters is described in Sec.~\ref{Sec:2ExpDesign}.}
\end{table}

A sequence defined by steps of the sort used in Ref.~\cite{Knill2008} requires $3$ steps before it has some probability of having generated every Clifford gate.  In fact, as noted in Sec.~\ref{Sec:2AppTwirl}, such sequences have a parity symmetry such that even and odd subsequences each have a non-zero probability of representing only one half of the gates in $\mathcal{C}_1$.  This $3$-step minimum indicates that sequences with lengths less than or near $3$ contain almost no well-twirled steps and cannot be expected to follow the depolarized-error assumptions described in Sec.~\ref{Sec:2AppTwirlFail}.  This justifies excluding lengths of a small multiple of $3$ or less from the model fitting, although no such cutoff was defined before the experiment.  Establishing such a cutoff after the experiment potentially introduces the sort of experimenter bias described in Sec.~\ref{Sec:2Goodness}.  

\begin{figure} [tbh]
    \begin{center}
	\includegraphics[width=100mm]{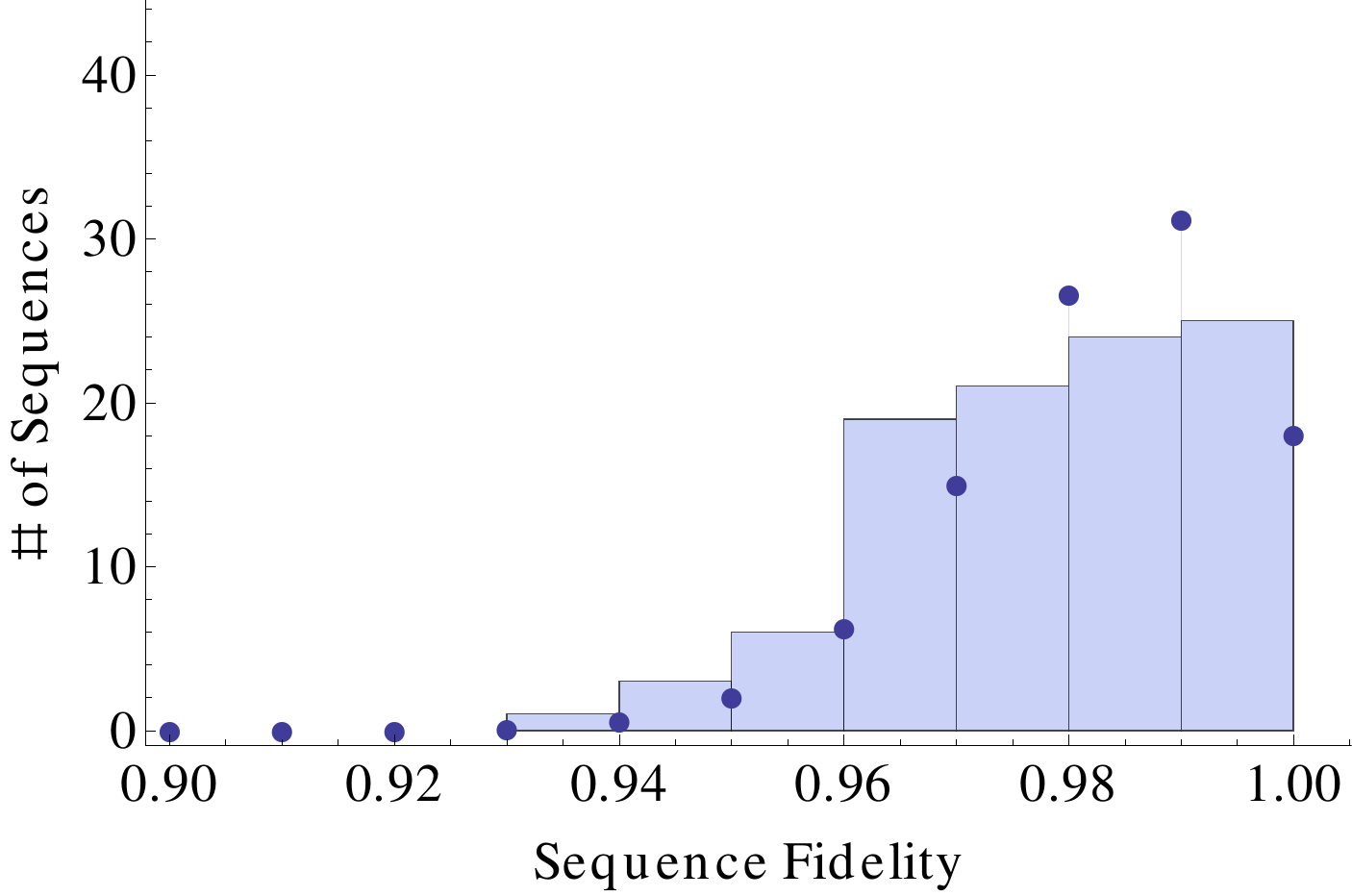}
    \end{center}
    \caption[Sequence-Fidelity Distribution for One-Qubit Benchmark]{
Depicted are the mean experimental sequence fidelities for the $100$ random sequence of length $1$ used for the experiment in Ref.~\cite{Brown2011}.  The fidelities are placed in bins of width $.01$.  It is clear that these fidelities are not normally distributed.  The points to the right of the bins correspond to the values of the best-fit binomial distribution (with parameters $n=100, p=.983$), rescaled by the total number of sequences.  This distribution appears to be a better model for the data. \label{Fig:2SeqDistKen}
}
\end{figure}

Being unsure of the types of gate error models, we chose to make no assumptions about the distribution of the sequence fidelities.  Fig.~\ref{Fig:2SeqDistKen} shows the spread in individual sequence fidelities for length $1$.  In this example, the sequences could not be normally distributed because of the proximity of the upper bound on fidelity. It is possible that a binomial (or perhaps beta) distribution would be a better fit.  For this reason, error bars based on resampling are expected to be more accurate than those based on normality assumptions, although the latter were used in the main report for simplicity.  We became aware of the perturbative analysis of Ref.~\cite{Magesan2011B} after the analysis using the usual RB model given in Eq.~\ref{Eq:2Main} was complete.  The results of applying the model of Eq.~\ref{Eq:2Magesan} are shown in Fig.~\ref{Fig:2AltFits}, and an explanation of these results is given in the next section.

\subsection{Results} \label{Sec:2ExpResults1}

\begin{figure} [tbh]
    \begin{center}
	\includegraphics[width=140mm]{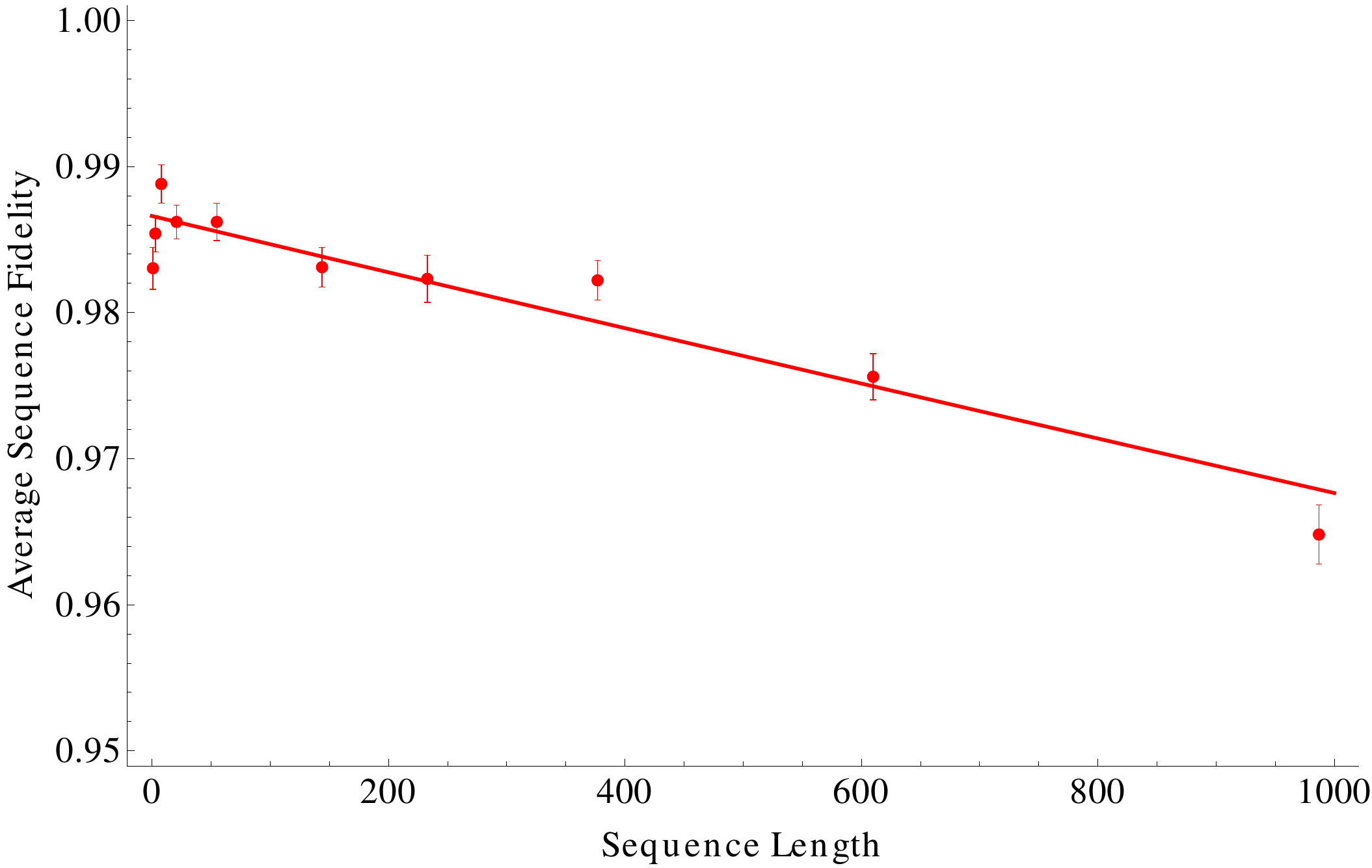}
    \end{center}
    \caption[Benchmark Results for the One Ion Experiment]{
Plot of the main benchmark results for the one-ion experiment.  The mean sequence fidelities for each of the sequence lengths are shown along with error bars that depict the standard errors of the means.  The best fit to Eq.~\ref{Eq:2Main}, corresponding to the reported benchmark parameters, is plotted as a solid curve.     \label{Fig:2ExpResults}
	}
\end{figure}

The primary analysis of the benchmarking data using Eq.~\ref{Eq:2Main} is depicted in Fig.~\ref{Fig:2ExpResults}.  Benchmark parameters of $\epsilon_s =1.99(20)*10^{-5}$ and $\epsilon_m = 1.3(1)*10^{-2}$ represent the best fit found by the fitting algorithm.  $\epsilon_s$ is significantly below the $10^{-4}$ goal.  If the errors for each pulse were truly depolarizing and the $Z$ and $I$ gates were errorless, then the error in each step would be the product of the depolarizing channels corresponding to the $X(\frac{\pi}{2})$ and/or $Y(\frac{\pi}{2})$ pulses, of which there were two on average.  Thus an estimate of the error per single $\frac{\pi}{2}$ pulse would be $\frac{\epsilon_s}{2} = .99*10^{-5}$, a full power of $10$ below the fault-tolerance threshold.  The value of $\epsilon_m$ is of less interest, although it is notable that $\epsilon_m$ is comparable to the estimated error in measurement, suggesting that this error dominates the SPAM error, as expected.

\begin{figure} [tbh]
    \begin{center}
	\includegraphics[width=140mm]{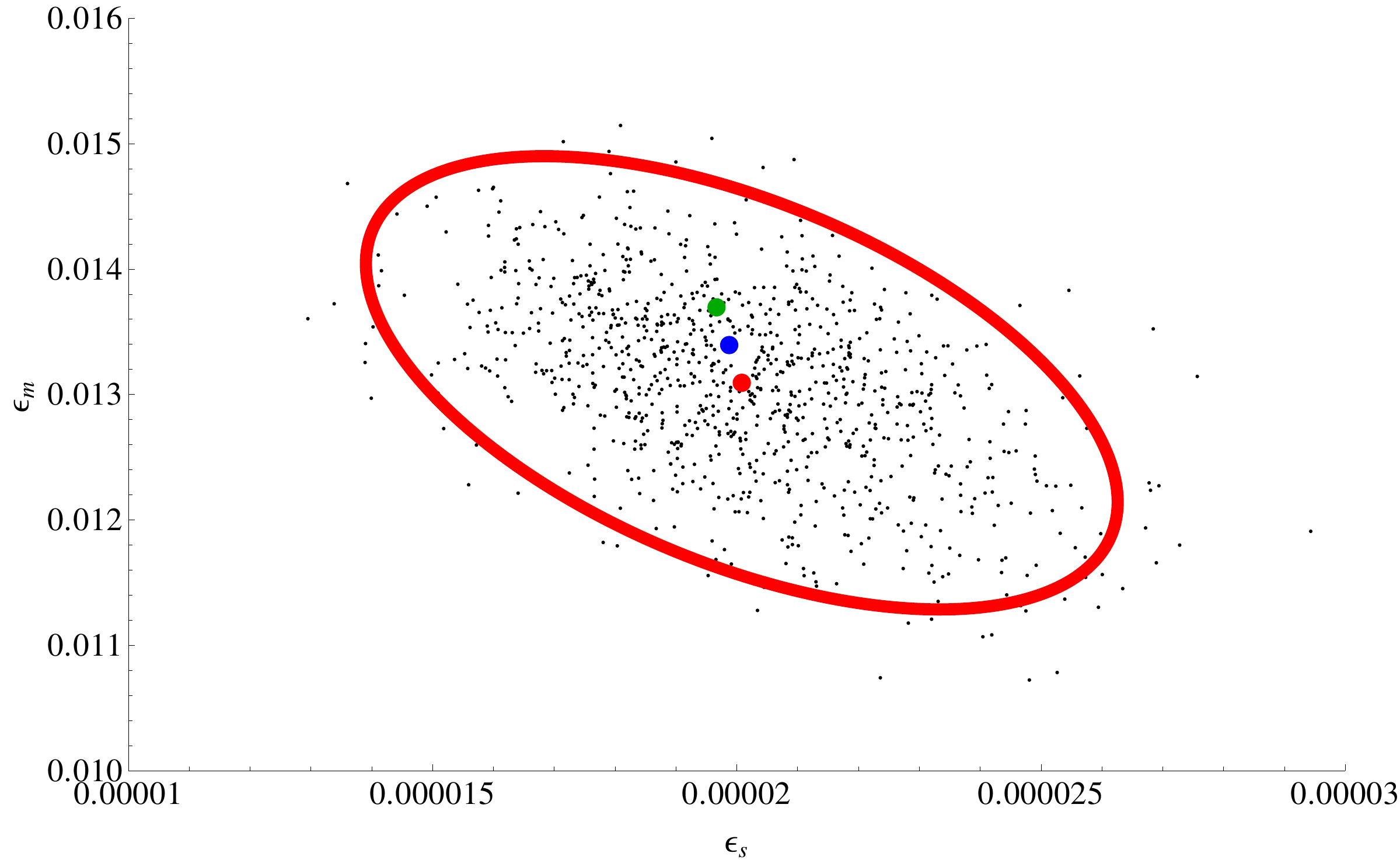}
    \end{center}
    \caption[Resampling Results for One Ion Experiment]{
Plot of the benchmarking parameters extracted from $1000$ resampled data sets.  An elliptical region containing $95 \; \%$ of the resampled parameters is depicted and estimates a $95 \; \% $ confidence region for the original parameter estimates.  Plotted as larger points are the locations of the benchmark parameters returned by the original fit (blue), the means of the resampled values (red), and the bias-corrected estimates (green).  Covariance of the two parameters can be observed in the tilt of the ellipse. \label{Fig:2Resamples}
	}
\end{figure}

Semi-parametric bootstrap resampling of the data is shown in Fig.~\ref{Fig:2Resamples}.  $1000$ resampled data sets are plotted, along with a $95 \; \%$ confidence ellipse.  All of the resampled values of the error per step fall below the threshold value of $10^{-4}$, increasing our confidence in the claim that the experiment surpassed the threshold.  As the figure indicates, the bias-correction procedure is deemed insignificant in this case because the corrected value is within $.25$ standard errors of the reported values.  

The $\chi^2$ statistic for the fit returns $X_\chi = 17.5$ with 
$8$ degrees of freedom, which is deemed significant at the $p = .95$ level.  However, this value is not significant at the weaker $p=.975$ level.  Despite the danger of bias in removing data points after the analysis has been completed, it is noteworthy that the same fit is acceptable at the $p=.95$ level if the data point corresponding to the length $1$ sequences is removed and the remaining $7$ degrees of freedom are used.  In fact, if the data points corresponding to the first $1$, $2$, $3$, $4$, or $5$ lengths are removed and the degrees of freedom are adjusted appropriately, the fit is always found to be acceptable at the $p=.95$ level.  While no length cutoff was introduced before the experiment, it is likely that at least the first two data points, which realize poor twirling, should not be included in the fitting.  In retrospect, the potential for misrepresenting the significance of the results could have been avoided by deciding which lengths to discard prior to the experiment by analyzing the behavior of the approximate twirl for these sequences; this issue is discussed in Sec.~\ref{Sec:2Goodness}.

\begin{figure} [tbh]
    \begin{center}
	\includegraphics[width=140mm]{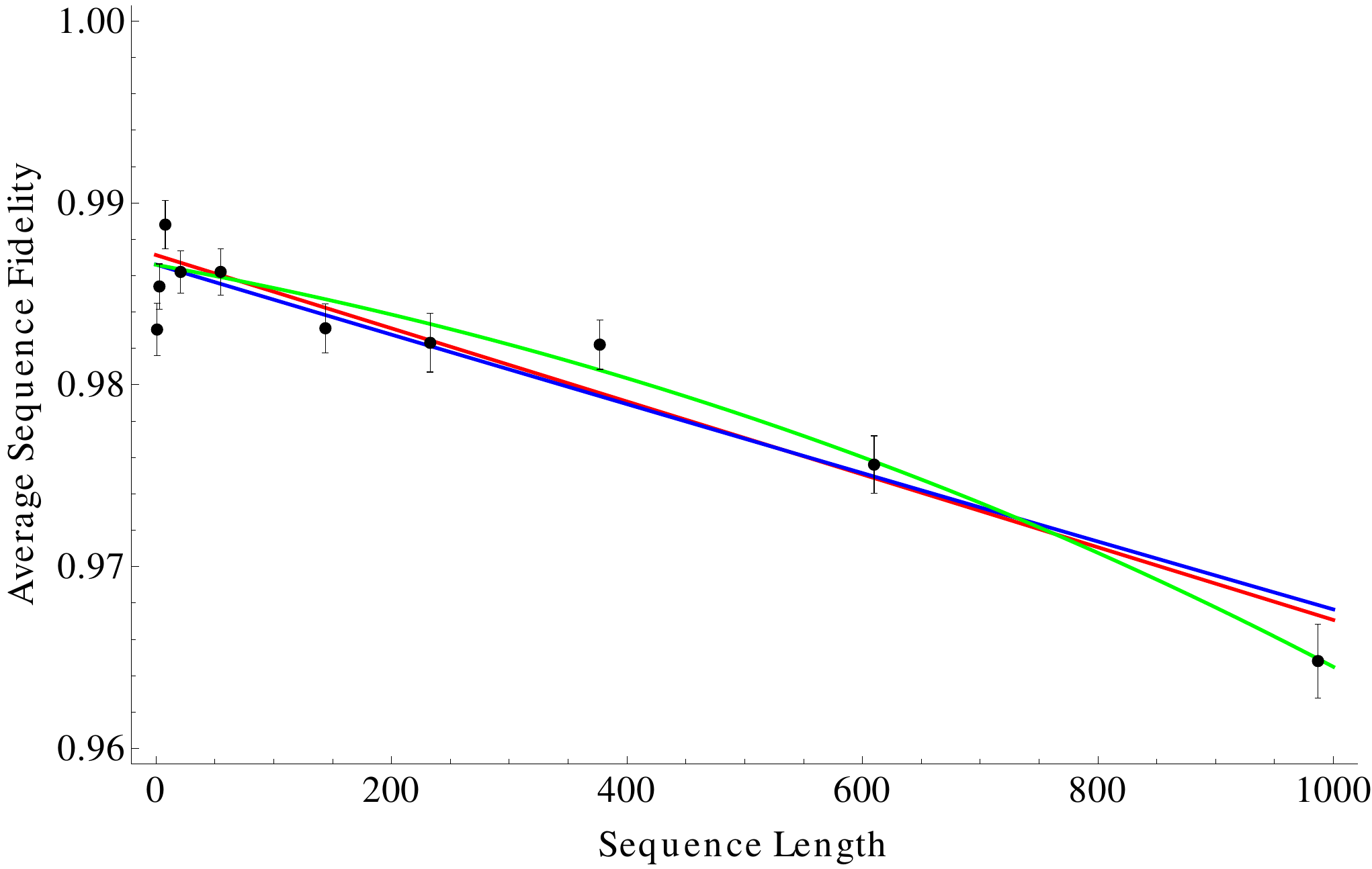}
    \end{center}
   \caption[Alternative Fits for the One-Ion Experiment]{
In this figure the average sequence fidelities are plotted along with the best-fit curves described by the main model and two alternative models for the benchmarking experiment.  The blue, red, and green lines correspond to the models of Eqs.~\ref{Eq:2Main}, \ref{Eq:2RB3}, and \ref{Eq:2Magesan}, respectively.  Interpretation of these fits is given in the text.    \label{Fig:2AltFits}
	}
\end{figure}

Several different fits to the same data using alternative models are depicted in Fig.~\ref{Fig:2AltFits}.  Although these fits appear good and in fact return comparable $\chi^2$ statistics, there is significant degeneracy in the fitting parameters.  Essentially, the data is fairly well fit by a straight line, causing any fit with initial linear behavior to appear valid.  For example, for the $3$ parameter fit corresponding to Eq.~\ref{Eq:2RB3}, the fit parameters returned are $\epsilon_s=4.7158*10^{-6}$, $\epsilon_m=-1.52$, and $C =-1.04$, where the last value implies that the asymptotic state would have negative fidelity, which has no physical interpretation.  However, this model has the same initial linear behavior as the two parameter model, which has a comparable  $\chi^2$ value and makes more sense physically.  This fitting degeneracy issue is not captured by our reliance on $\chi^2$, but considering prior knowledge leads us to reject the completely unphysical description in favor of the second, more meaningful description.  The fit to Eq.~\ref{Eq:2Magesan} suffers from a similar unphysical interpretation, as it returns a negative mean depolarizing strength of $-1.34*10^{-4}$.

\subsection{Error Sources} \label{Sec:2Systematics}

Analysis of the experiment using data extracted from the recalibrations and from, for example, Ramsey tests of the dephasing rate $T_2^*$ allows us to estimate the contributions to the error per step from various systematic sources.  The two largest estimated sources of error are fluctuating stray magnetic fields in the lab, which slightly shift the qubit transition frequency relative to the microwave used to perform the gates, and fluctuating microwave power, which causes over- or under-rotation of our intended $\frac{\pi}{2}$ rotations about the $X$ and $Y$ axes.  If a pulse is intended to perform a $\frac{\pi}{2}$ rotation, then it should have duration $t$ such that $\Omega t = \frac{\pi}{2}$, where $\Omega$ is the Rabi frequency of the microwave drive with respect to the transition.  (A different convention sometimes leads to writing the condition as $\Omega t = \frac{\pi}{4}$ in ion-trap literature.) If the microwave intensity has fluctuated such that $\Omega_e = (1 + \gamma) \Omega$ is the experimental Rabi frequency, then the effective rotation of the pulse applied for time $t$ is $\theta = \Omega_e t = \frac{\pi}{2}(1+\gamma)$.  Thus the error in such an operation is a coherent over-rotation of angle $\theta_e = \frac{\gamma \pi}{2}$.

Suppose the intended rotation is an $X$ rotation about a single qubit.  Then the (unitary) error operator has form $E_e = e^{-\imath \frac{\gamma \pi}{2}} X = \cos(\frac{\gamma \pi}{2}) I + \imath \sin(\frac{\gamma \pi}{2}) X$.  Assuming that $\gamma$ is small, $E_e \approx \left  (1-\frac{1}{2} \left (\frac{\gamma \pi}{2} \right )^2 \right ) I +  \frac{\imath \gamma \pi}{2}X$.  As a superoperator, this error takes $\rho \rightarrow \left ( 1- \left (\frac{ \gamma \pi}{2} \right )^2 \right ) I\rho I + \left (\frac{\imath \gamma \pi}{2} \right ) \left (X\rho I - I\rho X \right ) + \left (\frac{ \gamma \pi}{2} \right )^2 X\rho X$.  The depolarization of this error has depolarizing strength 
\begin{equation} \label{Eq:2IntensityDep}
p_d = \frac{4}{3}\left (\frac{\gamma \pi}{2} \right )^2
\end{equation}
on a single qubit, which makes a contribution of $\frac{2}{3}\left (\frac{\gamma \pi}{2}\right )^2$ to the error per step.  

Notably, errors from power fluctuations are completely coherent, so they are particularly susceptible to adding in unexpected ways in sequences with imperfect randomization.  For example, if two such $X(\frac{\pi}{2})$ rotations are performed in succession, and the error of the two is considered together, then the over-rotation angle is $\gamma \pi$. The depolarization of this aggregate superoperator has strength $\frac{2}{3}\left (\frac{2\gamma \pi}{2}\right )^2$, twice what would be expected for two individually depolarized errors.  This might cause us to over-estimate the error per step with respect to an experiment using perfectly randomized sequences.

Observations of the power fluctuations of the microwave drive revealed somewhere between $.1 \; \% - 1 \; \%$ drift in the power over $120s$, depending on whether the source had just turned on (the worst case) or had been continuously active for $20$ minutes (the better case).  These fluctuations translate to an error per step contribution between $.03*10^{-5}$ and $3*10^{-5}$.  It is unclear which case corresponds to the actual randomized-benchmarking experiment, where the drive is turned on and off rapidly.  Similarly, by recalibrating the microwave's frequency every $60$ s to the qubit resonance, it was possible to observe the drift in resonance frequency.  The RMS frequency drift in between calibrations was approximately $15$ Hz, leading to an estimated contribution to the error per step of $0.4*10^{-5}$.  However, estimates of the frequency drift using $T_2^*$ Ramsey experiments predict an error per step due to frequency-shifting magnetic field fluctuations of as high as $9*10^{-5}$.  Because the benchmark error per step is less than this estimate, we conclude that the Ramsey experiment is likely capturing drifts over longer intervals than are relevant for the benchmark experiment.

Coherent errors such as power fluctuations and dephasing due to slowly varying fields are particularly amenable to correction with self-correcting pulse sequences, dynamical decoupling, and decoherence-free subspaces (see Refs.~\cite{Lidar2012} and~\cite{Khodjasteh2010}).  However, these pulse sequences take several times longer to implement than a single, bare pulse, and they also require several pulses; it is not unusual that the increase in error (especially from decohering processes) caused by the additional pulses and time interval might overwhelm the reduction in error afforded by the correction.  As this experiment was being fine-tuned, it seemed possible that coherent $Z$-type errors due to slowly varying B-fields were the dominant error source.  These errors occur regardless of whether a pulse is applied or not.  For this reason, an $I$ gate implemented by $X(\frac{\pi}{2})X(\frac{-\pi}{2})$ (a very simple self-correcting sequence) might have shown smaller error than an $I$ gate implemented with no pulses.  In the end, the issue of dephasing was avoided in another way, by using a qubit transition less sensitive to magnetic fields; however, this optimization of gates through optimization of the implementing sequences is a valid strategy for randomized benchmarking and is discussed at more length in the next chapter.

\subsection{Conclusions} \label{Sec:2Conclusions}
This experiment was the first to demonstrate control of a single qubit (well) below the fault-tolerance threshold of $10^{-4}$ in error per step.  In some sense, such a result signals that the experiment has sufficiently good one-qubit gates and that it is time to work toward achieving the same goal for two-qubit gates. 

In the next chapter I discuss randomized-benchmarking experiments with two-qubit gates.  The strategy of approximate twirling used in this chapter is actually ill-suited to such a task; because experiments differ much more significantly in the kinds of two-qubit interactions they can employ, the concerns in Sec.~\ref{Sec:2AppTwirl} regarding comparison between benchmark parameters from different approximately twirled experiments are more pronounced.  In addition, as the number of qubits and the complexity of the control increase, it becomes desirable for experimenters to isolate the error contributions from different gates.  These concerns motivate a slightly different strategy for randomized benchmarking that both standardizes the interpretation of the benchmark results and allows for the isolation of errors from different gates.

\chapter{Exactly-Twirled Randomized Benchmarking}
\label{Chap:3}
In this chapter I describe an alternative strategy for Clifford randomized benchmarking based on implementing uniformly-random Clifford operators in each step of the randomized sequences.  This approach introduces some additional complexity to the design of the experiment, but it simplifies the analysis of the benchmark.  I describe the motivation for this approach before explaining the basic procedure and its advantages, including the ability to robustly benchmark a single gate.  Next I describe the main challenge of designing the benchmark experiment and how it can be approached.  Finally, I demonstrate the use of this procedure through an example from an experiment published as Ref.~\cite{Gaebler2012} that I helped to design, analyze, and interpret.  This experiment was the first to adopt this simplified procedure and the first to demonstrate randomized benchmarking of an individual gate.

\section{Weaknesses of Approximately-Twirled Benchmarks} \label{Sec:3AppTwirl}

As described throughout Chap.~\ref{Chap:2}, there are three significant disadvantages to using the approximate-twirling method for randomized benchmarking.  First, the error-per-step parameter has an interpretation that depends on the contents of a step; therefore, in order for two randomized-benchmarking experiments to be directly comparable, they must use the same gate distribution in each step.  Second, it is difficult to describe when two steps of an approximately-twirled experiment are separated enough for their errors to be considered independent.  In a related issue, it is unclear when sequences are long enough that the depolarizing assumption should be considered a good approximation.  Third, there is no prescribed way to extend an arbitrary approximately-twirled experiment to more qubits or to qutrits (for example), which is a problem for standardization purposes.

\subsection{Benchmark Interpretation} \label{Sec:3AppInterp}
A simple example suffices to introduce the interpretation difficulties involved in comparing approximately-twirled experiments.  Suppose there are two experiments each meant to benchmark two qubits.  The first experiment, A, can perform the usual one-qubit Clifford gates $X \left (\frac{\pm\pi}{2} \right)$ and $Y \left (\frac{\pm\pi}{2} \right )$ on each of the two qubits and a single two-qubit Clifford gate $\CX_{1,2}$ (the subscripts denote the control and target qubit indices, respectively).  The second experiment, B, can perform the same one-qubit gates, but the fundamental physics of the experiment only allow the $\CZ_{1,2}$ gate to be performed naturally as a two-qubit gate.  Both of these gate sets are sufficient to generate the Clifford operator group on two qubits.   The A experimenters define a step so that each of their gates $g$ is performed in each step with probability $q(g) \neq 0$, so errors are approximately twirled after some number of steps.  The extracted error per step describes the error per gate averaged over this gate distribution.  

Now suppose the B experimenters would like to directly compare their results to experiment A.  To do this, they must define their gate distribution in the same way as the first experiment, but this requires them to perform the $\CX$ gate with the same probability.  The $\CX_{1,2}$ gate can be exactly implemented using their gate set through the sequence $Y_2 \left (\frac{\pi}{2} \right ) X_2 \left (\frac{\pi}{2} \right )X_2 \left (\frac{\pi}{2} \right ) \CZ_{12} Y_2 \left (\frac{\pi}{2} \right ) X_2 \left (\frac{\pi}{2} \right )X_2 \left (\frac{\pi}{2} \right )$.  This gate sequence is likely to have a much higher error rate than the $\CZ$ gate by itself, so, even if the $\CX$ gate in experiment A and the $\CZ$ gate in experiment B had equal fidelities, the error per step of experiment B would be higher than that for experiment A.  Alternatively, if experiment B defines an error per step using $\CZ$ gates, then experiment A, in trying to replicate this definition, is likely to find a higher error per step than experiment B.  Using this analysis, it is difficult to decide which experiment actually has better control of their quantum systems because the step definition naturally favors one gate set over the other.  A coarse method for comparing the two experiments is introduced in Sec.~\ref{Sec:4StepDef}.

\subsection{Approximate Depolarization}

The second difficulty, described in more depth in Secs.~\ref{Sec:2AppTwirl}, \ref{Sec:2AppTwirlFail}, and \ref{Sec:4AppDep}, with approximately-twirled benchmarks relates to the known failure of the depolarization assumption in short approximately-twirled sequences.  To the extent that two errors (or an error and the measurement) are separated by a large number of steps, the aggregate random operator between them will be approximately uniform.  The issue of how many steps of separation are necessary for this approximation to be valid is addressed in Sec.~\ref{Sec:4AppDep}, but adjacent steps in a sequence are almost never well-enough separated in approximately-twirled benchmark experiments.  First, this can cause an issue with the interpretation of the results of the benchmark, as the parameter intended to represent the depolarizing strength of the average step actually describes some complicated, averaged interaction between steps.  Second, this makes the issue of which sequence lengths to use in the experiment difficult to solve.  Sequences so short that no errors are separated by enough steps are likely to be poorly described by the idealized model, and even slightly longer sequences can still exhibit significant non-depolarized behavior.

\subsection{Scaling} \label{Sec:3AppScaling}
As the number of qubits $n$ increases, the natural gates in different experiments diverge rapidly.  For example, even for experiments with natural $\CX$ gates, it is frequently the case that $\CX_{i,j}$ gates are not natural for all pairs $(i,j)$ for reasons related to geometry and connectivity.  Additionally, common gate sets typically require composition of many more gates to produce an arbitrary $n$-qubit Clifford operator than they do to produce an arbitrary $2$-qubit Clifford operator.  Therefore, for larger $n$, longer sequences are typically required before assumptions about depolarization hold reliably.  Both of these issues also arise when one moves from the discussion of qubit randomized benchmarking to, for example, qutrit randomized benchmarking, as the qutrit Clifford groups are larger and have different structures.  As I demonstrate in Sec.~\ref{Sec:3ClifDecomp}, there are some ways in which approximately-twirled benchmarks scale better than the exact benchmarks described in this chapter; there is a trade-off in different types of complexity between the two strategies.

\subsection{NMR Experiment} \label{Sec:3NMR}

Refs.~\cite{Ryan2009} and~\cite{Laforest2008} describe a three-qubit RB experiment performed using a liquid-state NMR apparatus.  The authors define a step in this experiment according to Table~\ref{Tab:3NMR}.  This choice of gates is sufficient, in that the gates generate the Clifford group on three qubits; however, the choice is also minimal, in that removing the possibility of performing any one of these gates destroys this property.  Because the generating set is minimal, the convergence to an exact twirl in this experiment is slow; nonetheless, the sequence fidelities are still well fit by an exponential decay function.  The authors report an error per step of $4.7(3) * 10^{-3}$.  If one assumes that the dominant contribution to the error came from two-qubit gates, as it does in ion-trap experiments, then this error per step would correspond to an error per $\CX$ gate of $1.4 *10^{-2}$.  This experiment set the precedent for multi-qubit benchmarking.

\begin{table}[h] 
\begin{center}
\begin{tabular}{|c|c|c|c|c|c|c|c|c|}
\hline
Gate & $H_1$ & $H_2$ & $H_3$ & $PH_1$ & $PH_2$ & $PH_3$ & $\CX{1,2}$ & $\CX{1,3}$ \\
\hline
Prob. & 1/9 & 1/9 & 1/9 & 1/9 & 1/9 & 1/9 & 1/6 & 1/6 \\
\hline
\end{tabular}
\end{center}
\caption[Step Definition for NMR RB Experiment]{This table presents the probabilities with which gates were performed in each step of the benchmarking experiment in Ref.~\cite{Ryan2009}.  The ``$PH_i$'' gate is equivalent to the product $Z \left ( \frac{\pi}{2} \right )_i H_i$. \label{Tab:3NMR}}
\end{table}  
\section{Exact-Twirl Benchmark}

The new Clifford randomized-benchmarking procedure I introduce is actually based on the historically older one described in Ref.~\cite{Emerson2005}, although there are some notable differences from the approach in that paper. The procedure is essentially the one described in Chap.~\ref{Chap:2} before approximate twirling is introduced.

\subsection{New Step Definition}

The sequences of random Cliffords constructed for this procedure differ from those of the old procedure in two ways.  First, each step of each sequence (except for the inversion step) is chosen uniformly at random from the Clifford operators on $n$ qubits.  In order to implement a typical Clifford operator, many individual gates are required in each step, and this number of gates varies between steps; this issue is discussed in Sec.~\ref{Sec:3ClifDecomp}.  
Second, whereas in the experiment of Sec.~\ref{Sec:2Exp1}, the Clifford inversion step at the end of the sequence only returned the system to a computational basis state, the inversion step in a sequence of the new procedure exactly inverts the composite Clifford operator preceding it, except for a final Pauli operator randomization.  One reason for this change is that such an inversion step can be seen to implement a marginally uniformly-random Clifford operator, so the expected error in this step is the same as the error in any of the other steps.  This sort of inversion also allows for deterministic measurements of each of the qubits in an $n$-qubit experiment, so the probability of measuring the correct results on all of the qubits decays asymptotically to $\frac{1}{2^n}$ as in Eq.~\ref{Eq:2Main} instead of to $\frac{1}{2}$ as in Eq.~\ref{Eq:2MainApp}.  Note also that this inversion step is different than that described in Refs.~\cite{Emerson2005} and \cite{Magesan2012} in its use of a random Pauli operator before measurement.  The motivation for this randomization is given in Sec.~\ref{Sec:2MeasRand}.

The greatest technical challenges of the new, exact-twirling approach are related to the size and structure of the Clifford group, about which much more is said in App.~\ref{App:A}.  In order to choose uniformly from all Clifford operators, it is necessary to either index all of the operators (and pick an index at random) or to develop an algorithmic technique for constructing Clifford operators that produces them with a uniform distribution.  See Table~\ref{Tab:AClifSize} for an idea of the sizes of various Clifford operator groups.  In this chapter I treat only experiments with small $n$ where the Clifford operators can be enumerated; for information about other cases, see Sec.~\ref{Sec:AGeneration} and Ref.~\cite{Magesan2012B}.  

If $n$ were greater than 20 or so, standard matrix inversion of the Clifford unitary (required to compute the operator for the inversion step) would also be a difficult task. It is here that the Clifford approach to randomized benchmarking would show its strength, as techniques described in Secs.~\ref{Sec:ADecomp} and \ref{Sec:AGeneration} can be used to invert a Clifford operator with polynomial, instead of exponential, classical computing time.  

\subsection{Interpretation} \label{Sec:3ExactInterp}

The result of this approach to randomized benchmarking is a benchmark error-per-step parameter that describes the error per Clifford operator averaged uniformly over the implementations of the Clifford operator group $\mathcal{C}_n$.  In some ways, this parameter seems arbitrary; in actual quantum algorithms (e.g. Shor's algorithm equipped with any particular fault-tolerance procedure), one typically performs a small set of Clifford operators a large number of times.  In order to better approximate the performance of such an algorithm, it might be more appropriate to describe the error per operator averaged only over this small set.  For now, however, it is unclear which set of Clifford operators will be utilized for a large implementation of Shor's algorithm, so I take a general approach instead.  Almost certainly, by working to improve the error averaged over all Clifford operators, one improves on the ability to perform any particular algorithm.  

The resulting error per step can also be thought of as a pure benchmark (without reference to a particular underlying error parameter), especially in the case where the gate error models are complicated and the relationship between experimental systematics and the benchmark parameter is unclear.  In this case, the error per step can be regarded simply as a robust, fair number to be compared to other experiments or to be used to judge improvement of a single experiment.  Further, the behavior of the benchmarking experiment can be useful for describing aspects of experimental control even when analysis indicates that Eq.~\ref{Eq:2Main} is not a good description.  For example, the presence of time-dependence, while disruptive to the standard analysis, is important to note for diagnostic purposes.

\section{Advantages of the Exact Clifford Twirl} \label{Sec:3ExactAdv} 
\subsection{Direct Comparison} \label{Sec:3ExactCompare}
The first and simplest advantage of defining a step by a uniform distribution of Clifford operators could be achieved by any fixed distribution.  By picking a fixed distribution of operators, all experiments computing an $n$-qubit benchmark can be compared directly.  One advantage of the uniform-distribution definition of a step in comparison with other possible distributions is that it is easy to extend to $n$ qubits or qudits (with $d>2$).  The Clifford operator groups are easy to define on such systems, so the uniform distributions over the groups are easy to define as well.  Using the structure of the Clifford groups, one might even hope to be able to compare RB experiments on different numbers of qubits directly.  In comparison,  although reasonable extensions can be made for certain structured distributions, there is no canonical way to extend an arbitrary operator distribution over $n$ qubits to an operator distribution over $n+k$ qubits.

Compared with the step definitions introduced in Refs.~\cite{Knill2008} and \cite{Ryan2009} , this choice of a distribution of operators has the advantage of not being intentionally biased toward any particular experimental set-up.  This definition also encourages experimenters to develop as many different gates as possible: For an experiment using this step definition, one might improve the error per step without actually improving any gates simply by introducing additional gates of comparable quality.  If the expanded gate set makes it so that a typical Clifford operator can be composed using fewer gates, then the error per step will be reduced.

\subsection{Simplified Assumptions} \label{Sec:3ExactSimple}

The use of exact twirling instead of approximate twirling in a randomized-benchmarking experiment loosens the assumptions that must be made about the gate errors.  The concerns of Sec.~\ref{Sec:2AppTwirlFail} are no longer an issue, as two errors that occur after adjacent steps are separated by a uniformly random Clifford operator, which decouples their errors on average.  This means, for example, that one may assume that only one step (the inverting step in my formalism) contributes to the $\epsilon_m$ parameter that describes SPAM errors as long as the gates are reasonably good.  In addition, while sequences with length $1$ are still suspicious because the inversion step and the one random step are strongly correlated, experiments of length $2$ and greater should generally be trustworthy with respect to independent-error assumptions. 

\subsection{Operator Typicality} \label{Sec:3ExactTypicality}
Required assumptions about gate-independent errors may also be relaxed slightly for this procedure.  Consider Table~\ref{Tab:ACNOT}, which portrays the number of $\CX$ gates required to implement Clifford operators as $n$ increases.  When $n$ is large, almost all of the gates require almost the same number of $\CX$ gates, i.e.~the fractional variation in the number of gates needed to compose an arbitrary Clifford operator decreases as $n$ increases.	 For $n$ very large, one expects that the variance of the Clifford operator errors also decreases, so the assumption of operator-independent errors is closer to being satisfied even if it is far from being satisfied for individual gates.  Formalizing this argument would lead to better understanding of the operator(gate)-independence assumption.

\subsection{Scaling} \label{Sec:3ExactScaling}

One appeal of randomized benchmarking in comparison to quantum process tomography is that randomized benchmarking is feasible for experiments with larger numbers of qubits.  A good discussion of the scaling of randomized benchmarking can be found in Ref.~\cite{Magesan2012B}.  The exact-twirling approach to RB has scaling advantages (and disadvantages) in comparison to the approximate-twirl approach.  

The distance at which gates must be separated in the approximately-twirled benchmarks in order to approximately satisfy independence assumptions grows with $n$.  Suppose that an experimental gate set consists of one-qubit operators and $\CX_{i,j}$ between arbitrary $(i,j)$.  Sec.~\ref{Sec:ADecomp} establishes that at least $\frac{n^2}{\log{n}}$ $\CX$ gates are required to implement some operators in $\mathcal{C}_n$.  Therefore, in order for there to be some probability of implementing any Clifford operator in between two errors, the errors must be separated by $\frac{n^2}{\log{n}}$ steps in an experiment implementing at most one $\CX$ gate per step.  Errors separated by fewer steps cannot be assumed to behave as independent depolarizing errors.  

In contrast, some aspects of the implementation and design of exactly-twirled benchmarking experiments scale more poorly with $n$ than for the approximately-twirled approach.  In particular, the decomposition of a Clifford operator in $\mathcal{C}_n$ into a generating gate set, which is necessary for the design of an exactly-twirled RB experiment, is shown in Sec.~\ref{Sec:ADecompAlg} to have (classical) computational complexity of order $n^3$.  This computation is not necessary for the approximately-twirled benchmark.  In a separate issue, the length of the steps (in terms of number of gates from a generating set) scales as $O \left (\frac{n^2}{\log{n}} \right )$ at worst.  In comparison, the steps in the approximately-twirled experiment may have constant length.  There is a direct trade off here between confidence in depolarization assumptions and the length of each step.   

\subsection{Invididual-Gate Benchmarking} \label{Sec:3GateBenchmark}

The simplicity of the exactly-twirled benchmark from a theoretical point of view makes extensions of the procedure fairly straightforward.  In this section I describe how it is possible to extract the gate fidelity of a single gate within the framework of randomized benchmarking.  This technique has a practical appeal to experimenters, as it allows for a robust, simple description of the performance of a single gate in the context of a long computation.  When a change in the experimental apparatus is made that impacts the gate, this parameter is a useful way to report whether the gate has improved or not.  The idea of individual-gate benchmarking was developed concurrently by the authors of Ref.~\cite{Magesan2012}, who call the process interleaving.

It is also possible to benchmark all of the individual Clifford gates used in an experiment, and thereby to determine which of the gates are the most in need of improvement.  The experimental time required for this process scales as $(m+1)T$, where $T$ is the time required for the standard benchmarking experiment and $m$ is the number of individual gates that one wishes to benchmark.  

This technique has the potential to improve estimations of the resources required for a computation; if one knows for each Clifford gate used in a computation how often it is used and its gate fidelity, then it is possible to make a more accurate estimation of the probability of failure of the computation.  There is a strong caveat to any such estimation using randomized-benchmark error parameters, and that is that the estimation is only reasonably accurate if the computation looks as randomized as the benchmark sequences, so that the errors do not add consistently or coherently.  If certain gate sequences are repeated frequently, then one must be concerned that the errors in these sequences add coherently instead of incoherently, as is the case for depolarized errors.  For this reason, it may be desirable to pseudo-randomize algorithms in order to avoid unwanted coherence of errors; for an algorithm performed in this way, RB error parameters would describe the behavior very well.

\subsubsection{Derivation} \label{Sec:3GateDerivation}
In order to perform the benchmarking of a gate $g$ represented by a superoperator $\hat{\Lambda}_g$, first perform an exactly-twirled RB experiment as usual.  The error of each step of the sequences in such an experiment has average, depolarized form $\hat{\Lambda}_e$ with depolarizing strength $p_e$, so that the probability of success for length $l$ sequences is 
\begin{equation} \label{Eq:3Main}
F(l) = (1-\alpha_n) + \alpha_n(1-p_m)(1-p_e)^l,
\end{equation}
where $p_m$ is the SPAM error strength and $\alpha_n$ is a constant, as in Eq.~\ref{Eq:2Main}.

Now construct a second randomized-benchmarking experiment in the same way as the first, except compose each step from a uniformly random Clifford operator followed by a $g$ gate.  The inversion step is performed in the same fashion as in the first benchmark experiment, inverting the random operations and $g$ gates that preceded it.  Crucially, the product of a uniformly random Clifford and any particular Clifford operator is still a uniformly random Clifford (because the Clifford operators form a group).  The implementation of a single step then has the form $\hat{\Lambda}_{e,g} \circ \hat{g} \circ \hat{\Lambda}_{e} \circ \hat{U} $, where $\hat{\Lambda}_{e,g}$ is the error operator for the $g$ gate defined in the same way as before.  Considering only one error at a time, each error is conjugated by uniformly random Clifford operators which enact an exact Clifford twirl.  Under the assumption that this depolarizes each error perfectly, the average probability of sequence success is
\begin{equation} \label{Eq:3MainGate}
F'(l) = \alpha_n + (1-\alpha_n)(1-p_m')(1-p_e')^l,
\end{equation}
where $(1-p_e') = (1-p_e)(1-p_g)$ according to the composition of depolarizing channels.  Note that, in general, $p_m'$ should be equal to $p_m$ if the SPAM errors are consistent and independent of the operator sequence as desired.  However, it is not necessary to assume that they are identical for the analysis to proceed, and observing their discrepancy might yield insight into the SPAM errors.

Finally, after this second benchmarking experiment has been performed, analysis of both experiments reveals $p_e$ and $p_e'$, which may then be used to solve for 
\begin{equation} \label{Eq:3Gate}
\epsilon_g  = \frac{1}{\alpha_n} \left (1-\frac{1-p_e'}{1-p_e} \right ),
\end{equation}
the error per $g$ gate.  As before, this is related to the fidelity of gate $g$ (considered as an $n$-qubit operator) by $F_g = 1- \epsilon_g$.

In order to be able to use the inversion algorithms described in Sec.~\ref{Sec:ARBSim} to calculate the appropriate inversion step, it is crucial that $g$ be a Clifford operator.  However, if the number of qubits is small so that the computations of the inversion operator and the correct result for each sequence are not prohibitive, then a similar technique could be used to calculate the gate fidelity of a non-Clifford gate.  This could potentially allow for a randomized-benchmark fidelity estimate for each of the gates in a generating gate set capable of universal quantum computation.

\subsubsection{Assumptions} \label{Sec:3GateAssumptions}
Although the exact Clifford twirling ensures that errors in $g$ gates are decoupled from each other and that errors in the random operators are decoupled from each other, it does not ensure that the error in each $g$ gate is decoupled from the error of the random Clifford operator that precedes it.  For this reason, the approximation that the gate and step errors are independently depolarized is not always warranted.  Ref.~\cite{Magesan2012} gives some bounds on the error in this approximation.

For a simple but adversarial example of this problem, consider the case where $g$ is a one-qubit $Z$ rotation (for example the Pauli $Z$ gate) and both $\hat{\Lambda}_{e,g}$ and $\hat{\Lambda}_{e}$ represent positive, small-angle rotations about the $Z$ axis.  Because these errors commute with $\hat{g}$, the two errors would add coherently in every step of every sequence.  Then the depolarization of the combined error $\hat{\Lambda}_{e,g}\hat{\Lambda}_{e}$ will have strength larger than the composition $1-(1-p_g)(1-p_e)$ of the depolarizations of the individual errors.  If the two errors have the same small rotation angle, then the observed error per step is twice what it would be if the errors were independently depolarized.  In this case the approximation overestimates $p_g$ by a factor of $2$.  A similar issue is described in Sec.~\ref{Sec:2Systematics}.  If the errors in this example were small-angle rotations in opposite directions about the same axis, then an error per step smaller than appropriate would be observed.  Coherence this extreme is not expected in actual experiments because $\hat{\Lambda}_{e}$ is not expected to be either consistent for all operators or entirely unitary, as in this example.

In contrast, if the errors of the random operators or $g$ gate are depolarizing (or stochastic Pauli) before Clifford twirling is applied, then the error in this approximation approaches zero.  In actual experiments, something in between is expected: random operators should have fairly well-randomized, partially depolarizing errors, especially when $n$ is large, but they are not expected to be exactly depolarizing.  

The effects of the separate Pauli and $\overline{\mathcal{C}_n}$ twirls are described in Sec.~\ref{Sec:2Clifford}.  If the Pauli operators have negligibly small error compared to the $g$ gate (as is the case described later in this chapter), then the required error assumptions can be relaxed slightly by performing the experiment with a random Pauli gate inserted before and after each $g$ gate.  This Pauli randomization twirls the errors of the step and $g$ gate into Pauli channels so that the worst-case addition of the errors is ameliorated.

\subsubsection{Extensions} \label{Sec:3GateExtensions}
After the standard benchmark has been performed, one additional repetition of the procedure of the previous section is needed to benchmark each individual gate.  This general strategy of extending the benchmarking procedure by performing a second benchmark and comparing the resulting error per step is not limited to benchmarking individual gates.  Ref.~\cite{Gambetta2012} describes a similar procedure where the goal is to benchmark the additional errors that occur when two gates are performed simultaneously.  This general issue, which the authors call ``crosstalk'' error, is crucial for understanding how modular components interact when they are combined in a potentially scalable architecture.  The procedure is as follows: perform benchmarking on a single qubit A, perform benchmarking on a single qubit B, perform one-qubit benchmarking on the two qubits, A and B, simultaneously.  The authors then derive expressions designed to extract the additional error that occurs when qubits have simultaneous interactions.  The table of benchmarking experiments in App.~\ref{App:B} includes a column indicating which experiments have attempted such an additional benchmarking-based procedure.

\section{Clifford Decomposition} \label{Sec:3ClifDecomp}
A difficult aspect of the implementation of the exact Clifford twirl is the need to use a small set of experimental gates to implement each of the operators in the set $\mathcal{C}_n$ of Clifford operators on $n$ qubits.  By definition, this is only possible if the ideal operators implemented by the experimental gates form a generating set for $\mathcal{C}_n$.  One preliminary task for designing a randomized-benchmarking experiment is to verify that the gates form such a generating set.  Next, it is necessary to compute compositions of these generating gates that implement each of the operators required for the sequences.  The reported error per step is related to the average operator fidelity over all experimental implementations of the Clifford operators.  If the implementations are inefficient in a way described in this section, then the error per step can be high even if the error per gate for each individual gate is low.  For this reason, there is much to be gained by searching for efficient decompositions of Clifford operators into generating gates.  Algorithms for finding these decompositions are described and discussed in more depth in App.~\ref{App:A}; in this section I focus on the uses for these algorithms and the motivation for improving them.

By calling a decomposition of an operator inefficient, I mean that its implementation is worse at satisfying some goal than another decomposition of the same operator.  The most common overarching goal is to implement the operator with low error, although not enough is typically known about the individual gates to approach this optimization directly.  A first proxy for this goal might be to minimize the number of gates in a decomposition.  For example, if one-qubit Clifford gates and $\CX$ gates are available, then a $\CZ_{12}$ gate can be implemented in many ways, but $H_2 \CX_{12} H_2$ requires a minimal number of gates.  Identities of this sort can be used to reduce the number of gates required in an implementation.  There might be experimental gates that are known to have higher error than others; in this case a second, more advanced goal might be to minimize the number of ``bad'' gates in addition to minimizing the number of gates overall.  An even more advanced goal might be to use self-correcting pulse sequences (see Ref.~\cite{Khodjasteh2010}), which describe sequences of gates designed to self-correct consistent gate errors, in order to find longer composition sequences might actually have higher fidelities.  In another approach entirely, one might wish to optimize Clifford decompositions so that they minimize time or some other computational resource of interest.
This subject has importance outside of the realm of randomized benchmarking, as it allows for efficient implementations of Clifford operators necessary for implementing quantum stabilizer codes and other fault-tolerance and communications procedures. 

\subsection{Exhaustive Search} \label{Sec:3DecompSearch}
One approach to Clifford operator decomposition is exhaustive search.  As described in detail in Sec.~\ref{Sec:ADecomp}, this is the process of enumerating all possible compositions of the generating gate set until a desired objective is met.  Whether this search terminates depends on the optimality criterion (objective function) that has been chosen to define the target of the search.  For example, if an optimal implementation of a fixed operator is defined to be the one that uses the smallest number of gates, then the enumeration iterates through all possible compositions of gates of length $1$, then $2$, etc., and the first composition it finds is an optimal implementation, terminating the search.  The experimental optimality criterion used later in this chapter concerns the number of two-qubit gates of a certain type that are needed in the implementation.  This search can also be constructed so that it terminates.  The crucial fact about these terminating examples is that the optimality criterion can be defined as a real valued function of the composition sequence that is non-decreasing as gates are added to the sequence.  Thus a valid implementation can never be improved by adding more gates.

If the optimality criterion is a more-complicated function of the sequences, such as the actual error of the composed operation, then it is possible that the objective function might be improved by adding more gates.  In this case, the search is not likely to terminate.  However, one may always generate composition sequences until a certain time has elapsed and pick the best composition implementing the desired gate.  This implementation might not be optimal, but much can be gained from the search in any case.

The problem with all such exhaustive search techniques, whether they terminate or not, is that the Clifford groups become very large as $n$ increases (see Table~\ref{Tab:AClifSize}).  In general, the size of the Clifford group scales as $4^{n^2}$, so it becomes completely untenable to search for optimal implementations for all of the Clifford operators when $n$ is larger than $5$.  In fact, even finding an optimal implementation of a single operator using the search described in Sec.~\ref{Sec:ADecomp} takes half as long on average as finding optimal implementations for all gates.  
More-advanced search algorithms can find a decomposition for a single group element more efficiently, but they are not enough to make exhaustive search a generally appealing strategy for a large number of qubits.

\subsection{Algorithmic Decomposition} \label{Sec:3DecompAlg}
The second approach described in Sec.~\ref{Sec:ADecomp} for decomposing a Clifford operator into generating gates is to use decomposition algorithms based on the structure of the Clifford group.  One such algorithm is described in detail in that section that decomposes a single Clifford element in $\mathcal{C}_n$ into $O(n^2)$ gates taken from the over-complete generating set consisting of one-qubit gates, $\CX$ gates, and $\CZ$ gates.  The algorithm takes time that scales only polynomially in $n$ to implement (on a classical computer) and is thus much more efficient than exhaustive search if implementations are only required for a small subset of the Clifford operators.  However, this algorithm gives an implementation of the operator that is not guaranteed to be optimal in any way.

Such algorithms can be designed with an optimality criterion in mind, but none is yet known that strictly optimizes the decomposition even for the simple criterion of minimum gate number.  More-complicated criteria such as total time required or minimum error per operator seem even more difficult to implement.  However, it is possible to take the results of any such algorithm and apply simple, circuit identities to make some improvements.  In the most important example, it is possible to take the output of my algorithm or the one used by Ref.~\cite{Aaronson2004} and rewrite it to use a different generating set of gates by expressing the gates used by the algorithm in terms of the new generating set.  This translation procedure does not work optimally, but it is typically satisfactory.

Because of the size of the Clifford groups, algorithmic decomposition is the only scalable way to design randomized-benchmarking experiments for large $n$.  As treated in Sec.~\ref{Sec:AGeneration}, selecting uniformly at random from all of the Clifford operators on $n$ qubits and computing the inverting Clifford operator can also be exponentially hard computational tasks for large $n$ if the structure of the Clifford group is not utilized.  The size of the Clifford group, while finite, again precludes listing all of the operators and approaching these problems with exhaustive techniques.  As with decomposition, algorithms relying on the algebraic structure of the Clifford group are available to solve individual instances of these problems in polynomially scaling time (examples of both are given in Sec.~\ref{Sec:AGeneration}).  All of these algorithms are important for the design of randomized-benchmarking experiments with $>5$ qubits, and not much effort has been applied toward optimizing any of them in this context.

\section{Two-Qubit Experiment} \label{Sec:3Exp2}
In this section I recapitulate an experiment performed by the NIST Ion-Storage group and published as Ref.~\cite{Gaebler2012} (with additional information in Ref.~\cite{Gaebler2012L}).  The exact Clifford twirling procedure, including the individual-gate benchmarking and the optimal-decomposition techniques for Clifford operators, was developed for this experiment.  There were several complementary goals for this experiment.  On the experimental side, the goals were to provide the first ion-trap benchmark of more than one qubit and also to investigate the use of a two-qubit gate called the M\o{}lmer-S\o{}rensen (MS) gate (first described in Refs.~\cite{Sorensen1999} and \cite{Molmer1999}), which had only recently been introduced to the experiment.  From a theoretical perspective, the goals were to test the new exact-twirling procedure as a more-scalable benchmarking procedure and to further investigate the statistical analysis and validity of the benchmarking equations on a more-complicated system.

\subsection{Experiment Basics} \label{Sec:3ExpBasics}
The essential layout and design parameters of the trap used for this experiment are found in Ref.~\cite{Home2009}.  One of the important aims of the design of this trap was to demonstrate the component techniques needed to implement a scalable ion-trap quantum-computing architecture of the kind described in Refs.~\cite{Wineland1998} and \cite{Kielpinski2002}.  These techniques include the usual interactions and protection from the environment needed for low-error gates, but they also include the ability to transport ions between different trapping zones without disturbing the states of the qubits.  This in turn necessitates advanced cooling techniques for the ions because the transport can heat them significantly and standard cooling techniques destroy quantum information.

Sympathetic cooling of the ions is used in order to preserve the ability to perform accurate gates when transport is required.  For sympathetic cooling, one may take advantage of the Coulomb coupling between the ion storing the information (qubit ion) and a second ion (cooling ion), usually of a different species (see Ref.~\cite{Barrett2003} for more information).   By standard laser cooling of the second ion far-off resonance from any relevant electronic transition in the qubit ion, the second ion can be cooled, which damps the coupled-motion mode and, in turn, the motion of the qubit ion.  Because the laser is far-off resonance from the qubit transition, it introduces a negligible perturbation to the qubit's state.  Notably, the two-qubit interactions to be described later also utilize coupled motional modes of the ions, so cooling is necessary before these gates but cannot be performed while the gates are being executed.

Although schemes that require only one additional cooling ion for a large number of qubit ions are possible (as in Ref.~\cite{Lin2009}), they typically do not allow all motional modes to be cooled sufficiently quickly, so our experiment utilized a dedicated cooling ion for each of the two qubit ions.  The qubit ions in this case were ${}^9\mbox{Be}^+$, as in the experiment of Sec.~\ref{Sec:2Exp1}, but the cooling ions were ${}^{24}\mbox{Mg}^+$.  When two-qubit gates were performed, the ions were ordered in the Be - Mg - Mg - Be configuration.  For the rest of this section, I ignore the cooling ions in discussions; however, they add significantly to the complexity of the laser pulses and trapping potentials needed for the experiment.

\begin{figure} 
    \begin{center}
	\includegraphics[width=80mm]{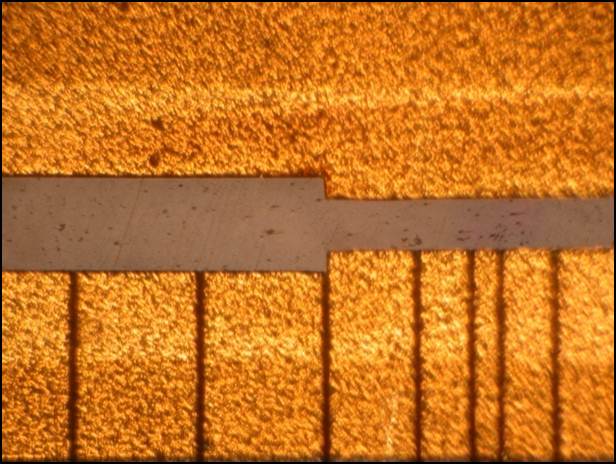}
    \end{center}
\caption[Ion Trap Image for the Two-Qubit Benchmark Experiment]{This image of the trap for the two-qubit benchmark experiment shows the electrodes used to create the different trapping zones.  The two trapping zones described in the text are above the electrodes that surround the thinnest electrode in the image. Image from John Gaebler, 2013.  Used with permission. \label{Fig:3TrapPic}} 
\end{figure}

The trap consists of six zones, or locations where qubits can be trapped, of which only two are mentioned here.  The electrode structure of the trap is shown in Fig.~\ref{Fig:3TrapPic}.  Because the two-qubit interaction is mediated by common motional modes of the qubits, it is desirable to have the qubits near each other when this gate is performed and further from each other at all other times.  This is the general motivation for the ability to move ions from one trap zone to another.  

Before every two-qubit gate, all of the ions are moved into the same trap zone and cooled.  After the gate, one qubit is once again moved to a separate zone, and both are cooled again.  When the ions are thus separated, they can be addressed individually with lasers, so that one-qubit gates are possible.  This movement was not expected to be a large source of error in the experiment in comparison with the gate operations themselves; however, the movement and cooling adds a great deal of time and complexity that indirectly affects the fidelity of the experiments \cite{Bowler2012}.  For example, the laser pulse that enacts the two-qubit MS gate lasts $\sim 20\;\mu$s, however the two-qubit gate with movement and cooling included takes $\sim 3$ ms.  While errors occurring during this time are not expected to make a large contribution to the error per step, the timing severely limits the number of experiments that can be carried out and likely contributes to the issue of ion loss described later.

Each qubit is encoded in the $\ket{0} = \ket{F=1,m_f=0}$ and $\ket{1} = \ket{F=2,m_f=1}$ electronic states of the valence electron of ${}^9\mbox{Be}^+$.  The transition frequency is $f_0 = 1.2$ GHz, but instead of direct microwave drive, as in Sec.~\ref{Sec:2Exp1}, two lasers with difference frequency $f_0$ are used in the stimulated Raman configuration to drive one-qubit $X$ and $Y$ rotations.  At the intensities and detunings used for the benchmark experiment, a $\pi$ pulse on this qubit resonance takes $9\;\mu$s.  This transition is first-order insensitive to magnetic field fluctuations about the applied static $12$ mT magnetic field.  As in Sec.~\ref{Sec:2ExpBasics}, $Z$ rotations can be implemented by adjusting the phase of subsequent pulses, and a cycling transition is used to measure each qubit.  Over a $250\;\mu$s measurement interrogation interval $\sim 30$ emitted photons are detected if the system was in the $\ket{0}$ state versus $\sim 1.5$ if it was in the $\ket{1}$ state.  

In order to implement an entangling two-qubit gate necessary to generate the Clifford group, a M\o{}lmer-S\o{}rensen (MS) type of interaction is used, as in Refs.~\cite{Molmer1999,Sorensen1999}.  In short, the gate operates by simultaneously driving the blue and red sidebands of the qubit resonances of the two qubits with respect to some motional mode (that is, the resonances that flip the qubit state while adding or removing one quantum of motional excitation).  The authors of Ref.~\cite{Monz2011}, using ${}^{40}\mbox{Ca}^+$ ions in a different trapping geometry with no transport or sympathetic cooling, demonstrated that this type of two-qubit gate could be used to create a Bell state with remarkable fidelity.  Two great advantages of the MS-type gate are that it can be implemented on qubits with a magnetic-field-insensitive transition and that it is insensitive, to some degree, to the initial excitation of the motional mode of the two qubits.  This motivated the development of the MS interaction for Be ions in the benchmarked trap.  

Ignoring some issues with relative phases, the MS gate has unitary $U_{MS,1,2} = e^{-\imath \frac{\pi}{4} \sigma_{x,1} \sigma_{x,2}}$ in this experiment.  If the MS gate is preceded and followed by one-qubit, $Y$-type rotations on each qubit, then it acts more like a familiar $\CZ$ gate.  The actual identity is given by
\begin{equation} \label{Eq:3GGate}
Y_1 \left (\frac{\pi}{2} \right )Y_2 \left (\frac{\pi}{2} \right )U_{MS,1,2}Y_1 \left (\frac{-\pi}{2} \right )Y_2 \left (\frac{-\pi}{2} \right ) = Z_1 \left (\frac{\pi}{2} \right )Z_2 \left (\frac{\pi}{2} \right )\CZ \equiv G_{1,2},
\end{equation}
up to a global complex phase that does not matter.  App.~\ref{App:C} should be used as a reference for Clifford operator definitions.  Using these identities, the $G$ gate and one-qubit rotations can be composed to implement a $\CZ$ or $\CX$ gate.  

The gate set $\{Y_{1} \left (\frac{\pi}{2} \right ), Y_2 \left (\frac{\pi}{2} \right ), X_1 \left (\frac{\pi}{2} \right ), X_2 \left (\frac{\pi}{2} \right ), G_{1,2}\}$ is sufficient to generate the two-qubit Clifford group $\mathcal{C}_2$.  The $G$ gate commutes with $Z$ rotations and is also insensitive to path-length changes in some of the laser beam lines used; these features motivate its use instead of the bare MS gate.  The total pulse time required to implement the five pulses (the first and last two in Eq.~\ref{Eq:3GGate} are concurrent) necessary for the $G$ gate is $\sim 110 \;\mu$s, but, as mentioned above, the addition of the separation, recombination, and cooling steps for the two qubits brings the total operation time to $\sim 3$ ms.

The $G$ gate was expected to be, by far, the largest source of error for this experiment, with somewhere between $10^{-1}$ and $10^{-2}$ error per $G$.  This estimate was based on simple calibrations of the gate, including an experiment where the gate was used to try to produce a Bell state and the fidelity of the resulting state with the ideal Bell state was estimated.  As demonstrated in the next section, the fact that the $G$ gate was the dominant error source motivated the design of the benchmarking experiment.  The long gate times also played a large part in the design, limiting the number and length of the individual sequences. 

\subsection{Experiment Design} \label{Sec:3ExpDesign}

There is no simple way to extend the step definition of the only previous multi-qubit RB experiment (Ref.~\cite{Ryan2009}, described in Sec.~\ref{Sec:3NMR}) to the gate set available for this experiment, so it was necessary to define a new step distribution.  In order to avoid the same problem in the future, the exactly twirled procedure was chosen; the standardized definition of a step allows it to be more extensible.  In the process of working through this technique, the experimenters raised the question of whether randomized benchmarking could help to describe the $G$ gate, which was the most important new aspect of the experiment, and this led to the idea of individual-gate benchmarking.  To benchmark a single gate, two benchmarking experiments are needed: the standard exactly-twirled benchmark (called the primary benchmark) and a second benchmark with an extra $G$ gate in each step (called the primary+$G$ benchmark).  The goal was to be able to complete this entire experiment in several hours, as the apparatus could not be expected to maintain consistent performance over days and a significant amount of calibration was needed before the benchmarking experiments began.

The first task was to explicitly determine gate-sequence implementations of each of the Clifford operators.  Because the Clifford group on two qubits has a manageable size, this was done using the exhaustive search techniques described in Secs.~\ref{Sec:3DecompSearch} and~\ref{Sec:ADecomp}.  The desired optimality criterion in the search was to minimize the error per step; as stated earlier, such a search might not terminate.  As a proxy, I optimized to find the implementation sequence that used the fewest $G$ gates first and the fewest one-qubit gates second.  This was expected to be an effective substitution because the error in the $G$ gates was so much higher than other gates.  Because a $G$ gate can be written as a product of one-qubit gates and a $\CX$ gate, the results of Table~\ref{Tab:ACNOT} apply to this generating set as well: 1.5 $G$ gates are needed on average, and $3$ $G$ gates are needed at most, to implement any operator in $\mathcal{C}_2$.  

In this decomposition, the Pauli-operator part of the randomization was separated from the $\overline{\mathcal{C}}_2$ part as described in Sec.~\ref{Sec:AClifDef}.  In practice, this means that a random Pauli operator was performed after the random operator from $\overline{\mathcal{C}}_2$, and the two were chosen independently.  Consequently, the number of one-qubit gates in each implementation was not actually optimized as a whole.  On average, $6.5$ effective $\frac{\pi}{2}$ gates were used per step ($\pi$ gates count as two effective $\frac{\pi}{2}$ gates, while $Z$ and $I$ gates use no effective $\frac{\pi}{2}$ gates). 
This separate Pauli randomization goes some way toward decoupling the errors in the step from the error of the $G$ gate.

The loading time of the pulse sequences into the apparatus was significant: $\sim 1$ s per step of the sequence.  In this case, $3$ ms per $G$ gate translates to $4.5$ ms per step, on average, or $7.5$ ms per step for the steps with an extra $G$ gate.  Preparation and initial laser stabilization require an extra $10$ ms per instance.  $100$ instances of a standard length $l$ sequence then require $(~.45*(l+1) + 1)$ seconds, roughly the same as the loading time.  For this reason and for ease of comparison to analysis of previous experiments, $n_e=100$ was chosen as the number of instances per sequence.  Sometimes, due to an error in the control software, only $99$ instances of a sequence were performed.  This was properly accounted for in the statistical analysis through the weighting in the fitting routine.

Because the expected error per $G$ gate was as high as $10^{-1}$, the expected error per step was also of order $10^{-1}$.  Sequences of length greater than $10$ would reveal little statistical information in this scenario, especially relative to the time required for their implementation.  In practice however, it was found that an unknown experimental complication was causing ions to be lost from the trap at a high rate for sequences of length greater than $6$.  Other consequences of this peculiarity are discussed below.  Thus, the chosen set of lengths was unfortunately limited to $\{1,2,3,4,5,6\}$.

At $~7s (8s)$ per $100$ instances of each standard (standard+$G$) sequence on average, it was reasonable to design $\sim 50$ sequences of each length for each sequence type.  The actual published experimental run was performed in a less organized way: because of concerns about ion loss for longer sequences, the experiment proceeded in stages where first sequences of length $1-3$ were performed, then $1-4$, $1-5$, and finally $1-6$.  The sequences used for the standard+$G$ benchmark are actually the same as the standard sequences, with the exception of the extra $G$ gate after each step, which also changes the required inverting operator.  The relative benefits of choosing the second set of sequences in this way as opposed to independently at random have not been thoroughly investigated.  At each stage a roughly equal number of sequences of each length were used, so $\{45, 55, 53, 39, 28, 15\}$ sequences were performed for each respective length (these numbers are similar for the standard+$G$ benchmarks, except for anomalies due to lost ions).  This is acceptable because the analysis procedure developed in Sec.~\ref{Sec:2Analysis} does not assume an equal number of sequences for each length.  The experimental parameter choices are summarized in Table~\ref{Tab:3Exp2}.

\begin{table}
\begin{center}
\begin{tabular}{|c|c|c|c|c|}
\hline
$T_t$ & $t_s$ & $n_e$ & $n_l$ & $\{l_i\}$ \\
\hline
1.75 hr & $4.5 (7.5)$ ms & $100$ & $\{45, 55, 53, 39, 28, 15\} (\{46, 54, 53, 38, 28, 15\})$ & $\{1,2,3,4,5,6\}$ \\
\hline
\end{tabular}
\end{center}
\caption[Experimental Design Parameters for the Two-Qubit Benchmark]{Presented are the experimental design parameters chosen for the two-qubit benchmark of \cite{Gaebler2012}.  Notation for the parameters is described in Sec.~\ref{Sec:2ExpDesign}.  When the parameters are different for the standard and standard+$G$ benchmarks, the latter value is in parentheses.  \label{Tab:3Exp2}}
\end{table}

In addition to the two two-qubit benchmarks, a simultaneous one-qubit benchmark was performed on both of the individual qubits.  I describe this experiment in less detail, as it was designed only to support the two-qubit benchmark.  The one-qubit benchmarks were designed using the same approximate-twirling benchmark procedure utilized in Sec.~\ref{Sec:2Exp1}, except that the computational gates were chosen from $\{X \left (\pm \frac{\pi}{2} \right ), Y \left (\pm \frac{\pi}{2} \right ), I\}$, each with probability $0.2$.  The benchmarks were performed at the same time, and each step included a recombination of ions into the same zone followed by a separation, so that the error per step also describes the error accrued in this movement process.  Again $n_e=100$, but this time $n_l \approx 12$ for each $l \in \{2,3,4,6,8,12\}$.

\subsection{Results} \label{Sec:3ExpResults}
\begin{figure} 
    \begin{center}
	\includegraphics[width=140mm]{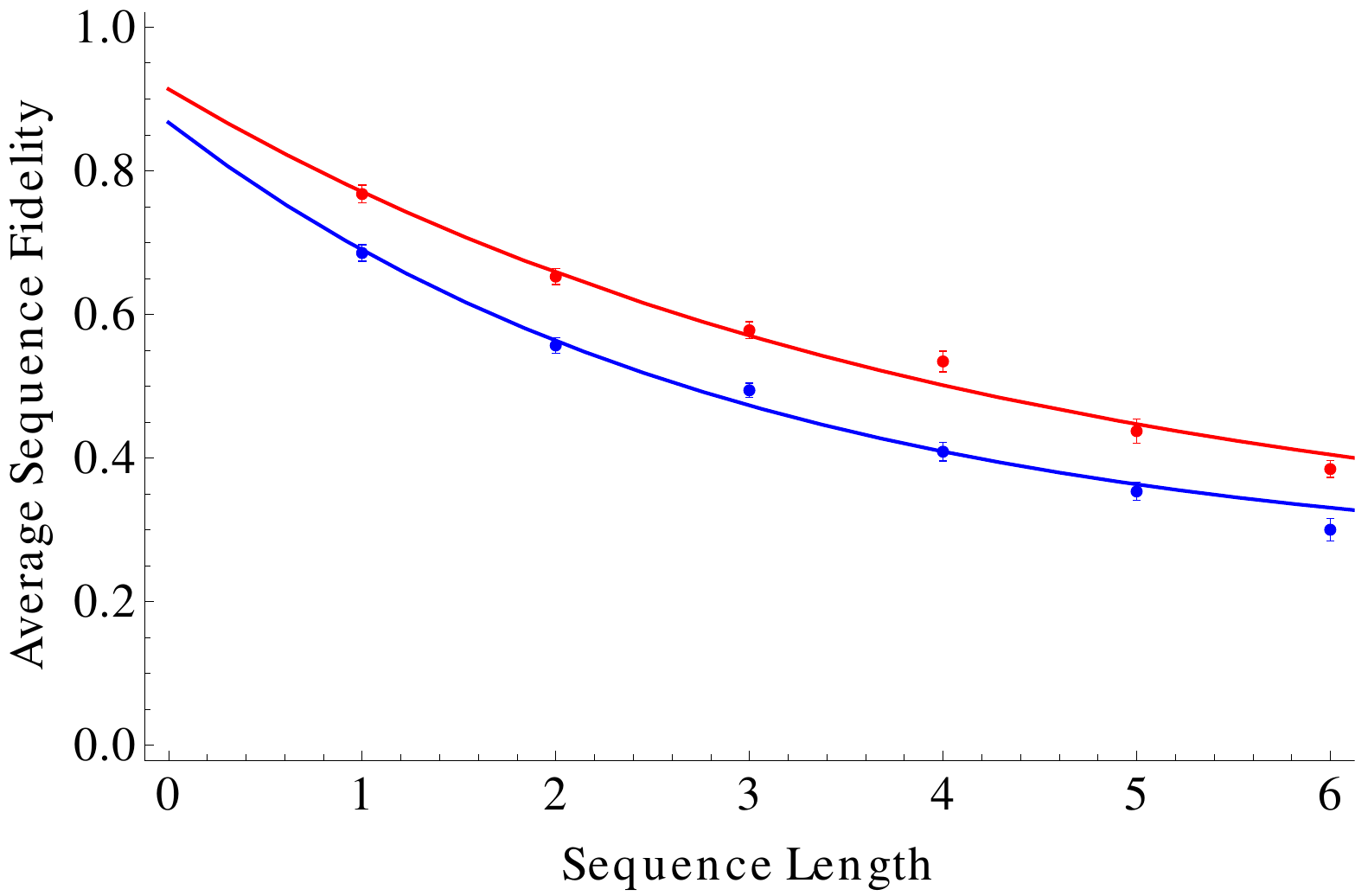}
    \end{center}
\caption[Benchmarking Results for the Two-Qubit Experiment]{Plot of the main benchmark results for the two-qubit experiment.  Both primary (red) and primary + $G$ (blue) experiments are depicted.  The mean sequence fidelities for each of the sequence lengths are shown along with error bars that depict the standard errors of the means.  The best fits to Eq.~\ref{Eq:2Main}, corresponding to the reported benchmark parameters, are plotted as solid curves. \label{Fig:3ExpResults}
} 
\end{figure}

The results of the two two-qubit randomized-benchmarking experiments are depicted in Fig.~\ref{Fig:3ExpResults}.  The benchmark parameters are $\epsilon_s = 0.162(8)$ and $\epsilon_m = 0.086(22)$ for the primary benchmark and $\epsilon_s' = 0.216(8)$ and $\epsilon_m' = 0.132(26)$ for the primary+$G$ benchmark.  Solving for the error per $G$ gate reveals $\epsilon_G = .069(17)$.  The values of the $\chi^2$ statistic for the two fits are $\chi^2 = 9.28$ and ${\chi^2}' = 9.48$, respectively, which are both (barely) insignificant at the $p=.95$ level for a fit with $4$ degrees of freedom.  

\begin{figure} 
    \begin{center}
	\includegraphics[width=140mm]{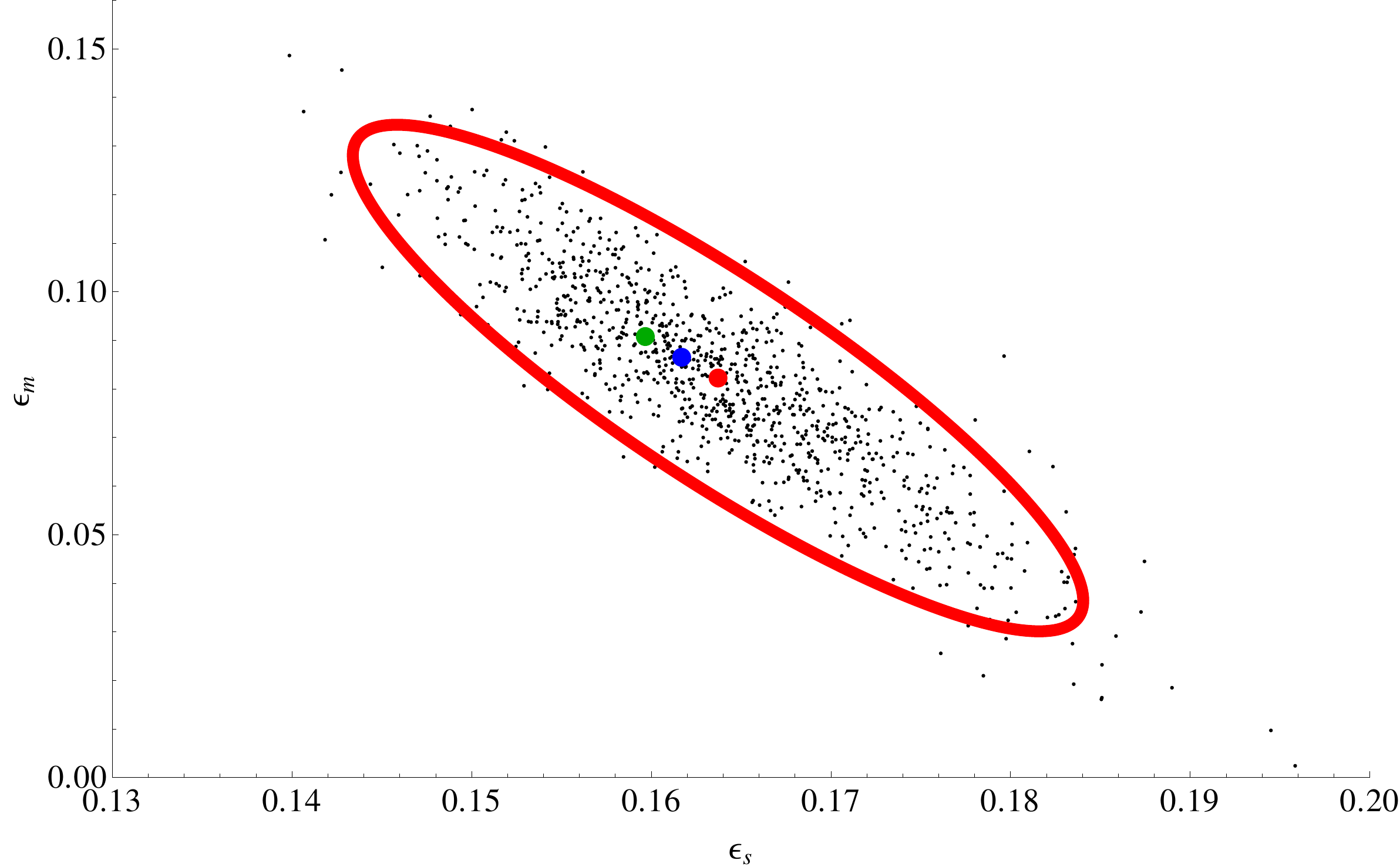}
    \end{center}
\caption[Resampling Results for the Two-Qubit Experiment]{
Plot of the benchmarking parameters for the primary benchmark on two qubits extracted from $1000$ resampled data sets.  An elliptical region containing $95\% $ of the resampled parameters is depicted and estimates a $95\% $ confidence region for the original parameter estimates.  Plotted as larger points are the locations of the benchmark parameters returned by the original fit (blue), the means of the resampled values (red), and the bias-corrected estimates (green).  Covariance of the two parameters can be observed in the tilt of the ellipse. \label{Fig:3Resamples}
}
\end{figure}

\begin{figure} 
    \begin{center}
	\includegraphics[width=120mm]{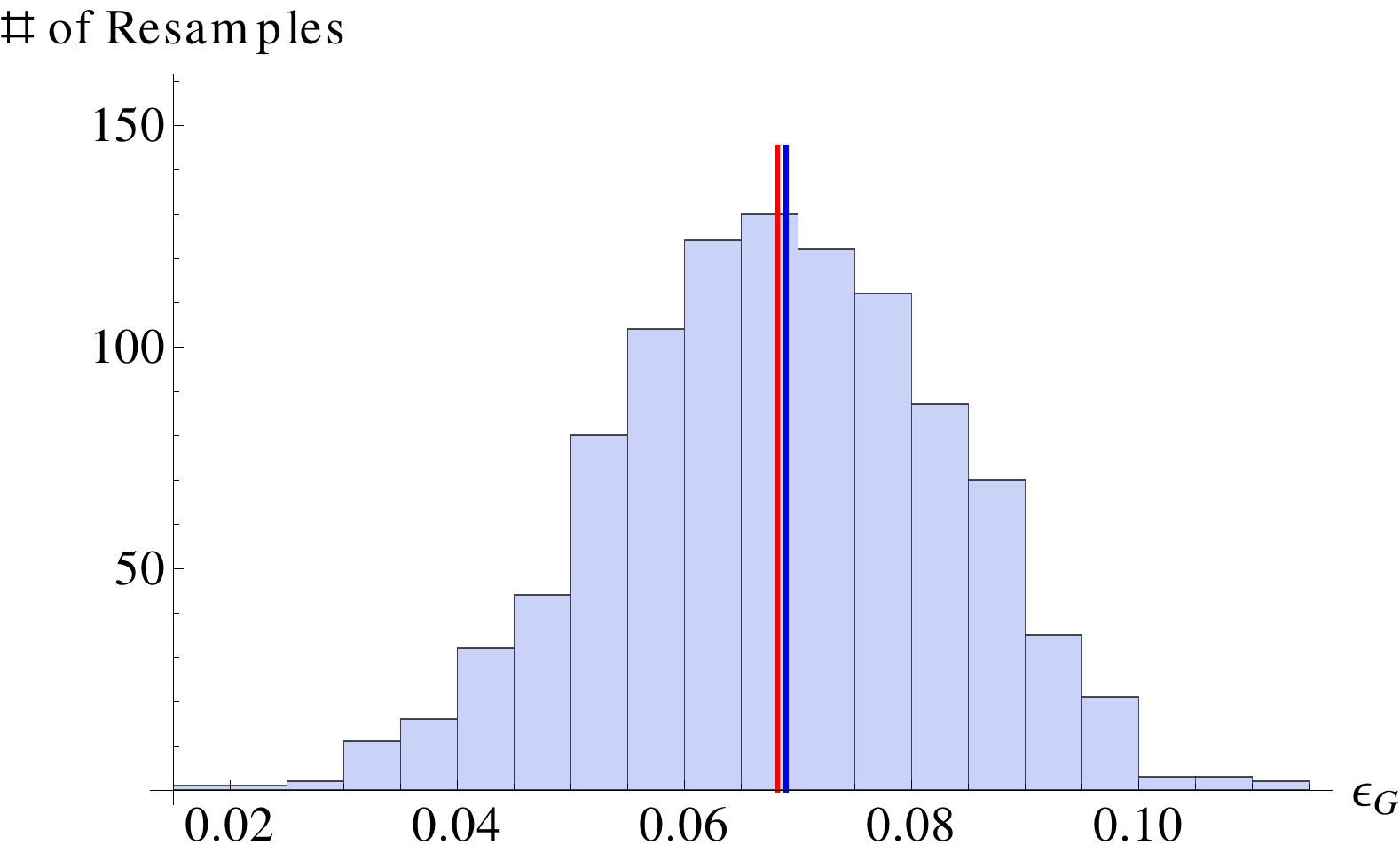}
    \end{center}
    \caption[Resampled $\epsilon_G$ for the Two-Qubit Experiment]{
Histogram of the estimates of the error per $G$ from resampled data set pairs.  This plot uses the same $1000$ resampled data sets as Fig.~\ref{Fig:3Resamples} paired with resampled data sets from the primary + $G$ experiment.  The best fit from the original data is marked by the red line, while the sample mean from the resamples is marked by a blue line. \label{Fig:3ErrorPerGHist}
	}
\end{figure}

Semi-parametric resampling helps to describe the expected errors on the parameter estimates.  Using the same technique described in Sec.~\ref{Sec:2Resampling} with $1000$ resampled data sets returns the data for the primary benchmark depicted in Fig.~\ref{Fig:3Resamples}.  Again there is a fairly strong covariance evident between the two parameters, and again the bias does not appear to be very significant.  By simultaneously resampling for the primary and primary+$G$ data sets, it is possible also to derive resampled estimates of $\epsilon_G$.  Fig.~\ref{Fig:3ErrorPerGHist} shows these estimates in histogram form.

\begin{figure} 
    \begin{center}
	\includegraphics[width=120mm]{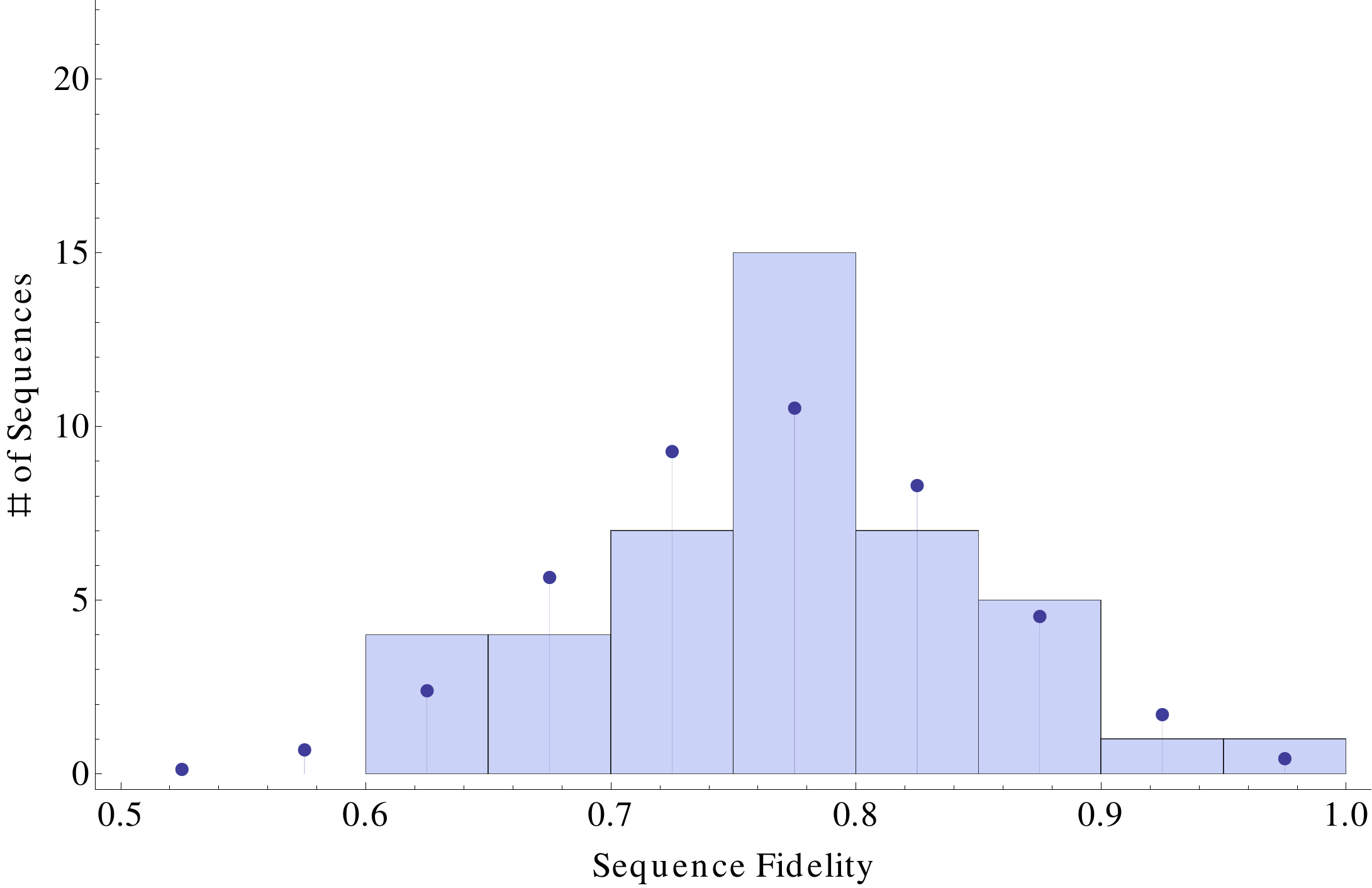}
    \end{center}
\caption[Sequence Distribution for the Two-Qubit Experiment] { \label{Fig:3SeqDistGaeb}
Histogram binning the number of length-$1$ sequences with observed sequence fidelity within a given range for the $2$-qubit experiment.  The points correspond to the integral over each bin of a Gaussian distribution times the number of sequences.  The distribution is much closer to Gaussian than that of Fig.~\ref{Fig:2SeqDistKen}.
}
\end{figure}

Fig.~\ref{Fig:3SeqDistGaeb} depicts the distribution of fidelities for individual length-$1$ sequences of the primary benchmark.  Compared to the one-qubit benchmark of Sec.~\ref{Sec:2Exp1}, this distribution does not show strong signs of non-Gaussianity, although it might be slightly more peaked than a Gaussian distribution.  Because the sequences described in this figure contain between zero and six $G$ gates, and the $G$ gates have the highest error rate, this distribution might be expected to show six peaks.  However, the optimal implementation of the inverse of a Clifford operator requires the same number of $\CX$ gates as the optimal implementation of the original operator, and so the possible numbers of $\CX$ gates in these sequences are $\{0,2,4,6\}$.  The distribution would then have $4$ equally spaced peaks with different amplitudes related to the numbers in Table~\ref{Tab:ACNOT}.  It is difficult to tell whether this is a notably better description of the data than a Gaussian model.

Despite the results of the goodness-of-fit tests, there are several reasons to be suspicious about whether the errors in this experiment were time-independent.  First, it is unusual, but not impossible, that the SPAM error $\epsilon_m$ is smaller than the error per step $\epsilon_s$ in both cases.  The SPAM error represents the error in the inversion step, among other things, so it should be at least as large as the error per step unless coherent errors in the step and measurement are canceling consistently.  Second, $\epsilon_m$ is expected to be close to $\epsilon_m'$ because the experiments are performed in a random order and the SPAM processes were the same for both experiments; there is no reason to think that the preparation, measurement, or inversion step (which does not include an extra $G$ gate) should be different for the second set of sequences.  Third, there is a visible trend toward the data from the long sequence lengths falling below the fit line consistently in both benchmarks reported here and in a previous data set not included in this analysis.  Finally, unlike the experiment described in Sec.~\ref{Sec:2Exp1}, this experiment has results that appear to be fit worse by Eq.~\ref{Eq:2Main} when the data from the shortest sequence lengths are ignored.

The fact that ion-loss events happen more frequently as the sequence lengths increase is a clear sign of time dependence in the experiment, although the source of this time dependence is unclear.  In this case, the relevant time interval is apparently the time since the preparation of the qubit, not a global time.  Many of the suspicious statistical signatures mentioned previously are also consistent with this sort of time dependence.  In fact, the simple model of time-dependent errors depicted in Fig.~\ref{Fig:2TimeDep} and described in the surrounding text replicates two of these signatures: the $y$-intercepts of the fit curves are inconsistent with the error-per-step (leading to an underestimation of the SPAM errors) and the longer-sequence data points fall below the fit curves.  In contrast, those curves also do not fit the short sequences well, whereas the short sequences are fit well for the actual experiment.  

If there is a time-dependence, it might be expected that any of its effects would be more prominent in the standard+$G$ benchmark, which takes longer for sequences of the same length due to the additional, slow $G$ gates.  This might explain why the SPAM error parameters are different for the two benchmarks.  

Finally, and perhaps most importantly, calculating the error per step using different subsets of the sequence lengths reveals that the shorter length sequences indicate a smaller error per step than the longer-length sequences, as shown in Table~\ref{Tab:3Truncations}.  The differences here are just about at the limit of what might be considered significant, especially since the data set is small to begin with.

In the end, it might be best in scenarios where time dependence is a potential problem to simply provide a range of errors per step.  The ion-loss events are a clear indication that some sort of equilibrium is not being achieved for some aspect of the qubits and apparatus.  Because the randomized benchmark is supposed to describe a steady-state behavior, this failure should be noted.  By providing the range of errors per step and errors per gate based on different truncations (as in Table~\ref{Tab:3Truncations}), we both acknowledge this limitation and provide some indication of the time dependence trends.

\begin{table}
\begin{center}
\begin{tabular}{|l|c|c|c|c|c|}
\hline
& 1-4 & 1-5 & 1-6 & 2-6 & 3-6 \\
\hline
$\epsilon_s$ & .144 & .151 & .162 & .165 & .185 \\
\hline
$\epsilon_m$ & .12 & .106 & .086 & .074 & -.037\\
\hline
$\epsilon_s$ & .201 & .209 & .216 & .224 & .275\\
\hline
$\epsilon_m$ & .16 & .15 & .133 & .101 & -.218\\
\hline
$\chi^2$ p-value & .70 & .80 & .94 & .97 & .94 \\
\hline
$\chi^2$ p-value & .79 & .83 & .94 & .96 & .34 \\
\hline
$\epsilon_G$ & .071 & .072 & .069 & .076 & .12\\
\hline
\end{tabular}
\end{center}
\caption[Benchmark Parameters for Truncated Sequences of Two-Qubit Benchmark]{Table of the benchmark parameters returned by fitting different subsets of sequence lengths for the two-qubit ion-trap experiment.  The values for both the primary and primary+$G$ experiments are included, as well as the $\epsilon_G$ inferred from these parameters.  The $p$ values corresponding to the $\chi^2$ statistics for the fits are also included; values above $0.95$ indicate a significantly bad fit of the data. \label{Tab:3Truncations}}
\end{table}

\begin{figure} 
    \begin{center}
	\includegraphics[width=140mm]{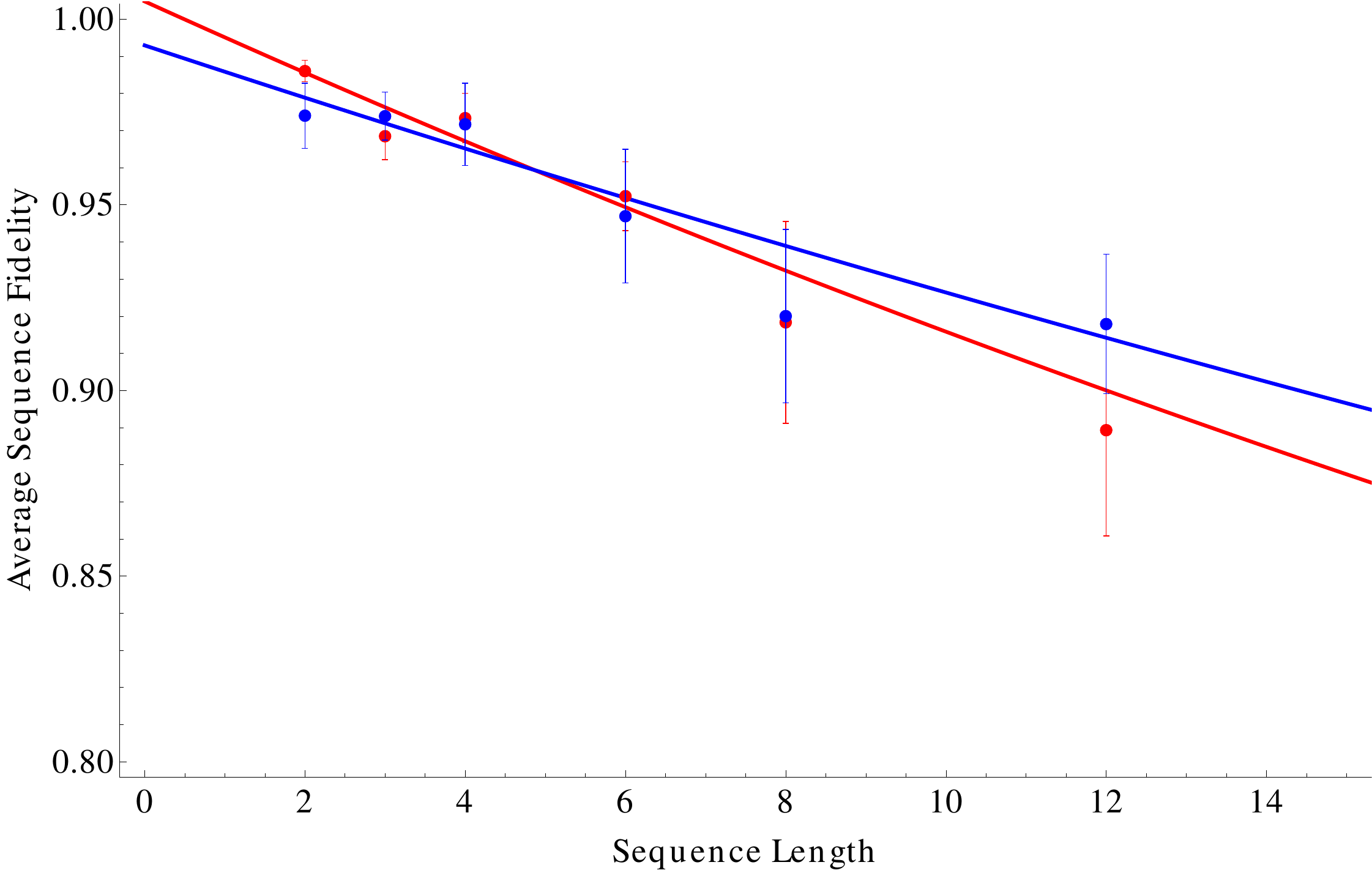}
    \end{center}
\caption[One-Qubit Benchmarks for the Two-Qubit Experiment]{Plot of the main benchmark results for the parallel one-qubit RB experiments performed with the two-qubit apparatus.  The red and blue data represent the first and second Be ions, respectively.  The mean sequence fidelities for each of the sequence lengths are shown along with error bars that depict the standard errors of the means.  The best fits to Eq.~\ref{Eq:2Main}, corresponding to the reported benchmark parameters, are plotted as solid curves. \label{Fig:3OneQubit}
} 
\end{figure}

The results of the parallel one-qubit benchmarks are depicted in Fig.~\ref{Fig:3OneQubit}.  The fit parameters are $\epsilon_{s,1} = .010(2)$ and $\epsilon_{m,1} = -.005(4)$ for the first qubit and $\epsilon_{s,2} = .07(2)$ and $\epsilon_{m,2} =.007(7)$ for the second qubit.  In comparison with the one-qubit control demonstrated in Sec.~\ref{Sec:2Exp1}, this system experiences significantly stronger errors except in the measurement, where it has higher fidelity.  This is not surprising: the added ion movement, in addition to the difficulties of laser compared to microwave control, make this experiment more complicated.

The distribution of gates in each step of the one-qubit benchmarks is different than the standard set by Ref.~\cite{Knill2008}.  If the error were purely a function of the number of effective $\frac{\pi}{2}$ gates in the steps, then the result of this experiment should be renormalized by a factor of $\frac{2}{1.8}$ to $\epsilon_{s,1}' = .011(2)$ and $\epsilon_{s,2}' = .008(2)$ for comparison with the standard one-qubit benchmark.  As mentioned at the beginning of this chapter, strict comparisons of the two benchmarks are not possible without such an assumption.  

An interesting check on the consistency of these experiments is provided by comparing the error per step for the two-qubit experiment to a sum of $\epsilon_G, \epsilon_{s,1}, $ and $\epsilon_{s,2}$ weighted by the number of $G$ gates and one-qubit ``steps'' that make up a step of the two qubit experiment.  However, this check is not necessarily valid if the errors are coherent and the Clifford operators are not randomly constructed, as the depolarization of the composition of errors is not likely to be the composition of their depolarizations in this case.  In order to make the comparison, I use the numbers of $G$ and one-qubit $\frac{\pi}{2}$ pulses, $1.5$ and $6.5$, respectively, to weight $\epsilon_G$ and $\frac{\epsilon_{s,1}+\epsilon_{s,2}}{2*1.8}$, where the latter value represents the averaged error per one-qubit step divided by the number of effective
$\frac{\pi}{2}$ gates in the step.  However, a depolarizing error on one qubit is not the same as a depolarizing error on two qubits.  The depolarization on $n$ qubits of a $k$-qubit depolarizing channel with strength $p_k$ has strength 
\begin{equation} \label{Eq:3DepConv} 
p_n = \frac{4^k-1}{4^k}\frac{4^n}{4^n-1}.
\end{equation}
Therefore, the final estimation of the error per step of the two-qubit experiment using these constituent parts is 
\begin{equation} \label{ConCheck}
 1.5*\epsilon_G + \frac{6}{5} * 6.5 * \frac{\epsilon_{s,1}+\epsilon_{s,2}}{2*1.8} = .136.
\end{equation}
By propagating through the standard error on the two error-per-step estimates, an error of $.02$ is estimated for this sum.  This estimate is notably lower than $\epsilon_s$ by about one standard error; if this is not simply statistical noise, it could be an indication of coherent errors or a difference in time-dependence between the one- and two-qubit experiments.

The dominant error sources for this experiment were difficult to isolate.  Spontaneous emission caused by the Raman lasers used to drive the gates contributes at approximately the $.013$ and $.001$ levels to the error per $G$ and error per one-qubit step, respectively.  This error is incoherent and corresponds to an amplitude-damping channel when the decay to the ground state is fast.  

Large sources of coherent error include the intensities and relative phases of the lasers, in addition to the mismatch between laser frequency and qubit transition frequency due to Stark shifts.  Estimates of these errors from calibration data indicate that their combined effect is at the $.03$ level for the error per $G$ gate. 

 These error estimates, and several others, can be summed incoherently to give an estimate of $.048$ for the error per $G$ gate; this estimate lies below the estimate provided by the randomized-benchmarking experiment by more than the resampled error estimates.  If this discrepancy is real, and not a product of statistical noise or un-noticed systematic effects, then it might be traced back to one or more of the following: time dependencies in the gates not included in diagnostic tests, error contributions that add coherently, or mis-estimation of the benchmark due to the violation of some of the benchmark assumptions.

\subsection{Conclusions} \label{Sec:3ExpConclusions}

The two-qubit benchmarking procedure demonstrated in this experiment is interesting in several ways.  First, it provides estimates of the error per random, two-qubit Clifford gate and error per $G$ gate that are approximately consistent with other (also imperfect) estimation techniques based on calibration data and simpler tests.  Refinement of the randomized-benchmarking experiment and the error estimates of individual sources would be necessary to decide which technique is more accurate in this case.  However, the procedure as a whole is considered a success, especially given its relative ease. 

Second, the statistical analysis developed for this experiment actually finds some reasonable, but not conclusive, evidence that the assumptions required for the ideal benchmark (specifically the assumption of time-independence) are not well-satisfied by the experiment.  This observation, and the possibly related issue of ion-loss, point to aspects of the experiment that were not understood at the time, and can therefore be seen as a worthwhile diagnostic.  Finally, in several ways the analysis of this experiment needed more data.  I take this as an indication that the number of lengths used and the number of sequences per length are near the lower bound of what is necessary for a successful randomized-benchmarking experiment.  Particularly, the limitation to sequences of length $6$ or less was vexing from both the experimental and analytic points of view; even one or two additional sequence lengths might have answered the open questions about time-dependence and the various consistency checks.

\chapter{Further Extensions}
\label{Chap:4}

Use of the exact-twirl approach described in Chap.~\ref{Chap:3} greatly simplifies the analysis of randomized-benchmarking experiments.  This simplification makes it easy to construct the individual-gate benchmarks described in Sec.~\ref{Sec:3GateBenchmark} and Ref.~\cite{Magesan2012}.  In this section I describe two other extensions of randomized benchmarking that use the framework of exact twirling.  First, I present the idea of RB of encoded qubits.  Encoding provides an approach to scaling RB to much larger numbers of qubits, but, more importantly, it also allows for benchmarking in the context of fault-tolerance architectures.  Second, I introduce a subgroup of the one-qubit Clifford group and a class of subgroups of the $(n>1)$-qubit Clifford groups that have the same twirling behavior as the entire Clifford group.  Such subgroups may be used to simplify individual-gate benchmarking.  Finally, I return to approximately-twirled benchmarks to place bounds on the comparison between the error per step in an approximately-twirled experiment and the error per step that would be found using the exact twirl.

\
\section{Encoded Randomized Benchmarks} \label{Sec:4EncodedBenchmark}

As the number $n$ of qubits grows large and the classical computation aspects of the design of standard RB experiments grow difficult, an alternative approach to benchmarking $n$ qubits is to encode them in quantum codes or decoherence-free subspaces and benchmark the encoded (logical) qubits.  There are two appealing aspects of this approach: First, by encoding $k$ logical qubits into $n$ physical qubits at a rate $\frac{k}{n}<1$, one reduces the size of the Clifford group $\mathcal{C}_k$ required for randomization, which reduces the difficulty of the classical computations involving the Clifford group.  Second, and more important, a description of the average performance of encoded qubits is a useful figure of merit for large-scale quantum computers.  When encoded qubits perform better than bare (unencoded) qubits, the system has passed a fault-tolerance threshold and is suitable for integration into a fault-tolerant quantum-computing (FTQC) architecture.  Before elaborating on these advantages and speculating on the resource requirements for a minimal encoded-benchmarking experiment, I introduce some relevant information about quantum error-correcting (QEC) codes and decoherence-free subspaces (DFS).

\subsection{DFS and QEC} \label{Sec:4DFSQEC}

For using decoherence-free subspaces, the goal is to find within an $m$-qubit Hilbert space (of dimension $2^m$) a $k$-qubit subspace that is protected from a certain type of error.  A canonical example is the one-qubit subspace of two qubits spanned by $\ket{\overline{0}} \equiv \ket{01}$ and $\ket{\overline{1}} \equiv \ket{10}$.  Consider an experiment that suffers only from global phase errors.  These (unitary) errors transform $\ket{0} \rightarrow e^{i \zeta} \ket{0}$ and $\ket{1} \rightarrow \ket{1}$ on each of the qubits, where $\zeta \in \mathbb{R}$ is the phase.   The encoded subspace can be seen to be unaffected by the errors; that is, any superposition of the encoded states is transformed as $\alpha\ket{01} + \beta\ket{10} \rightarrow e^{i \zeta} \left (\alpha \ket{01} + \beta\ket{10} \right )$.  Because the overall phase $e^{\imath \zeta}$ is not important, this error acts as the identity operator on the states in this subspace and does not affect the information they represent.  Treatments of decoherence-free subspaces with many more interesting examples can be found in Refs.~\cite{Lidar2012,Viola2000}.

In order to compute with qubits encoded in a decoherence-free subspace, it is necessary to find operators that generate the $k$-qubit unitary group on the encoded qubits.  These operators can be more complicated than the bare operators. In the simple example of the previous paragraph, a Pauli $Z$ gate is implemented by performing a bare $Z$ gate on the first physical qubit while a Pauli $X$ gate is implemented by performing bare $X$ gates on both of the physical qubits.  The degree to which these encoded gates are convenient to implement helps to determine the suitability of a particular DFS in practice.  If the encoded gates introduce errors that more than compensate for the error protection afforded by the DFS, then the DFS is not useful.  By benchmarking the performance of gates before and after encoding into a DFS, it would be possible to demonstrate the utility (or lack thereof) of the encoding. 

For error-correcting codes, the situation is slightly more complicated.  For simplicity, I discuss here only stabilizer codes for qubits; see Sec.~\ref{Sec:AStabilizer} or Ref.~\cite{Gottesman1997} for information about these codes.  Like a DFS, an error-correcting code is a distinguished $k$-qubit subspace of an $n$-qubit Hilbert space.  Instead of passively protecting the encoded information from a particular kind of error (as with a DFS), the error-correcting code is structured such that measuring certain operators (called stabilizers) detects whether one of a class of errors has occurred to the system.  Gates can then be used to correct the error and return the encoded qubit to its proper state.  

As with decoherence-free subspaces, in order to perform universal quantum computation with the encoded qubits, one must identify operations at the physical-qubit level that implement a generating set of unitary operators on the encoded qubits.  There are many strategies for performing these operators, but a common goal is to avoid having a Pauli operator error with support (non-identity action) on one physical qubit prior to the gate transform into a Pauli error with support on more than one physical qubit within the code through the implementation of the operator.  Performing operators in this way keeps the errors from ``spreading'' and turning into errors that cannot be corrected by the error-correction procedure.  This strategy is a fundamental part of FTQC.  Again, the encoded error rate introduced by these encoded gates must be lower than the error rate introduced by the physical-level gates if the error-correcting code is to be useful.  This could be demonstrated robustly by quantifying the fidelity of the bare and encoded gates using randomized benchmarking.

\subsection{Encoded-Benchmarking Procedure} \label{Sec:4EncBenchmarkProc} 

The fundamental structure of the randomized-benchmarking experiment is unchanged by qubit encoding.  For $k$ encoded qubits, sequences of random $k$-qubit Clifford operators are constructed.  Exact or approximate twirls might be used, but I focus here only on exact-twirl experiments.  Each Clifford operator must be decomposed as before into a generating set of encoded gates.  These encoded gates can be quite complicated, adding another layer of complexity to the experiment and increasing the required time. However, given implementations of a generating set of encoded Clifford operators, any Clifford operator can be constructed by use of Clifford decomposition algorithms. The actual measurement of an encoded qubit may be performed using a fault-tolerant measurement strategy or by simply decoding the state onto the first physical qubit and measuring it in the usual basis.   The measurement strategy need not be optimized for RB as long as it has a small enough error rate that the fidelity decay for increasing sequence lengths can be distinguished from the measurement noise.

The assumptions of time- and gate-dependence of errors are the same for the encoded benchmark as they are for the un-encoded benchmark.  However, the assumption that the errors preserve the (encoded) qubit subspace is more likely to fail in the encoded benchmark, especially for qubits encoded into a DFS.  Any error that mixes the encoded qubit subspace and the remainder of the Hilbert space violates this assumption.  Because no twirling is performed intentionally on the parts of the Hilbert space that are not identified with the encoded subspace, such errors cannot be assumed to be depolarized and can cause non-exponential fidelity decay.  This problem is less prevalent in QEC codes because each Clifford operator in such a scheme should be followed by an error-correction step.  This step returns the state into the encoded qubit subspace (to the extent that the correction is ideal), so that the aggregate error of the step will preserve the encoded qubit subspace.  However, many unitary errors on a DFS mix the encoded and un-encoded subspaces; the utility of the DFS depends on these errors being very improbable for physical reasons.

\subsection{Simple Encoded Experiments} \label{Sec:4MinimalExp}

The minimal resources required to implement an actual encoded-benchmarking experiment include not only the gates and qubits required for the QEC code or DFS, but also those required to implement the encoded gates and encoded measurements fault-tolerantly.  The $[[5,1,3]]$ code encodes one logical qubit in five physical qubits, and is the smallest code that allows for correction of any error with support on at most one physical qubit.  Ref.~\cite{Gottesman1997} provides a full description of the $[[5,1,3]]$ code and the terminology used to describe stabilizer codes. In order to implement the entire Clifford group fault-tolerantly on this code,  some operators are typically implemented using ancillary qubits.   However, a randomized-benchmarking experiment using the subgroup $\mathcal{T}$ of the Clifford group might be conveniently implemented for this code using the prescription of Sec.~\ref{Sec:4SmallDesigns}.  

The $[[7,1,3]]$, or Steane, code, while requiring two more physical qubits per encoded qubit, has the feature that all Clifford operators can be performed transversally.  This means that no additional qubits (ancillas) are needed to implement Clifford operators.  In order to perform the error-correction step for this (or any other) code, however, several measurements must be made that require ancillas.  For standard Steane error correction, error detection would require using one $\ket{\overline{0}}$ ancilla at a time, at a cost of $7$ additional physical qubits (see Ref.~\cite{Cross2007} for a description of error-detection strategies for different codes).  This number could be reduced by using a more tailored error-correction scheme.

If the simplest strategy of decoding the code and measuring is implemented, then an encoded benchmark using the Steane code could be implemented using $14$ qubits ($7$ for the code and $7$ for the detection ancillas).  In such an experiment, error detection and correction would be performed at the end of each step of the RB sequences.  This would require re-preparing the $\ket{\overline{0}}$ ancilla with high fidelity during each step, which could significantly add to the time required for the step.  

Similarly, a RB experiment with two encoded qubits could be implemented using $21$ physical qubits.  After each step, the detection operations for the two qubits can be performed serially, so the 7 qubits required for the logical ancilla can be reused for every detection.  The trade-off for requiring so few qubits is that the experiment must wait for the re-preparation of the ancillas.  This slows down the experiment and can introduce errors on the qubits if the fidelity of the storage (or identity) operation is not very high.  By using two encoded qubits, it would be possible to compare the bare $\CX$ gate fidelity to the encoded $\CX$ gate fidelity using an individual-gate benchmark; demonstrating an improvement in $\CX$ fidelity through encoding in this experiment would be an experimental milestone.  However, the bare error-per-$\CX$ threshold of $\sim \! 10^{-4}$ (see Ref.~\cite{Cross2007} for details) for achieving this improvement is beyond the reach of current experiments.  

If gate fidelity were the major limiting factor for an experiment, then a minimal encoded-benchmarking experiment might be performed using a small instance of the surface codes developed by the authors of Refs.~\cite{Raussendorf2007,Fowler2009} and others.  At gate error rates in the $10^{-3}$ regime, these codes break even at improving $\CX$ gates using a number of qubits in the $30-100$ range, depending on the particular surface-code implementation and error-correction scheme used.

It is more difficult to estimate the parameters needed to design a DFS-based encoded-benchmarking experiment.  At a minimum, only four qubits might be needed to encode two logical qubits in the toy DFS given above.  Whether such a DFS would be able to demonstrate improvement in $\CX$ gate error rate depends on both the global phase error (the error to which the protected subsystem is immune) being very high and all other errors (the errors that affect the subsystem) being very small.  Previous experiments using triple quantum dots \cite{Laird2010} might already be considered to operate on a DFS.  To demonstrate improvement from bare to DFS-encoded qubits, it must be possible to perform RB on the un-encoded qubits; this is difficult because control of the individual quantum dots has an unmanageable error rate.  Regardless, a successful RB experiment on such a system might be considered a demonstration of encoded RB.

\section{Clifford-Subgroup Twirls} \label{Sec:4SmallDesigns}

The crucial aspect of using Clifford operators for the randomized benchmark is the action of the Clifford twirl.  However, averaging over the full Clifford group is not necessary to achieve this effect.  In this section I present Clifford operator subgroups whose exact (uniform) twirls have the same effect on an error superoperator as the Clifford twirl, and I discuss how these subgroups might be used to simplify the benchmarking of individual gates.

\subsection{Required Twirl Property} \label{Sec:4TwirlProp}

Let $S$ be a super-superoperator.  The property that $S$ depolarizes any superoperator $\hat{\Lambda}$ can be written as
\begin{equation} \label{Eq:4TwirlProp}
S(\hat{\Lambda}) = (1-p)(I \otimes I) + \frac{p}{D^2} \sum_{P \in \mathcal{P}_n} (P \otimes P)
\end{equation}
for $p = \frac{D^2 - \tr{\hat{\Lambda}}}{D^2-1},$ as shown in Sec.~\ref{Sec:2Dep}.  In that section, I also show that the Clifford twirl can be decomposed into a Pauli twirl followed by a twirl over $\overline{\mathcal{C}}_n$ and  that this composition of twirls satisfies the property above.  

By itself, the Pauli twirl transforms any superoperator $\hat{\Lambda}$ into a superoperator of the form $\sum_{P \in \mathcal{P}_n} \gamma_P P \otimes P$.  Such a superoperator is sometimes called a (stochastic) Pauli channel, since it applies each Pauli operator to the system with some probability.  In the constructions that follow, a Pauli twirl will always be performed before any other twirl, so one can assume that any superoperator has the form of a Pauli channel before the second twirl is applied.

In order for a super-superoperator $S$ to transform a Pauli channel into a depolarizing channel, it is sufficient for $S(P_j \otimes P_j) = \frac{1}{D^2-1} \sum_{P_i \in \mathcal{P}_n, i \neq 0} (P_i \otimes P_i)$ for each $P_j \in \mathcal{P}_n, j \neq 0$.  Such an $S$ balances the weights $\gamma_P$ of all of the tensored Pauli terms (except for the identity), mapping a Pauli channel into a superoperator of the form in Eq.~\ref{Eq:4TwirlProp}.  I consider here only super-superoperators $S$ with the form $S(\hat{\Lambda}) = \frac{1}{|K|} \sum_{C \in K} \hat{C}^{\dag} \hat{\Lambda} \hat{C}$, that is, $S$ is a uniform twirl over a (sub)set of Clifford operators $K$.  In order to satisfy the Eq.~\ref{Eq:4TwirlProp}, it is sufficient for $S$ to satisfy
\begin{equation} \label{Eq:4ClifTwirlProp}
S(P_j \otimes P_j) = \frac{1}{|K|} \sum_{K} C^{\dag} P_j C \otimes C^{\dag} P_j C = \frac{1}{D^2-1} \sum_{P_k \in \mathcal{P}_n, k \neq 0} (P_k \otimes P_k),
\end{equation}
for all $P_j \in \mathcal{P}_n, j \neq 0$.  

The transformations $C^{\dag} P_j C$ in Eq.~\ref{Eq:4ClifTwirlProp} describe the conjugation action of a Clifford operator on a Pauli matrix as described in Sec.~\ref{Sec:AClifDef}, although the sign involved in that action is squared and therefore irrelevant.  Using the terminology of this action, a sufficient condition can be stated that guarantees that a twirl over a subset of Clifford operators (composed with a Pauli twirl) has the form of Eq.~\ref{Eq:4TwirlProp}:

\begin{lemma} Suppose $K$ is a subset of Clifford operators such that the number of $C \in K$ for which $C(P_j) = P_k$ or $-P_k$ is the same for all $j,k \neq 0$.  Then the composition of a $K$-twirl and a Pauli twirl depolarizes any superoperator as described in Eq.~\ref{Eq:4TwirlProp}. \label{Lem:4TwirlCondition}

\paragraph{Proof:}  Call the superoperator implementing the $K$-twirl $S_{K}$.  The action of  $S_{K}$ on any superoperator of the form $P \otimes P$ with $P \neq I$ is 
\begin{eqnarray} \label{Eq:4Lem3Proof}
S_{K}(P \otimes P) &=& \frac{1}{|K|}\sum_{C \in K} C P C^{\dag} \otimes CP C^{\dag} \\
&=& \frac{1}{|K|}\sum_{C \in K} C(P) \otimes C(P) \\
&=& \frac{1}{|\mathcal{P}_n|}\sum_{P_j \in \mathcal{P}_n, j \neq 0} P_j \otimes P_j,
\end{eqnarray}
where the last equality requires the property of $K$ given in the statement of the lemma.  This last expression shows that the twirl satisfies the condition of Eq.~\ref{Eq:4ClifTwirlProp}.  This condition was shown to be sufficient to prove that the composition of $S_{K}$ and a Pauli twirl depolarizes any superoperator. $\square$
\end{lemma}

This condition may be minimally satisfied if there is exactly one Clifford operator in the set that transforms a fixed $P$ into each of the $P_{j \neq 0}$, so the minimal size of a set satisfying this property is $|\mathcal{P}_n|-1 = D^2-1$.

When the goal is to implement a randomized-benchmarking experiment using such a  twirl, it is convenient for the set $K$ to constitute a group.  This guarantees that the exact inversion operator at the end of each sequence is contained in $K$.  
However, if the requirement of exact inversion is relaxed, then $K$ need not form a group:  Any product of operators from this set will be a Clifford operator, so the ideal state at the end of any such experiment will be a stabilizer state.  For any stabilizer $P$ of that state, the assumption of Lemma~\ref{Lem:4TwirlCondition} guarantees that there is at least one $C \in K$ such that $C(P) = Z_1$, the Pauli $Z$ operator on the first qubit.  If this operator is performed in the inversion step, then a measurement of the first qubit can be made that deterministically returns either $0$ or $1$ if no errors occur.  If the measurement step is performed in this way, then there is no reason $K$ needs to be a group.  In the two constructions that follow, however, $K$ does happen to form a group; the additional constraint of group structure helped to focus the search for sets with the desired property.

\subsection{One-Qubit Subgroup} \label{Sec:4OneQubit}

A minimal set whose twirl depolarizes one-qubit errors can be constructed from the Clifford operator $T$ defined by
\begin{equation} \label{Eq:4TDef}
T(X) = Y \;  , \; T(Y) = Z  \; , \; T(Z) = X.
\end{equation}
Note that this is not the same as the operator $Z(\frac{\pi}{4})$ that is often called ``T'' by other authors.  The set $\mathcal{T} = \{I,T,T^2\}$ satisfies the conditions of Lemma~\ref{Lem:4TwirlCondition}.  It also has minimal size $D^2-1$ for a set of this kind on one qubit ($D=2$).  The $\mathcal{T}$-twirl must be performed after a Pauli twirl, which averages over $D^2$ operators, so the total number of aggregate operators that might occur in one step (Clifford twirl + Pauli twirl) is $(D^2-1)D^2 = 12$.  The set $\mathcal{T}$ also happens to form a group.

The ``T'' terminology is used for this operator because its action (by conjugation) on the three principal Pauli axes in the one-qubit Bloch-sphere picture looks like a turnstile with three arms rotating $120$ degrees.  In other words, this operator cyclically permutes the principal axes in this picture; equivalently, its conjugation action cyclically permutes the Pauli operators.  If an operator of this kind existed that cyclically permuted the Pauli operators on $n>1$ qubits, then a corresponding minimal twirling set could be constructed from the powers of the operator; however, I do not know if such operators exist for any $n>1$. 

\subsection{$n$-Qubit Subgroup} \label{Sec:4NQubit}

An example based on a suggestion from E. Knill provides sets satisfying the assumption of Lem.~\ref{Lem:4TwirlCondition} for every $n>1$ that are proper subgroups of the Clifford groups on $n$ qubits.  Because the twirl over one of these sets is preceded by a Pauli twirl, the Pauli part of each Clifford operator (in the decomposition into a Pauli operator and a representative of a coset in $\overline{\mathcal{C}}_n,$ described in Sec.~\ref{Sec:2CliffordFacts}) is not important for this construction.  It is therefore convenient to construct these sets from elements of $\overline{\mathcal{C}}_n$.  For $[C] \in  \overline{\mathcal{C}}_n$, let $[C](P)$ denote the conjugation action of any Clifford operator representative of $[C]$ on the Pauli operator $P$, disregarding the sign of the result.  When implementing an element from one of the subsets described below in an algorithm for twirling purposes, any Clifford operator representative of the corresponding coset may be chosen.

Let $\vec{k} = \sum_{i=1}^n \alpha_i  \vec{b_i}$ be an element in the Galois field $\mbox{GF}[2^n]$, where each $\alpha_i \in \{0,1\} \cong \mbox{GF}[2]$ and $\{\vec{b_1} \cdots \vec{b_n}\}$ is a basis of $\mbox{GF}[2^n]$ considered as an $n$-dimensional vector space over $\mbox{GF}[2]$ chosen so that $\vec{b_1}$ is the identity of the field.  I use $*$-multiplication to explicitly denote multiplication over the field.  Define the linear functions $(\vec{k})_i \equiv \alpha_i$ with respect to this decomposition.  Let $\{\vec{\overline{b_1}} \cdots \vec{\overline{b_n}}\}$ be a second basis for the field which is the dual of the first with respect to $(\cdot)_1$ so that $(\vec{b_i}*\vec{\overline{b}_j})_1 = \delta_{i,j}$.  Note that $\vec{\overline{b}_1} = 1$.  I prove the existence of such a basis in Lem.~\ref{Lem:4DualThing}.

\begin{lemma} \label{Lem:4DualThing}
Given an additive basis $\{\vec{b_1} \cdots \vec{b_n}\}$ for $\mbox{GF}[2^n]$, there exists a second basis $\{\vec{\overline{b_1}} \cdots \vec{\overline{b_n}}\}$ such that $(\vec{b_i}*\vec{\overline{b}_j})_1 = \delta_{i,j}$.

\paragraph{Proof: }
There exists a dual basis of linear functions $f_i$ on the vector space $\mbox{GF}[2^n]$ such that $f_i(\vec{b_j}) = \delta_{i,j}$.  It is a standard result that the space of such linear functions has the same dimension ($n$) as the original space and so there are $n$ elements in the dual basis.  The function $(\cdot)_1$ is a linear function on $\mbox{GF}[2^n]$, so it is contained in this space.  In addition, for any fixed $\vec{p} \in \mbox{GF}[2^n]$, $(\cdot * \vec{p})_1$ is also a linear function on $\mbox{GF}[2^n]$, so it is contained in this space as well.

Suppose the linear functions associated in this way with $\vec{p}$ and $\vec{q}$ (both in $\mbox{GF}[2^n]$) are the same.  Then $(\vec{x}*\vec{p})_1=(\vec{x}*\vec{q})_1$ for all $\vec{x} \in \mbox{GF}[2^n]$.  By linearity, this means $((\vec{p}-\vec{q})*\vec{x})_1 = 0$ for all $\vec{x}$.  If $\vec{p}-\vec{q}$ were not equal to $0$, then a multiplicative inverse $\vec{x}=(\vec{p}-\vec{q})^{-1}$ would exist, and $(\vec{x}*(\vec{p}-\vec{q}))_1 = 1$ would not equal $0$ for this choice of $\vec{x}$.  Therefore $(\vec{p}-\vec{q})$ must equal $0$ and it must be that $\vec{p}=\vec{q}$.  Therefore, each element $\vec{p} \in \mbox{GF}[2^n]$ can be associated with a distinct linear function and there is a vector space isomorphism between the original space and the dual given by this association.  Call the element isomorphic to $f_i$ in this way $\vec{\overline{b}_i}$.  The construction dictates that $(\vec{b_i}* \vec{\overline{b}_j})_1 = f_j(\vec{b_i}) = \delta_{i,j}$, so $\{\vec{\overline{b}_1} \cdots \vec{\overline{b}_j}\}$ is a basis satisfying the original claim. $\square$
\end{lemma}

Assume $\vec{k} = \sum_{i=1}^n \alpha_i  \vec{b_i}$ can be written as $\vec{k} = \sum_{i=1}^n \beta_i \vec{\overline{b}_i}$ in the second basis, so that the coefficient $\beta_i$ of $\vec{\overline{b}_i}$ equals $(\vec{k}*\vec{b_i})_1$.   Note also that $(\vec{k}*\vec{\overline{b}_i})_1 = (\vec{k})_i = \alpha_i$.  

Define a vectorial exponentiation of a Pauli operator on $n$ qubits to the power $\vec{k} \in \mbox{GF}[2^n]$ as follows:  Let 
\begin{equation} \label{Eq:4Xpower}
X_j^{\vec{k}} \equiv \bigotimes_{i=1}^n X_i^{(\vec{k}*\vec{b_j})_i} = \bigotimes_{i=1}^n X_i^{(\vec{k}*\vec{b_j}*\vec{\overline{b}_i})_1}.
\end{equation}
In contrast, let 
\begin{equation} \label{Eq:4Zpower}
Z_j^{\vec{k}} \equiv \bigotimes_{i=1}^n Z_i^{(\vec{k}*\vec{\overline{b}_j}*\vec{b_i})_1}.
\end{equation}
Note that $X_j^{\vec{k}} = X_1^{\vec{b_j}*\vec{k}}$ and $Z_j^{\vec{k}} = Z_1^{\vec{\overline{b}_j}*\vec{k}}$ according to these definitions, so that any n-qubit Pauli operator can be written as $X_1^{\vec{a}}Z_1^{\vec{c}}$ for some $\vec{a},\vec{c} \in \mbox{GF}[2^n]$.  This is closely related to the qubit Pauli operator notation utilizing pairs of binary numbers that is used in Sec.~\ref{Sec:AStabMat}. Finally, for such a general $n$-qubit Pauli operator $P = X_1^{\vec{a}}Z_1^{\vec{c}}$ and any $\vec{k} \in \mbox{GF}[2^n]$, let $P^{\vec{k}} = X_1^{\vec{k}*\vec{a}}Z_1^{\vec{k}*\vec{c}}$.

\begin{theorem} \label{Thm:4OnlyTheorem}
Consider the set of operators in the Clifford quotient $\overline{\mathcal{C}}_n$ that satisfy an additional property: if $[C](P) = P'$, then $[C](P^{\vec{k}}) = (P')^{\vec{k}}$ for all operators $P \in \mathcal{P}_n$ and any $\vec{k} \in \mbox{GF}[2^n].$  The set of all such operators forms a subgroup $\mathcal{Q}_n$ and has order $2^n(4^n-1)$.  Furthermore, a twirl over this subgroup has the same action on superoperators as a twirl over $\overline{\mathcal{C}}_n$. 
\end{theorem}

The remainder of this section consists of a proof of this theorem.

Consider the possible Clifford transformation rules for an element $[C] \in \mathcal{Q}_n$: There are $D^2-1$ possible choices of Pauli operators that could be the result of $[C](X_1)$ and $\frac{D^2}{2}$ possible Pauli operators that could be the result of $[C](Z_1)$ and that anti-commute with the choice of $[C](X_1)$, as required for the action.  Once the action of $[C]$ on $X_1$ and $Z_1$ has been fixed, the action on all other Pauli operators is a consequence of the additional property (since $[C](X_i) = [C](X_1^{\vec{b_i}}) = ([C](X_1))^{\vec{b_i}}$, for example).  Lem.~\ref{Lem:AClifAction} establishes that the action of a Clifford operator on a generating set of Pauli operators uniquely specifies the operator, so it also uniquely specifies the coset of the operator.

To check whether any choice of these transformation rules describes an operator in $\overline{\mathcal{C}}_n$, one needs to check that $[C](X_i)$ and $[C](X_j)$ commute for all $i,j$, $[C](Z_i)$ and $[C](Z_j)$ commute for all $i,j$, and $[C](X_i)$ and $[C](Z_j)$ anti-commute for $i=j$ and commute for all $i\neq j$.  It is also necessary that the results $[C](X_i),[C](Z_i)$ of the action on the generating basis $\{X_1,Z_1,\cdots X_n,Z_n\}$ are all mutually independent (in that one cannot be written as a product of the others).  In order to prove these results, it is convenient first to find an expression describing the commutation relation of two Pauli operators.

\begin{lemma} \label{Lem:4Commutation}
Consider two Pauli operators $P_{(\vec{a},\vec{c})} \equiv X_1^{\vec{a}}Z_1^{\vec{c}}$ and $P_{(\vec{d},\vec{e})} \equiv X_1^{\vec{d}}Z_1^{\vec{e}}$, where $\vec{a},\vec{c},\vec{d},\vec{e}$ are all in $\mbox{GF}[2^n]$.  The operators commute if and only if $(\vec{c}*\vec{d})_1 + (\vec{a}*\vec{e})_1 = 0 \mod{2}$.

\paragraph{Proof:}
The operators $P_{(\vec{a},\vec{c})}$ and $P_{(\vec{d},\vec{e})}$ commute if and only if $P_{(\vec{a},\vec{c})}P_{(\vec{d},\vec{e})} = (-1)^s P_{(\vec{d},\vec{e})}P_{(\vec{a},\vec{c})}$ for their matrix representations and $s=0 \mod{2}$, so it remains to find an expression for $s$.  It is a basic fact of Pauli matrices that $X_iX_j = X_jX_i$, $Z_iZ_j = Z_jZ_i$, and $X_iZ_j = (-1)^{\delta_{i,j}}Z_jX_i$.  Using the decompositions of Eqs.~\ref{Eq:4Xpower} and \ref{Eq:4Zpower},
\begin{eqnarray}
P_{(\vec{a},\vec{c})}P_{(\vec{d},\vec{e})} &=& \left (\bigotimes_i X_i^{(\vec{a}*\vec{\overline{b}_i})_1} \right ) \left (\bigotimes_i Z_i^{(\vec{c}*\vec{b_i})_1} \right ) \left (\bigotimes_i X_i^{(\vec{d}*\vec{\overline{b}_i})_1} \right ) \left (\bigotimes_i Z_i^{(\vec{e}*\vec{b_i})_1} \right ) \nonumber \\
&=& (-1)^{\sum_i (\vec{c}*\vec{b_i})_1 (\vec{d}*\vec{\overline{b}_i})_1} \left (\bigotimes_i X_i^{(\vec{a}*\vec{\overline{b}_i})_1} \right )\left (\bigotimes_i X_i^{(\vec{d}*\vec{\overline{b}}_i)_1} \right )\left (\bigotimes_i Z_i^{(\vec{c}*\vec{b_i})_1} \right )\left (\bigotimes_i Z_i^{(\vec{e}*\vec{b_i})_1} \right ) \nonumber \\
&=& (-1)^{\left (\sum_i (\vec{c}*\vec{b_i})_1 (\vec{d}*\vec{\overline{b}_i})_1 \right ) + \left (\sum_i (\vec{e}*\vec{b_i})_1 (\vec{a}*\vec{\overline{b}_i})_1 \right )} P_{(\vec{d},\vec{e})}P_{(\vec{a},\vec{c})} \\
&=& (-1)^{(\vec{c}*\vec{d})_1 + (\vec{e}*\vec{a})_1}P_{(\vec{d},\vec{e})}P_{(\vec{a},\vec{c})} 
\label{Eq:4ProofArray1}
\end{eqnarray}
This derivation involves repeated applications of the Pauli commutation rules at each index, first for the inner $Z$ and $X$ operators and then for the outer $X$ and $Z$ operators.  The last equality can be derived in reverse by noting that $(\cdot)_1$ is a linear operator from $\mbox{GF}[2^n]$ to $\mbox{GF}[2]$ and expanding $\vec{c}$ and $\vec{d}$ in the $\{\vec{b_1}\cdots \vec{b_n}\}$ and $\{ \vec{\overline{b}_1} \cdots \vec{\overline{b}_n}\}$ bases, respectively.  This final expression characterizes the commutation relation of $P_{(\vec{a},\vec{c})}$ and $P_{(\vec{d},\vec{e})}$. $\square$ 
\end{lemma}

Consider the possible Clifford actions $[C](X_1) \equiv X_1^{\vec{a}}Z_1^{\vec{c}}$ and $[C](Z_1) \equiv X_1^{\vec{d}}Z_1^{\vec{e}}$.  First check the commutation relation between $[C](X_i) = X_1^{\vec{a}*\vec{b_i}}Z_1^{\vec{c}*\vec{b_i}}$ and $[C](X_j) = X_1^{\vec{a}*\vec{b_j}}Z_1^{\vec{c}*\vec{b_j}}$.  The two operators can be seen to commute using Lem.~\ref{Lem:4Commutation} because $(\vec{c}*\vec{b_i} * \vec{a}*\vec{b_j})_1 + (\vec{a}*\vec{b_i}*\vec{c}*\vec{b_j})_1 = (\vec{c}*\vec{a}*(\vec{b_i}*\vec{b_j} - \vec{b_j}*\vec{b_i}))_1 =  0 \mod{2}$.  An analogous application of the lemma suffices to show that $[C](Z_i)$ and $[C](Z_j)$ commute for all $i,j$.

Now consider the commutation of $[C](X_i)$ and $[C](Z_j)$.  Using the lemma again, one can see that the two commute if and only if $(\vec{c}*\vec{b_i}*\vec{d}*\vec{\overline{b}_j})_1 + (\vec{a}*\vec{b_i}*\vec{e}*\vec{\overline{b}_j})_1 = 0 \mod{2}$.  The linearity of $(\cdot)_1$ allows the left hand side of this equation to be written as
\begin{equation} \label{Eq:4HardRelation}
((\vec{c}*\vec{d} + \vec{a}*\vec{e}) * \vec{\overline{b}_j}*\vec{b_i})_1
\end{equation}
If $\vec{c}*\vec{d} + \vec{a}*\vec{e} = \vec{b_1} = 1$, then this expression evaluates to $1$ if $j=i$ and $0$ if $j \neq 1$.  These are the desired commutation rules.  If $\vec{c}*\vec{d} + \vec{a}*\vec{e} \neq \vec{b_1}$, then the expression cannot evaluate to these same values for all $i,j$, so some commutation rules for the Clifford action must be violated.

In conclusion, the commutation relations required for the transformation rules of an operator in $\overline{\mathcal{C}}_n$ and the additional exponentiation property for $\mathcal{Q}_n$ are satisfied by the operator defined by $[C](X_1) =X_1^{\vec{a}}Z_1^{\vec{c}}$ and $[C](Z_1) = X_1^{\vec{d}}Z_1^{\vec{e}}$ only if $\vec{c}*\vec{d} + \vec{a}*\vec{e} = \vec{b_1}$.  It is straightforward to check that this condition is also sufficient.  Therefore, the possible transformation rules $[C](X_1)$ and $[C](Z_1)$ that satisfy this condition are in one-to-one correspondence with the elements of $\mathcal{Q}_n$.

To see that the transformations $[C](X_i)$,$[C](Z_i)$ are independent when they are defined using the additional property for $[C] \in \mathcal{Q}_n$, suppose that $[C](X_i)$ could be written as a product of the other transformed operators: $ [C](X_i) = [C](Z_i)^{\beta_i} \prod_{j \neq i} [C](X_i)^{\alpha_j}[C](Z_i)^{\beta_j}$ for some $\alpha_j,\beta_j,\beta_i \in \{0,1\}$.  Given the commutation rules already shown, $[C](X_i)$ must anti-commute with $[C](Z_i)$.  However, 
\begin{eqnarray}
[C](X_i)[C](Z_i) &=& \left ([C](Z_i)^{\beta_i} \prod_{j \neq i} [C](X_i)^{\alpha_j}[C](Z_i)^{\beta_j} \right ) [C](Z_i) \nonumber \\
&=& [C](Z_i) \left ([C](Z_i)^{\beta_i} \prod_{j \neq i} [C](X_i)^{\alpha_j}[C](Z_i)^{\beta_j} \right ) \nonumber \\
&=& [C](Z_i)[C](X_i)
\end{eqnarray}
because all of the operators in the product have been shown to commute with $[C](Z_i)$.  The operators cannot both commute and anti-commute, leading to a contradiction of the supposition that $[C](X_i)$ could be written as such a product.

Given a transformation rule $[C](X_1) = X_1^{\vec{a}}Z_1^{\vec{c}}$ (where $\vec{a}$ and $\vec{c}$ cannot both be $0$ for a Clifford operator), the additional constraint imposed by Eq.~\ref{Eq:4ProofArray1} on the possible transformations $[C](Z_1)$ limits the possibilities to $\frac{|\mathcal{P}_n|}{2^n}$ operators for any choice of $[C](X_1)$.  Because there are $|\mathcal{P}|-1$ choices for the first rule and all other rules are determined by these two rules and the additional property, the total number of elements in $\mathcal{Q}_n$ is $\frac{|\mathcal{P}_n|}{2^n}(|\mathcal{P}|-1) = (4^n-1)(2^n)$.

To see that $\mathcal{Q}_n$ is a group, consider first the closure under products of its operators.  If $[C](P_i) = P_j$ and $[D](P_j) = P_m$, then $([D] [C])(P_i) \equiv ([D] \circ [C])(P_i) = P_m$.  To check the additional property, observe that $([D] \circ [C])(P_i^{\vec{k}}) = [D](P_j^{\vec{k}}) = P_m^{\vec{k}}$.  Because $[C]$ and $[D]$ were both assumed to be valid elements of $\overline{\mathcal{C}}_n$ , $[D] [C]$ must be one as well, so it is not necessary to check the commutation relations of the transformation rules for the operators in this coset.  Because $\overline{\mathcal{C}}_n$ is finite, the closure property suffices to prove that $\mathcal{Q}_n$ is a subgroup of $\overline{\mathcal{C}}_n$.

I next prove that $\mathcal{Q}_n$ satisfies the property of Lem.~\ref{Lem:4TwirlCondition} by showing first that there is an operator in   $\mathcal{Q}_n$ that transforms $P_i$ into $P_j$ for each $P_i,P_j \in \mathcal{P}_n$ with $i,j \neq 0$.  To begin with, I show that there exist $[C],[D] \in \mathcal{Q}_n$ such that $[C](X_1) = P_j$ and $[D](X_1) = P_i$:  Suppose without loss of generality that $P_i = X_1^{\vec{a}}Z_1^{\vec{c}}$ and $P_j = X_1^{\vec{c}}Z_1^{\vec{d}}$.  Then at least one of the operators defined by the rules ($[C](X_1) = P_i$ and $[C](Z_1) = X_1^{\frac{1-\vec{a}}{\vec{c}}} Z_1$) or the rules ($[C](X_1) = P_i$ and $[C](Z_1) = X_1^{\frac{1-\vec{c}}{\vec{a}}} Z_1$) and the additional property is guaranteed to be in $\mathcal{Q}_n$.  Similarly, at least one of the operators defined by the rules  ($[D](X_1) = P_j$ and $[D(Z_1) = X_1^{\frac{1-\vec{d}}{\vec{e}}} Z_1$) or the rules ($[D](X_1) = P_j$ and $[D](Z_1) = X_1^{\frac{1-\vec{e}}{\vec{d}}} Z_1$) and the additional property is guaranteed to be in $\mathcal{Q}_n$.  Because $\mathcal{Q}_n$ is a group, given this choice of $[D]$, $[D]^{-1}$ is contained in $\mathcal{Q}_n$ and $[D]^{-1}(P_i)=X_1$.  Then by the closure of the group, $[C][D]^{-1} \in \mathcal{Q}_n$, and $[C][D]^{-1}(P_i) = P_j$.  

The claim that there is an element of $\mathcal{Q}_n$ that acts to transform any non-identity Pauli operator to any other non-identity Pauli operator is the same as the statement that the action of $\mathcal{Q}_n$ on the Pauli group minus the identity is 1-transitive.  The argument of Cor.~\ref{Cor:ASymmetricAction} of App.~\ref{App:A}, which is actually a general statement about 1-transitive group actions, can be repeated verbatim to show that the number of elements of $\mathcal{Q}_n$ that transform $P_i$ into $P_j$ is independent of $i$ and $j$ (as long as both operators are not the identity).  One may then conclude that a uniform twirl over $\mathcal{Q}_n$ (preceded by a uniform twirl over the Pauli operators) satisfies the condition of Lem.~\ref{Lem:4TwirlCondition} and therefore depolarizes superoperators in the same way as a uniform twirl over the entire Clifford group.  This concludes the proof of Theorem \ref{Thm:4OnlyTheorem}. $\square$

Repeating the preceding arguments shows that a uniform twirl over any subgroup of $\mathcal{C}_n$ (or $\overline{\mathcal{C}_n}$) depolarizes superoperators as long as at least one element of the subgroup transforms $P_i$ into $P_j$ for each $i,j \neq 0$.  This very restricted condition might be convenient for searching for other useful Clifford subgroups with the desired twirling behavior.

\subsection{Uses of Alternative Twirling Sets} \label{Sec:4AltTwirlUses}

For any Clifford subset satisfying the twirling condition, one can implement a randomized-benchmarking experiment whose steps are defined by a uniform distribution over the subset.  The errors on the operators will be depolarized in the same way as in the full Clifford-twirl experiment, and the analysis will proceed in the same way as well.  However, it is not clear why one would wish to know the error per step for steps defined in this way.  An obvious analog of the argument made for the full Clifford group indicates that this error per step would be useful if an important computational routine was known to utilize only operators from the Clifford subset.  While this is possible in principle, it seems somewhat contrived.

A more useful application of the subset twirls comes from the individual-gate benchmarking experiment described in Sec.~\ref{Sec:3GateBenchmark}.  A subset-twirl individual-gate benchmark could proceed as follows:  Perform a first randomized-benchmark experiment where a step is defined by a uniform distribution over the operators from the subset.  Perform a second randomized benchmark where each step consists of a uniformly random Clifford from the subset followed by some distinguished gate $G$.  Note that, if $G$ is not a member of the subset, the measurement step for this benchmark might be different from the first experiment.  By comparing the error-per-step parameters from the two benchmarks, the error per $G$ can be extracted as in Eq.~\ref{Eq:3Gate}.  The analysis for this experiment has essentially the same limitations and required assumptions as the analysis for the full Clifford group; the only potential complication comes from designing the measurement step efficiently.  The key criterion that decides whether such a subset would be useful in this experiment is whether its operators are more convenient to experimentally implement on average than are those of the entire Clifford group.

Error-correcting codes provide a specific instance where a twirl over such a subset might be convenient.  For any code, there is a set of gates that can be implemented transversally.  That is, in order to implement a gate $G$ on the encoded qubit, one simply implements the gate $G$ on each of the physical qubits making up the code.  Transversal gates are a particularly simple way to implement encoded gates without spreading errors from one physical qubit to a second physical qubit within the code.  A significant aspect of the study of suitability of quantum codes for quantum computing tasks is the description of which gates are transversal for each code.  Ref.~\cite{Eastin2009} proved that the set of transversal gates cannot contain a generating set of gates for universal quantum computation (i.e.~a generating set for the unitary operators).

Nevertheless, codes exist for which a subset sufficient for twirling can be implemented transversally.  For example, the $[[7,1,3]]$ Steane code has transversal implementations for all Clifford gates and can conveniently implement any Clifford twirl.  I know of no codes for which one of the $(n>1)$-subgroup twirls is transversal but the full Clifford group is not.  However, Ref.~\cite{Zeng2007} indicates that the gate called $T$ in this thesis is transversal for the $[[5,1,3]]$ code.  Performing encoded randomized benchmarking on one encoded qubit for this code (as described in Sec.~\ref{Sec:4EncodedBenchmark}) would be significantly easier if only the $\mathcal{T}$ subgroup were used for twirling.

\section{Using Approximate Twirls} \label{Sec:4AppTwirl}
One of the biggest obstacles to using approximate twirling for RB comes from a lack of understanding about how much randomness is required in the gates separating two errors before they may be approximated by independent depolarizing channels.  For example, it should be possible to perform an individual-gate benchmarking experiment using approximate twirls as follows: Perform a standard approximate-twirling RB benchmark using some distribution of gates $p(g)$ to define a step.  Perform a second approximately-twirled RB experiment with the same $p(g)$, but insert an additional $G$ gate after every $m$ randomly chosen steps.  If the error in the $G$ gates is depolarized, it is straightforward to extend Eq.~\ref{Eq:3Gate} in such a way that the depolarizing strength of this error can be extracted.  The issue that remains is how big $m$ must be before these $G$ gates have errors that are sufficiently depolarized for the measurement step to detect them as expected.

A distinct but related issue, mentioned in Sec.~\ref{Sec:2AppTwirl}, is the desire to compare two approximately-twirled RB experiments for which two different distributions $p_A(g)$ and $p_B(g)$ have been used to define the steps.  The problem arises especially if one wishes to compare the relative merits of the two experiments by describing how well experiment A would perform if it were using the same step definition as experiment B.  I restrict this problem somewhat by assuming that experiment B uses the uniform distribution.  If tight bounds could be placed on an estimate of the error per step (as defined by $p_B$) for experiment A based on the known error per step (as defined by $p_A$), then approximate twirling would be a more extensible technique.

These two problems actually relate to different aspects of the gate distribution defining a step.  For the latter problem, it is desirable for $p_A(g)$ (or $p_{l,A}(g)$, the distribution of gates in the composition of $l$ steps) to be as close to the uniform distribution over Clifford operators as possible.  To the extent that $p_A(g)$ is not uniform, the benchmark under- or over-represents some gates that might have higher or lower depolarizing error strengths than the others.  This forces larger bounds on the estimation of the error per step defined by the uniform step definition.  In contrast, in order to guarantee the depolarizing nature of the twirl prior to the measurement, Sec.~\ref{Sec:4SmallDesigns} proved that it is not necessary to randomize uniformly over all Cliffords, so a step with $p(g)$ far from the uniform distribution over the Cliffords can still twirl effectively.  In this section I attempt to bound the relationships between estimates of the error per step defined by $p(g)$ and the error per step defined by the uniform distribution as influenced by these two effects.

\subsection{Step Definition} \label{Sec:4StepDef}

If the main focus of a randomized-benchmarking experiment is to extract the error per gate for a single gate $G$ using the techniques in Sec.~\ref{Sec:3GateBenchmark}, then the step definition need not be uniform as long as it twirls the errors and it is consistent between the primary benchmark and the benchmark with extra $G$ gates.  However, if the main focus of a randomized-benchmarking experiment is, instead, to provide a benchmark parameter describing the ability to perform average Clifford operators, then the exact definition of a step becomes important.  In order to compare the error per step of experiment A to that of experiment B (especially the standard, exact-twirl experiment), it is necessary to place bounds on an estimate of the error per step experiment A would exhibit if it used the step definition from experiment B.  

To study this comparison in the simplest (nontrivial) way possible, I assume that all gates $g$ have depolarizing errors of different strengths $s_{g}$.  In such a scenario, the twirling behavior of the randomized benchmark is not necessary, as each error is already independent and depolarizing.  I also assume that SPAM errors are non-existent.  The correct result of the analysis of an exactly twirled benchmarking experiment with these errors would be an error per step 
\begin{equation} \label{Eq:4PerfectTwirl}
\epsilon_{s,B} = \alpha_n \sum_{g \in \mathcal{C}_n} \frac{1}{|\mathcal{C}_n|} s_{g}.
\end{equation}

Suppose each step of experiment A is defined instead by the distribution $p_A(g)$, and let $p_{l,A}(g)$ be the distribution of aggregate operators after $l$ consecutive steps.  I assume that $p_A(g)$ is the same for each step, so $p_{l,A}(g)$ is the same for any subsequence of $l$ steps.  The error per step in such an experiment is proportional to the weighted mean of the depolarizing strengths of the different errors that might occur in a step,
\begin{equation} \label{Eq:4ImpTwirl1}
\epsilon_{s,A,1} = \alpha_n \sum_{g \in \mathcal{C}_n} p_A(g) s_{g}.
\end{equation}

One can estimate $\epsilon_{s,B}$ by the value $\epsilon_{s,A,1}$ extracted from experiment A, but it is necessary to bound the error in this estimate, $\delta_1 \equiv |\epsilon_{s,B} - \epsilon_{s,A,1}|$.  I maximize the value of $\delta_1$ subject to the known values of $\epsilon_{s,A,1}$ and the distribution $p_{A}(g)$.  

The value of $\epsilon_{s,A,1}$ forces a constraint on the values $s_g$ because
\begin{equation} \label{Eq:4Constraint1}
\sum_g p_A(g)s_g = \frac{\epsilon_{s,A,1}}{\alpha_n}.
\end{equation}
Another set of constraints based on the definition of a depolarizing channel is given by $0 \leq s_g \leq \frac{D^2}{D^2-1}$ for all $g$.  Finally, the desired upper bound on $\delta_1$ can be calculated by maximizing
\begin{equation} \label{Eq:4Objective}
\delta_1 = \alpha_n \sum_g \left (\frac{1}{|\mathcal{C}_n|} - p_A(g) \right ) s_g
\end{equation}
by varying the $s_g$ subject to these constraints.  An equivalent lower bound can be found by maximizing $-1$ times this sum.  In order to solve these maximization problems, a linear program may be used.  

Because the assumptions underlying these bounds are not likely to be exactly true in practice, I intend these bounds only as a guideline for the types of variation that might be found.  This toy model also demonstrates why it is desirable for the definition of a step in an approximately twirled experiment to be as close to a uniform distribution as possible.  Note, in particular, that the sum of the terms in parentheses in Eq.~\ref{Eq:4Objective} is upper bounded by twice the total-variation distance described in Sec.~\ref{Sec:2GateChoice}.  Therefore, for gate distributions with small total-variation distance, this upper bound is generally small as well.  It is also important to note that gates with $p_A(g) = 0$ are not constrained at all by Eq.~\ref{Eq:4Constraint1}.  For this reason, the terms in Eq.~\ref{Eq:4Objective} corresponding to such gates will dominate the maximization.

In order to find a tighter comparison between the A and B experiments, one can consider grouping $k$ steps of the experiment A into a new definition of a step.  The aggregate distribution of operators in this step is $p_{A,k}(g)$, and the error-per-step parameter for the aggregate step is now $\epsilon_{s,A,k} = \frac{1}{\alpha_n} \left (1-(1-\alpha_n\epsilon_{s,A,1})^{k} \right )$.  Notably, for this aggregate step, the constraint that corresponds to Eq.~\ref{Eq:4Constraint1} is 
\begin{equation} \label{Eq:4Constraintk}
\sum_g p_{A,k}(g)s_g = \frac{\epsilon_{s,A,k}}{\alpha_n}.
\end{equation}
For any generating set of gates and a step definition that does not lead to periodicity (see Sec.~\ref{Sec:2AppTwirl}), $p_{A,k}(g)$ is greater than $0$ for sufficiently large $k$, so eventually this equation constrains all of the $s_g$.

It is possible to combine all of the constraints given by Eq.~\ref{Eq:4Constraintk} with different $k$ values.  For simplicity, I describe here using each constraint equation for $k$ alone in a new linear program to maximize $\delta_k$ given by:
\begin{equation} \label{Eq:4Objectivek}
\delta_k = \sum_g \left (\frac{1}{|\mathcal{C}_n|} - p_{A,k}(g) \right ) s_g.
\end{equation}
It may be possible to find a tighter upper bound for the estimate of $\epsilon_B$ by using $\epsilon_{s,A,k} + \delta_k$ with the $\delta_k$ obtained in this way.

The bounds provided above are still likely to be pessimistic, motivating the use of the common definition of a step based on the exact twirl that is advocated in Chap.~\ref{Chap:3} whenever one intends to compare two benchmarks.  However, it is possible to describe much tighter bounds if some simplifying physical assumptions are made regarding the errors for different gates.  For example, in Sec.~\ref{Sec:3Exp2} I use the assumptions that $s_{g}$ depends only on the number of $\frac{\pi}{2}$ pulses (in the one-qubit experiments) and on the number of $G$ gates (in the two-qubit experiments).  These assumptions allow $s_{g}$ to be estimated for operators even if they are not actually implemented in a single step, so the pessimistic $s_g < \frac{D^2}{D^2-1}$ constraint need not be used.  
\subsection{Incomplete Depolarization} \label{Sec:4AppDep}

Regardless of whether an individual-gate benchmark or an error per step for some distribution is intended, there is a problem with imperfect depolarization when using approximate twirls.  When the steps that separate two unitary errors do not have the possibility of implementing a random Clifford operator, the two errors can add coherently as described in Sec.~\ref{Sec:2AppTwirl}.  Similarly, if an error at a certain place in a sequence is not separated by enough steps from the measurement, then it may not appear as a depolarizing channel with respect to the measurement.  In this section, I present a very simplified class of error models that demonstrate the possible effects of imperfect depolarization on an individual-gate RB experiment.  For these error models, the depolarizing strength of the errors inferred from a standard analysis of the benchmark results differs from the true strength of the depolarization of the errors, so I bound the difference between the true and inferred values.


In terms of taking consistent small-angle unitary errors (which I call coherent errors here) and ensuring that they do not add coherently, the most crucial randomization is the Pauli twirl.  In Sec.~\ref{Sec:2PauliTwirl} I demonstrate that the Pauli twirl acts to transform any superoperator into a (stochastic) Pauli channel, which has the form $\sum_{P \in \mathcal{P}_n} \gamma_P P \otimes P$, with $\sum_{P \in \mathcal{P}_n} \gamma_P = 1$.  Pauli channels do not add coherently in the usual sense.  This superoperator can be interpreted as one in which a Pauli operator $P$ is applied to the system with probability $\gamma_P$.  Because the remaining gates after an error in a RB sequence are Clifford gates, a Pauli operator $P$ at one location in the sequence is equivalent to a (possibly different) Pauli operator $P'$ at the end of the sequence.  This error operator either causes an incorrect measurement result (if $P'$ anti-commutes with the measurement operator) or causes a correct measurement result (if $P'$ and the measurement operator commute).  

In order to demonstrate in a simple way the effects of a non-uniform twirl on the imperfect depolarization of errors, I consider the following idealized RB experiment:  Suppose all Clifford gates used for twirling are error-less, so that the standard RB experiment reveals an error per step of $\epsilon_s = 0$.  The step in this experiment is described by a probability distribution $p(C)$, and the aggregate of $k$ gates is described by a distribution $p_k(C)$.  I assume that $p_k(C)$ approaches a uniform Clifford distribution as $k$ increases, but it is only necessary that it approaches some distribution that has the twirling behavior described by Eq.~\ref{Eq:4TwirlProp}.  The inversion step for a length-$l$ sequence is not performed exactly, as is typical for approximately twirled RB experiments.  Instead, the inversion step is described by a distribution $m_l(C)$ over Clifford operators, where there is potential for dependence on $l$ because the distribution of aggregate Clifford operators from the rest of the sequence $p_l(C)$ depends on the length of the sequence $l$.  Assume that a single binary measurement of a Pauli operator is made after the inversion.  Also, assume that there are no SPAM errors. 

Suppose further that there is a single gate $G$ (not used in the steps of the standard RB experiment) that has non-zero error.  A second randomized benchmark is performed in the same way as the first, except that an extra $G$ gate is performed after each step.   I assume that $p(C)$ and $m_l(C)$ are such that Pauli randomization is performed perfectly in each step and before the measurement.  Because of the assumption that the Pauli randomization is performed perfectly, it is acceptable to assume that the error of $G$ is described by a Pauli channel $\hat{\Lambda}_G = \sum_{P \in \mathcal{P}} \gamma_P P \otimes P$ acting after the gate.  Because of the approximate twirl, this error may not be perfectly depolarized before the measurement.  

\paragraph{Example:} As a simple demonstration of imperfect depolarization of a Pauli channel in this RB context, consider a one-qubit RB experiment using the step distribution and inversion operators described in Sec.~\ref{Sec:2Exp1}.  Consider further performing individual-gate RB of a gate $G$ intended to act as the identity using these steps.  Then the contribution of the Pauli channel error of the last $G$ gate (immediately preceding the inversion) to the sequence infidelity depends on more than just the strength of the depolarization of the Pauli channel, which can cause a mistake in the estimation of the depolarized strength of the channel:  Consider a sequence with only one random step, so that the Clifford part of the inversion operator is $X \left (\frac{\pi}{2} \right )$ or $Y \left (\frac{\pi}{2} \right )$ with equal probability.  If a Pauli operator of the $X$ or $Y$ type is applied by the error channel, then it is detected by the measurement $\frac{1}{2}$ of the time, depending on the inversion operator.  However, a $Z$ error is always transformed by the inversion operator into an error that anti-commutes with the measurement and is always detected.  So changing the relative weight of the $Z$ part of the Pauli error channel for $G$ (without changing the strength of its depolarization overall) changes the error per step that an experimenter would infer.  In other words, the standard analysis of the experiment may mis-estimate the strength of this Pauli channel.  In this section I derive bounds for how far off this estimate can be for gates with Pauli channel errors.

Returning again to the general case, consider the effect of a non-identity Pauli operator acting after the $k^{th}$ random step from the end of a sequence.  In an exactly twirled experiment, this error is twirled by the rest of the random sequence so that it appears as a $\frac{1}{D^2-1} \sum_{i \neq 0} P_i \otimes P_i$ superoperator prior to the measurement.  If a single Pauli measurement is made on each qubit, $\frac{D^2}{2}$ terms in this sum anti-commute with the binary measurement and $\frac{D^2}{2}-1$ commute.  Therefore, if the error is perfectly depolarized, it causes an incorrect measurement with probability $\frac{D^2}{2(D^2-1)}$.  However, in the imperfectly-twirled experiment, the distribution of random aggregate Cliffords for the remaining steps is not guaranteed to transform each Pauli operator into a uniformly weighted mixture of Pauli operator as before, i.e.~the depolarization of this error is not perfect.

Let $Q_{C}$ be the set of non-identity Pauli operators that are the pre-image under the action of a Clifford operator $C$ of the $\frac{D^2}{2}$ operators that commute with the measurement.  If the $k^{th}$ step from the end of a sequence is followed by steps (including the inversion step) whose aggregate operator is $C$, then a Pauli operator error from the set $Q_{C}$ in this step will not be detected by the measurement, while any Pauli operator not in the set will be detected.  For the approximately twirled RB experiment under consideration, call $p_k'(C)$ the probability that an error after the $G$ gate in the $k^{th}$ step is separated from the measurement by the aggregate operator $C$.  Note that $p_k'(C)$ can be calculated from the step definition $p(C)$ and the inversion operator distribution $m_k(C)$ (and with the inclusion of the interleaved $G$ gates).

Because the twirl is approximate, the $p'_k(C)$ are not equal for all $C$, and so the probabilities with which the various Pauli operators lie in the pre-image of the remaining aggregate operator $C$ are not all the same.  For this reason, a particular Pauli operator error can make a contribution to the observed infidelity of the sequence that is greater (or less) than it would make in an exactly-twirled benchmark.

I now show how an imperfectly depolarized Pauli channel affects the result of the measurement in the RB experiment.  For each Pauli operator $R \neq I$, let $q_{k}(R) = \sum_{\substack{C \in \mathcal{C}_n : R \in Q_{C}}} p_k'(C)$.  Now $\max_{R \in \mathcal{P}_n} q_{k}(R)$ is the maximum probability that an error described by a Pauli operator will go undetected in the measurement and $\min_{R \in \mathcal{P}_n} q_k(R)$ is the minimum probability that such an error will go undetected.  Call the Pauli operators that achieve these extreme values $R_{k,\max}$ and $R_{k,\min}$, respectively.  Now, for a Pauli channel with a fixed depolarized strength, the two extreme cases for mis-estimating the strength of the depolarization of the error occur if the error is described by the stochastic Pauli channels
\begin{eqnarray} \label{Eq:4Extreme1}
\hat{\Lambda}_{k,\max}(\gamma) \equiv (1-\gamma) \left (I \otimes I \right ) + \gamma \left ( R_{k,\max} \otimes R_{k,\max}\right ) \mbox{\; \; or }\\
\hat{\Lambda}_{k,\min}(\gamma) \equiv (1-\gamma) \left (I \otimes I \right ) + \gamma \left (R_{k,\min} \otimes R_{k,\min} \right ). \label{Eq:4Extreme2}
\end{eqnarray}

If the gate $G$ that occurs after the $k^{th}$ step from the end of the sequence has error $\hat{\Lambda}_{k,\max}(\gamma)$ and no other step has an error, then, averaged over all sequences, the measurement reports an error with probability $e \equiv \gamma (1-q(R_{k,\max}))$.  In order to have this effect on a measurement, a perfectly depolarizing channel would need strength $2e$, so the standard analysis of the RB experiment would infer that the depolarization of the error on $G$ had this strength.  However, the actual depolarized strength of $\hat{\Lambda}_{\max}(\gamma)$ is $\frac{D^2}{D^2-1} \gamma$.  Therefore the upper bound on the difference between the estimated depolarized strength and the true depolarized strength is 
\begin{equation} \label{Eq:4KSolution}
\kappa_{k,\max} = \left ( \frac{D^2}{D^2-1}\gamma - 2e \right ).
\end{equation}

Eq.~\ref{Eq:4KSolution} describes one worst case for how badly the standard benchmark analysis can mis-estimate the true depolarizing strength of the error.
In order to bound the possible magnitude of this difference using only the parameters available in the experiment, a substitution for $\gamma$ must be used in Eq.~\ref{Eq:4KSolution}.  In this simplified scenario, $e \geq \gamma (1-q_k(R_{k,\max}))$ is the minimum error that could be detected for a Pauli channel with non-identity Pauli-probability $\gamma$, where the equality is achieved for $\hat{\Lambda}_{\max}$.  Using the boundary value for $\gamma$ as a substitute, but noting that $\gamma \leq 1$ by construction, the result is
\begin{equation} \label{Eq:4KFinal}
 \kappa_{k,\max} \leq \left ( \min \left \{1,\frac{e} {1-q_k(R_{k,\max})} \right \} \frac{D^2}{D^2-1} - 2 e \right ).
\end{equation}
This estimation error is maximized when $q_k(R_{k,\max}) \rightarrow 1$.  Intuitively, when the Clifford operators of the aggregate steps are imbalanced so that a Pauli error goes undetected with high probability, then the discrepancy between the inferred and true parameters is maximized. 

In a more realistic scenario where errors can occur on any of the $l$ $G$ gates interleaved in the sequence, the contributions from all of these gates must be summed.  I assume that the product $l \gamma$ is small so that the probability of more than one error occurring can be neglected.  Then the probability of detecting an incorrect measurement result is lower-bounded by $e \geq \sum_{k=1}^{l} \gamma (1-q_k(R_{k,\max}))$.  This equation pessimistically allows the type of Pauli channel on $G$ to vary at each step to maximally take advantage of the distribution of gates in the remainder of the sequence.  This is an unrealistic model that could be improved by solving a linear program, but it allows for a simpler bound to be expressed.  Because multiple-error events are negligible (allowing higher order terms to be ignored), the standard analysis of the experiment would infer $ 1 - 2 e = (1-p_d)^l \approx (1-l p_d)$, so that the inferred average depolarizing strength is $p_d = 2 e/l$.  As before, the true depolarized strength of the error is $\frac{D^2}{D^2-1} \gamma$.  Using the bound of $\gamma \leq \frac{e}{\sum_{k=1}^{l} 1-q_k(R_{k,\max})}$ to substitute, the difference between the estimated and true strengths of the depolarized error is 
\begin{equation}\label{Eq:4Final1}
\kappa_{\max} \leq \left ( \min \left \{1,\frac{e}{\sum_{k=1}^{l} \left (
1-q_k(R_{k,\max}) \right )} \right \} \frac{D^2}{D^2-1} - 2 \frac{e}{l} \right ). 
\end{equation}
This discrepancy grows as the mean over all step numbers of the deviation between the step distribution and a uniform distribution grows.

First-order lower-bound estimates on the true depolarizing strength can be found in a similar way.  By defining $\kappa_{\mbox{min}} \equiv -\kappa_{\mbox{max}}$ to be the difference between the true depolarizing strength and the estimated strength, it possible to bound 
\begin{eqnarray} \label{Eq:4Final2}
\kappa_{\min} &=& \left (2 \frac{e}{l} - \gamma \frac{D^2}{D^2-1} \right ) \nonumber \\
 &\leq& \left (2 \frac{e}{l} - \frac{e}{\sum_{k=1}^{l} 1- q_k(R_{k,\min})} \frac{D^2}{D^2-1} \right ).
\end{eqnarray}
Again, the discrepancy is larger when the gate distribution is far from uniform.  

For both bounds, the terms $1-q_k(R_{k,\min /\max})$ are expected to approach $\frac{D^2}{2(D^2-1)}$ quickly as $k$ grows, so that the largest contributions to this discrepancy come from small $k$.  Then the discrepancy can be seen to approach $0$ as $l$ increases.  This leads to non-exponential behavior in the scaling of the short-sequence fidelities for approximately-twirled experiments that becomes closer to exponential as lengths increase.  By examining the behavior of the sum $\sum_{k=1}^{l} \left (1-q_k(R_{k,\max}) \right )$ as a function of $l$, one could perhaps better understand the shortest sequence lengths that should be included in the analysis of an RB experiment, although the approximations used here are only valid when $l$ is small compared to the error per step. 

I stress that the two results given in Eq.~\ref{Eq:4Final1} and~\ref{Eq:4Final2} are based on the assumption that at most one error occurs in any sequence and are also based on a very restricted and simplistic gate error model.  Higher-order corrections could be calculated based on the correlations of multiple errors and the twirling behavior of the gates separating them.  However, these first-order estimates for an admittedly-simple model are enough to demonstrate the basic effects of approximate twirls in the worst case.  In particular, these bounds demonstrate that as total-variation distance between the approximate-twirl distribution and the exact-twirl distribution decreases, the possible estimation error decreases.  Also, the results indicate that short sequences demonstrate larger relative deviations from the perfect depolarizing model.  Therefore, when approximate twirls are used, short sequences should be regarded with suspicion.  A third way to minimize this estimation error is to ensure that the distribution of Clifford operators in the inversion step is as uniform as possible, so that the $q_k(R)$ do not take extreme values.  This supports the argument for randomized measurements made in Sec.~\ref{Sec:2MeasChoice}.

The same analysis can be applied to a standard (as opposed to individual-gate) RB experiment.  To first order, one may assume that all steps are perfect except for the $k^{th}$ and sum over the different indices $k$ corresponding to the steps.

In general, if the Pauli twirl is not performed well, because the Pauli operators have large, unitary errors or because the Pauli randomization is not uniform, for example, then errors can add coherently on average and cannot be treated semi-independently as the Pauli channels are in this section.  Treatment of approximate twirls in the case of errors other than Pauli channels is more difficult.  However, there is reason to hope that as experiments advance, the implementations of Pauli operators (which are always tensor products of one-qubit gates) will have low error, so that the relative effect of the errors in the Pauli twirl will be dwarfed by the size of the errors in the multi-qubit gates.  In this scenario, the assumption of Pauli channel errors would still hold reasonably well, and the error bounds provided in this section would be approximately correct.

\chapter{Conclusions}
\label{Chap:5}
Randomized benchmarking has found a place in the broader study of how to quantify the behavior of quantum-information protocols.  Speaking generally, there are three purposes for such a quantification tool.  First, one might wish to simply quantify how well a task is being performed in the current experiment.  Second, one might wish to deduce some more fundamental parameters of an experiment; this is especially common when the goal is to troubleshoot an experiment.  Third, one might wish to infer how well the experiment might perform some task, such as a computation on a larger scale, to which it has not yet been applied.  Randomized benchmarking has the capability to assist with all three of these goals, but it has been especially relevant for inferring how an experiment might perform large-scale computations.  This task is the one for which the most broadly used quantification protocol, quantum process tomography, is the least well suited, and thus RB fills an important void in the quantification of quantum information protocols.

\section{Quantification} \label{Sec:5Quantification}

Demonstrating that the capabilities of quantum-computing efforts are being improved through new techniques and technologies is important both internally (within a group) and externally (to progress the field).  For example, suppose that a new laser is installed in an ion-trap experiment.  It could be that this laser has higher power but worse stability; this can improve certain aspects of quantum control but worsen other aspects.  It is then important to perform a benchmark that accurately reflects the experimental goals in order to determine whether the new laser has a positive or negative effect overall.  If the purpose of an experiment is to perform universal fault-tolerant quantum computation, then I have argued that randomized benchmarking of Clifford operators is a worthwhile way to quantify progress.  This is tied to the prevalence of Clifford operators in fault-tolerant quantum computation and the importance of observing the consistency and average behavior of gates in the context of a long computation.

In order to improve the accuracy and error estimates of the parameters reported by randomized benchmarking, I introduced the use of semi-parametric resampling.  This technique allows for better understanding of the confidence regions for the results of RB in the frequently found case where the statistics of the sequence fidelities are known to be non-Gaussian.  In addition, resampling can reveal information about the bias in the inference procedure, which is known to be a crucial problem for the analysis of quantum process tomography.  When approximate-twirling techniques are used for RB, bounds should be placed on the discrepancy between the true and estimated error per step caused by this approximation.  In Sec.~\ref{Sec:4AppDep} I described a simplified model that leads to first-order estimates of these bounds.

In order to compare benchmark results from different experiments, it is important to ensure that the reported parameters have a common interpretation.  I have placed loose bounds on the comparison between benchmarks defined using approximately-twirled steps and benchmarks defined using exactly-twirled steps.  More importantly, I have argued that for most purposes it is preferable to perform the benchmark using an exact twirl; this not only standardizes the experiment, making comparisons easier, but also reduces the complexity of the analysis.

Even in the cases where the assumptions, such as time independence and exact twirling, introduced in Sec.~\ref{Sec:2GateError} do not hold, randomized benchmarking is still useful for describing the status of an experiment.  If time dependence is observed in the decay of sequence fidelities, for example, then progress can be demonstrated by decreasing the dependence in future experiments.  Three different approaches to dealing with these assumptions have been demonstrated in this thesis: First, I tried to analyze (especially for the assumptions of approximate twirling and gate-dependent errors) when these assumptions are necessary and how their violation affects the results of the benchmark.  Second, I described modifications to the design and analysis of the RB experiments that allow the experimenter to sidestep these assumptions to some degree.  Finally, I developed a statistical analysis procedure that allows the experimenter to evaluate when the RB experiment is not well described by these strict assumptions.  It is this last effort that is most important for quantifying progress in experiments; by testing for signatures of the violations of these assumptions, one can demonstrate how they change from experiment to experiment.

I have also described an extension of the randomized-benchmarking procedure that allows for the benchmarking of individual gates.  This enables the quantification of the difficult and crucial task of demonstrating a two-qubit entangling gate with fidelity above a fault-tolerance threshold.  Further extensions, such as the benchmarking of cross-talk errors introduced by Ref.~\cite{Gambetta2012}, will enhance the breadth of tasks for which randomized benchmarking can be used. 

\section{Deducing Systematics} \label{Sec:5Deducing}
For most parameters that might be of interest in an experiment, randomized benchmarking is not the ideal diagnostic tool.  While RB is useful for deducing the overall fidelity of a gate, it is not well suited to deducing the probability that the gate suffers from a particular sort of error.  Typically, the tool best suited to such a task is either quantum process tomography (as a general tool) or a protocol specifically designed to isolate that parameter.  For example, a Rabi-oscillation experiment is ideal for the task of investigating the Rabi frequency of a qubit.

The great appeal of quantum process tomography is that by performing the protocol once, one is able to extract a large number of experimental parameters.  Particularly, it is not necessary to know in advance which parameters are of interest; QPT can be performed first, and the data can be post-analyzed to decide which parameters are important and then to quantify them.  In contrast, the techniques used to extract parameters such as cross-talk error through randomized benchmarking require the experiment to be specifically redesigned with this goal in mind.  

There are several difficulties with using QPT to deduce experimental parameters.  The first, discussed at length in Chapters \ref{Chap:1} and \ref{Chap:2}, is the difficulty of separating errors in state preparation and measurement from the parameters of interest.  This is particularly problematical when the values of the parameters are small, as is the case when the goal is to quantify the infidelity of a very good gate.  This issue can be ameliorated to some degree by using the ``overkill'' tomography of Ref.~\cite{Merkel2012}.  The second problem is that there is no obvious way to describe the performance of a gate in the context of a large computation. In the QPT procedure, the gate is always performed shortly after the preparation protocol.  This is not necessarily the context in which the gate would be performed in a large computation.

The final difficulty with using quantum process tomography is the adverse, exponential scaling that occurs as the number of qubits involved increases.  Refs.~\cite{Flammia2011} and~\cite{Cramer2010} introduced some strategies to remove the exponential scaling of QPT, but these techniques have not yet been tested extensively in experiments.  For the few parameters that can be estimated using randomized benchmarking in a straightforward way, I have attempted to reduce further the resource requirements of the experiment by improving the statistical analysis for the situation when a small number of sequences are used.  I have also discussed experiment-design choices that allow one to use experimental time effectively.  For the restricted problem of discerning individual gate fidelities, I introduced subgroups of the Clifford group that can be used to implement randomized benchmarks with reduced control requirements and, possibly, reduced overhead.  For any kind of Clifford randomized benchmark, algorithms for sampling from, computing with, and inverting Clifford operators can simplify the classical computation requirements significantly and allow for the design of more efficient experiments. 

\section{Inferring Large-Scale Performance} \label{Sec:5Inferrence}

Attempts to infer the performance of a quantum control architecture on a large-scale computation are prevalent in literature but are quite difficult to construct carefully.  There are several ways that reporting an experiment in this way is appealing.  First, it suggests progress toward a quantum computer.  More importantly, however, attempts to describe the performance of a hypothetical large-scale computation using small-scale experiments force the experimenter to think about aspects of computer architecture and fault tolerance.  This connection to theory provides benefits to those studying architectures, as they are given specific error models to address.  The connection also provides a key benefit to the experimenter: by taking a larger view of the components necessary for a large-scale computer, the experimenter is less likely to be distracted by short-term fixes that are impractical in the long term.  The loss of interest in liquid-state NMR quantum computing is an extreme example of this process; although liquid-state NMR can have very high-fidelity gates, increasing focus on the lack of scalability of this strategy led to the abandonment of this platform as a prospect for large-scale quantum computation.

The ultimate goals of attempts to infer the behavior of large-scale computation are, first, to decide whether the qubits and gates in the small-scale experiment would be sufficient to implement a general fault-tolerant quantum computation and, second, to quantify how many resources (gates, qubits, time, etc.) would be required to implement a specific algorithm.  To count these resources with high accuracy, one needs accurate specifications of the errors on all of the physical processes in the computer and also detailed understanding of how to design algorithms and fault-tolerant architectures that take advantage of the properties of these physical processes.  In practice, both of these requirements are too difficult to accomplish without a large time commitment.  Instead, simplified analyses are used that consider reduced descriptions of the physical processes and standard, non-optimized architecture and algorithm assumptions.  

A simple proxy for the question of whether experimental control is sufficient to implement fault-tolerant computations is to ask whether it has a set of Clifford gates with depolarizing error below a given fault-tolerance threshold.  Even with this simplification, there is still debate about which fault-tolerance threshold to use; in this thesis I have chosen to use the conservative estimate of $10^{-4}$ in error per gate for one- and two-qubit Clifford gates. 
Measurement and state preparation errors, while important, are often ignored in order to simplify the problem because fault-tolerance theory indicates that their requirements are not as stringent.  

With these simplifications, the problem is rephrased to ask whether the experiment is capable of implementing a set of gates with errors below a threshold in a context that approximates typical fault-tolerant computation.  If so, then one might predict that a scaled-up version of the experiment with many qubits might be able to execute an interesting quantum computation.  However, even this last inference relies on the assumption that a larger-scale experiment would have gates that behave in the same way as the small-scale experiment.  Because of interactions between gates and qubits, this assumption is not usually valid.  Analysis of cross-talk is a crucial step toward making the inference to larger computations more rigorous.

Randomized benchmarking is an appealing tool for making inferences about long computations because it inherently addresses the issue of placing the gates in a computational context.  Although RB is limited in the precise parameters it can quantify, the parameters useful for comparison with typical fault-tolerance thresholds are conveniently accessible using the standard RB protocol and the individual benchmark extensions.  The limitation of describing random sequences of Cliffords seems not useful at first, but pseudo-randomizing quantum algorithms in various ways can provide an important way to avoid coherent errors with little downside.  If an algorithm is pseudo-randomized with respect to its Clifford gates, then the errors per step and gate returned by RB protocols are ideal figures of merit. 

Two limitations of randomized benchmarking for the purpose of this inference do exist.  First, the crucial assumption that errors are time-independent is too restrictive.  Although a quantum computation must have limited time-dependence in its errors if it is to compute successfully, there is no reason to think that a computation with some short-term or low-level time dependence in its errors could not work.  The same argument applies to a lesser degree to the assumption that all errors are trace-preserving; there are known protocols (in optical quantum computing especially) that deal successfully with non-trace-preserving errors.  Second, the results of the usual formulation of randomized benchmarking do not provide any information about what will happen as more qubits are introduced into the system and cross-talk of various kinds becomes an issue.  One could imagine a suite of tests in which the qubit number is increased from experiment to experiment to establish some sort of consistent (hopefully constant) scaling in gate fidelities as a function of qubit number.  In such a way, a more valid inference could be made to the large-scale behavior of the system.  No such test has been established, but this seems like a valuable avenue for future research.

Finally, as experiments begin to adopt specialized fault-tolerant architectures, the simplification of only reporting depolarizing strength will be of limited use.  It is well-known that the effectiveness of a quantum code is dependent on the types of errors it must correct.  Typically, only errors with small weight (that is, errors that only affect a small number of qubits) can be corrected.  Some codes can also preferentially protect against errors of the $Z$ type, for example.  In order to select an effective code, it is important to know more about the errors on the system than randomized benchmarking typically reveals.  The techniques of Refs.~\cite{Moussa2011} and~\cite{Emerson2007} use different kinds of twirls in a randomized-benchmarking-like framework to reveal other useful information about the errors in a process.  I think that such techniques, once further developed, will provide information useful for selecting better from fault-tolerant architectures.  Once specific experiment and architecture information is available, the task of inferring the computational overhead for a useful computation can be solved in a more meaningful way (see Ref.~\cite{Jones2012} for an example based on quantum dots).

An experiment performing an encoded randomized benchmark would be an exciting demonstration of quantum control.  Although challenging and interesting by itself, the greater goal for such an experiment is to perform both bare and encoded benchmarks on the same system in order to show improvement in operator fidelities through the use of encoding.  I have introduced some of the requirements for such an experiment and described why it would be an important milestone.

\section{Open Problems}

As an experimental technique and theoretical procedure, randomized benchmarking is still relatively young and under-developed, especially compared to quantum process tomography.  At this point, one-qubit RB experiments have already been performed in most of the systems considered serious prospects for quantum computation, with the exceptions of quantum dots and photonics.  In contrast, multi-qubit benchmarks have only been performed by three groups (see Refs.~\cite{Ryan2009},~\cite{Gaebler2012}, and~\cite{Corcoles2012}).  These experiments are considerably harder and have varied more significantly in their implementations of RB.  Multi-qubit procedures are likely to undergo several more improvements before a standard is established for the community.

One of the most important open questions in randomized-benchmarking theory is the poor understanding of the behavior of benchmarking experiments in the absence of the strong time- and gate-dependence assumptions.  Numerical simulations, as seen in Ref.~\cite{Magesan2011B}, for example, show that the analysis problems induced by violating these assumptions are significant but nowhere near as bad as the available pessimistic bounds predict.  In other words, there is a lack of understanding about why randomized benchmarking seems to work well, given that the assumptions, especially that of gate-independent error, are almost certainly violated in practice.

Assumptions about gate error models are especially crucial for individual-gate, or interleaved, benchmarks.  The simplified results presented here are only strictly valid when either the step or the gate has depolarizing error, but this is not a good assumption in general.   However, numerical analysis performed by the quantum computing group at IBM has indicated that interleaved benchmarking still works reasonably well most of the time even when this assumption is violated \cite{Gambetta2013}.  In order to use interleaved benchmarks formally to compare different experiments, deduce specific experimental systematics, or extrapolate to large-scale computing, it would be necessary to better characterize the role of these assumptions.

There is good reason to believe that time-dependent errors can cause major problems for the benchmarking analysis of actual experiments.  The most basic approach to this problem in this thesis has been to attempt to provide a test for time-dependence so that the benchmark can be regarded with proper confidence.  However, a much more attractive approach would be to modify the benchmark experiment so that it could quantify the time dependence of the average operator fidelity.  In Ref.~\cite{Gaebler2012L} we introduced an ad hoc attempt to do just this, but a formal, scalable treatment would be much more valuable.

Errors that do not preserve the qubit subsystem are perhaps the least-well studied of all of the problems for randomized benchmarking.  While it is not difficult to come up with adversarial errors of this kind, I know of no published efforts to study the sorts of realistic errors that one might encounter in common systems.  These can be loosely divided into loss-type errors, where the state is permanently removed from the subsystem, and memory-type errors, where correlation with the environment or temporary excitation causes the errors to exhibit a finite memory lifetime.  Loss-type errors are probably not hard to incorporate into the description of a benchmark experiment, as their interactions with other errors are simple.  Memory-type errors, which share some overlap with time-dependent errors, have the potential to be much more subtle and difficult.

A promising mathematical approach to new randomized-benchmarking procedures involves studying the effects of twirling over different unitary operator groups.  I have described Clifford subgroups that have the same twirling effects as the twirl over the entire Clifford group, but there might be smaller subgroups or entirely different groups that exhibit this behavior.  In addition, Ref.~\cite{Emerson2007} introduced a procedure that uses a different kind of twirl to reveal different interesting parameters of gate errors.  It is likely that there is a large class of similar procedures based on different groups that could be of use for characterizing errors.

Another algebraic issue warranting attention is the comparison between errors per step on different-sized registers, as between RB experiments on two and three qubits.  Because the Clifford groups on any number of qubits greater than one are all generated by the same kinds of gates, it should be possible to establish some predictive scaling that uses the error per step on a small number of qubits to bound the error per step that would be possible on a larger number of qubits.  This will require some assumptions about the consistency of gates and the scaling of the cross-talk errors as the size of the register changes.  Comparing RB experiments with different numbers of qubits is valuable by itself for quantifying progress, but understanding this scaling is also important at a more fundamental level for the design and analysis of large-scale quantum computers.

By improving algorithms for manipulating Clifford operators, one can improve the design and performance of RB protocols.  Ref.~\cite{Amy2013} gives a related example of how to find a large computational-overhead reduction by optimizing the way $Z \left (\frac{\pi}{4} \right )$ and $\CX$ gates are used in a circuit.  Similarly, by designing algorithms that allows Clifford operators to be decomposed into gates in different ways, one can implement arbitrary Clifford operators more efficiently in RB experiments.  These same algorithms can also help to design more-efficient encoding and decoding algorithms for quantum codes. The Clifford decomposition algorithms currently available and applicable to RB do not take advantage of parallelism in control, and they make assumptions about the convenient gate set.  Both of these restrictions deserve attention, as does the general computational efficiency of the algorithms.

Finally, at the same time as specific benchmarking procedures are standardized, the quantum computing community should converge on a suite of procedures and benchmark parameters to use to describe their quantum control capabilities.  This suite might include reporting of the results of the standard RB procedure for different-sized subsets of qubits along with individual-gate benchmarks for the most crucial gates.  Ref.~\cite{Gambetta2012} has advocated reporting cross-talk errors as well; this seems like an important addition in the common situation where scalability is a major concern.  On the other hand, there is a danger of reporting too many parameters and spending too much time on their calculation.  As with quantum process tomography, a compromise must be made between calculating and reporting all parameters that could be of use and finding a set of parameters small and meaningful enough to be understandable.

\bibliographystyle{plain}	
\bibliography{ThesisBib}		

\appendix

\chapter{Clifford Simulation}	
\label{App:A}

Because the standard randomized benchmarking procedures are based on Clifford operators, it is advantageous to have an efficient way of simulating the evolution of quantum states under the action of Clifford operators.  Fortunately, the Gottesman-Knill theorem \cite {Gottesman1998} provides a framework that allows exactly this: by specifying states using their stabilizing operators, one can simulate computations that involve only Clifford operations in time that scales polynomially with the number of qubits involved.  This ability to simulate efficiently Clifford operations is the most convincing reason to use Clifford twirling instead of full unitary twirling in the randomized benchmarking protocol.  In this appendix, I give some background on Clifford operations and the Gottesman-Knill theorem before describing a Clifford simulator written in collaboration with E. Knill that was used to facilitate and study randomized-benchmarking experiments.

\section{Clifford Operations} \label{Sec:ACliffords}

The set of computational steps consisting of preparation of qubits in the $\ket{0}$ or $\ket{1}$ states, Clifford operators, and measurement of the $Z_i$ operators generates (by composition) the set of Clifford operations (or stabilizer operations).  Notable quantum computer operations that are not Clifford operations include the one-qubit operations corresponding to rotations about the $X,Y,Z$ axes by $\frac{\pi}{4}$, the Toffoli gate, measurements of non-Pauli Clifford operators,  and preparation of so-called magic states \cite{Bravyi2005}.  In fact, because the group of unitary operators is infinite and the group of Clifford operators is finite (for any fixed $n$), there is a sense in which Clifford operations are sparse in the full set of quantum operations.  

\subsection{Pauli Operators} \label{Sec:APauliOps}
The importance of Clifford operators stems directly from that of the Pauli matrices and their associated operators.  Define the group of $n$-qubit Pauli matrices $P_{\vec{a}} = \imath^{\zeta} \bigotimes_{i=1}^{n} P_{a_i}$, where $\vec{a}$ is a list of integers $\mbox{mod } 4$, $\otimes$ is the Kronecker product, $\zeta \in \{0,1,2,3\}$, and 
\begin{equation} \label{Eq:APauliDef}
P_{0} = I = \left( \begin{array}{cc}
1 & 0 \\
0 & 1 \end{array} \right),
P_{1} = X = \left( \begin{array}{cc}
0 & 1 \\
1 & 0 \end{array} \right),
P_{2} = Z = \left( \begin{array}{cc}
1 & 0 \\
0 & -1 \end{array} \right),
P_{3} = Y = \left( \begin{array}{cc}
0 & -\I \\
\I & 0 \end{array} \right).
\end{equation}

I define the group of Pauli operators on $n$ qubits to be isomorphic to the quotient of the group of Pauli matrices on $n$ qubits with the scalar multiples $\imath^{\zeta}\otimes_{i=1}^{n} I_i$ of the identity matrix.  This definition of the Pauli operators (as distinct from the Pauli matrices) is somewhat non-standard.  In shorthand, I frequently refer to any group of such multi-qubit Pauli operators as a Pauli group, although previous literature frequently reserves this terminology for the one-qubit group.  I denote the Pauli operator group on $n$ qubits with $\mathcal{P}_n$ and its elements using the same notation as for the Pauli matrices (but without the scalar).  

By choosing the Pauli matrix with $\zeta=0$, I define a standard matrix representative of each Pauli operator coset.  It is important to note that the elements of this set of representatives of Pauli operators are orthogonal, i.e., $\tr{P_{\vec{a}} P_{\vec{b}}} = 2^n \delta_{\vec{a},\vec{b}}$.  The Kronecker product on the matrices translates to a tensor product structure on the operators indexed by the individual qubits, so that $P_{\vec{a}}$ acts as $P_{a_i}$ on the $i^{th}$ qubit.  When the qubit indices of the Pauli operators on multiple qubits are not important, they are written as $P_k$, where the index $k$ is an integer between $0$ and $4^n-1$ ($k=0$ always identifies the identity operator).  If $G$ is an operator acting on a specified ordered set of $m$ qubits, then its action can be extended to any larger set of qubits containing the specified set by tensoring with the identity on the other qubits.  This extended operator is also denoted by $G$ when the extension is obvious from context.

The most important feature of the standard matrix representatives of the Pauli operators for the purposes of this thesis is that they form an additive basis for the complex vector space of all $2^n \times 2^n$ matrices.  That is, any $2^n \times 2^n$ matrix $M$ can be uniquely written as $M = \sum_{\vec{a}} c_{\vec{a}} P_{\vec{a}},$ with each $c_{\vec{a}} \in \mathbb{C}$.  Sets of matrices with this property are called error bases in Ref.~\cite{Knill1996} because they are used in the study of quantum operations to decompose error superoperators on $n$ qubits into constituent unitary parts whose effects on a computation can be studied independently.  

Given two Pauli matrices $P$ and $Q$, their product $PQ$ is equal to $\pm QP$.  If the sign in this equation is positive, the matrices (and their corresponding operators) commute; otherwise, they are said to anti-commute.  The single-qubit Pauli operators $X,Y,Z$ pairwise anti-commute.  Two Pauli operators commute when their non-identity tensor parts do not overlap in indices. Together with the group structure, these rules allow the computation of the commutation relation of any two multi-qubit Pauli operators.  Another related, often-used feature of the standard representatives of the Pauli operators is that they are self-inverse.  As a consequence their eigenvalues can only be $\pm 1$.

Groups of matrices and operators with similar properties to the qubit Pauli operators have been found for quantum systems with Hilbert spaces of any dimension $k$.  Ref.~\cite{Knill1996} presents as an example a construction that leads to a group of matrices on such Hilbert spaces forming an error basis.  I refer to the corresponding operator groups as the Pauli operator groups of dimension $p$ when the dimension $p$ is a prime.  In particular, the qutrit Pauli operator group is the group formed in this way for systems with Hilbert-space dimension $3$.  Multi-qutrit Pauli operators can be formed from tensor products of qutrit Pauli operators in the same way as for qubit Pauli operators.  Uses for these groups are described briefly at the end of Sec.~\ref{Sec:AClifDef}. 

\subsection{Clifford Group Definition} \label{Sec:AClifDef}
A matrix $C \in U(2^n)$ is an $n$-qubit Clifford matrix if, for any Pauli matrix  $P_{\vec{a}}$ (with $\zeta=0$), there exists a Pauli matrix $P_{\vec{b}}$ (with $\zeta=0$) such that $C P_{\vec{a}} C^{\dag} = \pm P_{\vec{b}}$.  
The Clifford matrices form a subgroup of $U(2^n)$.  This rule for the conjugation by a Clifford matrix of a Pauli matrix implies that the subgroup of Clifford matrices is the normalizer of the subgroup of Pauli matrices in $U(2^n)$.
Notably, the Pauli matrices are also Clifford matrices, as $PQP^{\dag} = \pm Q P P^{\dag} = \pm Q$ for all Pauli matrices $P,Q$.  The sign in this equation describes whether the Pauli matrices commute or anti-commute, as described in the previous section.

I identify the $n$-qubit Clifford operators with elements of the quotient of the group of Clifford matrices on $n$-qubits with scalar multiples $e^{\imath \phi} I$ of the $n$-qubit identity matrix.  I denote this group with $\mathcal{C}_n$. Because the Pauli matrices are a subgroup of the Clifford matrices, there is a natural embedding of the Pauli operators in $\mathcal{C}_n$ implied by this identification.   Table~\ref{Tab:AClifSize} presents the sizes of the Clifford and Pauli operator groups for different $n$; both groups are finite for all $n$, despite the fact that the groups of Clifford matrices are all infinite due to the unconstrained complex scalar.  The elements $C \in \mathcal{C}_n$ are defined entirely by the conjugation action of their representative matrices on the Pauli matrices (as shown in Lem.~\ref{Lem:AClifAction}): if $C_U$ is a unitary matrix representing the Clifford operator and $P$ is a Pauli matrix, then I define the Clifford operator action $C(P) \equiv C_UPC_U^{\dag}$.   

\begin{lemma} \label{Lem:AClifAction}

A Clifford matrix $C$ is uniquely defined up to an overall phase (complex scalar) by its action by conjugation on the group of Pauli matrices.

\paragraph{Proof:}   Consider the set of matrices  $E_{i,j}$ that have zeroes in the standard basis except for a $1$ in the $(i,j)$ location.  Because each $E_{i,j}$ can be written as a sum over Pauli matrices and the Clifford action distributes across sums, it is possible to calculate $A_{i,j} = CE_{i,j}C^{\dag}$ given the knowledge of the Clifford action on the Pauli matrices.  The $(l,m)$ entry of $A_{i,j}$ is $(A_{i,j})_{l,m} = (C)_{l,i}(C)_{m,j}^*$.  At least one entry of $C$ must be non zero since its action on Pauli operators does not transform them to the $0$ operator.  Without loss of generality, assume $(C)_{a,b} \neq 0$.  Then $(A_{a,a})_{b,b} = |(C)_{a,b}|^2 \neq 0$. For this $(a,b)$ pair, $(C)_{a,b} = \gamma |\sqrt{(A_{a,a})_{b,b}}|,$ where $\gamma$ is a unit-magnitude complex number.  Now, for any other element, $(C)_{c,d} = \frac{(A_{d,b})_{c,a}}{(C)_{a,b}^*} = \frac{\gamma (A_{d,b})_{c,a}} {|\sqrt{(A_{a,a})_{b,b}}|}$.  Using this formula, any element of the matrix $C$ can be calculated in the standard basis up to the consistent factor $\gamma$.  Therefore, the unitary matrix that has the given action is uniquely specified up to an overall phase. $\square$
\end{lemma}

The proof of this lemma demonstrates how to find a unitary matrix representation of a Clifford operator specified by its action.  The phase ambiguity in the definition specifies a Clifford matrix up to its Clifford operator coset; therefore, the Clifford action specifies a Clifford operator without ambiguity.    It is also straightforward to calculate the transformation rules for a Pauli operator if a unitary matrix representation of $C$ is known.  Note that not all sets of transformations of Pauli elements can define a Clifford operator.

There are several important properties of this action; together they prove that the action of each Clifford operator on the Pauli matrices is an isomorphism that fixes $\imath^{\zeta} I$ for any $\zeta \in \{0,1,2,3\}$.  
\begin{enumerate}
\item 
Action on the identity matrix: $C(\imath^{\zeta} I) = \imath^{\zeta} CIC^{\dag} = \imath^{\zeta} I$ for $\zeta \in \{0,1,2,3\}$.
\item
Action on non-identity matrices: Because $C(P) = I \Leftrightarrow CPC^{\dag}=I \Leftrightarrow P = I$, the action of any $C$ on a non-identity Pauli matrix leaves a non-identity Pauli matrix. 
\item Homomorphism: The action of a Clifford operator on the group of Pauli matrices is a homomorphism.
\begin{equation}
C(P_{\vec{a}}P_{\vec{b}}) = CP_{\vec{a}}P_{\vec{b}}C^{\dag} = CP_{\vec{a}}C^{\dag}CP_{\vec{b}}C^{\dag} = C(P_{\vec{a}})C(P_{\vec{b}})
\end{equation}
\item
Preservation of commutation relations: If $P_{\vec{a}}P_{\vec{b}} = (-1)^d P_{\vec{b}}P_{\vec{a}}$, then previous properties indicate that $C(P_{\vec{a}})C(P_{\vec{b}}) = (-1)^d C(P_{\vec{b}})C(P_{\vec{a}})$, so the action preserves commutation relations of Pauli matrices.
\item
Injectivity: 
$C(P_{\vec{a}}) = C(P_{\vec{b}})$ implies that $CP_{\vec{a}}C^{\dag} = CP_{\vec{b}}C^{\dag}$, so $P_{\vec{a}} = P_{\vec{b}}$.
\item
Inversion: 
$C^{\dag}(C(P)) = P$, so the map given by the action of $C^{\dag}$ is the inverse of the map given by the action of $C$.
\end{enumerate}

Because there are a finite number of Pauli matrices for a fixed $n$, the injectivity and homomorphism properties are enough to show that the map provided by the action of any Clifford operator is an isomorphism.  The consequences of this algebraic structure are important.  First, the action of a Clifford operator on any Pauli matrix can be specified by its action on a (multiplicative) generating set of Pauli matrices.  Second, preservation of commutation relations implies that the action of any Clifford operator maps a set of mutually-commuting Pauli matrices to a set of mutually-commuting Pauli matrices.  This fact, in combination with Eq.~\ref{Eq:AClifUpdate}, implies that if $\ket{\psi}$ is a simultaneous eigenstate of the matrix representatives of $k$ distinct Pauli operators, then $C\ket{\psi}$ is also a simultaneous eigenstate of the matrix representatives of $k$ distinct Pauli operators.  When a state is described as being an eigenstate of a Clifford operator, I mean that it is an eigenstate of any matrix representative of the operator.

Any Clifford operator $C$ can be written as the product $C = P\tilde{C}$ of a Pauli operator $P$ and a Clifford operator $\tilde{C}$ such that $\tilde{C}X_i\tilde{C}^{\dag} = P_{\vec{a},i}$ and $\tilde{C}Z_i\tilde{C}^{\dag} = P_{\vec{b},i}$ for every $i$ and some Pauli matrices $P_{\vec{a}/\vec{b},i}$ (with scalar $1$).  Here $X_i$ and $Z_i$ refer to the Pauli operators which are tensor products of the $X$ or $Z$ operators on qubit $i$ with identity operators on the other qubits.  The existence of this decomposition is a consequence of the action of Pauli operators considered as Clifford operators given above: the Clifford action of a Pauli operator on a Pauli matrix can only change its sign, allowing for the separation between the part of the operator changing the type of Pauli matrix and the part changing its sign.  The restricted class of Clifford operators satisfying the condition for $\tilde{C}$ can be thought of as labeling (right) cosets of the Pauli group in the Clifford group, and these cosets are the elements of a quotient group $\overline{\mathcal{C}}_n = \mathcal{C}_n / \mathcal{P}_n$.  Practically, considering this quotient group instead of the entire Clifford group in derivations and calculations involving the Clifford operators separates the effects of Pauli operators from the interesting effects of Clifford operators and makes the group size more manageable.  In fact, it is also possible to define double cosets of the form $\mathcal{P}_n C \mathcal{P}_n$.  Considering the set of double cosets for problems such as Clifford decomposition (see Sec.~\ref{Sec:ADecomp}) greatly reduces the number of elements that need to be considered.

In terms of experimental implementation, the group of Pauli operators is often thought of as the group generated by $\pi$ rotations about the $X,Y,$ and $Z$ axes of the Bloch sphere for each qubit.  Then the group of Clifford operators is the group generated by the $\frac{\pi}{2}$ rotations about the $X,Y,$ and $Z$ axes on each qubit and $\CX$ gates between pairs of qubits.  Because of the relative ease of one-qubit operators and historical (and experimental) focus on the $\CX$ gate, implementations of the Clifford operators are frequently well-developed in quantum control experiments.  The $\CX$ in this construction may be replaced with several common experimental two-qubit gates, including the $\CZ$ gate and the M{\o}lmer-S{\o}rensen gate.  See App.~\ref{App:C} for definitions of these gates.  Note that the Clifford group does not contain the $\frac{\pi}{4}$ rotations about the $X,Y,Z$ axes (or any smaller-angle rotations).  

Clifford operators have attracted considerable interest from the quantum information community because of the middle ground they appear to occupy between classical and universal-quantum computation.  The Gottesman-Knill theorem described in the next section implies that a computation that consists only of Clifford operators and measurement and preparation in the fiducial ($\ket{0},\ket{1}$) basis can be efficiently simulated using a classical computer.  This means that Clifford operators must not be sufficient to achieve any super-polynomial quantum-computing speed-ups.  On the other hand, Clifford operators are enough to generate entanglement, encode/decode quantum stabilizer codes, perform quantum teleportation, and achieve superdense coding; these tasks are all frequently seen as important non-classical capabilities of quantum computing.

Although they are not described in depth in this thesis, it can be useful to consider the normalizers of the $p$-level system Pauli matrix groups mentioned in the previous section.  The operators corresponding to the matrices in this normalizer are the $p$-level-system Clifford operators, and, with care, qubit Clifford operator results can typically be extended in a fairly straightforward way to work for these operators as well.  In particular, stabilizer coding, Clifford simulation, and randomized benchmarking can be performed using $p$-level-system Clifford operators.  Table \ref{Tab:AClifSize} includes the sizes of some multi-qutrit Clifford groups for reference.

\begin{table} 
\begin{center}
\begin{tabular}{|c|c|c|c|}
\hline
& $|\mathcal{P}_n|$ & $|\mathcal{C}_n|$ & $|\overline{\mathcal{C}}_n|$ \\
\hline
1 & 4 & 24 & 6 \\
\hline
2 & 16 & 11520 & 720 \\
\hline
3 & 64 & $~9*10^7$ & 1451520 \\
\hline
4 & 256 & $~1*10^{13}$ & $~5*10^9$ \\
\hline
5 & 1024 & $~2*10^{19}$ & $2*10^{16}$ \\
\hline
1 (qutrit) & 9 & 216 & 24 \\
\hline
2 (qutrit) & 81 & 4199040 & 51840 \\
\hline
3 (qutrit) & 729 & $~7*10^{12}$ & $~9*10^{9}$ \\
\hline
\end{tabular}
\end{center}
\caption[Clifford and Pauli Group Sizes]{Presented are the sizes of the Pauli groups $\mathcal{P}_n$, the Clifford groups $\mathcal{C}_n$, and their quotients $\overline{\mathcal{C}}_n = \mathcal{C}_n / \mathcal{P}_n$ for $n=1-5$ qubits and $n = 1-3$ qutrits. \label{Tab:AClifSize}}
\end{table}
 
\subsection{Stabilizer States} \label{Sec:AStabilizer}
In this section, I discuss states that are specified uniquely by the operators for which they are eigenstates.  If a state $\ket{\psi}$ is an eigenstate of the Pauli operator $P_{\vec{a}}$ with eigenvalue $\gamma_{\vec{a}} \in \{1,-1\}$ and $C(P_{\vec{a}})=\pm P_{\vec{b}}$ for some Clifford operator $C$, then the state $C\ket{\psi}$ is an eigenstate of the Pauli operator $P_{\vec{b}}$:
\begin{equation} \label{Eq:AClifUpdate}
P_{\vec{b}}C\ket{\psi} = \pm CP_{\vec{a}}\ket{\psi} = \pm \gamma_{\vec{a}}  C\ket{\psi}.
\end{equation}
Calculations of this form can allow for efficient computation of the result of a Clifford operator applied to a state through use of knowledge of the Clifford action and the operators that stabilize the state. 

Theorem 1 of Ref.~\cite{Aaronson2004} establishes that if a vector for a pure state $\ket{\psi}$ is a simultaneous $+1$ eigenvector of a group of $2^n$ $n$-qubit Pauli matrices, none of which is a scalar multiple of another, then it is unique in this sense.  The corresponding state is called a (pure) stabilizer state and the operators corresponding to the Pauli matrices are called stabilizers and form a group.  As proven in Lem.~\ref{Lem:AGenSet}, the group of stabilizers is generated by $n$ Pauli operators; these generators and the eigenvalues of the generators on a state then specify the stabilizer state uniquely.   Because each Pauli operator $P_{\vec{a}}$ on $n$ qubits can be indexed by a length-$n$ list of integers $\{0,1,2,3\}$, a stabilizer state can be uniquely specified with $n(2n+1)$ bits, where the extra $n$ bits specify which eigenstate of each operator the state occupies.  In comparison, a general $n$-qubit quantum state requires roughly $2^n$ complex numbers to specify uniquely (overall phase ambiguity and normalization conditions reduce this slightly).  Therefore stabilizer states can be represented much more efficiently than general quantum states in a classical computer.  I show later in this section that computations involving only stabilizer states can also be performed more efficiently than general quantum computations. 

\begin{lemma}
\label{Lem:AGenSet}

A group $\mathcal{G}$ of $2^n$ mutually-commuting Pauli operators on $n$ qubits contains a minimal generating set of size $n$.

\paragraph{Proof:} Produce a set of independent Pauli operators as follows by iterating through the elements of $\mathcal{G}$ in any order: Let $\langle \{Q_i'\}_i \rangle$ denote the group of Pauli operators generated (multiplicatively) by the set $\{Q_i'\}_i$ of generators marked so far.  Choose the next element $Q$ from $\mathcal{G}$. If $P \notin \langle \{Q_i'\}_i\rangle$, then add it to $\{Q_i'\}_i$. After $\{Q_i'\}_i$ has been constructed in this way, assume it has size $k$. 

Suppose $k<n$.  Then all possible elements in $\langle \{Q_i'\}_i \rangle$ can be written as $\prod_{i=1}^k Q_i'^{a_i}$, where $a_i \in \{0,1\}$.  There are only $2^k$ elements of this form, so at least one element of $\mathcal{G}$ is not contained in $\langle \{Q_i'\}_i \rangle$, contradicting the construction. 

Now, suppose $k>n$.  Suppose that for an element $P \in \mathcal{G}$ one can write $P = \prod_{i=1}^k Q_i'^{a_i} = \prod_{i=1}^k Q_i'^{b_i}$, then $I = \prod_{i=1}^k Q_i'^{a_i-b_i}$ implies $a_i = b_i$, since each generator cannot be written as a product of other generators.  Therefore, each product $\prod_{i=1}^k Q_i'^{a_i}$ represents a unique element in $\langle \{Q_i'\}_i \rangle$, and $|\langle \{Q_i'\}_i \rangle| = 2^k > 2^n$, contradicting the claim that $\mathcal{G}$ is a group of size $2^n$. Therefore, $k=n$. $\square$

\end{lemma}

The minimal generating set of Pauli operators for $\mathcal{G}$ in the preceding lemma can be thought of as a basis for a space of Pauli operators.  An isomorphism that formalizes this intuition exists from the Pauli operators to a vector space over $GF[2]$ of dimension $2n$.  This extends to a matrix representation over the same field for $\overline{\mathcal{C}}_n$.  I use this vector space notation only informally in this appendix, although several proofs, including the preceding one, could be more simply expressed by first establishing and then utilizing the full formalism.

A corollary of the lemma above is that the result of applying a Clifford operator to a stabilizer state is a stabilizer state.  This is the essential technique used to simulate computations consisting only of Clifford operators; as long as the initial state is a stabilizer state, only the space of stabilizer states needs be considered in such a simulation until a measurement is reached.

Computing the result of a measurement of a Pauli operator $P$ on a stabilizer state is also necessary for such simulations.  In the following discussion, I show that such a measurement also results only in stabilizer states, enabling its efficient simulation.  To simplify the problem, I show in Lem.~\ref{Lem:AMeasTrans} that it is sufficient to be able to compute the result of measuring the $Z_1$ operator on any stabilizer state, where $Z_1$ indicates the tensor product of the Pauli $Z$ operator on the first qubit with identity operators on all other qubits..  In order to introduce the process of measuring a Pauli operator on a stabilizer state, I first lay out the process for measuring $Z_1$ on any stabilizer state.

Let $S$ be the set of stabilizers for $\ket{s}$.  If $Z_1 \in S$, then the result of the measurement of the operator $Z_1$ will be the eigenvalue of $Z_1$ corresponding to $\ket{s}$ and the resulting state will be $\ket{s}$.  

If $Z_1$ does not stabilize $\ket{s}$, then $Z_1$ must anti-commute with some operator $P \in S$.  Then 
\begin{equation} \label{Eq:AOrthogonal}
\bra{s}Z_1\ket{s} = \bra{s}PZ_1\ket{s} = -\bra{s}Z_1P\ket{s} = -\bra{s}Z_1\ket{s},
\end{equation}
so $\bra{s}Z_1\ket{s} = 0$, and $Z_1 \ket{s}$ must be orthogonal to $\ket{s}$.  The states $\ket{\pm s'} = \frac{1}{\sqrt{2}}(\ket{s} \pm Z_1\ket{s})$ have $Z_1\ket{\pm s'} = \frac{1}{\sqrt{2}}(Z_1\ket{s} \pm \ket{s}) = \pm \ket{\pm s'}$.  In addition, $\ket{s} = \frac{1}{\sqrt{2}}(\ket{s'} +\ket{-s'})$.  For these reasons, measurement of the operator $Z_1$ will project $\ket{s}$ into one of the two states $\ket{\pm s'}$, each with probability $\frac{1}{2}$.  In order to efficiently represent the resulting state, one must calculate the stabilizing generators of $\ket{\pm s'}$, as follows.

Let $S_g$ be a minimal set of Pauli operators that generates $S$.  I claim that one can assume without loss of generality that at most one operator in $S_g$ anti-commutes with $Z_1$.  If more than one generator anti-commutes with $Z_1$, choose the first such operator $P \in S_g$.  For any another generator $Q \in S_g$ that anti-commutes with $Z_1$, replace it in the generating set with $PQ$.  This operator commutes with $Z_1$, and the new set still generates the same group of stabilizers.  

Now consider any generator $R \in S_g$ such that $R \neq P$; by assumption, $R$ commutes with $Z_1$.  The effect of $R$ on the states $\ket{\pm s'}$ is $R \ket{\pm s'} = \frac{1}{\sqrt{2}}(R\ket{s} \pm R Z_1\ket{s}) = \ket{\pm s'}$ due to the commutation relation.  So the generating set of operators $S' \equiv Z_1 \cup (S - P)$ are stabilizers of the new state $\ket{\pm s'}$.  In fact, because $Z_1$ cannot be written as a product of the other generators by assumption, these stabilizers are independent generators of a group of stabilizers of size $2^n$.  Therefore, they completely specify the state $\ket{\pm s'}$, which must also be a stabilizer state by definition.  Calculating this generating set of stabilizers for the result of the measurement along with their eigenvalues allows one to efficiently compute and specify the result of a $Z_1$ measurement on a stabilizer state.

\begin{lemma} \label{Lem:AMeasTrans}
The result of a measurement of a Pauli operator $P$ on a stabilizer state $\ket{s}$ is the same as the measurement result of $Z_1$ on $C\ket{s}$ for a Clifford operator $C$ such that $C(P)=Z_1$.  The resulting state of the measurement of $P$ on $\ket{s}$ is the same as $C^{\dag}$ applied to the result of the measurement of $Z_1$ on $C\ket{S}$. 

\paragraph{Proof:}
Find $C \in \mathcal{C}_n$ such that $C(P) = Z_1$; Lem.~\ref{Lem:ACliffordSym} proves that such a Clifford operator exists.  Suppose first that $P$ stabilizes $\ket{s}$, i.e. $P\ket{s}=\lambda \ket{s}$.  Then $\lambda \ket{s} = P\ket{s} = C^{\dag}Z_1 C \ket{s}$, so $Z_1 C \ket{s} = \lambda C \ket{s}$.  So a measurement of the $Z_1$ operator on the state $C\ket{s}$ returns the result $\lambda$ as desired, and applying $C^{\dag}$ to the resulting state returns $\ket{s}$ as claimed.

If $P$ does not stabilize $\ket{s}$, then the result of measurement of $P$ on $\ket{s}$ is binary random as in the previous discussion of the measurement $Z_1$.  Using the logic of the previous paragraph in reverse, it must be that $Z_1$ is not a stabilizer of $C\ket{s}$, so the result of the measurement of $Z_1$ on that state is binary random as well.  However, one must still check that the resulting states of these two measurements are consistent conditioned on the measurement.  When the measurement of $P$ projects the state $\ket{s}$ into a new state $\ket{s'}$, the new state is stabilized by $P$ with eigenvalue corresponding to the (random) measurement result $\lambda$ and all previous stabilizers $S$ that commuted with $P$.  Likewise, the result of the measurement of $Z_1$ on $C \ket{s}$ leaves a state $\ket{r}$ that is stabilized by $Z_1$ with a (random) eigenvalue $\phi$ and all previous stabilizers $C S C^{\dag}$ that commuted with $Z_1$.  Then the state $C^{\dag} \ket{r}$ is stabilized by $C^{\dag} Z_1 C = P$ and all $S$ where $S$ satisfies the previous condition.  Because the two states are stabilized by the same operators with the same eigenvalues, it must be that $C^{\dag} \ket{r} = \ket{s'}$ up to the (undetermined) eigenvalue of the measurement operator corresponding to the measurement result, proving the claim. $\square$
\end{lemma}

Thus, any computation only involving Clifford operators and projective Pauli measurements can be rewritten so that it contains only Clifford operators and measurement of one-qubit $Z$ operators.  These operations can be simulated efficiently by tracking a generating set of stabilizers for the state and their associated eigenvalues.

\section{Gottesman-Knill Theorem} \label{Sec:AGK}

The preceding section established that it is possible to specify the state at any step of a computation involving only Clifford operations by specifying $n$ Pauli operators which stabilize the state along with their eigenvalues when applied to the state.  This means that the computation can be simulated in a way that is space efficient: the amount of memory required to track the computation scales as $O(n^2)$.  However, in order to efficiently simulate a computation involving stabilizer states, it must also be possible to simulate each elementary operation of the computation in an amount of time that scales polynomially with $n$.  This is the claim of the Gottesman-Knill theorem \cite{Gottesman1998,Nielsen2004}.  In fact, the previous section has laid all of the groundwork for a sketch of the constructive proof of this claim, which follows.  In Sec.~\ref{Sec:AClifGSSF} I describe in more detail the algorithms used to perform this simulation by the particular Clifford simulator used in this thesis.

First, consider simulating the result of a Pauli operator measurement on a stabilizer state.  In the previous section, I demonstrated how to rewrite an arbitrary Pauli operator measurement as a measurement of the $Z_1$ operator on a related state. The Clifford operator necessary to transform the computation can be efficiently computed using the algorithm of Lem.~\ref{Lem:ACliffordSym}.  In order to simulate the effects of a measurement of the operator $Z_1$ on a state using the calculations in the previous section, one must first check whether $Z_1$ commutes with all of the recorded stabilizer generators.  If it does, then some product of these generators is equal to $\pm Z_1$.  By a Gaussian-elimination-like protocol similar to the one described in depth in Sec.~\ref{Sec:AClifGSSF}, one can calculate this product efficiently.  Then the result of the measurement is $\pm 1$ times the product of the eigenvalues of those operators.  If $Z_1$ does not commute with all of the stabilizer generators, then the result of the measurement is $1$ or $-1$ with equal probability, and the generating operators of the resulting state must be updated as described previously.  Because the number of generators is $n$ and each has length of order $n$, these tasks can all still be done efficiently.

In order to simulate the effects of a Clifford operator $C$ on a stabilizer state, one must update all of the stabilizer generators and their eigenvalues according to the action of the Clifford operator.  There are only $n$ stabilizer generators, so this can be done efficiently as long as the update of each stabilizer generator only requires polynomial time.  In fact, each stabilizer generator $P_j'$ consists of a tensor product of $n$ one-qubit Pauli operators.  Take the three non-identity Pauli operators on each of the $n$ qubits as a (not minimal) generating set for the Pauli operators on $n$ qubits.  Then in order to specify the operator $C$, it is sufficient to specify its action on $3n$ generators.  To calculate $C(P_j')$, decompose $P_j'$ as a product of these generators, transform each of the generators, and compute their product.  This product can be computed in the general case in time that scales as $n^2$: each of the $n$ tensor parts of the result can be computed separately by calculating a product of one-qubit Pauli operators $n$ times.
Because each generator must be transformed in this fashion, the updates required to simulate a Clifford operator in this simple way scale as $O(n^3)$. 

In order to simulate the preparation of a new qubit indexed by $i$ in the state stabilized by the $Z_i$ operator, one extends all of the recorded generators for the stabilizing group by adding a tensored $I$ operator at index $i$.  Then one adds the $Z_i$ operator to the list of stabilizers.  In order to simulate the preparation of a qubit stabilized by some other single-qubit Pauli operator, one simply changes which new operator is added to the list of stabilizers.  This simulation takes time scaling as $O(n)$.

Together, the algorithms for preparation, measurement, and Clifford operator simulation, which require time scaling polynomially in the number of qubits, sketch a constructive proof of the Gottesman-Knill theorem. 

The general algorithms used in the proof of the Gottesman-Knill theorem can be extended to the normalizers of other finite subgroups of the unitary group that satisfy similar properties to the qubit Pauli groups with respect to generating sets (see Ref.~\cite{Clark2007}, for example).  In order to define scaling, there must be a clear notion of how these subgroups extend to different numbers of qubits; for these algorithms to scale polynomially with qubit number, the size of a minimal generating set for the subgroup must scale polynomially with qubit number.

Particularly, the groups of Pauli matrices on multiple (prime) $p$-level systems have generating sets that scale in the same way as the qubit Pauli matrices, and their normalizers, the $p$-level-system Clifford operators, can be simulated efficiently as a consequence.  It is also possible to compute efficiently with certain proper subgroups of the $p$-level-system Clifford groups; some interesting groups describing $p^d$-level systems can be considered in this way \cite{Knill1996,Knill1996B,Ashikhmin2001}.  These groups are used to design alternative RB protocols for qubits in Sec.~\ref{Sec:4SmallDesigns}.

Clifford operation computations can be extended in limited ways without sacrificing their efficient simulation. Refs.~\cite{Aaronson2004,VanDenNest2010} and others have noted that because stabilizer states contain a basis for all quantum states, any pure initial state of an $n$-qubit system can be written as a sum over stabilizer states.  If an initial state can be written as a stabilizer-state sum with a small number of terms, then the evolution of each summand can be tracked independently as a Clifford computation and re-summed before a measurement.  This allows efficient simulation of Clifford computations in which the system is initialized into such a state.  A similar trick allows the efficient simulation of computations containing a constant number of certain classes of non-Clifford operators.  After each such operator, the state of the system is rewritten as a sum of stabilizer states and the computation can be tracked on each constituent part.

Although the scaling of Clifford simulations is known to be polynomial in $n$, there is much to be gained from optimizing the simulation so that this polynomial is as small as possible.  In the most extreme example, tracking of Pauli errors on large fault-tolerant architectures is a task suited to Clifford computation.  Such architectures are likely to contain much greater than $n=10^6$ qubits, so a naive Clifford simulator which operates with scaling $O(n^3)$ would still be hopelessly incapable of this task, even if memory overhead were the only issue.  However, published Clifford simulators in Refs.~\cite{Aaronson2004,Garcia2012} are capable of performing the simulation of a single 1- or 2-qubit operator in $O(n)$ time and a measurement in $O(n^2)$ time.  Their memory overhead scales as $O(n^2)$ as well.  

The simulator written for and used in this thesis scales slightly differently.  Define the support of a stabilizer to be the set of qubit indices at which it has non-identity tensor Pauli parts.  The Clifford operator and measurement steps take time that is proportional to $\overline{s_i} \; \overline{s_i \log{s_i}},$ and memory proportional to $n \overline{s_i}$, where $s_i$ is the size of the support of the $i^{th}$ stabilizing operator and the bar denotes the mean over all of the recorded stabilizer generators.  In the worst case (which is also the typical case for random Clifford computations), $\overline{s_i} \propto n$; however, in many cases of interest it does not depend strongly on $n$.  Examples of this scenario include computations involving concatenated codes, states with structured, local entanglement, or blocks of systems of constant size that are never entangled with each other.  These scenarios are of frequent-enough interest that the large overhead reduction in these cases can be worth the somewhat-poorer scaling in the worst case.

While many Clifford simulation tools assume a basic set of computational gates, the algorithms used in our simulator are more agnostic with respect to fundamental gate sets.  This not only makes the simulator more intuitively useful in experimental scenarios with unusual gate sets, it also allows the algorithms to be naturally extended to Clifford simulations on the Clifford groups and subgroups for $p^d$-level systems mentioned above, although I do not describe such extensions in this thesis.

\section{Graph-State Standard Form} \label{Sec:AGSSF}
\subsection{Stabilizer Matrix} \label{Sec:AStabMat}
All efficient Clifford simulators to date except that of Ref.~\cite{Anders2006} make use of matrix algorithms (especially Gaussian elimination) to realize their speed-ups.  The matrix on which these algorithms operate is formed as follows: Denote each one-qubit Pauli operator with a pair of elements of the Galois field $GF[2]$.  In practice this is done by identifying each operator with a pair of binary numbers representing the powers of $X$ and $Z$ in a decomposition of the operator.  An equivalent notation using the integers $\{0,1,2,3\}$ is introduced in Sec.~\ref{Sec:APauliOps}.  The Galois field notation is introduced here only because it allows the simulator to be extended to operator groups indexed by other Galois fields in a somewhat-straightforward fashion.  Then each stabilizer operator a state of $n$ qubits can be represented as an ordered string of pairs of binary numbers or, equivalently, as a pair of ordered strings of binary numbers.  Create an array by writing in each row the representation of a stabilizer generator as such a string.  This leaves an $n \times n$ array of pairs of binary numbers or an array consisting of two $n \times n$ blocks of binary numbers; for illustrative purposes, I write the array as an $n \times n$ matrix of one-qubit Pauli operators.  Such an array of stabilizer generators is called a stabilizer matrix.

The stabilizer matrix notably does not capture the information about the eigenvalue associated with each $n$-qubit Pauli generator.  This information is typically stored in a separate ``sign'' list (each index associated with a row of the matrix).  It is sufficient to store each eigenvalue as an element of $GF[2]$ as well, since each qubit Pauli operator has only two possible eigenvalues.  Together, the stabilizer matrix and sign list specify a unique stabilizer state, although freedom in the choice of stabilizer generators means that the stabilizer matrix is not unique for a given stabilizer state. 

A product of rows (which is a composition of stabilizing Pauli operators) can be computed as follows: the pair of binary numbers denoting the Pauli operator at any index in the resulting operator is the component-wise binary sum of the pairs of binary numbers at that index for the two original rows.  The (binary) sign of the new row is the binary sum of the signs of the old rows plus a term which is $0$ if the $X$ part of the second operator and the $Z$ part of the first operator commute and $1$ if they do not.  

The different choices of stabilizer generators are manifested in the stabilizer matrix in the following way: by replacing any row with its product with some different row (and updating the sign list) one creates a different stabilizer matrix describing the same state.  That this operation (which looks like a row sum and is sometimes described in this language) can be performed without changing the meaning of the matrix is what makes stabilizer-matrix algorithms amenable to Gaussian elimination.  For example, the algorithms in the remainder of this section rely on a reduction to a standard form which is accomplished by a Gaussian-elimination-like procedure. General Gaussian-elimination algorithms require $O(n^3)$ element additions, since each of $n$ rows must be added to $n$ other rows, and they have length $n$.  The only way to beat this scaling is to ensure that only a limited number of rows needs to be added, and this is only possible if the matrix has some exploitable structure before the algorithm is performed.

\subsection{Definition of Graph-State Standard Form} \label{Sec:AGSSFDef}
The structure I impose on the stabilizer matrix is called graph-state standard form (GSSF) to indicate its relation to the ideas of Refs.~\cite {Elliott2008, Elliott2009}.  The usage of graph-state formalism to speed up stabilizer simulation was introduced by the authors of Ref.~\cite{Anders2006}, whose ideas feature prominently in our simulator.  In the first two papers, the authors work through some transformation rules for a graphical representation of stabilizer states subjected to Clifford operators and measurements.  The essential idea of the graph-state representation is that all stabilizer states can be experimentally constructed as follows: First, prepare each qubit in the state $\ket{+} = \frac{1}{\sqrt{2}}( \ket{0} + \ket{1})$.  This can be done by applying a Hadamard ($H$) gate to each qubit initialized in the $\ket{0}$ state.  Next, apply $\CZ$ gates between some pairs of qubits.  This gate has a symmetric action on the two qubits and commutes with other $\CZ$ gates, so it is useful to list the connections formed between qubits in this way via an adjacency matrix or (equivalently) a graph, where qubits are represented by vertices and two qubits are adjacent (share an edge) if a $\CZ$ has been performed on them.  Finally, apply one-qubit Clifford operators to each qubit.  In the graph notation, this final step is denoted by adornments on the graph that indicate which single-qubit operator has been performed.  

Because the gates used in this construction are all Clifford operators, any graph state must be a stabilizer state.  I show in Sec.~\ref{Sec:AReduction} than any stabilizer state can also be written as a graph state, so the two ideas are equivalent.  Consider the stabilizers of the states constructed according to the prescription of the previous paragraph.  The stabilizers of a tensor state of $\ket{+}$ on each qubit are generated by the $n$ operators $X_{i}$ for  $i \in \{1 \cdots n\}$.  The stabilizer matrix for such a state can be written as a diagonal matrix with $X$ on the diagonal (see Fig.~\ref{Fig:AGSSF}).  The action of $\CZ$ on two qubits is defined by \begin{equation} \label{Eq:ACZ}
\CZ(Z_1I_2)=+Z_1I_2, \quad \CZ(I_1Z_2)=+I_1Z_2,  \quad \CZ(X_1I_2)=+X_1Z_2, \quad \CZ(I_1X_2)=+Z_1X_2.
\end{equation}  If a graph state has adjacency matrix with $M_{i,j} = 1$, then a $\CZ$ is applied between qubits $i$ and $j$, so a $Z$ appears in the new stabilizer matrix at positions $(i,j)$ and $(j,i)$.  Finally, the one-qubit Clifford operators on each qubit act on each column of the matrix, transforming $X$ operators and $Z$ operators consistently in the column.

\begin{figure}[tbh] 
\begin{center}
\begin{picture}(400,60)(0,0)
\put(0,20){
	\makebox(0,0)[l]{ 
		$ \left ( \begin{array}{cccc}
			X & I & I & I \\
			I & X & I & I \\
			I & I & X & I \\
			I & I & I & X \\
		\end{array} \right )$}}
\put(125,20){
	\makebox(0,0)[c]{\Large $\rightarrow$}}
\put(150,20){
	\makebox(0,0)[l]{ 
		$\left ( \begin{array}{cccc}
			X & I & Z & Z \\
			I & X & I & I \\
			Z & I & X & Z \\
			Z & I & Z & X \\
		\end{array} \right )$}}
\put(275,20){
	\makebox(0,0)[c]{\Large $\rightarrow$}}
\put(300,20){
	\makebox(0,0)[l]{ 
		$\left ( \begin{array}{cccc}
			X & I & Y & X \\
			I & Y & I & I \\
			Z & I & X & X \\
			Z & I & Y & Z \\
		\end{array} \right )$}}
\end{picture}
\end{center}
\caption[Creation of a Graph-State Stabilizer Matrix]{Stabilizer-matrix perspective on the creation of an example graph state from four qubits, each prepared in the $\ket{+}$ state.  The graph adjacency matrix has $1$ entries at $(1,3),(1,4), (3,4)$ and their symmetric counterparts.  In the last step, single-qubit Clifford gates were applied as follows: identity applied to qubit 1, $Z \left (\frac{\pi}{2} \right )$ applied to qubit 2, $X \left (\frac{\pi}{2} \right )$ applied to qubit 3, Hadamard applied to qubit 4.  The sign list is omitted for simplicity. \label{Fig:AGSSF}}
\end{figure}

The defining features of a stabilizer matrix that is in GSSF are:
\begin{enumerate}
\item It has operators on the diagonal of each column that are not identity (a consequence of the preparation and ordering of the rows).
\item Non-diagonal, non-identity entries in each column can only be a distinguished operator that anti-commutes with the column's diagonal operator.  This operator is defined separately for each column and is called the neighbor operator for the column.
\item It is symmetric in the location of identity entries (a consequence of the adjacency matrix symmetry).
\end{enumerate}
I show in the next section that this last feature is actually a consequence of the first two and the additional constraint of stabilizer matrices that forces rows to commute.  Using this prescription it is possible to specify any stabilizer matrix in GSSF by giving an adjacency matrix, a list of neighbor and diagonal operators for each column, and a sign list.

The GSSF representation of a stabilizer state is not unique.  By changing the neighbor and diagonal operators and applying the reduction algorithm in the next section, a GSSF matrix can usually be transformed into a new GSSF matrix for the same state.  This non-uniqueness introduces ambiguity about how efficiently states are actually stored in a stabilizer matrix, as one GSSF representation can be much more sparse than another for the same state.

\subsection{Reduction to GSSF} \label{Sec:AReduction}
In this section, I show that any stabilizer matrix $S$ can be put into GSSF by briefly describing the reduction algorithm that is actually used.  The algorithm, which bears some resemblance to Gaussian elimination, acts iteratively, fixing one column and row per iteration.  At the end of each iteration, a set $C$ of columns will satisfy the two column conditions of GSSF.  Each column in $C$ corresponds to a row (with the same index) that is also removed from consideration in future iterations.

In each iteration, pick the first row $r$ whose index is not yet in $C$.  For the elements in this row, find the first index $d$ that is not in $C$ marking a non-identity element.  Such an index must exist.  Suppose, to the contrary, that no such index exists.  In this case, either: The row contains only identity elements, in which case the stabilizer matrix must not represent an independent generating set of stabilizers.  Otherwise, all non-identity elements in the row have indices in $C$; call the first such index $f$ (note that $f$ cannot be the same as $r$ in this supposition).  Consider the commutation rules of the stabilizers in the two rows $r$ and $f$.  Row $r$ only has non-identity elements whose tensor index is in $C$.  In the column indexed by $f$, row $f$ has a diagonal operator and row $r$ has a neighbor operator; these two elements anti-commute by construction.  For every other column, these two rows must either have identity elements or neighbor operators; these elements commute by construction.  Therefore, the two rows anti-commute at one tensor index and commute elsewhere, so they overall anti-commute.  This contradicts the assumption that the matrix represented a set of (mutually commuting) stabilizer operators for a state, and so contradicts the supposition that the index $d$ defined above does not exist.

Because $d \notin C$, the rows $r$ and $d$ can be swapped without disturbing any row that had previously been considered.  Now there is a non-identity operator in the $(d,d)$ location.  Call this the diagonal operator for column $d$, and iterate through the other non-identity operators in the column.  Pick the first such operator which does not commute with the diagonal operator and call it the neighbor operator; if no such operator exists, define the neighbor operator arbitrarily.  If there is a non-identity operator at index $(a,d)$ in row $a$ and it is not the neighbor operator, then replace the entire stabilizer operator in row $a$ with its product with the operator in row $d$.  The stabilizer matrix formed in this way has a generating set of stabilizers if it did before the replacement.  

This replacement of row $a$ also preserves the GSSF column conditions on the columns in $C$:  Consider a column $d' \in C$ with $a \neq d'$; the entry at index $(a,d')$ is the product of the old entry at $(a,d')$ and the entry at $(d,d')$, both of which could only be the neighbor operator or the identity.  Therefore, the new operator at that index is the neighbor operator or the identity, preserving the GSSF condition.  If $a=d'$, then the old entry at this index was the diagonal operator.  The diagonal operator and the neighbor operator for row $d'$ must anti-commute, so the product of the two anti-commutes with both the original neighbor operator and diagonal operator.  Call this new operator the new diagonal operator for the column, and the GSSF conditions are preserved.

After adding row $d$ to all the necessary rows in this way, column $d$ has a non-identity diagonal operator at index $(d,d)$ and identity or neighbor operators at every other index.  This column now satisfies the GSSF conditions for columns, so its index is added to the list $C$, and the algorithm is iterated again.  

After all column indices are contained in $C$, each column satisfies the column conditions of GSSF.  I now show that the matrix also satisfies the symmetry property of GSSF.  Consider two rows $r$ and $r'$; the only entries of these rows whose relationship is fixed by the symmetry are $(r,r')$ and $(r',r)$.  The two rows must overall commute, and entries in the same column commute except if one of the entries is the diagonal entry.  There are then two possibilities: It might be that the two rows anti-commute in both the $r$ and $r'$ columns, where one of the columns has a diagonal operator and the other has a neighbor operator.  In this case, $(r,r')$ and $(r',r)$ must both have a neighbor operator and therefore be non-identity.  The other possibility is that the two rows commute in both the $r$ and $r'$ columns.  In this case, because the $r$ column has a diagonal operator for row $r$, row $r'$ must contain an identity operator in that column in order to commute.  The same argument applies to column $r'$.  Therefore, both $(r,r')$ and $(r',r)$ contain the identity.  In either case, the presence of identity operators at these locations in the matrix is symmetric. 

Because the reduction requires $n$ steps in which a row might be added to $n$ other rows, it resembles a Gaussian-elimination algorithm; the asymptotic time  required scales as $O(n^3)$.  In order to achieve better scaling for a Clifford simulation, it is necessary to avoid implementing the entire reduction algorithm after each computational step.

\subsection{Clifford Simulation in GSSF} \label{Sec:AClifGSSF}
There are three central algorithms for Clifford simulation.  The first and simplest is the simulation of the preparation of a new qubit in the logical $\ket{0}$ state.  Suppose the old stabilizer matrix had $r-1$ rows and columns; the preparation simulation is achieved by adding a new sign list element, stabilizer (row), and qubit (column) to the stabilizer matrix.  The new column and row both contain only the identity element except at the $(r,r)$ index, which contains a $Z$ operator.  The state described by this stabilizer matrix is stabilized by all of the old stabilizers (which have a tensor identity part on the new qubit) and a new stabilizer which has no action on the old qubits.  This indicates that the state is a bipartite product state, with the new qubit being a separate part.  Inspection reveals that the state on the new qubit is $\ket{0}$ when the sign associate with the row is $0$.  The matrix formed in this way has GSSF if it did before the simulation of the preparation of the qubit.

The second crucial algorithm is the simulation of a Clifford operator.  Assume that the Clifford operator has non-trivial action on $m$ qubits $\{k_1, \cdots k_m\}$ and that the stabilizer matrix is in GSSF before the simulation of the Clifford operator.  Knowledge of the Clifford action on a generating set of Pauli operators allows efficient computation of the Clifford action on any Pauli operator.  Consider the submatrix formed from only the columns $\{k_1, \cdots k_m\}$, in that order.  Transform each row of the submatrix according to the action of the Clifford operator; this is a more-efficient way of transforming the entire stabilizer matrix using knowledge of the support of the operator.  Replace the transformed submatrix columns into the stabilizer matrix.  This part of the algorithm acts once on every entry of the submatrix and takes time scaling as $O(nm)$.  Note that Clifford operators preserve commutation of stabilizers.

The matrix is generally not in GSSF after such a transformation, although it is if the Clifford operator was a tensor product of one-qubit operators.  Up to $m$ columns no longer satisfy the GSSF column conditions.  In order to restore GSSF, perform the reduction algorithm described above, starting with $C_f = \{1, \cdots n\} - \{k_1, \cdots k_m\}$.  The same arguments and algorithm apply exactly.  Because the main step of the algorithm must only be iterated $m$ times, this procedure has asymptotic complexity in $O(n^2m)$; $m$ is typically considered to be a constant in descriptions of such algorithms, although it could potentially be as large as $n$.

The final algorithm simulates the effects of a measurement of the $Z_j$ operator (on a single qubit indexed by $j$).  This algorithm is the most complicated, so I describe the three necessary steps in more detail:

\begin{enumerate}
\item
First, check whether the diagonal operator for column $j$ is $Z$.  If it is, then check whether there are any other entries in the column.  If there are not, then the state described by this stabilizer matrix is the product of the state on the $j^{th}$ qubit and the state on all other qubits.  The state of qubit $j$ is stabilized by $Z$, so the result of the measurement is the sign associated with row $j$ and the state is unchanged by the measurement.  In all other cases, the state is not stabilized by $Z$ and so the measurement outcome will be random; however, the resulting state must still be simulated.  

If there are other entries in the column $j$, then pick the first such; suppose it has index $(i,j)$.  Swap rows $i$ and $j$ (and their corresponding sign list elements) and perform the reduction algorithm starting with $C_f = \{1, \cdots, n\} - \{i,j \}$.  This reduction does not change the state described by the stabilizer matrix, and it takes time scaling as $O(n^2)$.  At this point, the new diagonal operator of column $j$ is the old neighbor operator of column $j$, so it anti-commutes with $Z$.

\item
Now the diagonal operator for column $j$ is not $Z$.  Check whether the neighbor operator for the column is $Z$.  If it is not, then the structure of the one-qubit Pauli group necessitates that it must be a product of the diagonal operator and $Z$.  For each non-identity entry in column $j$ at $(i,j)$, with $i \neq j$, replace row $i$ with the product of row $j$ and row $i$ and adjust the sign list accordingly.  As before, this does not change the state stabilized by the matrix; however, it does change the neighbor operator for column $j$ to be $Z$ while preserving GSSF.

\item
Now the neighbor operator for column $j$ is $Z$.  At this step, the stabilizer at row $j$ clearly anti-commutes with the measured operator, so the result of the measurement will be random.  The resulting state will be an eigenstate of $Z_j$, and must not be stabilized by any operators that anti-commute with $Z_j$, so replace row $j$ with the row representing the operator $Z_j$ and change the sign list element to reflect the random measurement result.  Because the neighbor operator of column $j$ is $Z$, it is easy to check that all other rows commute with this new stabilizer, so they will still stabilize the new state.  Because all of the other rows commuted with the old row $j$, no product of those rows could be equal to the operator $Z_j$, which anti-commuted with the old row $j$.  So the new row $j$ is independent of the other rows, and the rows of the stabilizer matrix are an independent generating set for all stabilizers of the new state.  

Column $j$ does not yet satisfy the GSSF column conditions.  In order to restore GSSF, for each row $r$ with non-identity entry $(r,j)$, replace the operator at row $r$ with the product of the operators at rows $r$ and $j$ and adjust the sign list.  Now column $j$ has all identity operators except at the diagonal, so it satisfies the GSSF conditions.
\end{enumerate}

The algorithm above simulates the effects of a measurement of the $Z$ operator on qubit $j$ and leaves the set columns of the stabilizer matrix in GSSF.  The final GSSF condition, symmetry of identity operator locations, is satisfied automatically, as shown in the proof of the reduction algorithm.  The slowest step of the algorithm scales as $O(n^2)$.

Together, these three algorithms allow for the simulation of any stabilizer circuit.  The slowest algorithms are the Clifford operator simulation that scales as $O(mn^2)$ and the measurement simulation that scales as $O(n^2)$.  As noted above, if $m$ is upper-bounded, because the only Clifford operators considered are 1- and 2-qubit operators that generate the Clifford group, for example, then the Clifford operator algorithm scales as $O(n^2)$ as well.  However, if circuits with arbitrarily-large Clifford operators are considered, then this algorithm can scale as $O(n^3)$; such Clifford circuits typically need very few steps, so this scaling may still be tolerable.

\subsection{GSSF-based Proofs} \label{Sec:AGSSFAlgs}

In addition to enabling better scaling for Clifford simulations, GSSF can be used to simplify many constructions involving Clifford circuits and stabilizer states.  In this section I prove several useful properties of Clifford operators using the GSSF construction.

\begin{lemma} \label{Lem:AStabTransform}
For any two stabilizer states $\ket{s}$ and $\ket{s'}$ on $n$ qubits, there exists $C \in \mathcal{C}_n$ such that $C\ket{s} = \ket{s'}$.

\paragraph{Proof:} A gate sequence which transforms one state to the other can be formed as follows: Consider the stabilizer matrices for the two states.  Reduce both matrices to GSSF; this requires no gates, only a rearranging of the stabilizer matrix.  Using the construction method for graph states given above, generate the sequences of Clifford operators $(C_i)_i$ and $(C_j')_i$ necessary to transform an initial $\bigotimes \ket{0}$ state into each of the two stabilizer states, respectively.  These steps can be organized into blocks of Hadamard gates, $\CZ$ gates, and one-qubit gates, in that order.  Now the Clifford $C$ transforming $\ket{s} \rightarrow \ket{s'}$ can be constructed as the product of the sequence of gates $(C_i^{\dag})_i$ in reverse order (taking the state $\ket{s} \rightarrow \bigotimes \ket{0}$) followed by the sequence of gates $(C_j')_i$ (taking $\bigotimes \ket{0} \rightarrow \ket{s'}$). $\square$
\end{lemma}

These sequences can be decomposed into blocks of one-qubit gates, $\CZ$ gates, Hadamard gates, $\CZ$ gates, and one-qubit gates, in that order.  A similar algorithm for decomposing Clifford operators into products of one- and two-qubit gates will be described later in the chapter.

The general protocol used to construct graph states using Clifford operators can also be used to prove two convenient results about the conjugation action of the Clifford group on the Pauli group.  

\begin{lemma} \label{Lem:ACliffordSym}
Given any two Pauli operators $P,Q \in \mathcal{P}_n$ both not equal to the identity operator, there exists a Clifford operator $C \in \mathcal{C}_n$ such that $C(P) = CPC^{\dag} = Q$.

\paragraph{Proof:} Let $L$ be an initially empty list of Clifford operators.  Choose the first qubit indices $l$ and $m$ on which $P$ and $Q$ (respectively) have support (i.e. non-identity tensor part).  Append to $L$ the gate $\SWAP{l}{m}$.  Also append to $L$ a one-qubit Clifford gate that transforms the part of the Pauli at the $l$ index of $P$ to the $X$ operator.  For each other index on which $P$ has support, append a one-qubit Clifford gate to $L$ that transforms this tensor part to $Z$.  Acting with the operators in $L$ on $P$ at this point transforms it into a Pauli operator $P'$ with the same support except for a potential swap of $l$ and $m$ and with $X$ at index $m$ and $Z$ at all other supported indices.

For every index $a$ at which the supports of $P'$ and $Q$ differ, append a $\CZ_{m,a}$ operator.  Acting with these operators on $P'$ to make $P''$ removes $Z$ operators at undesired locations and adds them at desired locations, so that $P''$ has the same support as $Q$.  Finally, append to $L$ one-qubit gates at each supported index of $P''$ to transform the operator at that index to the operator at the same location in $Q$.

The product of the operators in $L$ transforms the Pauli operator $P$ into $Q$, and it is a Clifford operator since all of the elements in $L$ are Clifford operators.  This result can be summarized by saying that the conjugation action of the Clifford group on the set of non-identity Pauli operators is 1-transitive. The number of gates that must be performed scales as $O(n)$, so the computation as a whole scales as $O(n^2)$ because each gate has support on only one or two qubits.  $\square$
\end{lemma}

\begin{corollary} \label{Cor:ASymmetricAction}
The number of Clifford operators $C \in \mathcal{C}_n$ for which $C(P_i) = P_j$ is the same for all $P_i,P_j \in \mathcal{P}_n$ both not equal to $I$.

\paragraph{Proof:} Consider four arbitrary, non-identity Pauli operators $P_1,P_2,P_3,P_4 \in \mathcal{P}_n$ Let $\mathcal{D}$ be the set of Clifford operators $D$ such that $D(P_1) = P_2$.  Let $\mathcal{E}$ be the set of Clifford operators $E$ such that $E(P_1)=P_3$.  By the previous lemma, there exists at least one Clifford operator $C_{2,3}$ such that $C_{2,3}(P_2) = P_3$.  Therefore the set of distinct operators $C_{2,3}\mathcal{D} \subseteq \mathcal{E}$, and so $|\mathcal{D}| \leq |\mathcal{E}|$.  The same argument can be made in reverse to show that $|\mathcal{D}| \geq |\mathcal{E}|$; therefore $|\mathcal{D}| = |\mathcal{E}|$.  Because the Clifford operators form a group, each element of $\mathcal{E}$ has a distinct inverse, and the set $\mathcal{F}$ of operators $F$ for which $F(P_3)=P_1$ contains the inverses of the elements of the set $\mathcal{E}$.  A similar argument then proves that $|\mathcal{E}| = |\mathcal{F}|$.  Finally, the same steps can be repeated yet again to show that the set $\mathcal{G}$ of operators $G$ for which $G(C_3) = C_4$ is equal in size to all the previous sets; in particular, this implies that $|\mathcal{G}| = |\mathcal{D}|$, which proves the claim.  These results are a direct result of the one-transitivity of the conjugation action. $\square$
\end{corollary}
    
\section{Clifford Simulator}

The Clifford simulator I wrote with E. Knill uses GSSF and the algorithms of the previous sections as its fundamental parts.  It is coded in C++.  This simulator is available for use upon request.  The asymptotic scaling of the algorithms implemented differs from that of the previous section due to our use of sparse data structures.  Define $s_i$ to be the size of the support of a stabilizer-matrix row $i$.  Then $\overline{s_i}$ is the mean size of the support averaged over all rows of the stabilizer matrix and quantifies the sparseness of the matrix.  In the worst case where the rows are not sparse at all and $\overline{s_i}=n$, these sparse-data-structure algorithms introduce a multiplicative $\log{n}$ factor into the cost of each row addition (row operator multiplication) that comes from searching binary tree data structures.  This factor increases the asymptotic scaling of the measurement and Clifford operator algorithms described in Sec.~\ref{Sec:AClifGSSF} because they are all fundamentally based on row addition.

\begin{figure} [ht] 
    \begin{center}
	\includegraphics[width=130mm]{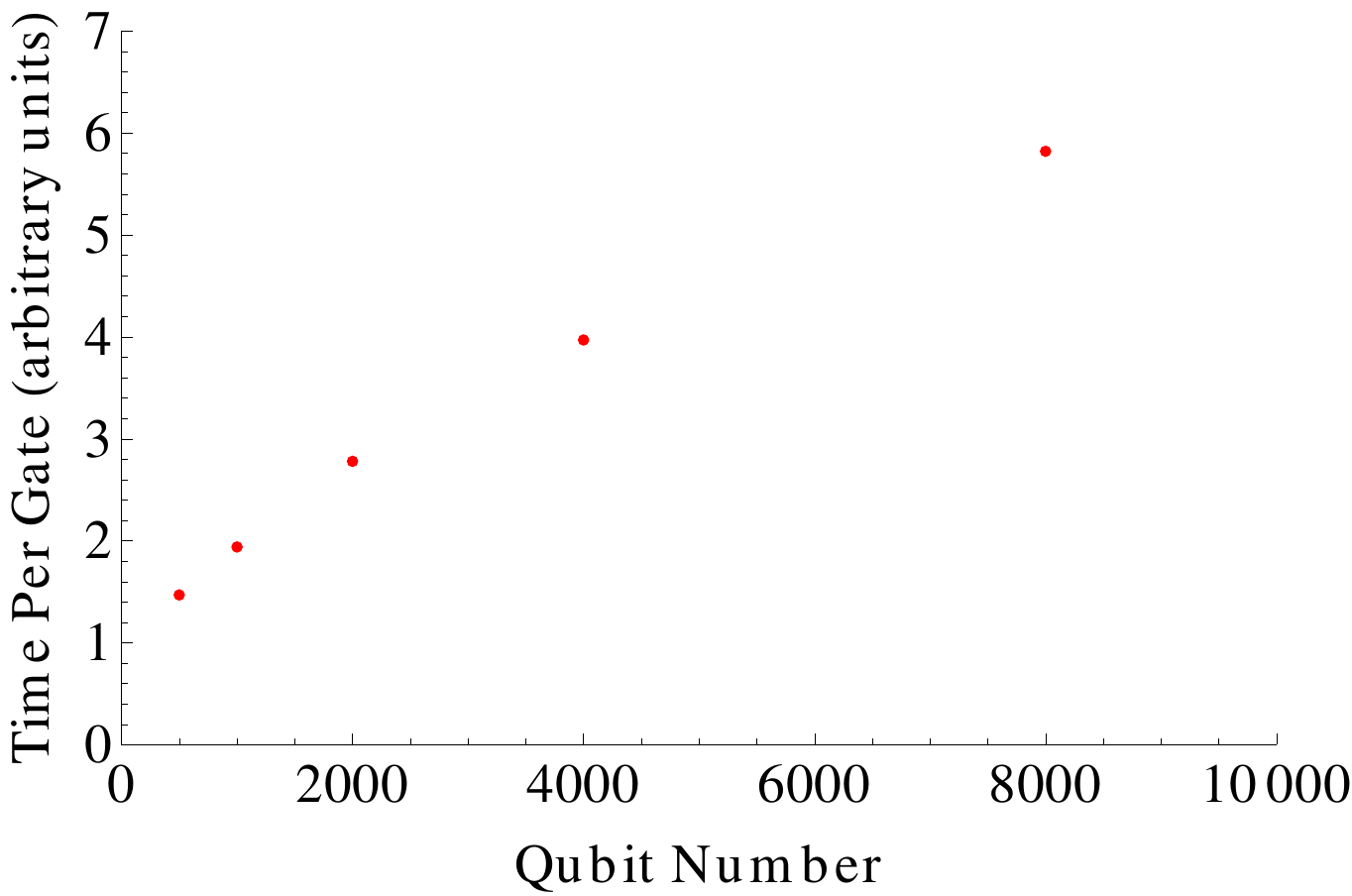}
    \end{center}
    \caption[Time Per controlled-NOT gate as a Function of Qubit Number]{
This figure depicts the time required to simulate a single $\CX$ gate as a function of the number of qubits in the simulation of a random stabilizer matrix, as computed in the C++ profiler \textbf{gprof}.  In these simulations, the parameter $d$ describing the mean support of each row is fixed at $50$ as the qubit number varies.  Ideally this function should be constant, but a sub-linear dependence on $n$ seems to be present.  The time units should be considered arbitrary. \label{Fig:ATimePerGate} }
\end{figure}

\begin{figure} [hb]
    \begin{center}
	\includegraphics[width=130mm]{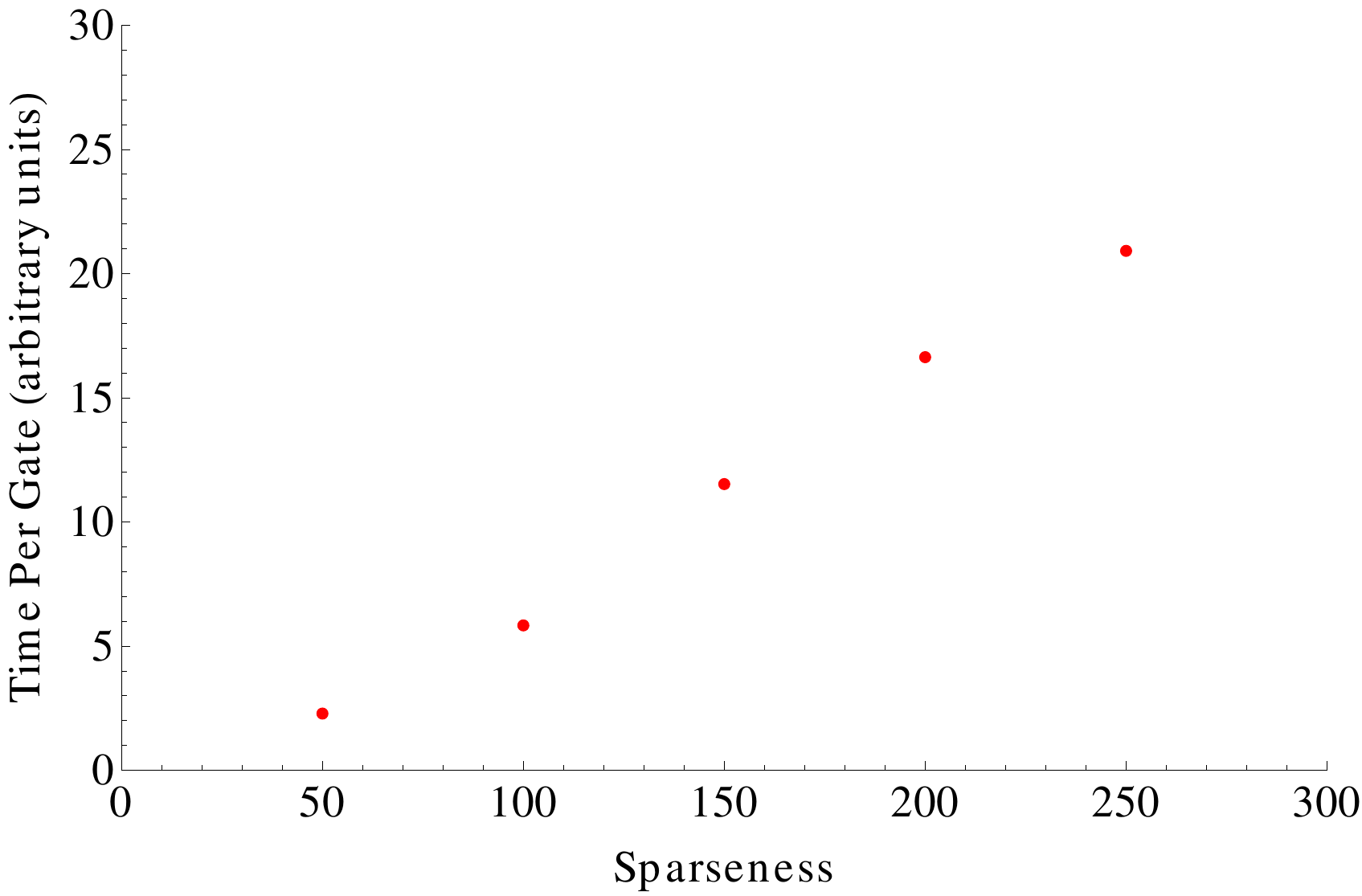}
    \end{center}
    \caption[Time Per controlled-NOT gate as a Function of Sparseness]{
This figure depicts the time required to simulate a single $\CX$ gate as a function of the mean size of the support of the rows of random stabilizer matrices in the simulator, as computed in the C++ profiler \textbf{gprof}.  In these simulations, the qubit number $n$ is fixed at $1000$ as the mean support varies.  This function is expected to grow asymptotically as $d^2 \log{d}$ in the worst case when $d$ is small compared to $n$ but is expected to flatten as $d$ approaches $n/2$.  The time units should be considered arbitrary. \label{Fig:ATimePerSparse} }
\end{figure}

As a trade-off for the logarithmic factor in the worst case, row addition can be performed with these data structures in such a way that only non-identity entries need ever be considered.  This allows a single row addition to be performed in time scaling on average as $\overline{s_i \log{s_i}}$.  For simplicity, I assume throughout that the distribution of $s_i$ is sharply peaked at $\overline{s_i} = d$ so that this time scaling is approximately $d \log{d}$.  Looking back at the algorithms of Sec.~\ref{Sec:AClifGSSF} reveals that, because the mean number of non-identity entries in each column is also $d$, only $O(d)$ row additions need to be performed for the measurement and Clifford operator algorithm on average, so the scaling of the algorithms as a whole is contained in $O(d^2 \log(d))$.
In addition, the memory overhead for the stabilizer matrix in this structure scales as $O(nd)$, and resizing the matrix (for example to introduce a new qubit) has lower cost.  In practice, I have been unable to avoid some weak dependence on the number of qubits $n$ in these algorithms.  Figs.~\ref{Fig:ATimePerGate} and \ref{Fig:ATimePerSparse} depict the actual scaling for the Clifford operator algorithm as functions of $n$ and $d$, respectively.

Because a stabilizer matrix can be equivalent to another stabilizer matrix with different mean $\overline{s_i}$, it is hard to make guarantees about when these data structures will allow large speedups.  This is true even when the stabilizer matrix is forced into GSSF.  It is clear that a graph state that can be constructed with very few $\CZ$ gates corresponds to a stabilizer matrix with very small $d$ in the best case.  For example, a stabilizer state with no entanglement can be represented by a stabilizer matrix with $d=1$.  In fact, GSSF requirements for the stabilizer matrix of such a state will protect this property.  However, the stabilizer matrices for more complicated states can change in sparseness significantly during the course of a Clifford simulation.  No attempt has yet been made to optimize the simulator to increase the sparseness of the stabilizer matrix chosen at each step; it would not be surprising if such algorithms were very difficult to implement optimally. 

\subsection{Stabilizer Subspaces} \label{Sec:AStabSpace}
If a set of stabilizer generators for $n$ qubits contains only $n-m$ operators, the set specifies a subspace of stabilizer states, called a code or stabilizer space, instead of a unique state.  Tracking the evolution of such an incomplete set of stabilizers through a Clifford circuit is a useful task for the study of stabilizer quantum codes, so I have included this capability for the simulator.  In stabilizer coding, the state of one or more qubits is encoded in the subspace of states consistent with the incomplete stabilizer set.  By measuring the stabilizer operators, it is possible to detect whether some (incomplete) set of errors has occurred to the state; the goal of stabilizer coding is to create a set of stabilizer generators such that the errors that cannot be detected in this way are unlikely to occur for physical reasons.  The measurement of any stabilizer operator in this scenario requires only Clifford operations, and the results of this procedure can be usefully simulated even if the overall state of the qubits is not a stabilizer state. 

The notion of GSSF must be modified slightly for stabilizer subspaces.  A matrix of the stabilizers for this subspace has $n$ columns (representing qubits), but only $n-m$ rows.  Attempting to perform the GSSF reduction on such a matrix leaves $m$ columns that do not satisfy the GSSF column restrictions.   The symmetry condition of GSSF, which was a result of the column conditions to begin with, is not possible for the non-square matrix.  Algorithmically, the simulations of Clifford operators and preparation generally look the same for stabilizer subspaces, although some work is required to ensure that a version of GSSF can be maintained efficiently.  Measurement steps include an additional difficulty: it is possible that an operator might be measured that is not in the set of stabilizers but commutes with all of the generators.  In this case, a new stabilizing operator is added to the set of generators, the dimension of the stabilized subspace is reduced, and the measurement result cannot be determined from the information contained in the stabilizer matrix.  Consequently, our simulator is not capable of actually simulating the result of a measurement of this kind, but such measurements are never used in protocols related to error correction (since they destroy the encoding).


\section{Uses of Clifford Simulation for Randomized Benchmarking} \label{Sec:ARBSim}

\subsection{Clifford Decomposition} \label{Sec:ADecomp}

As discussed in Sec.~\ref{Sec:3ClifDecomp}, one computationally intensive task necessary for randomized benchmarking of many qubits is the decomposition of arbitrary Clifford operators into an experimentally available generating set.  In fact, this task has ramifications for the reported benchmark.  By optimizing this decomposition, each random Clifford operator can be performed with fewer generating gates or with gates having higher fidelities.  The result of such optimization is that an experiment can improve its reported benchmark without improving any of the fundamental gates by optimizing the way the gates are used together.

The goal of improving Clifford decomposition has meaning beyond the structure of randomized benchmarking as well.  Stabilizer-code encoding and decoding operations use Clifford operations and benefit from this optimization.  Although much research is focused on conceptually-simple codes for which these operations are easily described, it is likely that large, complicated stabilizer codes with desirable properties exist as well.  The utility of any stabilizer code is tied to the efficiency of its implementation in a given quantum control architecture, so being able to optimize the implementation of the encoding and decoding unitaries is helpful for translating such large codes into practical algorithms.

The first approach to optimal decomposition of Clifford unitaries I took was a brute force search.  Table~\ref{Tab:AClifSize} shows the number of elements of the Clifford quotient $\overline{\mathcal{C}}_n$ on $n=1-4$ qubits.  These groups are of manageable size only when $n$ is very small.  However, for these small groups, it is feasible to approach the problem by enumerating sequences of gates drawn from some generating set.  The composition of the gates in each such sequence describes a Clifford operator, and, by comparing different sequences corresponding to the same operator, one can attempt to optimize in various ways the implementation each Clifford operator.  This task is particularly tractable when the optimization involves minimizing the number of times gates from some distinguished subset appear in the composition sequence; in this case the search halts in a finite time.  In order to implement such a search, one orders the investigation of the various sequences first by the number of gates from the distinguished subset that they contain and second by the total number of gates; one must also take care to replace subsequences of gates that are non-optimal implementations of Clifford operators already discovered in the search.  The first sequence found by such a search that implements a given Clifford operator is optimal.  This search procedure was used for the experiment described in Sec.~\ref{Sec:3Exp2}, where it was desirable to minimize the number of $G$ gates used to implement a two-qubit Clifford operator.  

There is a convenient way to write Clifford operators that utilizes some of the algorithms developed for the Clifford simulator.  Because a Clifford operator is defined by its action on a generating set of Pauli operators, one can store information about the Clifford operator in a matrix of Pauli elements as described in Fig.~\ref{Fig:ACJMatrix}.  In the left half of the matrix are stored a list of generating Pauli operators for the $n$ qubit Pauli group (ordered by $X_1,\cdots, X_n, Z_1, \cdots, Z_n$).  To the right of each generator is the transformation of the Pauli operator under the conjugation action of the Clifford operator.  Information about the sign of the resulting operator is stored in a list that associates a binary number with each row.  These transformation rules for the generating set allow one to calculate the conjugation action of the Clifford operator on any Pauli operator.

Such a matrix can also be thought of as a stabilizer matrix on $2n$ qubits. That the commutation relations required for a stabilizer matrix are enforced can be seen as follows: consider the commutativity of the first and second halves of two $2n$-element rows $i$ and $j$ separately.  The structure of the matrix forces the first halves to anti-commute if and only if $i = j \mod{n}$.  However, the Clifford conjugation action preserves commutation relations, so the second halves of the rows also anti-commute if and only if $i = j \mod{n}$.  Therefore the two whole rows (each of which is a tensor product of the half rows) commute for any $i,j$.

\begin{figure} [tbh]
\begin{center}
\[
\begin{pmat}[{..|}]
X & I & I & C(XII) \cr
I & X & I & C(IXI) \cr
I & I & X & C(IIX) \cr\-
Z & I & I & C(ZII) \cr
I & Z & I & C(IZI) \cr
I & I & Z & C(IIZ) \cr
\end{pmat}
\]
\end{center}
\caption[Stabilizer Notation for Clifford Gate]{Depicts the $6\times 6$ stabilizer-matrix notation for representing a Clifford operator on 3 qubits.  The submatrices on the left hand side contain a generating set of Pauli operators on three qubits, partitioned into Pauli operators with $X$ first and then $Z$.  The submatrices on the right hand side contain the transformation under $C(P) = CPC^{\dag}$ of the Pauli operators on the left side. Within each submatrix, the $3$-qubit Pauli operators are independent and commuting; however, the operators in different submatrices do not commute.  The rows of the matrix taken as a whole commute with each other, justifying the stabilizer-matrix description.  A separately stored list of signs indicates the scalar factor $\pm 1$ of the transformed Pauli matrix. \label{Fig:ACJMatrix}}
\end{figure}

This method of representing a Clifford operator is directly related to the Choi-Jamiolkowski isomorphism \cite{Choi1975}, which establishes a bijection between a unitary operator $U$ on $n$ qubits and a state of $2n$ qubits constructed by acting with $I \otimes U$ on $n$ pairs of qubits each in the Bell state (the tensor notation is meant to indicate a bipartition that takes half of each Bell pair). The stabilizer state for the initial tensor product of $n$ Bell pairs is specified by the generating stabilizer set $\{X_i\otimes X_i , Z_i \otimes Z_i\}_i$ for $i \in [1,n]$.  After transforming such a state by acting with $I \otimes C$ for an arbitrary Clifford operator $C$, the right half of each row of the resulting stabilizer matrix describes the result of the conjugation action of $C$ on the Pauli operator in the left half of the row.  In other words, the matrix representation I use for a Clifford operator is actually the stabilizer matrix for the Choi-Jamiolkowski mapping of the operator.

The product of two Clifford operators, $C \circ D,$ is also specified by its action on a generating set of Pauli operators.  $C \circ D (X_i) = C (D (X_i))$, where $D(X_i)$ can be extracted in a straightforward way from the representation of $D$ as a matrix described above.  $D(X_i)$ is a Pauli matrix, so it can be decomposed into a scalar $\imath^{\zeta}$ and a tensor product of one-qubit Pauli operators, each of which can again be decomposed into an $X$ part and a $Z$ part, so that $D(X_i) = \imath^{\zeta_i} \bigotimes_j X_j^{\alpha_{i,j}}Z_j^{\beta_{i,j}}$.  Finally, 
\begin{equation} \label{Eq:AClifProduct}
C \circ D (X_i) = \imath^{\zeta_i} \bigotimes_j C(X_{j})^{\alpha_{i,j}} C(Z_{j})^{\beta_{i,j}},
\end{equation}
where each $C(X_i),C(Z_i)$ can be extracted from the stabilizer-matrix representation of that operator.  The transformations $C \circ D(Z_i)$ of the other Pauli generators $Z_i$ can be computed in the same way. This construction allows the products of Clifford operators to be specified and computed compactly in the stabilizer-matrix notation already utilized in the Clifford simulator.  Note that the GSSF and sparse-matrix algorithms are not of any particular use here.  Because the product can be seen as applying a Clifford operator to a stabilizer matrix, previous algorithms allow the product to be computed in time scaling as $O(n^2)$, where $n$ is the number of qubits on which the larger operator acts.

Without employing any other clever tricks, the crucial information about such a matrix representing a Clifford operator can be represented by a binary number of length $2n\times n \times 2 + 2n$, where each bit represents the exponent of the $X$ or $Z$ operator on the $i^{th}$ qubit in the right half of the $j^{th}$ row of the matrix (the extra $2n$ digits are used to specify the sign list for the row).  So the best case for this scenario is that a Clifford operator can be uniquely specified in $4n^2 + 2n$ bits, corresponding to a byte per one-qubit operator or $\sim 50$ bytes per $10$-qubit Clifford.  In the same way, checking equality of two Clifford operators represented in this way requires at most $4n^2 +2n$ binary number comparisons.  These inform the computational limits of any attempt to enumerate or sort large numbers of Clifford elements.

In order to search for the shortest implementation of a Clifford operator for a small number of qubits $n$, compute the operators generated by compositions of gates from the chosen generating set $T$ of Clifford gates.  This is equivalent to traversing a directed Cayley graph for the Clifford group, which is a graph with vertices corresponding to Clifford operators and directed edges corresponding to generators which take one Clifford operator to another by the multiplication defined above.  Starting with the identity element, at each step of the traversing the graph, $|T|$ new elements are potentially accessible.  The search for the shortest composition sequence for a given operator is then analogous to finding the shortest path from the identity vertex to a vertex representing the operator.  If the optimality criterion is something other than the shortest path, then a weighting can be used for the edges.  For example, I chose in Sec.~\ref{Sec:3ExpDesign} to search for the operator implementation using the fewest $\CX$ gates first and, secondarily, the fewest one-qubit gates.  This corresponds to weighting the $\CX$ edges of the graph as $1$ and all other edges as some $\epsilon << 1$.  

The greatest challenge of such a search comes from the large number of Clifford operators.  In order to find the shortest implementation for all operators, one must store edge-path information for $|\mathcal{C}_n|$ operators and perform $O(|\mathcal{C}_n|)$ Clifford multiplications (operator compositions in the simulator).  By exploiting some symmetries of the Clifford group based on the double cosets mentioned in Sec.~\ref{Sec:AClifDef}, I found optimal $\CX$ decompositions for $n\leq 4$, but the super-exponential scaling of the Clifford group size will doom any such attempt for $n$ significantly greater than 5.  The double cosets are useful because every element of such a double coset requires the same number of $\CX$ gates to implement.  Table~\ref{Tab:ACNOT} contains the results of a study of the number of $\CX$ gates required to generate elements of $\mathcal{C}_n$ for $n=1-4$.  

\begin{table}[tb] 
    \begin{center}
    \begin{tabular}{l|c|c|c|c|c|c|c|c|c|c|r}
	$n \! \backslash \! \CX \!$\# & 0 & 1 & 2 & 3 & 4 & 5 & 6 & 7 & 8 & 9 & Mean \\ \hline
	1 & 6 & & & & & & & & & & 0 \\ \hline
	2 & 36 & 324 & 324 & 36 & & & & & & & 1.5\\ \hline
	3  & 216 & 5832 & 93312 & 601344 & 657012 & 93312 & 432 & & & & 3.51 \\ \hline
	4  & 1296 & $7 \! * \! 10^5$ & $2 \! * \! 10^7$ & $6 \! * \! 10^8$ & $1 \! * \! 10^9$ & $1 \! * \! 10^{10}$ & $3 \! * \! 10^{10}$ & $6 \! * \! 10^9$& $4 \! * \! 10^8$ & 7776 & 5.86
	\end{tabular}
	\end{center}
    \caption[Number of controlled-NOT gates required to generate elements of $\overline{\mathcal{C}}_n$]{ This table presents a partition of the Clifford quotient $\overline{\mathcal{C}}_n$ according to the minimum number of $\CX$ gates required in the decomposition of a representative of each element.  Each decomposition is allowed to contain only $\CX$ gates and one-qubit gates.  The mean number of $\CX$ gates required for the decompositions of the group elements is also presented.  Because elements of this group are equivalence classes of Clifford operators distinguished only by a Pauli operator multiplication, each Clifford in the equivalence class requires the same number of $\CX$ gates.  The number of Clifford operators in $\mathcal{C}_n$ requiring the given number of $\CX$ gates can then be calculated through multiplication by $|P_n|$.\label{Tab:ACNOT}}
\end{table}

Table~\ref{Tab:ACNOT} indicates that the difficulty of performing Clifford operators with a reasonable generating set does not scale badly with $n$, even though the number of Clifford operators scales super-exponentially.  Specifically, Ref.~\cite{Patel2008} proves that the mean number of $\CX$ gates required to implement an operator in $\mathcal{C}_n$ should scale as $O(n^2 / \log{n})$ at worst.  I find that the operators that require a maximal number of $\CX$ gates for each $n=2-4$ implement cyclic permutations of the $n$ qubits.  This trend must not continue for higher $n$, or else the max number of $\CX$ required would be $3(n-1)$, and the average number of $\CX$ needed could not continue its $O(n^2 / \log{(n)})$ scaling.  Being able to calculate and predict which operators require the maximal number of gates to implement can direct experimenters toward knowing which additional control would benefit them the most; this is a worthwhile open challenge.

\subsubsection{Clifford Decomposition Algorithm} \label{Sec:ADecompAlg}

For $n$ sufficiently large, the brute-force search for optimal implementations of Clifford operators becomes prohibitively computationally difficult.  However, decomposition algorithms are already available that split a Clifford operator on $n$ qubits into elementary gates.  For example, Aaronson and Gottesman \cite{Aaronson2004} showed an algorithm for decomposition of Clifford operators into $H$,$P$, and $\CX$ gates with a constant number of blocks of each gate type.  This result, combined with the result of Ref.~\cite{Patel2008}, promises that any decomposition requires at most $O(n^2 / \log{(n)})$ gates and provides the upper bound used in the previous paragraph.

In this section, I describe a different algorithm loosely based on GSSF that decomposes any Clifford operator into blocks of one-qubit gates, $\CZ$ gates, and $\CX$ gates.  The use of $\CZ$ gates in addition to $\CX$ gates means that the scaling of the algorithm cannot benefit from the trick of Ref.~\cite{Patel2008}, so the overall scaling is $O(n^2)$.  Note that if $\CX$ is the only experimentally available two-qubit gate, a $\CZ$ operator can always be enacted by replacing it with the identity $\CZ_{1,2} = H_2 \CX_{1,2} H_2$ using the Hadamard and $\CX$ gates.

Consider a Clifford operator $C$ that is represented by a Bell-state stabilizer matrix $M_C$ as described earlier in this section.  The goal of this algorithm is to transform this matrix into the form corresponding to the identity operator using a sequence of Clifford gates $(G_i)$.  If the sequence $G_m \cdots G_1 C= I$, then $C = G_1^{\dag} \cdots G_m^{\dag}$ is a decomposition of $C$.  The construction of $M_C$ lends itself to discussing the four quadrants of $M_C$ separately (labeled clockwise from upper-left).  I denote the stabilizer matrix at step $j$ of the algorithm as $M_j$ and the $3^{rd}$ quadrant of this matrix as $M_j^3$, for example.  A graphical summary of the algorithm is given in Fig.~\ref{Fig:ARedAlg}.

Whenever Clifford operators or row additions are performed on the stabilizer matrix, the sign list should be updated.  I do not state this explicitly in each step of the algorithm.  If one does not care about which Clifford operator in a Clifford coset $[C]$ in $\overline{\mathcal{C}}_n = \mathcal{C}_n/\mathcal{P}_n$ is implemented, then the last step of the algorithm is not necessary and tracking the sign list is not actually necessary for computing the decomposition.

\begin{enumerate} \label{ABellReductionAlgorithm}
\item
\paragraph{Fix GSSF on quadrant 3}
$M_0$ has: $M_0^1$ with $X$ operators on the diagonal and $I$ elsewhere and $M_0^4$ with $Z$ on the diagonal and $I$ elsewhere.  Note that, by construction, each quadrant is itself a stabilizer matrix (of full rank), and this will be preserved at each step.  This is the only initial constraint on $M_0^2$ and $M_0^3$.  First, reduce $M_0^{3}$ into GSSF.  This step involves no gates, only row additions that rearrange the stabilizer matrix.  In particular, row additions on $M_0^3$ are also row additions on $M_0^4$, so $M_1^4$ will contain only $I,Z$ but will no longer be diagonal.

\item
\paragraph{Use one-qubit gates to regularize quadrant 3}
$M_1^3$ is now in GSSF with arbitrary neighbor and diagonal operators.  Append to the (initially empty) sequence of gates $(G_i)$ the one-qubit gates necessary to change the diagonal operator to $X$ and the neighbor operator to $Z$ for each column of $M_1^3$.  These gates commute, so their order is unimportant.  This step also changes $M_1^2$, but no assumptions have yet been made on its form.

\item
\paragraph{Diagonalize quadrant 3 using $\CZ$ gates}
Each off-diagonal entry $[M_2^3]_{k,l}$ in quadrant 3 is either $I$ or $Z$ after the previous step.  If it is $Z$, then append to the sequence of gates $(G_i)$ the gate $\CZ_{k,l}$, which transforms $M_2^3$ to change this off-diagonal entry to $I$.  All such $\CZ$ gates commute, so again the order does not matter.  This step also changes $M_2^2$, but no assumptions have yet been made on its form.  After this step, $M_3^3$ is diagonal with $X$ operators on the diagonal.

\item
\paragraph{Fix GSSF on quadrant 4}
$M_3^4$ contains only $I,Z$ entries but is no longer diagonal.  Because this sub-matrix is still full rank, row additions can be used to diagonalize it.  These row additions can change the third quadrant so that $M_4^3$ is no longer diagonal, but it will still only have $I,X$ entries.

\item
\paragraph{Diagonalize quadrant 3 using $\CX$ gates}
$M_4^3$ has full row and column rank if considered as a binary matrix ($0 \sim I$ and $1 \sim X$).  For each row $k$: First use three $\CX$ gates to implement a $\SWAP{}{}$ gate, if necessary, to swap the diagonal column with a column of higher index so that the diagonal entry is $X$.  Append these gates to the sequence of gates $(G_i)$.  Second, for each non-diagonal entry $[M_4^3]_{k,l}$ of the new submatrix, apply a $\CX_{k,l}$ gate to change this entry to $I$; append this gate to $(G_i)$.  Performing the gates in this order will ensure that the corrections to previous rows are not undone.  The full column rank of the binary matrix guarantees that this algorithm can be performed to completion.  These gates will also change $M_4^2$, but no assumptions have yet been made on that matrix's form.

\item
\paragraph{Regularize quadrants 2 and 3 using one-qubit gates}
At this point, quadrants 1,3,4 are all diagonal.  In order for the overall stabilizers at rows $k$ and $k+n$ to commute, $[M_5^2]_{k,k}$ must anti-commute with $[M_5^3]_{k,k}$; this is because the two rows anti-commute in the quadrants 1 and 4, and they cannot anti-commute at any other entries of quadrant 3.  Use a one-qubit gate to transform $[M_5^3]_{k,k}$ from $X$ to $Z$ and $[M_5^2]_{k,k}$ to $X$.  Append this gate to the gate sequence $(G_i)$.  Repeat for each column; the order does not matter.  At this point, quadrant 3 is diagonal with $Z$ on the diagonal and quadrant 2 has $X$ on the diagonal and $Z$ or $I$ elsewhere.  This form is enforced by commutation rules of rows.

\item
\paragraph{Diagonalize quadrant 2 using $\CZ$ gates}
For each non-diagonal entry $[M_6^2]_{l,k}$, if the entry is $Z$, apply the gate $\CZ_{l,k}$ to change the entry to $I$.  Append this gate to the sequence $(G_i)$. Crucially, this gate does not affect $M_6^3$, since it has only $I,Z$ entries.  After this step, all four quadrants are diagonalized, and the stabilizer matrix is that of a Bell state although the list of signs might not be all $0$.

\item
\paragraph{(Optional) Use one-qubit gates to change all signs to $0$}
If the sign of a row $k$ (or n+k) in the top (or bottom) half of the matrix is non-zero, it can be changed to zero by applying a $Z$ (or $X$) gate to qubit $k$.  This gate changes nothing else about the stabilizer matrix.  Append these gates to the gate sequence $(G_i)$.
\end{enumerate}

\begin{figure} [tbh]
\begin{eqnarray}
\begin{pmat}[{|}]
M^1 & M^2 \cr\-
M^4 & M^3 \cr
\end{pmat} : 
\begin{pmat}[{|}]
D(X) & U() \cr\-
D(Z) & U() \cr
\end{pmat} \rightarrow
\begin{pmat}[{|}]
D(X) & U() \cr\-
U(Z) & G() \cr
\end{pmat} \xrightarrow{\mbox{1-q}}
\begin{pmat}[{|}]
D(X) & U() \cr\-
U(Z) & G(X,Z) \cr
\end{pmat} \xrightarrow{\CZ}
\begin{pmat}[{|}]
D(X) & U() \cr\-
U(Z) & D(X) \cr
\end{pmat} \rightarrow  \nonumber \\
\begin{pmat}[{|}]
D(X) & U() \cr\-
D(Z) & U(X) \cr
\end{pmat} \xrightarrow{\CX}
\begin{pmat}[{|}]
D(X) & U() \cr\-
D(Z) & D(X) \cr
\end{pmat} \xrightarrow{\mbox{1-q}}
\begin{pmat}[{|}]
D(X) & G(X,Z) \cr\-
D(Z) & D(Z) \cr
\end{pmat} \xrightarrow{\CZ}
\begin{pmat}[{|}]
D(X) & D(X) \cr\-
D(Z) & D(Z) \cr
\end{pmat} \nonumber
\end{eqnarray}
\caption[Clifford Decomposition Algorithm Schematic]{A graphical depiction of the transformation of the Clifford operation in the algorithm for its decomposition.  At each step, the Clifford operation is represented by the stabilizer matrix of a state Choi-Jamiolkowski isomorphic to it.  The form of the submatrices in each of the four quadrants of this stabilizer matrix is denoted as follows: If the submatrix is diagonal, it is denoted with a D.  If it is in GSSF, it is denoted with a G.  If it is unknown, it is denoted with a U.  The non-identity elements in the matrix are denoted in parentheses if they are constrained; otherwise the parentheses are empty.  If gates are used to transform the matrix in each step, the type of gates used is noted above the transformation arrow (1-q indicates one-qubit gates). \label{Fig:ARedAlg} }
\end{figure}

Because $\CX$ and $\CZ$ gates are self-inverse, the inverse sequence $G_1^{\dag} \cdots G_m^{\dag}$ contains blocks of: one-qubit gates, $\CZ$ gates,  $\CX$ gates,one-qubit gates, $\CZ$ gates, and (optional) one-qubit gates (in that order).  This gate sequence describes how to decompose the original Clifford operator $C$ into $O(n^2)$ elementary gates.  Crucially, all of these gates act only on the right half of the stabilizer matrix; operators acting on the left half or both halves of this representation of a gate would require a different interpretation in this context.  The calculation of this decomposition takes time scaling as $O(n^3)$ due to the GSSF reduction steps.  The optional step need not be used in randomized benchmarking experiments (and is not particularly useful) if a separate Pauli randomization is included in each step.

In general there are efficiency gains to be found by improving Clifford decomposition algorithms.  My algorithm is not particularly efficient in the number of $\CZ$ or $\CX$ gates it uses and could probably be improved to match the scaling of Ref.~\cite{Patel2008}.  In addition, if $\CX$ or $\CZ$ gates are not natural gates for a particular experiment, then these decompositions will not work as efficiently in practice.  However, Clifford operator equivalences and symmetries allow this algorithm and others to be restructured to use other basic gates within some limits.  As a separate issue, global control and parallel gates are not explicitly utilized at all in current algorithms although they are possible in many experiments.  Particularly, the relevant scaling of an operator decomposition may not be the total number of gates in the decomposition, but rather the number of time steps required when certain gates are allowed to act simultaneously in each time step.  Optimizing under this relaxed constraint is preferable when computation time is the crucial resource.

For example, suppose that any number of $\CX$ gates can be performed in parallel when they have a common control qubit (see Ref.~\cite{Lin2009} for an example of an architecture where this might be possible), and suppose that any number of $\CZ$ gates can be performed in parallel regardless of their supports (because they commute).  Additionally, any number of one-qubit gates can be performed in parallel for the same reason.  Then a decomposition into $O(n)$ time steps can be constructed from the decomposition algorithm above as follows: Algorithm steps that involve only one-qubit gates or only $\CZ$ gates take $O(1)$ time step each.  The algorithm step involving $\CX$ gates iterates over rows, and for each row it performs a single SWAP gate followed by $\CX$ gates sharing a common (diagonal) qubit, so the gates for each row except for the SWAP can be performed in parallel in $O(1)$ time step.  The $\CX$ part of the algorithm then requires $O(n)$ time steps.   Therefore the overall decomposition will require at most $O(n)$ time steps, even though it still requires $O(n^2)$ gates.

\subsection{Sequence Generation} \label{Sec:AGeneration}

Generating the sequences of Clifford operators that make up a randomized benchmarking experiment is a natural job for a Clifford simulator.  The first part of this task is to randomly choose the first $l$ Clifford operators for each sequence from the chosen operator distribution.  If the $|\mathcal{C}_n|$ Clifford operators have already been enumerated and listed, then this is as simple as choosing $l$ random integers in the range $\{1 \cdots |\mathcal{C}_n|\}$ with weights according to the distribution.  If there are too many elements in the group for this list to be stored in its entirety and the goal is to sample uniformly from all Clifford operators, then this can be accomplished by generating the Pauli elements that describe the Clifford action of each operator.  An exponentially slow, but functional approach is to pick Pauli operators completely at random and discard (and re-pick) those that do not satisfy the required commutation rules.  A smarter approach utilizing solutions to linear systems of equations, which would be necessary for large $n$, is alluded to in Ref.~\cite{Magesan2012B}.  I explicitly lay out one version of this approach in Lemma~\ref{Lem:ARandClif}.

\begin{lemma} \label{Lem:ARandClif}
A uniformly random Clifford operator from $\mathcal{C}_n$ can be constructed in time that scales as $O(n^3)$.

\paragraph{Proof:} 
I prove this scaling by describing an algorithm that constructs such a Clifford operator.  The algorithm works iteratively, producing at each step $k$ the transformations $P_{X,k}$ and $P_{Z,k}$ by the operator $C$ of the $X_k$ and $Z_k$ operators.  Lem.~\ref{Lem:AClifAction} proves that these transformation rules uniquely specify a Clifford operator.  At each step, a sequence of Clifford operators $L = (C_{X,1},C_{Z,1}, \cdots , C_{X,k},C_{Z,k})$ is updated that is necessary for the construction.  In shorthand, I write $L_k(P)$ to indicate $C_{Z,k-1} \circ C_{X,k-1} \circ \cdots C_{Z,1} \circ C_{X,1}(P)$ and $L_k^{-1}$ to indicate the inverse of the Clifford operator $L_k$.   The operator $L_k^{-1}$ has the same transformation rules as the operator $C$ eventually constructed by the algorithm for the Pauli operators $X_1,\cdots,X_k,Z_1,\cdots,Z_k$, and so $L_k$ can be thought of as a partial inverse for the operator $C$.  Finally the composition of gates leading to $L_{n}^{\dag} = C$ is the result of the algorithm.  At first, let $L$ be empty.  

For step $k$ of the algorithm: Construct a random $n$-qubit Pauli operator $(-1)^{h_{k}} P_{X,k}$ that is the transformation by $C$ of $X_k$ as follows: Such an operator must commute with all previously generated operators $P_{X,l},P_{Z,l}$ for $l<k$.  However, this property is automatically satisfied if $L_{k}(P_{X,k})$ commutes with all $L_k(P_{X,l}) = X_l , L_k(P_{Z,l}) = Z_l$ for $l<k$, due to the fact that Clifford operators preserve commutation relations.  For this reason, it is easier to construct $L_k(P_{X,k})$ first.  Inspection reveals that such an operator that commutes with all $X_l$ and $Z_l$ for $l<k$ must not have support on the first $k-1$ qubits.  For each remaining index $j$, choose a uniformly random integer $a_j$ from $0-3$ and let the $j^{th}$ tensor part of the operator, $(L_k(P_{X,1}))_j$, equal $P_{a_j}$, where $P_{a_j}$ is indexed using the usual convention for qubit Pauli operators.  $L_k(P_{X,k})$  can not be the all-identity operator, so if $a_j$ = 0 for all $j$, discard all $a_j$ and re-choose the integers.  Now choose a random integer $h_j \in \{0,1\}$ to be the sign of the matrix $P_{X,k}$ after transformation by $C$.  This accounts for all possible operators satisfying the required commutation relations so far.  Note that there are $2*(4^{n-k+1}-1)$ uniformly likely possibilities for this operator regardless of previous operators chosen in the algorithm. Finally, the actual result $(-1)^{h_k} P_{X,k}$ of $C(X_k)$ can be found by applying $L_k^{-1}$ to the chosen operator $L_k(P_{X,k})$ and multiplying by the chosen sign $(-1)^{h_k}$.

Now calculate the Clifford operator $C_{X,k}$ that transforms $L_k(P_{X,k}) \rightarrow X_k$ and does not affect any $X_l,Z_l$ for $l<k$.  This can be accomplished using the algorithm of Lem.~\ref{Lem:ACliffordSym}, which uses a composition sequence of $n$ gates.  That this algorithm does not affect the previous operators can be seen by observing that the operators $L_k(P_{X,k})$ and $X_k$ do not have support on the indices $l<k$ and that the algorithm does not act on indices with no support for either of the two operators.  This calculation requires time scaling as $O(n^2)$.  Add $C_{X,k}$ to the list $L$.

The transformation by $C$ of $Z_k$ must anti-commute with $P_{X,k}$ and commute with all $P_{X,l},P_{Z,l}$ for $l<k$.  The nature of Clifford operators ensures that this is true for an operator $C(Z_k) = (-1)^{g_k}P_{Z,k}$ if $C_{X,k} \circ L_k(P_{Z,k})$ anti-commutes with $X_k$ and commutes with all $X_l, Z_l$ for $l<k$.  Construct $C_{X,k} \circ L_k(P_{Z,k})$ as follows: As before, this operator must not have support on the first $k-1$ indices.  For the $k^{th}$ tensor part, let $(C_{X,k} \circ L_k(P_{Z,k}))_k = X \mbox{ or } Y$ with equal probability.  Let the other tensor parts $(C_{X,k} \circ L_k(P_{Z,k}))_j$ for $j > k$ be chosen uniformly at random from all single qubit Pauli operators as before.  Choose the sign $(-1)^{g_k}$ as before also.  The result $C_{X,k} \circ L_k(P_{Z,k})$ thus defined is guaranteed to anti-commute with $X_k$, so $P_{Z,k} = L_k^{-1} \circ C_{X,k}^{-1} (C_{X,k} \circ L_k(P_{Z,k}))$ is guaranteed to anti-commute with $P_{X,k}$.  In fact, this procedure selects uniformly from all possible Pauli operators that satisfy the required commutation rules with $C_{X,k} \circ L_k (P_{X,l})$ for $l\leq k$ and $C_{X,k} \circ L_k (P_{Z,l})$ for $l<k$, so the transformations of these operators by $L_k^{-1} \circ C_{X,k}^{-1}$ select uniformly from the operators that satisfy the required commutation relations on $P_{X,l}$ for $l \leq k$ and $P_{Z,l}$ for $l<k$ .  There are $2*2*4^{n-k}$ possible choices in this step independent of previous choices.

Now compute (and add to $L$) a Clifford operator $C_{Z,k}$ that transforms $C_{X,k} \circ L_k(P_{Z,k}) \rightarrow Z_k$ and does not affect $X_l,Z_l$ for $l<k$ or $X_k$.  The algorithm of Lem.~\ref{Lem:ACliffordSym} actually produces an operator with the latter conditions almost by default when attempting to satisfy the first.  As before, note that the lack of support of the new operator and $Z_k$ on the first $(k-1)$ indices means that the algorithm produces an operator that does not have any affect on $X_l,Z_l$ for $l<k$.  To satisfy the commutation relation with $X_k$, note first that no $\mbox{SWAP}$ gates are necessary in the algorithm of Lem.~\ref{Lem:ACliffordSym} because both operators have support on the $k^{th}$ qubit, which can be used as the control for the $\CZ$ gates.  In order to transform $Z_k$ to $X_k$ in the one-qubit portion of the algorithm, a $H_k$ gate can be used.  This gate transforms $X_k \rightarrow Z_k$, and none of the subsequent $\CZ$ gates have any affect on $Z_k$.  Finally, the algorithm can use $H_k$ at the end to transform $X_k$ back to $Z_k$, simultaneously restoring $Z_k \rightarrow X_k$.  The aggregate operator then has no affect on the $X_k$ operator.  Add this operator to $L$ and update $L_k$ and/or $L_k^{-1}$.  These updates require two Clifford operator products, which can be accomplished in time scaling as $O(n^2)$. 

After completing the $n^{th}$ iteration, this algorithm produces all of the transformation rules for $L_n^{\dag}=C$, thus uniquely defining the operator.  The transformation rules are chosen uniformly at random in each step from a set whose size is independent of the previous choices, and each Clifford operator is uniquely defined by these rules, so $C$ is a uniformly random Clifford operator. The computations of the two new operators in each iteration require $O(n^2)$ time each, so the overall algorithm requires $O(n^3)$ time. $\square$   

\end{lemma}

\begin{corollary} \label{Cor:AClifSize} The size of the Clifford group on $n$ qubits is $\prod_{k=1}^n (2*(4^{n-k+1}-1))(4*(4^{n-k}) = 2^{n^2+2n} \prod_{k=1}^{n} (4^{n-k+1} - 1)$.
\end{corollary}

After the $l$ random Clifford operators have been generated, one can use a Clifford simulator to simulate the experiment up through the final, inverting step.  There is a peripheral benefit to using this algorithm to construct Clifford operators for randomized benchmarking experiments: there is no need to separately calculate a decomposition of the operator, as the construction technique contains the details for a decomposition of the inverse which can be translated to a decomposition for the gate itself.  This decomposition is not likely to be very efficient, however.  

The techniques described in the previous section can be used to calculate efficiently the Clifford operator that is the product of the $l$ random Clifford operators in a randomized benchmarking sequence.  At this point, there are several options for how to compute the inverting step corresponding to different degrees of randomization and computational complexity.  If $n$ is large, brute force or matrix inversion techniques are both prohibitively slow.  In this case, one can instead find the inverse by first performing a Clifford operator decomposition as in Sec.~\ref{Sec:ADecomp}.  Because the gates in the decomposition are all one-qubit gates or self-inverse gates, it is easy to turn this into the composition sequence for the inverse operator.  Calculating the product of this new composition sequence then reveals the inverse Clifford in time that scales as $O(n^3)$, since each of $O(n^2)$ gates in the decomposition can be applied to the operator stabilizer matrix in time scaling as $O(n)$ using the algorithms of Ref.~\cite{Aaronson2004}.

Some approaches to randomized benchmarking use inversion steps that do not exactly invert the overall Clifford operator.  The experiments described in Sec.~\ref{Sec:3Exp2} and Refs.~\cite{Corcoles2012,Magesan2012} include a measurement step where the inversion operator is exact (except for a random, known $n$-qubit Pauli operator performed before the inversion operator in the first case).  This inversion can be approached through the algorithmic approach described above.  The experiments in Sec.~\ref{Sec:2Exp1} and Refs.~\cite{Knill2008, Biercuk2009, Chow2009, Olmschenk2010, Chow2010, Ryan2009, Brown2011} do not invert the Clifford operator, they only return the state at the end of the $l$ measurement steps into the measurement basis.  This is convenient for the approximate 2-design approach to randomized benchmarking and is described in more depth in Sec.~\ref{Sec:2MeasChoice}.  In single-qubit experiments, this approach is trivial to implement, but, in experiments of this type with larger $n$, such an inverting operator could be calculated and decomposed using the Clifford-simulator-based strategy described in Lem.~\ref{Lem:ACliffordSym} to find an elementary gate sequence that transforms the stabilizer state after $l$ steps to the initial state.

\subsection{Error Simulation} \label{Sec:AError}

In addition to simulating the behavior of a Clifford computation under ideal conditions, Clifford simulators are capable of simulating these computations in the presence of certain limited error models.  It is particularly natural to simulate error models for which Clifford operators are applied probabilistically.  Even easier to simulate are error models for which only Pauli operators are applied probabilistically, often called stochastic Pauli channels.  These error models are natural in stabilizer error-correction protocols, where successful error detection projects errors into a Pauli basis.  However, they are not sufficient for the experiments described in this thesis because, among other things, they do not model coherent errors, which are common experimental issues.  Nonetheless, by modeling Pauli error models in which the error probabilities vary in time or differ based on the intended Clifford operator, it is possible to investigate the violation of some of the gate error model assumptions made in Sec.~\ref{Sec:2GateError} for randomized benchmarking experiments. 

The Pauli error model I consider is one in which each ideal operator $g$ on $m$ qubits is actually implemented as one of $P \in \mathcal{P}_m$, and the probability of $P g$ is $p(P,g,t)$, where $t$ is time or step number.  One Monte-Carlo strategy for simulating stabilizer circuits that contain Pauli errors is to simulate the circuit many times, inserting Pauli errors according to their probabilities.  The results of such simulations are often called quantum trajectories, and they represent a possible result of the experiment.  By sampling many times from such quantum trajectories, one recovers a data set that should behave the same as an actual experimental data set (assuming that the error model is correct), and so it is possible to study features of the statistical analysis of the experiment.

This sampling approach can also be used effectively for error models with probabilistic Clifford operators.  Ref.~\cite{Gutierrez2012} describes using these error models to simulate even more general types of errors.  In that paper, error models which include probabilistic stabilizer measurements are also simulated in the stabilizer framework.  While general errors affecting current 1- and 2-qubit randomized benchmarking experiments can be fully simulated without using stabilizer techniques, scaling considerations will eventually force approximate and/or limited simulations such as those described in this section.
   
If there are $n$ qubits, then the sign list can only take $2^n$ possible values, so it is possible, for $n$ not much greater than $30$, to track every possible value of the sign list simultaneously.  In the simulation of a Pauli operator error, nothing about the stabilizer matrix changes except for the sign list.  Therefore, stochastic Pauli errors can be simulated if the simulator tracks every possible sign list and assigns to each a probability.  Before each intended operator, the simulator calculates how each Pauli error would transform each sign list and updates the probabilities for all of the sign lists accordingly.  For example, before the first step no errors have occurred, so all of the probability is concentrated on one ``correct'' sign list.  Each Pauli error transforms this sign list to a new sign list, so the probability weight of that sign list after the error step is increased by the probability of the Pauli error.  

When performing such an error simulation, Clifford operators are simulated as in Sec.~\ref{Sec:AClifGSSF}, except that all sign lists must be updated after each operator. A measurement of a stabilizer partitions the sign lists into two sets according to the observed measurement result for that stabilizer.  These sets are labeled by the measurement result and separated for the remainder of the simulation.  Measurement of a Pauli operator that anti-commutes with a stabilizer returns a random result, so the simulator measurement procedure is carried out as always except that the probability of each old sign list is split evenly between two new sign lists representing the two possible measurement results. 

The result of such an error simulation is a set of probabilities that defines both the likelihoods of measurement outcomes and the weights of different orthogonal states in the resulting mixed states for each potential measurement outcome combination at the end of the computation.  By using this complete distribution of results instead of sampling from possible results, one can exactly calculate expectation values of measurements and ignore concerns with under-sampling, which arise in the Monte-Carlo (trajectory) approach.  Although limited error simulations were performed for the experiment described in Sec.~\ref{Sec:2Exp1}, this application has not been investigated in depth.

This kind of error simulation of Clifford computations is closely related to a crucial classical computing requirement for FTQC.  In fault-tolerant architectures, it is desirable to track errors but to not actually correct them until designated points in the computation (typically before the next non-Clifford gate).  Because ideal error-detection for stabilizer codes projects any error onto a Pauli error, this task is the same as propagating Pauli errors through a Clifford circuit.  In practice, a Clifford simulator can accomplish this task by recording the measurement results as a separate sign list alongside the ideal sign list (for an error-free computation) and propagating both, along with a stabilizer matrix, in parallel with the experiment's computation.  The simulator can then be used to compare the two sign lists and determine a Pauli operator that would transform the actual sign list into the ideal sign list, correcting any errors that have occurred.  At the present, this computation is trivial for small demonstrations of FTQC.  However, to perform this type of error-correction on a large-scale quantum computer, it would be necessary for a Clifford simulator to be able to manage many millions (or more) of qubits.  At this scale, either Clifford simulators must be made significantly more efficient or the error-tracking algorithm must be decomposed in some way so that it need not track all of the qubits in the same stabilizer matrix.  In the long term, this is probably the most crucial application for an efficient Clifford simulator.

\chapter{Randomized-Benchmarking Experiments} \OnePageChapter
\label{App:B}
\begin{table}[h]
\centering
\begin{tabular}{|l|c|c|c|c|p{1.7in}|}
\hline
Ref. & $\epsilon_s$ & Qubit \# & Lengths & Qubit & Notes\\
\hline
\cite{Knill2008} & $4.82(17) *10^{-3}$ & 1 & 2-96 & Be Ion Trap & \\
\hline
\cite{Chow2009}  & $1.1(3) *10^{-2}$ & 1 & 1-196  & Transmon  & \\
\hline
\cite{Biercuk2009} & $8(1) *10^{-4}$ & 1 & 1-200  & Be Ion Crystal & \\
\hline
\cite{Ryan2009,Laforest2008} & $4.7(3) *10^{-3}$ & 3 & 1-120  & Liquid State NMR & Unique gate set\\
\hline
\cite{Ryan2009,Laforest2008} & $1.3(1) *10^{-4}$ & 1 & 1-192  & Liquid State NMR & \\
\hline
\cite{Olmschenk2010} & $1.4(1) *10^{-4}$ & 1 & 1-995  & Neutral Rb & simultaneous experiments\\
\hline
\cite{Chow2010} & $7(5) * 10^{-3}$    & 1 & 1-96 & Transmon & \\
\hline
\cite{Brown2011} & $2.0(2) *10^{-5}$ & 1 & 1-987 & Be Ion Trap & \\
\hline
\cite{Gambetta2012} & $.0029$ & 1 & 1-512 & Transmon & Cross-Talk Estimation\\
\hline
\cite{Gaebler2012} & $.162(8)$ & 2 & 1-6 & Be Ion Trap & Exact Twirl, \newline Individual-gate RB\\
\hline
\cite{Gaebler2012} & $.0085(20)$ & 1 & 1-12 & Be Ion Trap & \\
\hline
\cite{Corcoles2012} & $0.0936(58)$ & 2 & 1-20 & Transmon & Exact Twirl, \newline Individual-gate RB\\ 
\hline
\cite{Corcoles2012} & $4.4(1) * 10^{-3}$ & 1 & 1-512 & Transmon & Cross-Talk Estimation\\
\hline
\end{tabular}
\caption[Randomized-Benchmarking Experiment List]{List of all published experiments that use randomized-benchmarking protocols.  Also listed are the reported error per step, number of qubits, sequence lengths, and qubit type for each experiment.  Notes beside each experiment describe significant differences from the standard protocol (first experiment) and additional protocols implemented. \label{Tab:BRBExp}}
\end{table}

\chapter{Clifford Operator Conventions}\OnePageChapter
\label{App:C}

Detailed descriptions of the Clifford and Pauli operators are given in Secs.~\ref{Sec:2Clifford} and \ref{Sec:ACliffords}.  This appendix collects some consistently used notation for faster reference.
\begin{itemize}
\item
The Pauli operator group on $n$ qubits is denoted by $\mathcal{P}_n$.
\item
The Clifford operator group on $n$ qubits is denoted by $\mathcal{C}_n$.  
\item
The conjugation action of a Clifford operator $C$ on a Pauli operator $P$ is denoted by $C(P)$.  This action is sometimes simply referred to as the Clifford action.
\item
The quotient of the Clifford and Pauli operator groups is denoted by $\overline{\mathcal{C}}_n = \mathcal{C}_n/\mathcal{P}_n$.  The coset in $\overline{\mathcal{C}}_n$ containing a Clifford operator $C$ is denoted by $[C]$, and the distinguished representative of the coset is $\tilde{C}$.
\item
Subscripts on specific Pauli or Clifford operators are used to indicate the qubits on which they act.  If an operator has explicit support only on a strict subset of the qubits, it is assumed to act as the identity on all other qubits.  When a Clifford operator has multiple indices, the order can be important and must be understood by reference to its definition (e.g. in Table \ref{Tab:CCommonGates}).
\end{itemize}

\begin{table}[t] \label{Tab:CCommonGates}
\centering
\resizebox{!}{3.5in}{
\begin{tabular}{|l|c|p{2.5in}|}
\hline
Name(s) & Unitary Matrix & Clifford Action \\
\hline
$I$ (for one qubit) & $\left( \begin{array}{cc}
1 & 0 \\
0 & 1 \end{array} \right)$ & $I(X)=X$, $I(Z)=Z$\\
\hline
$X,X(\pi)$ & $\left( \begin{array}{cc}
0 & 1 \\
1 & 0 \end{array} \right)$ & $X(X)=X$, $X(Z)=-Z$ \\
\hline
$Z,Z(\pi)$ & $\left( \begin{array}{cc}
1 & 0 \\
0 & -1 \end{array} \right)$ & $Z(X)=-X$, $Z(Z)=Z$ \\
\hline
$Y,Y(\pi)$ & $\left( \begin{array}{cc}
0 & -\imath \\
\imath & 0 \end{array} \right)$ & $Y(X)=-X$, $Y(Z)=-Z$ \\
\hline
$X \left (\frac{\pi}{2} \right )$ & $\frac{1}{\sqrt{2}} \left( \begin{array}{cc}
1 & -\imath \\
-\imath & 1 \end{array} \right)$ & $X \left (\frac{\pi}{2} \right )(X)=X$, $X \left (\frac{\pi}{2} \right )(Z)=-Y$ \\
\hline
$Z \left (\frac{\pi}{2} \right )$, $P$, $S$ & $\left( \begin{array}{cc}
1 & 0 \\
0 & \imath \end{array} \right)$ & $Z \left (\frac{\pi}{2} \right )(X)=Y$, $Z \left (\frac{\pi}{2} \right )(Z)=Z$ \\
\hline
$T, PH$ & $\frac{i}{\sqrt{2}}\left( \begin{array}{cc}
1 & 1 \\
\imath & -\imath \end{array} \right)$ & $T(X)=Y$, $T(Z)=X$ \\
\hline
$H$, Hadamard & $\frac{1}{\sqrt{2}} \left( \begin{array}{cc}
1 & 1 \\
1 & -1 \end{array} \right)$ & $H(X)=Z$, $H(Z)=X$ \\
\hline
$\CX_{1,2}$, CNOT & $\left( \begin{array}{cccc}
1 & 0 & 0 & 0 \\
0 & 1 & 0 & 0 \\
0 & 0 & 0 & 1 \\
0 & 0 & 1 & 0  \end{array} \right)$ & $\CX(XI)=XX$, $\CX(ZI)=ZI$, \newline $\CX(IX)=IX$, $\CX(IZ)=ZZ$ \\
\hline
$\CZ_{1,2}$, CPHASE & $ \left( \begin{array}{cccc}
1 & 0 & 0 & 0 \\
0 & 1 & 0 & 0 \\
0 & 0 & 1 & 0 \\
0 & 0 & 0 & -1  \end{array} \right)$ & $\CZ(XI)=XZ$, $\CZ(ZI)=ZI$, \newline $\CZ(IX)=ZX$, $\CZ(IZ)=IZ$ \\
\hline
$MS_{1,2}$ & $ \left( \begin{array}{cccc}
1 & 0 & 0 & \imath \\
0 & 1 & \imath & 0 \\
0 & \imath & 1 & 0 \\
\imath & 0 & 0 & 1  \end{array} \right)$ & $MS(XI)=XI$, $MS(ZI)=-YX$, \newline $MS(IX)=IX$, $MS(IZ)=-XY$ \\
\hline
$G_{1,2}$ & $\frac{i}{\sqrt{2}}  \left( \begin{array}{cccc}
1 & 0 & 0 & 0 \\
0 & \imath & 0 & 0 \\
0 & 0 & \imath & 0 \\
0 & 0 & 0 & 1  \end{array} \right)$ & $G(XI)=YZ$, $G(ZI)=ZI$, \newline $G(IX)=ZY$, $G(IZ)=IZ$ \\
\hline
\end{tabular}
}
\caption[Commonly Used Clifford Operators]{List of some commonly used Clifford operators.  Each operator is identified by its name, a unitary matrix representation in the standard basis, and its action by conjugation on a generating set of Clifford operators.  The naming convention for gates includes an index that describes which qubit(s) the gate acts on; for two-qubit gates the order of these indices may be important.  }
\end{table}

\end{document}